\documentclass[useAMS,usenatbib]{mn2e}

\usepackage{graphicx}
\usepackage{natbib}
\usepackage{hyperref}
\usepackage{enumerate}
\usepackage{longtable}
\usepackage{deluxetable}
\usepackage{txfonts}
\usepackage{times}

\newcommand{\hi}{{\sc H\,i}}
\newcommand{\mhi}{$M$(\hi)}
\newcommand{\mhtwo}{$M$(H$_2$)}
\newcommand{\rhi}{$R$(\hi)}
\newcommand{\lk}{{$L_\mathrm{K}$}}
\newcommand{\mk}{{$M_\mathrm{K}$}}
\newcommand{\mhil}{$M$(\hi)/\lk}
\newcommand{\Nhi}{$N$(\hi)}

\newcommand{\msun}{{M$_\odot$}}

\newcommand{\mlsun}{{M$_\odot/$L$_\odot$}}

\newcommand{\ltsima} {$\; \buildrel < \over \sim \;$}
\newcommand{\gtsima} {$\; \buildrel > \over \sim \;$}
\newcommand{\lta} {\lower.5ex\hbox{\ltsima}}
\newcommand{\gta} {\lower.5ex\hbox{\gtsima}}

\newcommand{\atlas}{ATLAS$^{\rm 3D}$}
\newcommand{\atlashi}{\atlas\ \hi}
\newcommand{\mhilim}{\mhi$_\mathrm{lim}$}
\newcommand{\sn}{$\mathrm{\left(S/N\right)_{max}}$}

\title[The \atlas\ project -- XIII. \hi\ in early-type galaxies]{The \atlas\ project -- XIII. Mass and morphology of \hi\ in early-type galaxies as a function of environment.}
\author
[Paolo Serra et al.]{\parbox{\textwidth}{Paolo Serra,$^{1}$\thanks{E-mail:\texttt{serra@astron.nl}}
Tom Oosterloo,$^{1,2}$
Raffaella Morganti,$^{1,2}$
Katherine Alatalo,$^{3}$
Leo Blitz,$^{3}$
Maxime Bois,$^{4,5}$
Fr\'ed\'eric Bournaud,$^{6}$
Martin Bureau,$^{7}$
Michele Cappellari,$^{7}$
Alison F. Crocker.$^{7,8}$
Roger L. Davies,$^{7}$
Timothy A. Davis,$^{7}$
P. T. de Zeeuw,$^{4,9}$
Pierre-Alain Duc,$^{6}$
Eric Emsellem,$^{4,5}$
Sadegh Khochfar,$^{10}$
Davor Krajnovi\'c,$^{4}$
Harald Kuntschner,$^{11}$
Pierre-Yves Lablanche,$^{4,5}$
Richard M. McDermid,$^{12}$
Thorsten Naab,$^{13}$
Marc Sarzi,$^{14}$
Nicholas Scott,$^{7}$
Scott C. Trager,$^2$
Anne-Marie Weijmans,$^{15}$\thanks{Dunlap Fellow}
and Lisa M. Young,$^{16}$}\vspace{0.4cm}\\ 
\parbox{\textwidth}{
$^{1}$Netherlands Institute for Radio Astronomy (ASTRON), Postbus 2, 7990 AA Dwingeloo, The Netherlands\\
$^{2}$Kapteyn Astronomical Institute, University of Groningen, Postbus 800, 9700 AV Groningen, The Netherlands\\
$^{3}$Department of Astronomy, Campbell Hall, University of California, Berkeley, CA 94720, USA\\
$^{4}$European Southern Observatory, Karl-Schwarzschild-Str. 2, 85748 Garching, Germany\\
$^{5}$Universit\'e Lyon 1, Observatoire de Lyon, Centre de Recherche Astrophysique de Lyon and Ecole Normale Sup\'erieure de Lyon, 9 avenue Charles Andr\'e, F-69230 Saint-Genis Laval, France\\
$^{6}$Laboratoire AIM Paris-Saclay, CEA/IRFU/SAp -- CNRS -- Universit\'e Paris Diderot, 91191 Gif-sur-Yvette Cedex, France\\
$^{7}$Sub-Dept. of Astrophysics, Dept. of Physics, University of Oxford, Denys Wilkinson Building, Keble Road, Oxford, OX1 3RH, UK\\
$^{8}$Department of Astrophysics, University of Massachusetts, 710 North Pleasant Street, Amherst, MA 01003, USA\\
$^{9}$Sterrewacht Leiden, Leiden University, Postbus 9513, 2300 RA Leiden, the Netherlands\\
$^{10}$Max-Planck Institut f\"ur extraterrestrische Physik, PO Box 1312, D-85478 Garching, Germany\\
$^{11}$Space Telescope European Coordinating Facility, European Southern Observatory, Karl-Schwarzschild-Str. 2, 85748 Garching, Germany\\
$^{12}$Gemini Observatory, Northern Operations Centre, 670 N. A`ohoku Place, Hilo, HI 96720, USA\\
$^{13}$Max-Planck-Institut f\"ur Astrophysik, Karl-Schwarzschild-Str. 1, 85741 Garching, Germany\\
$^{14}$Centre for Astrophysics Research, University of Hertfordshire, Hatfield, Herts AL1 9AB, UK\\
$^{15}$Dunlap Institute for Astronomy \& Astrophysics, University of Toronto, 50 St. George Street, Toronto, ON M5S 3H4, Canada\\
$^{16}$Physics Department, New Mexico Institute of Mining and Technology, Socorro, NM 87801, USA
}}
\begin{document}

\date{Accepted 2011 November 16. Received 2011 October 03; in original form 2011 October 03}

\pagerange{\pageref{firstpage}--\pageref{lastpage}} \pubyear{2010}

\maketitle

\label{firstpage}

\clearpage

\begin{abstract}
We present the \atlas\ \hi\ survey of a volume-limited, complete sample of 166 nearby early-type galaxies (ETGs) brighter than \mk$=-21.5$. The survey is mostly based on data taken with the Westerbork Synthesis Radio Telescope, which enables us to detect \hi\ down to $5\times10^6$ - $5\times10^7$ \msun\ within the survey volume.



We detect \hi\ in $\sim40$ percent of all ETGs outside the Virgo galaxy cluster and in $\sim10$ percent of all ETGs inside it. This demonstrates that it is common for non-cluster ETGs to host \hi. The morphology of the detected gas varies in a continuous way from regular, settled \hi\ discs and rings to unsettled gas distributions (including tidal- or accretion tails) and systems of gas clouds scattered around the galaxy. The majority of the detections consist of \hi\ discs or rings (1/4 of all ETGs outside Virgo) so that if \hi\ is detected in an ETG it is most likely distributed on  a settled configuration. These systems come in two main types: small discs (\mhi$<10^8$ \msun), which are confined within the stellar body and share the same kinematics of the stars; and large discs/rings (\mhi\ up to $5\times10^9$ \msun), which extend to tens of kpc from the host galaxy and are in half of the cases kinematically decoupled from the stars.

Neutral hydrogen seems to provide material for star formation in ETGs. Galaxies containing \hi\ within $\sim1R_\mathrm{e}$ exhibit signatures of on-going star formation in $\sim70$ percent of the cases, $\sim5$ times more frequently than galaxies without central \hi. The ISM in the centre of these galaxies is dominated by molecular gas. In ETGs with a small gas disc the conversion of \hi\ into H$_2$ is as efficient as in spirals.

The ETG \hi\ mass function is characterised by $M^*\sim2 \times 10^9$ \msun\ and by a slope $\alpha\sim-0.7$. Compared to spirals, ETGs host much less \hi\ as a family. However, a significant fraction of all ETGs are as \hi-rich as spiral galaxies. The main difference between ETGs and spirals  is that the former lack the high-column-density \hi\ typical of the bright stellar disc of the latter.

The ETG \hi\ properties vary with environment density in a more continuous way than suggested by the known Virgo vs. non-Virgo dichotomy. We find an envelope of decreasing \mhi\ and \mhil\ with increasing environment density. The gas-richest galaxies live in the poorest environments (as found also with CO observations), where the detection rate of star-formation signatures is higher. Galaxies in the centre of Virgo have the lowest \hi\ content, while galaxies at the outskirts of Virgo represent a transition region and can contain significant amounts of \hi, indicating that at least a fraction of them has joined the cluster only recently after pre-processing in groups. Finally, we find an \hi\ morphology-density relation such that at low environment density (measured on a local scale) the detected \hi\ is mostly distributed on large, regular discs and rings, while more disturbed \hi\ morphologies dominate environment densities typical of rich groups. This confirms the importance of processes occurring on a galaxy-group scale for the evolution of ETGs.

\end{abstract}


\section{Introduction}
\label{sec:intro}

In the classification scheme proposed by \cite{1936rene.book.....H} galaxies are arranged on a sequence going from ellipticals to lenticulars and, from these, to spirals of progressively later type (Sa to Sc). Ellipticals are dominated by a stellar bulge while the spiral sequence is essentially one of decreasing bulge-to-disc ratio. The intermediate position of lenticular galaxies in this scheme has lead to the common idea that all early-type galaxies (ellipticals and lenticulars; hereafter ETGs) have higher bulge-to-disc ratio than spirals.

The other difference between ETGs and spirals is that the former lack the blue spiral arms typical of the latter \citep{1926ApJ....64..321H}. It was early recognised that this corresponds to a lack of star formation in ETGs, leading to the simplified picture that their stellar populations are uniformly old.

In contrast with this traditional view, ETGs exhibit a large variety of shapes and some authors suggest that their bulge-to-disc ratio can in fact be as low as that of Sc, disc-dominated spirals \citep{1951ApJ...113..413S,1970ApJ...160..831S,1976ApJ...206..883V}. Furthermore, following early insights by \cite{1981ApJ...249...48G} and \cite{1982ApJ...261...85R}, it is now established that a large fraction of ETGs are forming small amounts of stars or have done so in their recent past \citep[e.g.,][]{1993PhDT.......172G,2000AJ....119.1645T,2005ApJ...619L.111Y,2007ApJS..173..619K,2010MNRAS.404.1775T}.

Support to these ideas comes from recent studies of nearby ETGs using integral-field spectroscopy \citep{2002MNRAS.329..513D}. These show that most ETGs host a rotating, kinematically cold component \citep{2008MNRAS.390...93K} whose stars are usually younger and more metal-rich than those in the bulge \citep{2010MNRAS.408...97K}.

These results are placed on a firm statistical ground by \atlas, a multi-wavelength study of a volume-limited sample of 260 morphologically selected ETGs \citep[hereafter Paper I]{2011MNRAS.413..813C}. We find that as many as 80 percent of all ETGs in the nearby Universe consist of nearly axisymmetric, fast rotating stellar systems (\citealt[herafter Paper II]{2011MNRAS.414.2923K}; \citealt[hereafter Paper III]{2011MNRAS.414..888E}), most of which resemble spiral galaxies with the arms removed \citep[hereafter Paper VII]{2011MNRAS.tmp.1249C}.

The presence of a substantial disc and the occurrence of star formation in ETGs imply that cold gas has played an important role in their evolution. Indeed, \cite{2011MNRAS.tmp.1500K} suggest that most ETGs grow a stellar disc following gas cooling. In this respect, two fundamental lines of research are the direct observation of neutral hydrogen gas (\hi) and molecular gas (H$_2$, observed through the radiation emitted by CO molecules). In spirals, \hi\ is an essential constituent of the disc, and is the material from which H$_2$ and subsequently new stars form. Understanding the \hi\ and H$_2$ properties of ETGs is therefore crucial to investigate the origin of their structure and star formation history,  how they continue to evolve at $z=0$, and why they are so different from spiral galaxies.

Results from the \atlas\ CO survey are presented in \cite[hereafter Paper IV]{2011MNRAS.414..940Y}. Here we present an \hi\ survey of 166 nearby ETGs belonging to the \atlas\ sample. Thanks to its unprecedented combination of sample size, sample selection, depth and resolution of the \hi\ data, and availability of multi-wavelength data this survey represents a significant improvement over previous studies.

The study of \hi\ in ETGs dates back to the end of the 1960's. Early work already showed that ETGs have lower \mhi$/L_\mathrm{B}$ than spirals, consistent with their redder colour \citep[e.g.,][]{1969A&A.....3..281G,1970A&A.....6..453B,1972A&A....21..303B,1972AJ.....77..568G,1973A&A....22..281B,1975ApJ...202....7G,1976ApJ...209..710B,1977A&A....60..361B,1977AJ.....82..106K,1978ApJ...222..800K,1979ApJ...234..448K,1979A&A....75....7B,1979ApJ...227..776K,1980ApJ...242..931S,1982A&AS...50..451R}. Based on these data, \cite{1985AJ.....90..454K} and \cite{1986AJ.....91...23W} analysed the \hi\ content of $\sim150$ ellipticals and $\sim300$ lenticulars, respectively. They detected \hi\ in about 15 percent of all ellipticals and 25 percent of all lenticulars for a typical \mhi\ detection limit of a few times $10^8$ \msun. They also found a lack of correlation between \mhi\ and $L_\mathrm{B}$ and interpreted it in terms of gas external origin.

Early studies were mostly carried out with pointed single-dish observations aimed at determining global quantities like integrated \hi\ mass and velocity width of ETGs. Recent progress in this kind of analysis has been made possible by single-dish blind surveys of large areas of the sky such as HIPASS (\citealt{2001MNRAS.322..486B}; see \citealt{2002ASPC..273..215S} for an \hi\ study of ETGs based on these data) and ALFALFA \citep{2005AJ....130.2598G}. In particular, the latter  pushes the \mhi\ detection limit below $10^8$ \msun\ for galaxies within a few tens of Mpc from us, allowing a significant increase in the number of detected ETGs. ALFALFA data  show that the ETG \hi\ detection rate is a strong function of environment density, from just a few percent inside the Virgo galaxy cluster \citep{2007A&A...474..851D} to about 40 percent outside it (\citealt{2009A&A...498..407G}; see also \citealt{1986A&A...165...15C}). This result fits with the idea that \hi\ is easily stripped from galaxies inside clusters \citep[e.g.,][]{1983AJ.....88..881G}, and with the fact that recent episodes of star formation occur mostly in ETGs in poor environments \citep[e.g.,][]{2010MNRAS.404.1775T}.

While single-dish observations have the advantage of reaching good sensitivity in relatively short integrations, they lack the angular resolution needed to study the detailed \hi\ morphology and kinematics. Starting from the 1980's an increasing number of galaxies was observed at higher angular resolution with interferometers to perform such analysis \citep[e.g.,][]{1980A&A....82..314S,1981ApJ...246..708R,1984A&A...138...77K,1984MNRAS.210..497S,1986AJ.....91..791V,1987ApJ...314...57L,1987A&A...175....4S,1988ApJ...330..684K,1991A&A...243...71V,1994ApJ...423L.101S,1995ApJ...444L..77S}. These and many later observations revealed a surprisingly large diversity of \hi\ morphologies, ranging from very extended (tens of kpc), low column-density discs and rings to unsettled gas distributions indicative of recent gas accretion, gas stripping, or galaxy interaction and merging (see the review by \citealt{1997ASPC..116..310V} and \citealt{2001ASPC..240..657H}). These results revealed \hi\ as a fundamental tracer of the assembly history of ETGs, and prompted work aimed at increasing the size of a sample with deep, homogeneous high-resolution \hi\ data.

A significant step in this direction was made by the combined study of \cite[herafter M06]{2006MNRAS.371..157M} and  \cite[herafter O10]{2010MNRAS.409..500O} of 33 galaxies in the SAURON sample. Using data taken with the Westerbork Synthesis Radio Telescope they detected about 10 percent and 2/3 of all galaxies inside and outside the Virgo cluster, respectively (going down to \hi\ masses of a few times $10^6$ \msun). Exploiting the high resolution of their data they show that the \hi\ is distributed on regular discs and rings in about half of the detections (1/3 of all galaxies), with radius ranging from $\sim1$ to tens of kpc \citep[see also][]{2007A&A...465..787O}.

\begin{figure*}
\includegraphics[width=18cm]{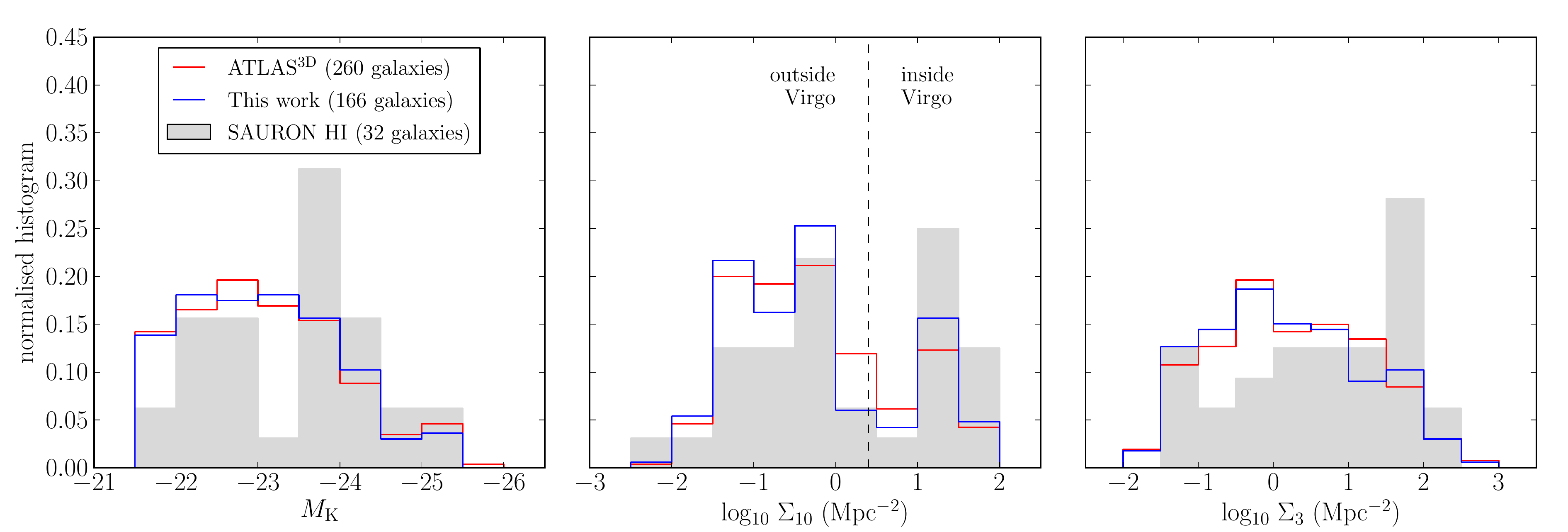}
\caption{Distribution of $M_\mathrm{K}$ (left), $\Sigma_{10}$ (middle) and $\Sigma_3$ (right) for galaxies in the \atlas\ sample (red line), \atlashi\ sample (blue line) and \hi\ SAURON sample (grey filled histogram). See the text for the definition of $\Sigma_{10}$ and $\Sigma_3$.}
\label{fig:sample}
\end{figure*}

Their analyses demonstrate that high sensitivity is crucial to detect the faintest signatures of \hi\ accretion in ETGs, and that these are present in most detected galaxies. Furthermore, high angular resolution and the availability of multi-wavelength data make it possible to connect the \hi\ to stars and multi-phase interstellar medium within the host galaxy. This is crucial to understand the role of \hi\ in replenishing ETGs with cold gas, which could then fuel star- and disc formation. For example, M06 and O10 found that all ETGs surrounded by settled \hi\ distributions host ionised gas within $1 R_e$, and that the two gas phases share the same kinematics. Furthermore, all galaxies with \hi\ within $1 R_e$ are detected in CO, and these systems are more likely to be detected in radio continuum too, indicating that some star formation is occurring (see O10 for a discussion of the origin of the radio emission).

The main limitation of M06 and O10 is that the SAURON sample is small, hampering a statistically strong study of the relation between \hi\ and other galaxy properties. For example, the number of galaxies was not sufficient to establish or rule out trends between \hi\ and ETG dynamics or stellar populations, nor to study trends with environment beyond the simple Virgo vs. non-Virgo dichotomy. Also, the uncertainty on the fraction of ETGs hosting regular, rotating \hi\ systems (undoubtedly a very interesting class of objects) was very large.

Another limitation of the SAURON sample is that galaxies were selected to be evenly distributed on the $M_\mathrm{B}$-ellipticity plane rather than to follow the ETG luminosity function, so that the interpretation of the results in terms of galaxy evolution is not straightforward. Works based on the HIPASS and ALFALFA surveys do not suffer from this limitation since both are blind surveys of large regions of the sky. However, they lack the resolution and sensitivity necessary to provide a picture as revealing as that emerging from M06 and O10 (and, in the case of ALFALFA, the sample studied so far is only 40\% larger than the SAURON sample within their common $M_\mathrm{B}$ range).

To overcome these limitations we extend here the study of M06 and O10 to a complete, volume-limited sample of 166 nearby ETGs ($\sim5$ times more galaxies) belonging to the \atlas\ sample, while maintaining approximately the same sensitivity and resolution of their observations. The large size and better selection of the sample studied here allow us to establish the detailed \hi\ properties of ETGs in the local Universe on a firm statistical basis. The availability of a large range of multi-wavelength data taken as part of the \atlas\ project is another element of continuity with work on the SAURON sample, and a major step forward relative to other studies.

The aim of this article is to present the \hi\ properties of galaxies in the sample, how they depend on galaxy luminosity and environment, and their relation to signatures of star formation in the host galaxy. Later work will explore how \hi\ in ETGs relates to other galaxy properties derived from different datasets available within the \atlas\ project, and the connection to simulations. This article is structured as follows. In Sec. \ref{sec:sample} we introduce the galaxy sample. In Sec. \ref{sec:data} we describe radio observations and data reduction. In Sec. \ref{sec:morph} we describe the \hi\ morphology of the detected galaxies and introduce a classification scheme based on it. We also discuss the signatures of star formation in galaxies with different \hi\ morphology, and the relation between \hi\ and H$_2$. In Sec. \ref{sec:detlim} we calculate upper limits on \mhi\ of undetected galaxies. In Sec. \ref{sec:mf} we discuss the \hi\ mass function of ETGs. In Sec. \ref{sec:spir} we compare the distribution of \mhi, \mhil\ and \hi\ column density of ETGs and spirals. In Sec. \ref{sec:env} we investigate the relation between \hi\ properties, galaxy luminosity and environment. Finally in Sec. \ref{sec:summ} we summarise our findings and draw conclusions. \rm

%
% ...

%


\section{Sample}
\label{sec:sample}

We study the \hi\ properties of galaxies in the volume-limited \atlas\ sample, which includes 260 ETGs within 42 Mpc and brighter than \mk$=-21.5$ (Paper I). ETGs are selected from a parent sample as galaxies without spiral arms in available optical images.

Our \hi\ study is based on data taken with the Westerbork Synthesis Radio Telescope (WSRT). For observability reasons we select only the 170 galaxies with declination $\delta \geq 10$ deg. We also exclude all 4 galaxies closer than 15 arcmin to the centre of the Virgo galaxy cluster (NGC~4476, NGC~4478, NGC~4486 and NGC~4486A; observations close to Virgo~A do not provide \hi\ data of sufficient quality). Therefore, we study here a sub-sample of 166 \atlas\ ETGs, 61 percent of the total. We refer to this as the \atlashi\ sample. Of these galaxies, 39 reside inside the Virgo galaxy cluster and 127 outside it (Paper I). All 166 galaxies are listed in Table \ref{tab:sample} together with properties derived in the present work.

Figure \ref{fig:sample} shows the distribution of galaxy absolute magnitude $M_K$, large-scale environment density $\Sigma_{10}$ and local environment density $\Sigma_{3}$ for the \atlashi\ sample and the full \atlas\ sample. These parameters are given in Paper I ($M_K$) and in Paper VII ($\Sigma_{10}$ and $\Sigma_{3}$; these are defined as the number density of galaxies contained within a 600-km/s-deep cylinder whose radius is equal to the distance from the tenth and third closest galaxy, respectively). Fig. \ref{fig:sample} shows that the distribution of these parameters is essentially the same for the \atlashi\ sample and the full \atlas\ sample.

This \hi\ survey expands the study performed by M06 and O10 on the SAURON sample. All but one of the 33 galaxies studied by M06 and O10 belong to the \atlas\ sample. Figure \ref{fig:sample} shows the distribution of galaxies common to the two samples. SAURON galaxies are not distributed very differently if one takes into account statistical uncertainties. The main improvement of the present study are the selection of the sample (which, as discussed in the previous section, is unbiased in galaxy luminosity) and sample size, so that we now have a more representative sampling of the ETG properties. \rm 

\section{\hi\ data}
\label{sec:data}

In this section we describe radio observations and data reduction. For some galaxies data were taken as part of earlier studies. Reference to the corresponding papers is given in the text below and  in Table \ref{tab:sample}.

\subsection{Observations}

We follow two different observational strategies for galaxies inside and outside the Virgo cluster. We observe all but one of the 127 galaxies outside Virgo with the WSRT (the only exception is NGC~4762, for which we use ALFALFA data). WSRT data were taken as part of previous projects for 18 of these 127 objects \citep[M06 and O10]{2009A&A...494..489J}. Another 3 fall within the field of view of these earlier observations. All remaining 105 galaxies are observed for \atlas. Both old and new data are taken using the WSRT in the maxi-short configuration. We observe using a bandwidth of 20 MHz sampled with 1024 channels, corresponding to a recessional velocity range of $\sim4000$ km/s and a channel width of $\sim4$ km/s for the \hi\ emission line. The only exception is NGC~2685, observed by \cite{2009A&A...494..489J} with 1024 channels over a 10-MHz bandwidth.

We observe each galaxy for 12 h. For some galaxies, earlier data are deeper than our 12 h integration (see Table \ref{tab:sample}, column 8) and we use those in our analysis.

Data are less homogeneous for the 39 galaxies inside Virgo. We take WSRT data with the same array configuration and correlator set-up described above for 3 galaxies. Equivalent data are available for 13 more galaxies in O10. A field including 2 more \atlas\ galaxies was observed with the WSRT by Oosterloo and collaborators with the same WSRT configuration and correlator set-up described above, and we use their data in our analysis. We use the ALFALFA database to look for \hi\ emission in the remaining 21 Virgo galaxies\footnote{\url{http://arecibo.tc.cornell.edu/hiarchive/alfalfa}.}. We find 2 galaxies with a possibly associated ALFALFA \hi\ detection: NGC~4694 and NGC~4710. We make use of VLA data taken for the VIVA project to study the former \citep{2009AJ....138.1741C}; and we observe the latter with the WSRT using the same array configuration and correlator set-up as above. We use ALFALFA spectra to derive \mhi\ upper limits for the remaining 19 galaxies (see Sec. \ref{sec:detlim}).
 
\subsection{WSRT data reduction and products}

We reduce the WSRT data in a standard way using a dedicated pipeline based on the Miriad reduction package \citep{1995ASPC...77..433S}. The standard pipeline products are \hi\ cubes built using robust=0 weighting and 30-arcsec-FWHM tapering (see \citealt{1995AAS...18711202B} for an explanation of the robust parameter). The discussion below is based on the analysis of these cubes unless otherwise stated. The pipeline is validated by comparing its products to those obtained with manual reduction by M06 and O10 for galaxies in common.

The typical angular resolution of the cubes is $35\times35/\sin\left(\delta\right)$ arcsec$^2$, where $\delta$ is the declination. All cubes are made at a velocity resolution of 16 km/s (after Hanning smoothing). The noise  ranges between 0.4 and 0.7 mJy/beam. The median noise is 0.5 mJy/beam, and 90 percent of the cubes have noise below 0.6 mJy/beam.  At the median noise level the 5$\sigma$ column density threshold within one velocity resolution element is $\sim3.6\times10^{19}\times \sin\left(\delta\right)$ cm$^{-2}$. The \atlashi\ sample covers the $\delta$ range 10 - 60 deg, corresponding to \Nhi\ in the range 0.6 - 3.1 $\times10^{19}$ cm$^{-2}$.

For each galaxy we build a total \hi\ image by summing along the velocity axis all emission included in a mask constructed as follows. A pixel belongs to the mask if the absolute value of its flux density is above 4$\sigma$ in at least one of the following cubes: \it (i) \rm the original cube; \it (ii) \rm the three cubes obtained smoothing the original cube in velocity with a Hanning filter of FWHM 32, 64 and 112 km/s; \it (iii) \rm the original cube smoothed with a 60-arcsec-FWHM Gaussian beam; \it (iv) \rm the three cubes obtained by Hanning smoothing the 60-arcsec cube as in point \it (ii)\rm. This allows us to be sensitive to \hi\ emission over a wide range of angular and velocity scales. The binary masks are then enlarged by convolution with the synthesised beam, and only mask pixels whose value is above 50 percent of the convolved mask peak are kept. This removes a large fraction of spurious detections with small angular size. It also ensures that we do not miss faint gas just below the detection threshold at the edge of the \hi\ distribution.

Total \hi\ images are obtained as the zero-th moment of the masks built following the above procedure. These images contain also pixels with negative surface brightness because we apply the detection algorithm to the absolute value of the flux density. We use these ``negative'' pixels to estimate the column density detection threshold of the images. In practice, we determine the surface brightness threshold $-S_\mathrm{th}$ below which only 5 percent of the negative pixels are retained. The final total \hi\ image includes only pixels with $|S|\geq +S_\mathrm{th}$. The typical value of $S_\mathrm{th}$ corresponds to an \hi\ column density of a few times $10^{19}$ cm$^{-2}$, the exact value varying from image to image.

We also build \hi\ cubes with robust=1 weighting (close to natural weighting) and no tapering. Below we refer to this weighting scheme as $R01$. The noise in $R01$ cubes is a factor of $\sim1.5$ lower than in the standard cubes. However, its pattern is very patchy and the overall image quality is lower. Therefore, we use these cubes only when the detection algorithm described above reveals low-surface-brightness emission missing from the standard cubes (see Table \ref{tab:sample}, column 7). In these cases, the \hi\ mass given in Table \ref{tab:sample} is derived from the $R01$ cube.

We note that the $R01$ cubes have slightly better angular resolution than the standard cubes: $25\times25/sin(\delta)$ arcsec$^2$. Therefore, the formal column density sensitivity is $\sim30$ percent worse.

\section{\hi\ morphology and its relation to the host galaxy}
\label{sec:morph}

We detect 53/166 = 32 percent of all ETGs in the \atlashi\ sample. Unlike the detection rate presented in M06 and O10, this does not include a number of cases where \hi\ is detected only in absorption or where we cannot securely assign \hi\ to the observed ETG because of confusion with neighbouring galaxies (see below).

The \hi\ detection rate depends strongly on environment density, being 4/39=10 percent inside Virgo and 49/127=39 percent outside it (see Sec. \ref{sec:env} for a comparison to previous results). We postpone a full discussion of environmental effects to Sec. \ref{sec:env}. In this section we describe the \hi\ morphology of all detected galaxies regardless of their environment.

\subsection{\hi\ morphological classes}

We show total \hi\ images of all detected ETGs in Fig. \ref{fig:gallery}. The figure suggests the existence of a few broad types of \hi\ morphology, which we describe with the following \hi\ morphological classes:

\begin{itemize}
\item $D$ (large discs) -- most of the \hi\ rotates regularly and is distributed on a disc or ring larger than the stellar body of the galaxy.
\item $d$ (small discs) -- most of the \hi\ rotates regularly and is distributed on a small disc confined within the stellar body of the galaxy.
\item $u$ (unsettled) -- most of the \hi\ exhibits unsettled morphology (e.g., tails or streams of gas) and kinematics.
\item $c$ (clouds) -- the \hi\ is found in small, scattered clouds around the host galaxy.
\end{itemize}

\noindent After examining morphology and kinematics of the detected \hi\ we assign each galaxy to one of these classes. The \hi\ class of detected objects is given in Table \ref{tab:sample} and indicated in Fig. \ref{fig:gallery}. Further notes on the \hi\ morphology and kinematics of individual galaxies are given in Table \ref{tab:sample}, column 7.

\begin{table}
\centering
\caption{Number of galaxies in the various \hi\ classes as a function of Virgo membership (see Sec. \ref{sec:env}) and the presence of CO and dust/blue features (see Sec. \ref{sec:h2}).}
\begin{tabular}{rrrrrr}
\hline
 \hi\ & all            & inside & outside &  CO                 & dust/blue\\
 class & galaxies & Virgo  & Virgo     &det.   & features\\
 \hline
 all                & 166 &39 &127 & 38 & 36 \\ 
undet. & 113 & 35 &78 & 15   & 13\\
det.      &   53 &  4  &  49  & 23  & 23 \\
  \\
$D$              & 24   & 1   & 23 & 10  & 10 \\
                      & (10) &  (1)   & (9)    &   (1)   &  (2) \\
$d$               & 10  & 1    & 9   & 8    & 10 \\
$u$               & 14  & 2    & 12 & 5    & 3 \\
                      & (4)  & (0)    & (4) & (0) & (0) \\
$c$               & 5     & 0    & 5   & 0    & 0\\
\hline
\end{tabular}

Numbers in parenthesis indicate the number of galaxies within that class for which \hi\ is not found in the centre of the host galaxy. All $d$'s and none of the $c$'s host \hi\ in their centre.
\label{tab:summary}
\end{table}

This classification scheme is an attempt to simplify the variety of \hi\ morphologies seen in Fig. \ref{fig:gallery}. It also reflects our view that different \hi\ morphologies give different indications on the assembly and gas-accretion history of the host galaxy. For example, $D$'s and $d$'s are galaxies with a relatively quiet recent history on a time-scale proportional to the orbital time of the rotating gas (this is obviously larger in $D$ than $d$ systems; as an example, it is $\sim1$ Gyr for the $D$ galaxy NGC~6798, and $\sim200$ Myr for the $d$ NGC~5422). On the contrary, galaxies in class $u$, where most of the gas is not rotating regularly around the stellar body, have recently experienced tidal forces or episodes of gas accretion/stripping involving a large fraction of their gas reservoir. In other words, their \hi\ content is evolving, and for some of them the current \hi\ properties may be just a transient phase. In contrast, in $D$ objects the \hi\ has likely been part of the galaxy (and orbiting around the stellar body) for many gigayears. Finally, galaxies in class $c$ are surrounded by \hi\ not obviously associated to them, and we regard them as objects with similar \hi\ properties as undetected ETGs.

\begin{figure*}
\includegraphics[width=43mm]{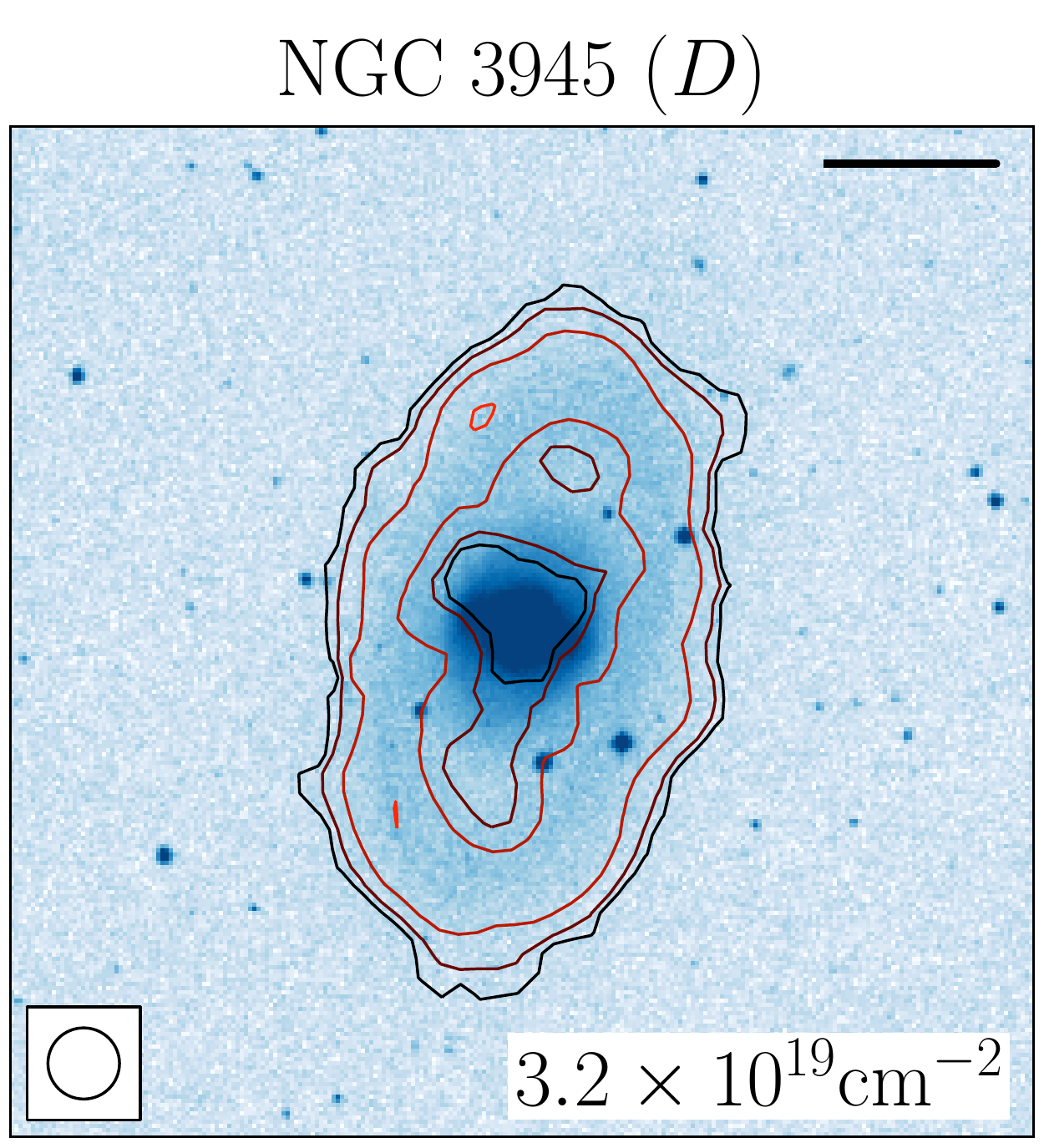}
\includegraphics[width=43mm]{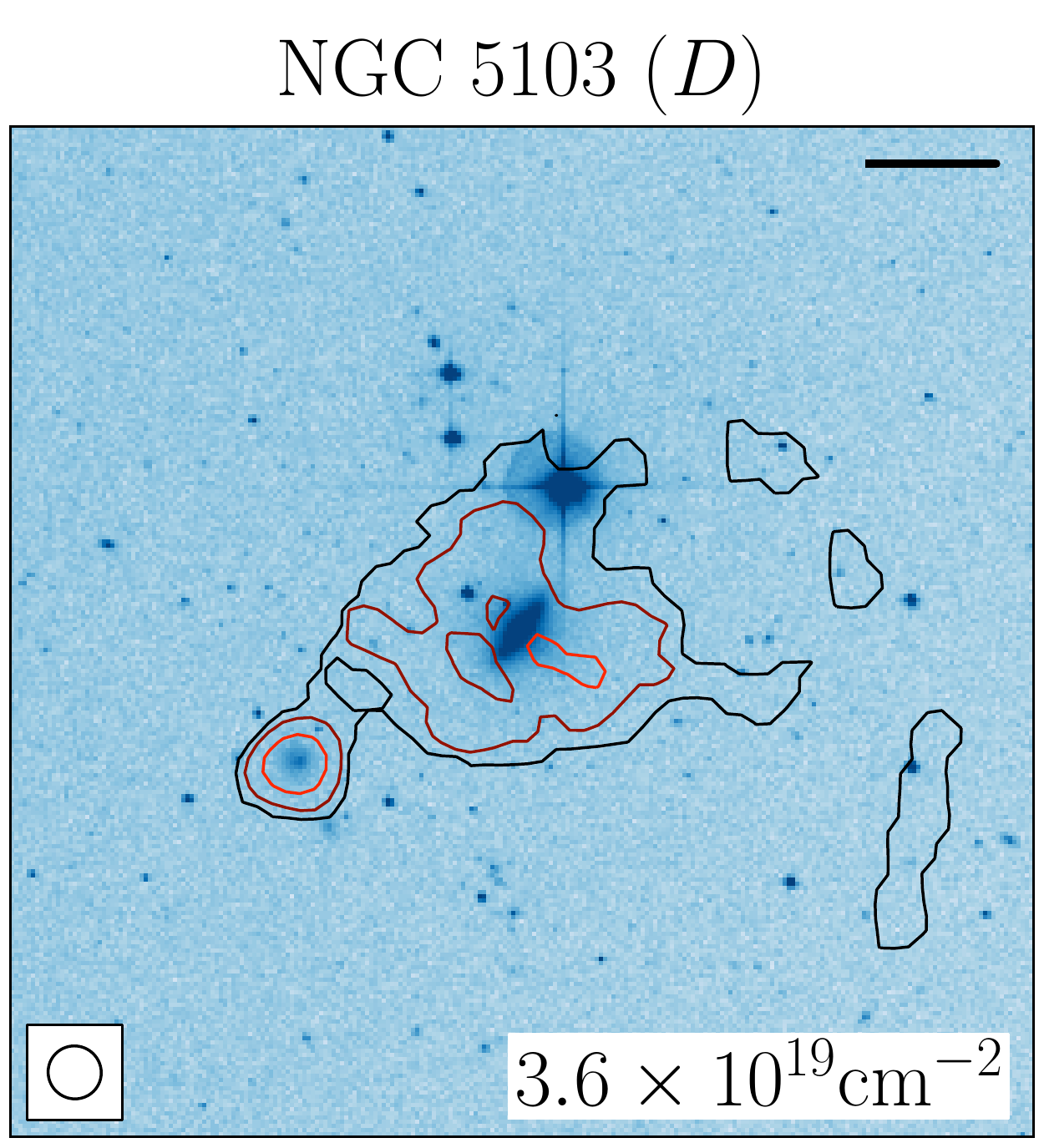}
\includegraphics[width=43mm]{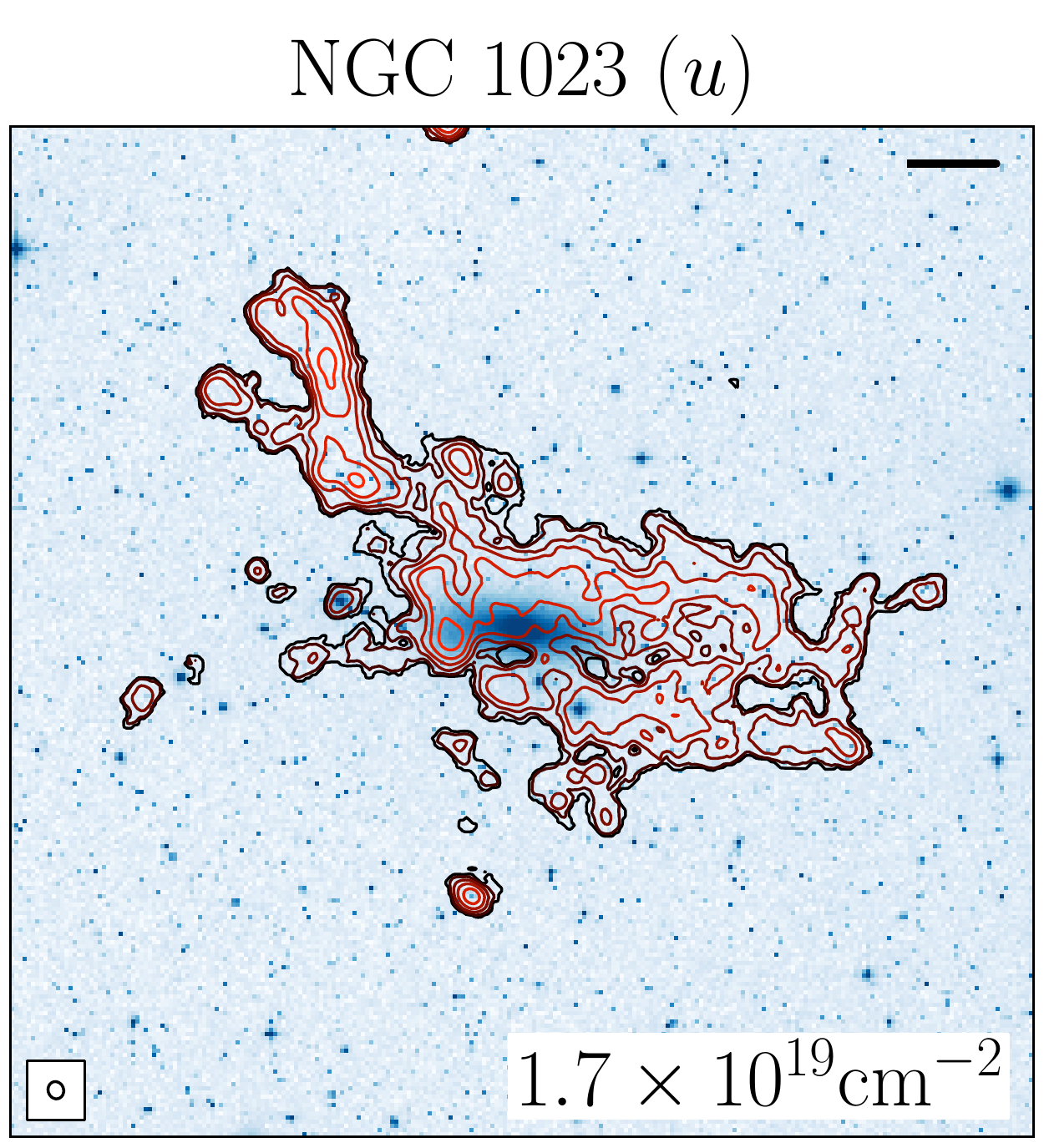}
\includegraphics[width=43mm]{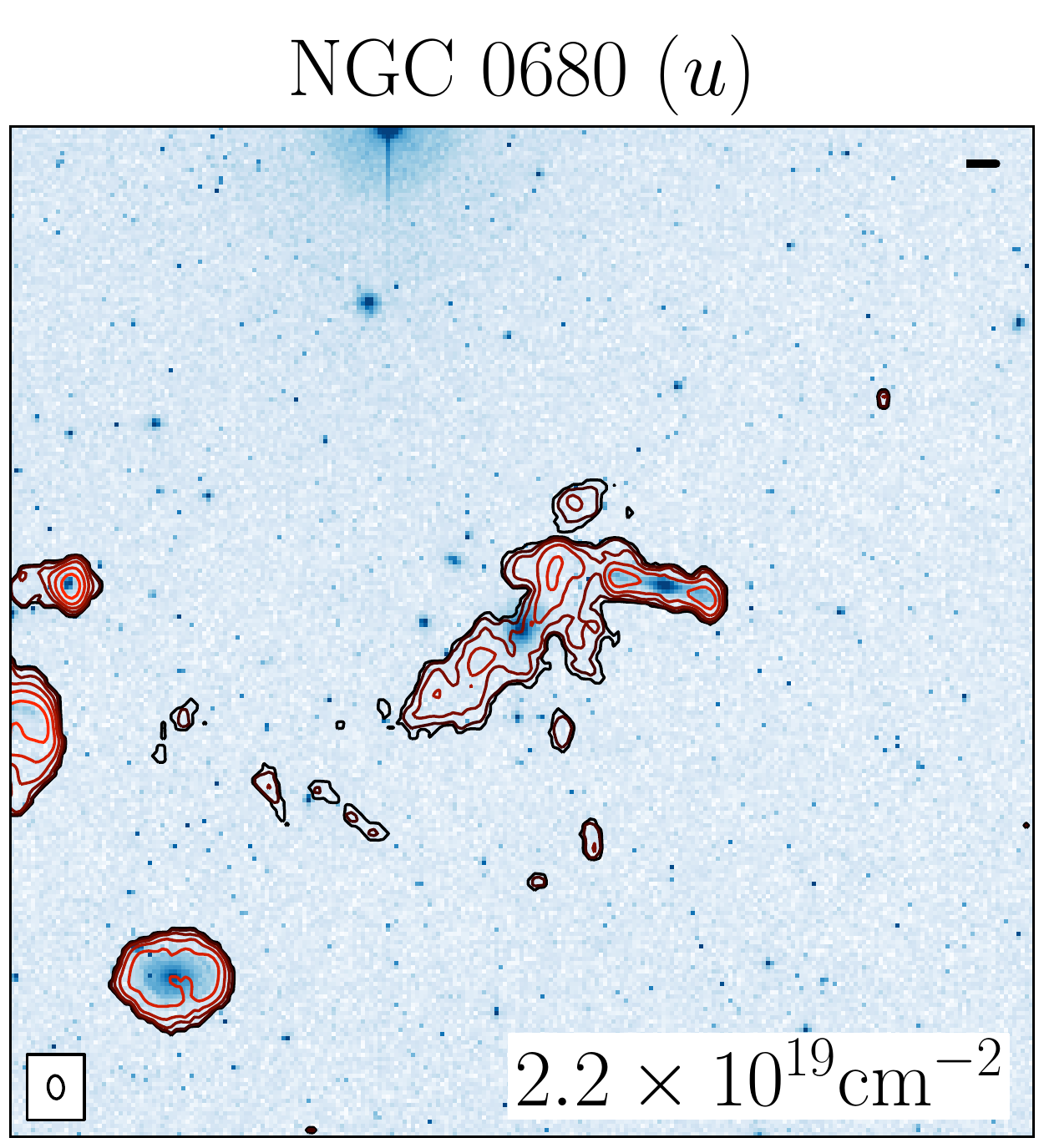}
\includegraphics[width=43mm]{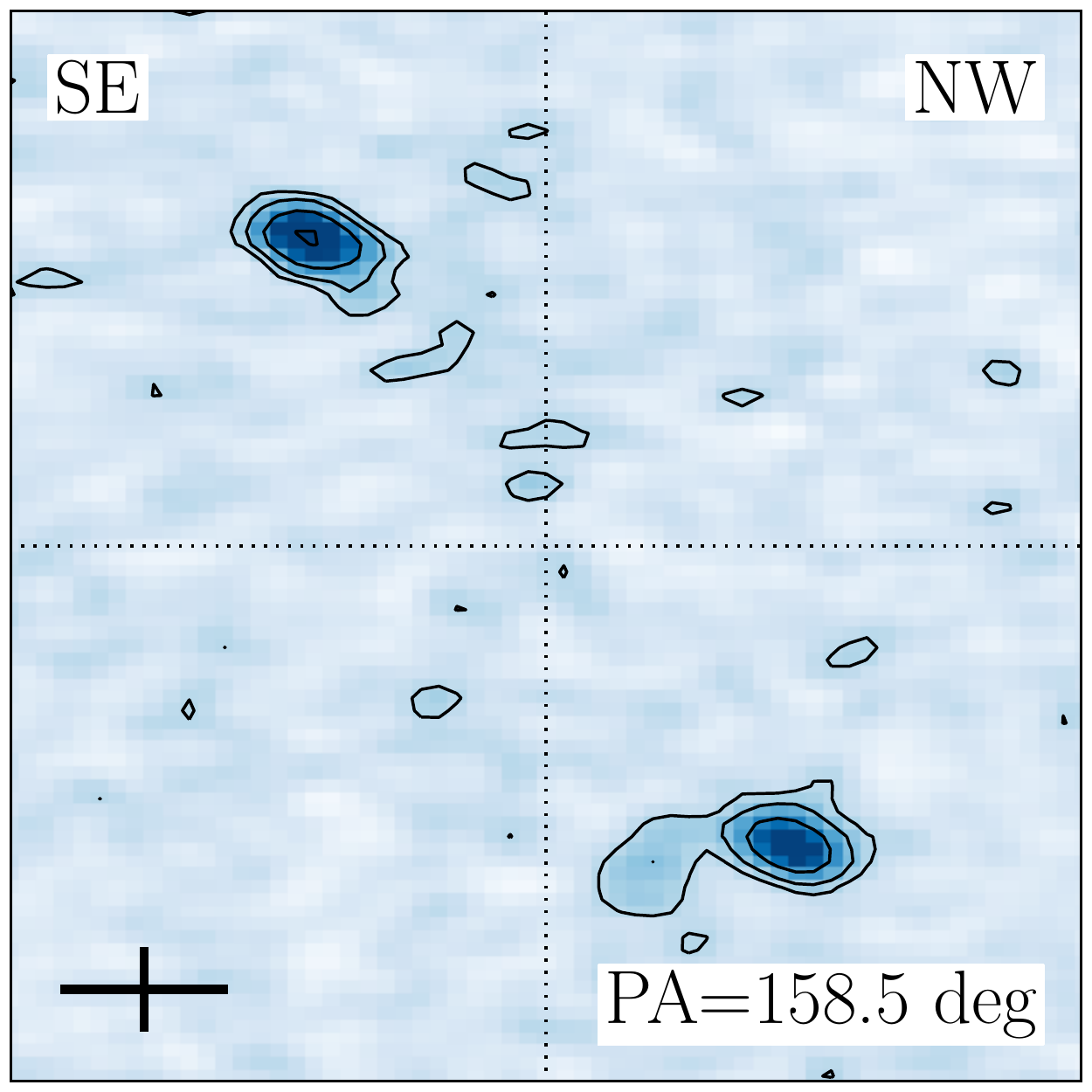}
\includegraphics[width=43mm]{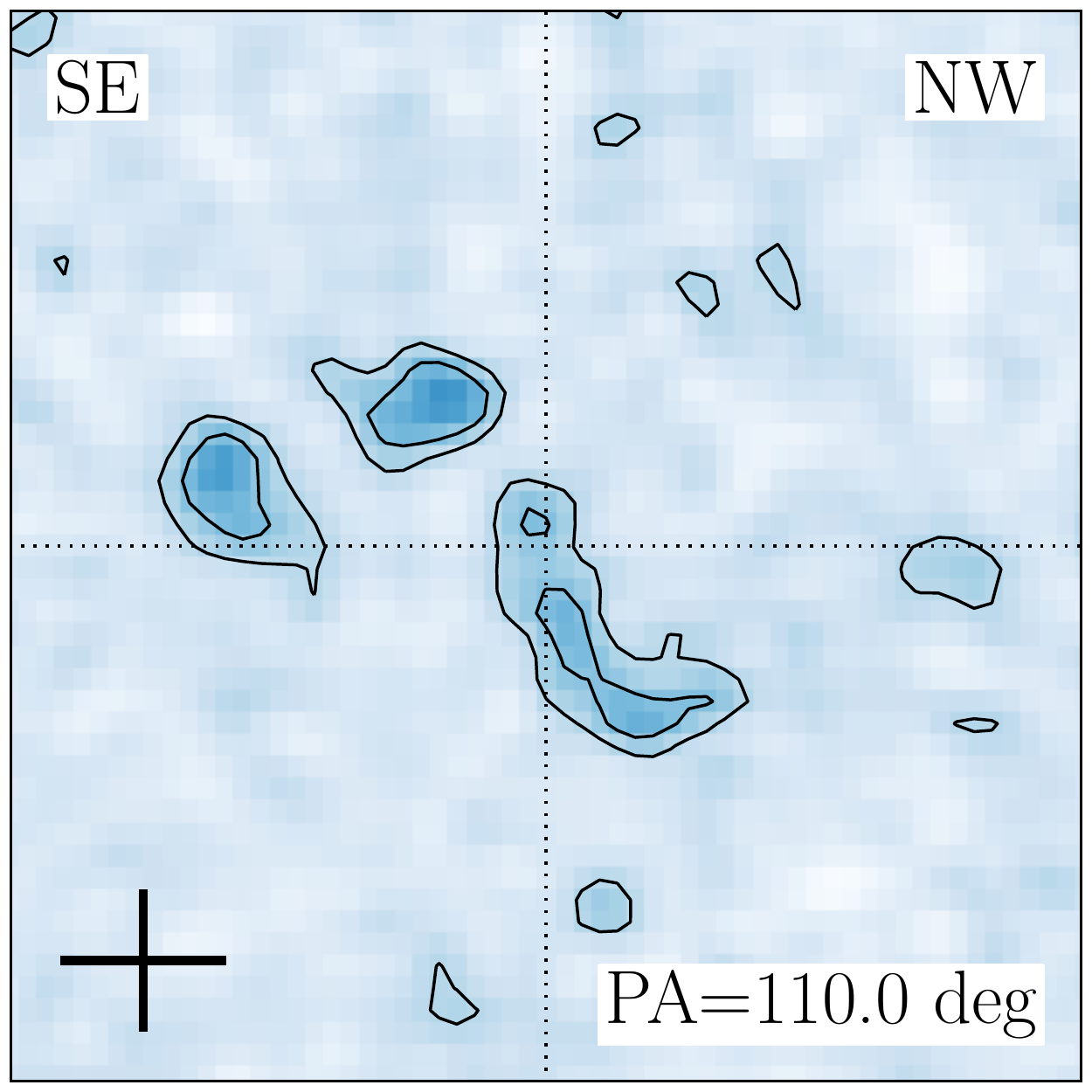}
\includegraphics[width=43mm]{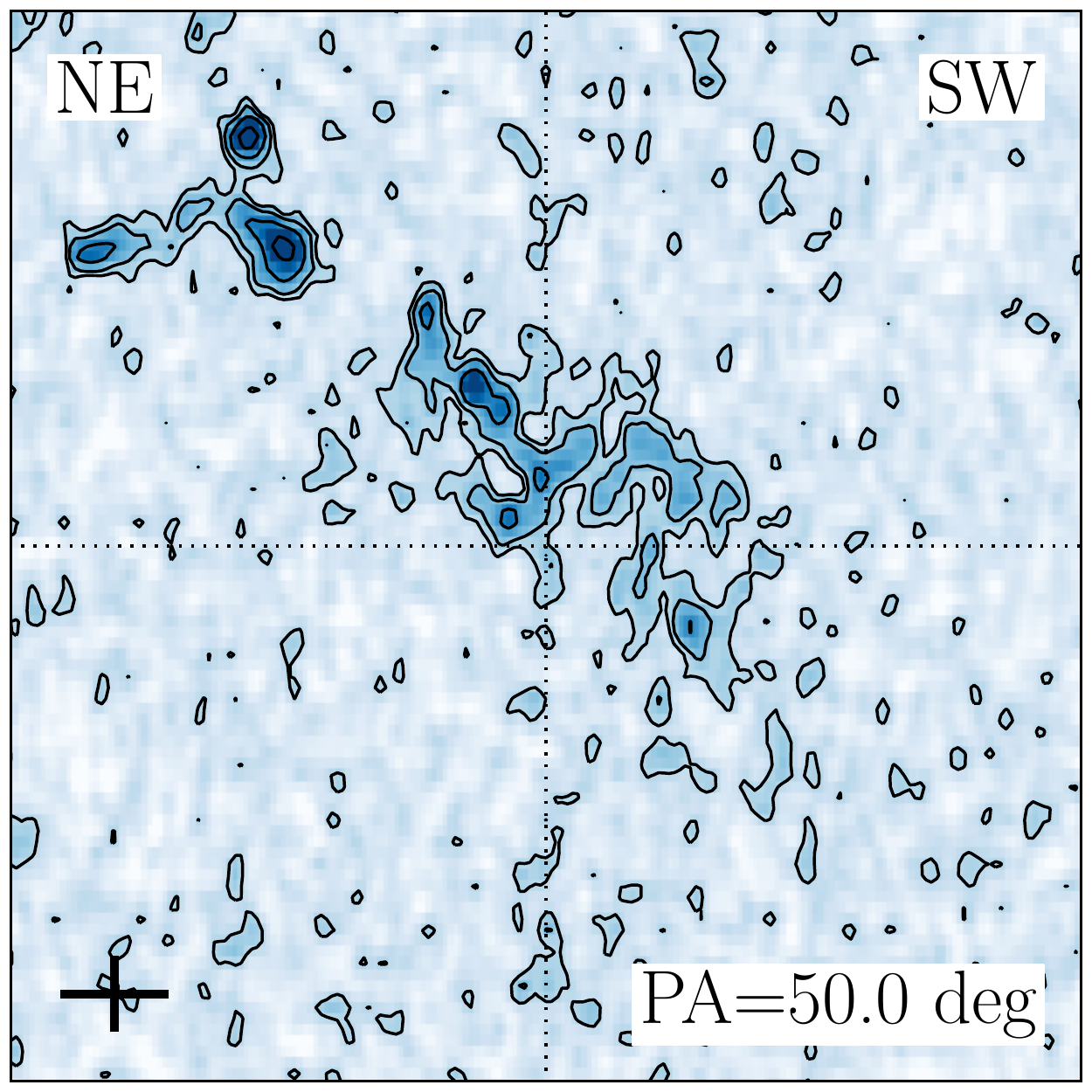}
\includegraphics[width=43mm]{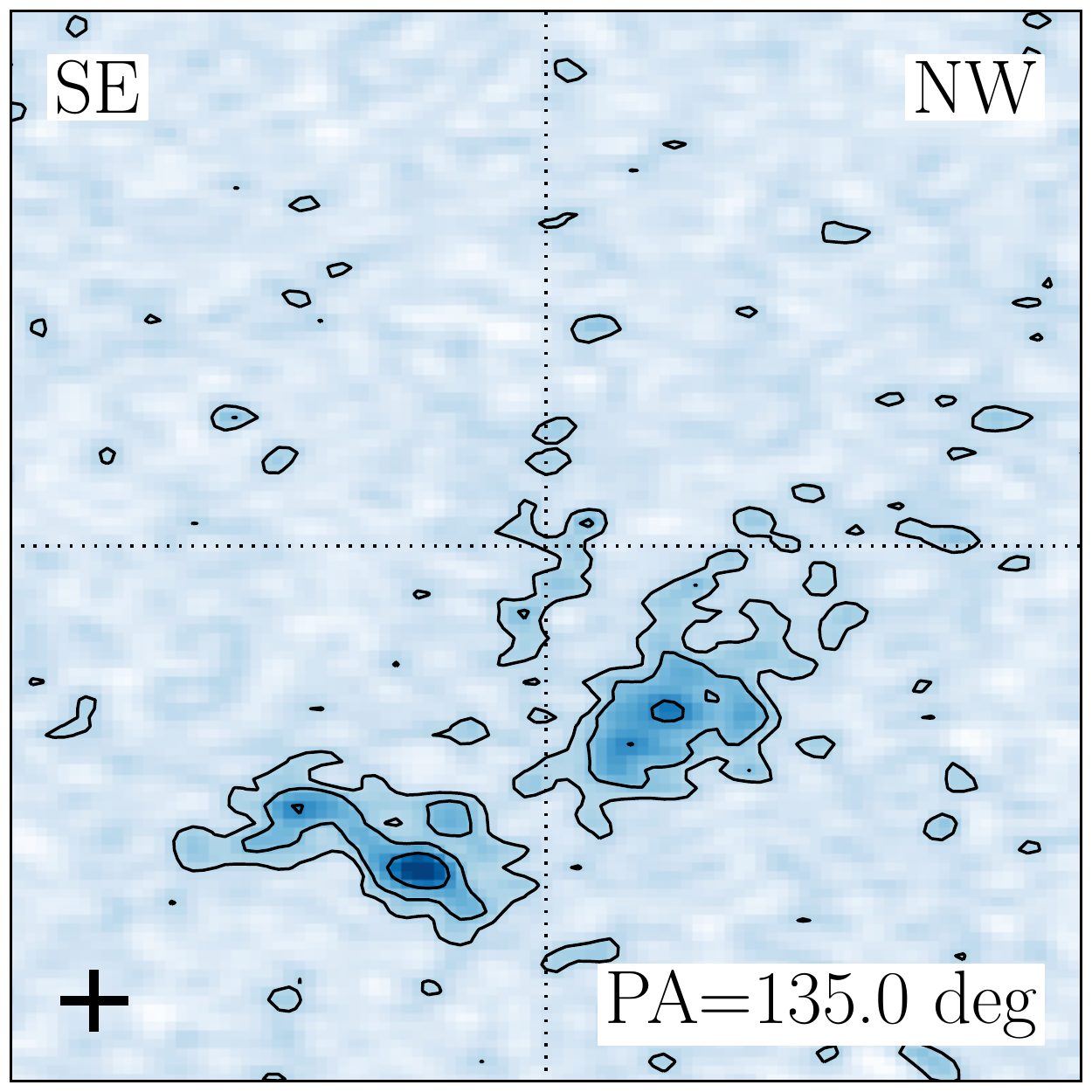}
\caption{A sequence of \hi-rich ETGs with increasingly less regular gas configurations (left to right). The sequence shows galaxies with very large \hi\ distributions relative to the stellar body. The top row shows total-\hi\ contour images. We refer to the caption of Fig. \ref{fig:gallery} for a description of the content of each image. In this figure the top-right scale bar indicates 10 kpc at the galaxy distance. The bottom row shows position-velocity diagrams of galaxies in the top row drawn along an axis which highlights relevant features in the \hi\ kinematics (see text). We plot angular and velocity offset along horizontal and vertical axis, respectively. The diagrams are drawn along an axis whose position angle is indicated on the bottom right of each diagram (north to east). Contours are drawn at $1.0\times2^n$ mJy beam$^{-1}$, $n=0,1,2,...$. The cross on the bottom left indicates 10 kpc along the horizontal axis and 50 km s$^{-1}$ along the vertical axis.}
\label{fig:sequenced}
\end{figure*}

Similarly, the distinction between $D$ and $d$ galaxies, which is entirely based on the size of the \hi\ distribution, is motivated by our view that they are fundamentally different objects (see Sec. \ref{sec:4.3}). In order to make this classification as objective as possible we define the \hi\ radius \rhi\ as the maximum distance of the \Nhi=$5\times10^{19}$ cm$^{-2}$ isophote from the galaxy centre (deconvolved with the beam size). We then compare \rhi\ to the optical effective radius $R_e$ given in Paper I. We adopt \rhi$=3.5\times R_e$ as the dividing line between small and large \hi\ systems (for a de Vaucouleur profile this radius includes 90 percent of the total light).\footnote{The only galaxy whose classification would not fit in our picture is NGC~3414, which we regard as hosting a large \hi\ system although its \rhi\ is smaller than 3.5$\times R_e$. However, the gas in this galaxy is distributed on a polar configuration, so that it may be more appropriate to compare \rhi\ to $R_e\times(1-\epsilon)$ where $\epsilon$ is the optical ellipticity given in Paper II. When this is done the ratio between \hi\ and optical radius is larger than 3.5 so that we classify this galaxy as $D$.}

Given this classification scheme, $D$ galaxies have \mhi\ in the range $10^8$-$10^{10}$ \msun, and the gas extends out to many tens of kpc from the centre of the galaxy (in some cases more than 10 $R_\mathrm{e}$). In contrast, \mhi\ in $d$ systems is typically below $10^8$ \msun, and the size of the gas distribution is up to a few kpc. Galaxies in group $u$ contain between a few times 10$^7$ and 10$^{10}$ \msun\ of \hi, often stretching to many tens of kpc from the galaxy. Finally, galaxies in group $c$ have \mhi\ below a few times 10$^7$ \msun, comparable to the \hi\ mass upper limits described in Sec. \ref{sec:detlim}.

\subsection{A continuum of \hi\ morphologies}
\label{sec:4.2}

We find 24, 10, 14, and 5 galaxies in group $D$, $d$, $u$ and $c$, respectively, corresponding to 14, 6, 8 and 3 percent of the sample. Only 1 $D$, 1 $d$ and 2 $u$'s belong to the Virgo cluster. Outside Virgo 18, 7, 9 and 4 percent of all galaxies belong to group $D$, $d$, $u$ and $c$, respectively. About 40 percent ($10/24$) of all $D$'s are rings (or discs with a central \hi\ hole). We summarise these results in Table \ref{tab:summary}.

The above figures show that within the \atlas\ volume as many as 1/5 of all ETGs (and 1/4 of all ETGs outside Virgo) host rotating \hi\ distributions (groups $D$ and $d$). These objects represent the majority of our detections, 64 percent, in agreement with M06 and O10. Therefore, \it if \hi\ is detected in an ETG it will most likely be distributed on a disc or a ring. \rm

Within the adopted classification scheme it is not always easy to assign a galaxy to a given class. In fact, we find a number of objects intermediate (or in transition) between classes. For the sake of simplicity we will make use of this classification in the rest of the article, but the \atlashi\ sample reveals rather a continuum of \hi\ morphologies going from settled, rotating systems to progressively more and more disturbed ones, and from the latter to systems of scattered \hi\ clouds which may or may not be related to the nearby ETG. We guide the reader through this continuum of morphologies in the rest of this section.

Figure \ref{fig:sequenced} shows a sequence of increasingly less regular \hi\ distributions (left to right) whose size is large relative to that of the host galaxy. The sequence shows the continuity of \hi\ morphology going from the most regular $D$ objects (left) to the most irregular, large $u$'s (right). It begins on the left with an \hi\ ring, likely a resonance related to a bar (see also the stellar ring coincident with the \hi\ distribution). The morphology (top row) and kinematics (bottom row) of the \hi\ ring are extremely regular. The sequence continues on the second panel with a disc whose rotation and morphology are somewhat less regular. In this galaxy a tail of gas is visible to the west, and a small neighbour to the east might be interacting with our target. Although the \hi\ kinematics exhibits an overall ordered rotation (but misaligned relative to the stellar kinematics), it is clear that some of the gas has not settled yet within the galaxy potential (or has been disturbed recently). Systems like this provide a link between $D$'s and galaxies in group $u$, where most (but not necessarily all) of the gas is unsettled.

The third panel shows a galaxy that we classify as $u$. However, there are hints that at least part of the \hi\ is rotating around the stellar body (see the position-velocity diagram on the bottom row and the description of this galaxy in \citealt{1984MNRAS.210..497S} and M06). Systems like this could move towards the $D$ class if their evolution proceeds without significant merger/accretion events for a few gas orbital periods. Finally, the fourth panel shows completely  unsettled \hi\ (\citealt{2011arXiv1105.5654D}, herafter Paper IX, discuss the optical signature of a recent merger in this galaxy). It is possible to see the sequence in Fig. \ref{fig:sequenced} as one of decreasing time passed since the last major episode of gas accretion or stripping (relative to the gas orbital period).

\begin{figure*}
\includegraphics[width=43mm]{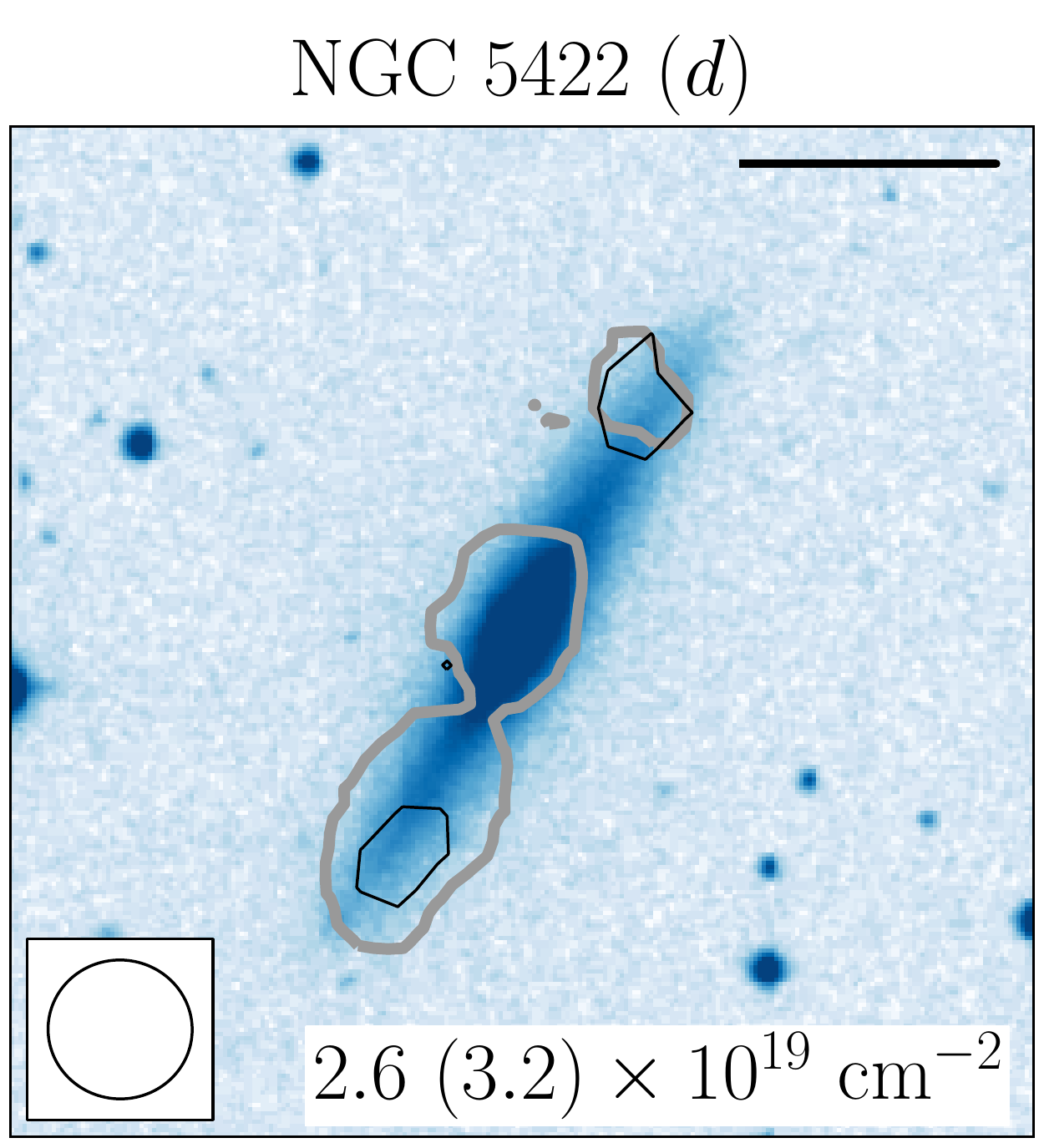}
\includegraphics[width=43mm]{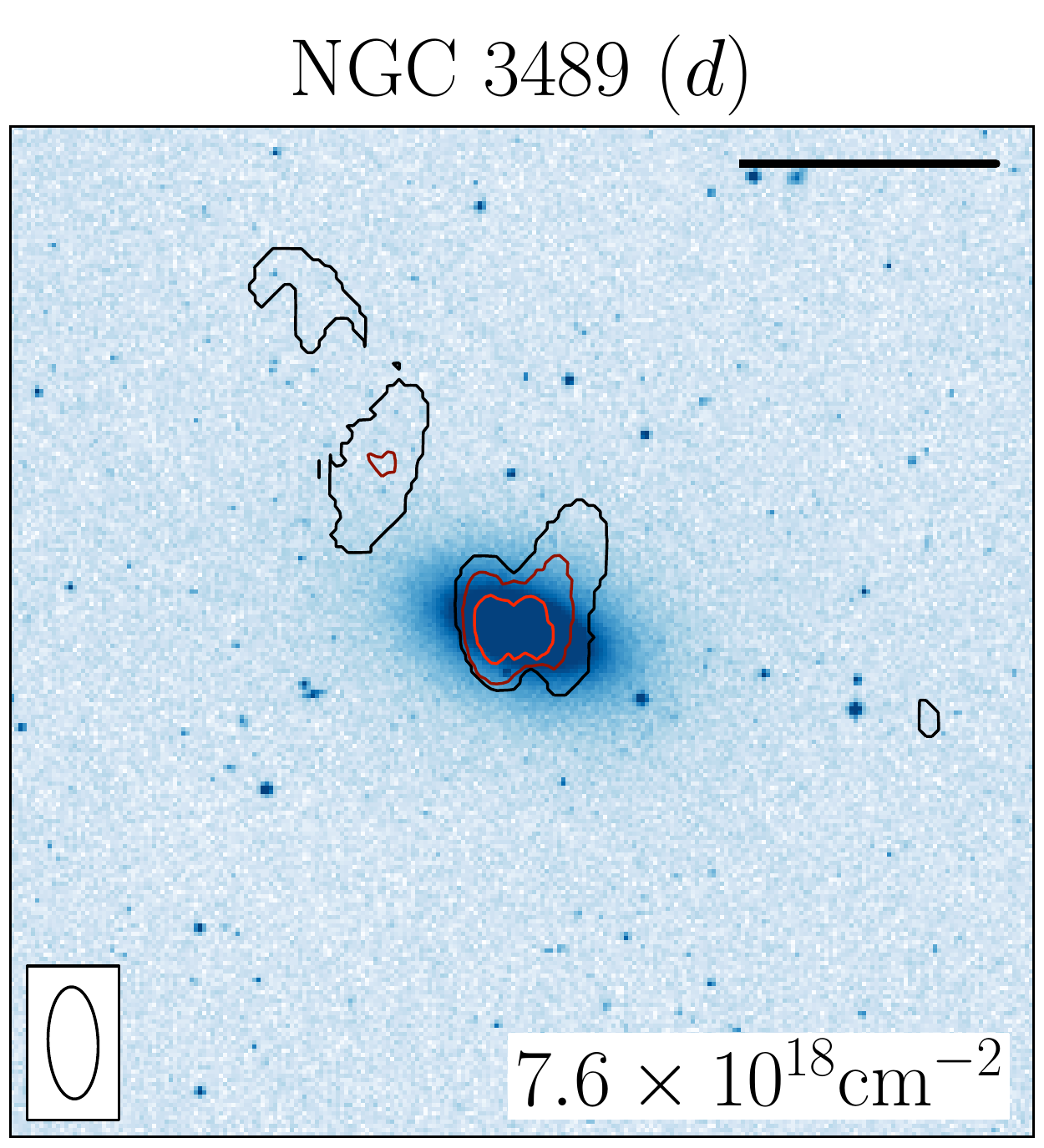}
\includegraphics[width=43mm]{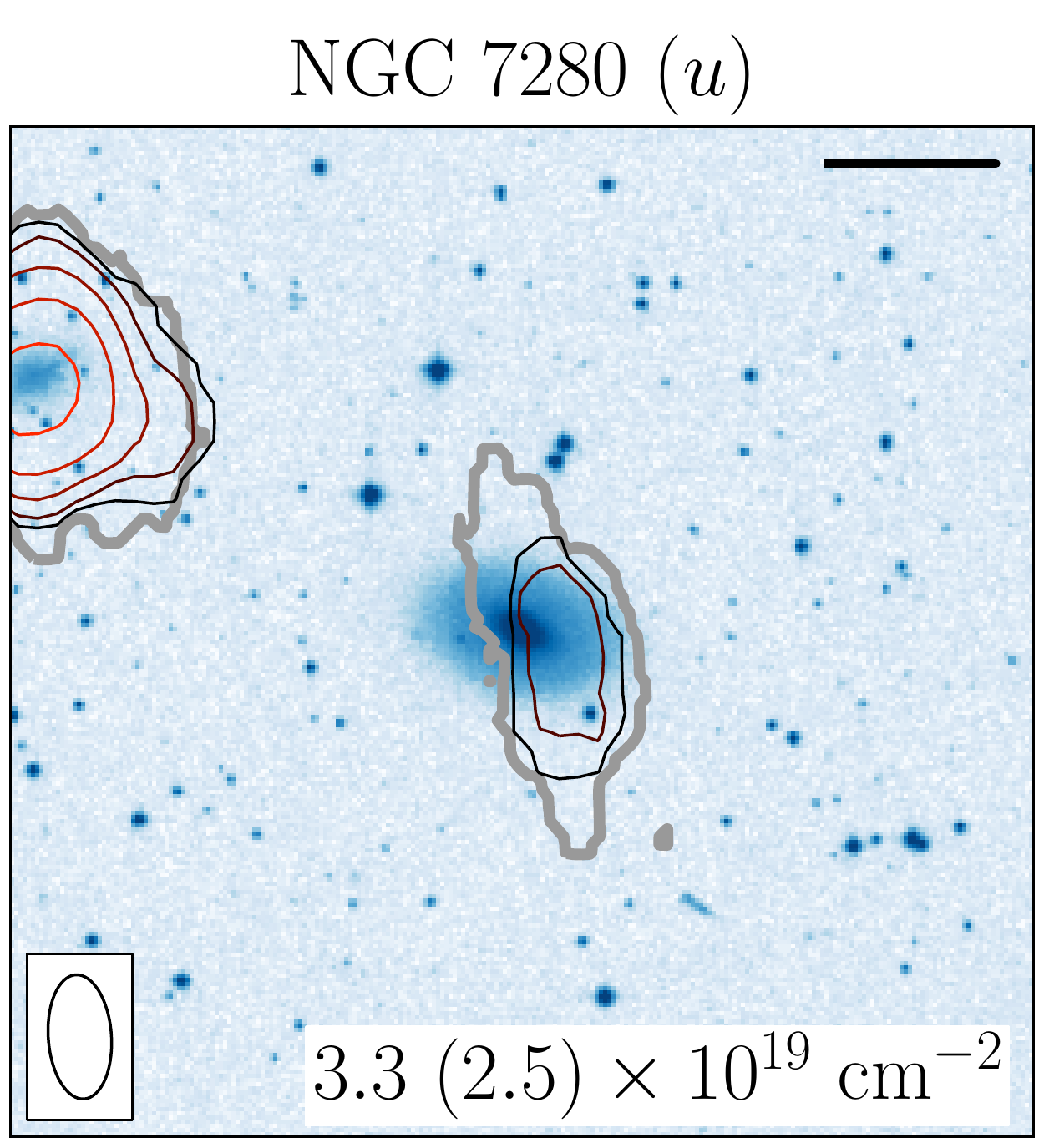}
\includegraphics[width=43mm]{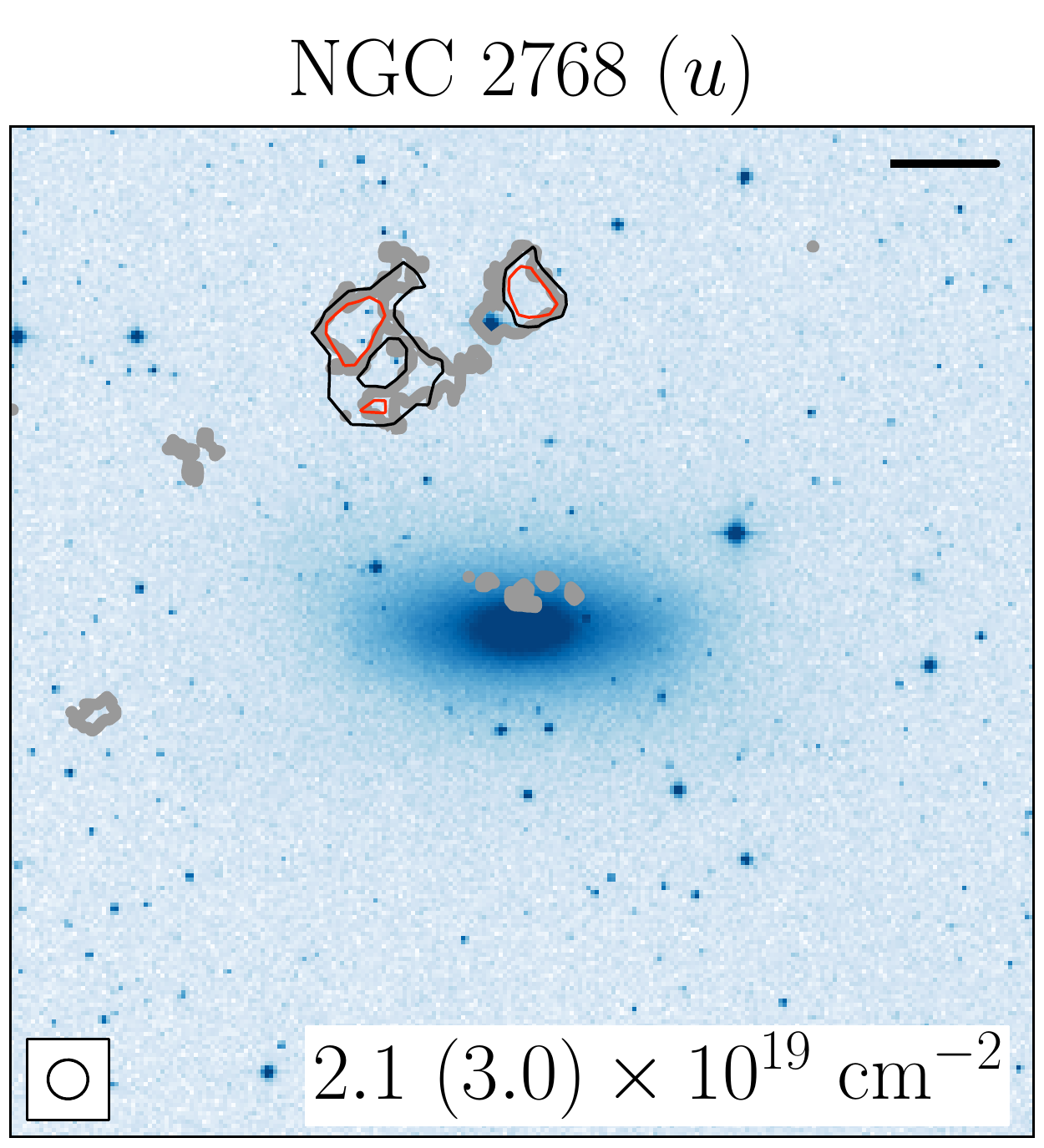}
\includegraphics[width=43mm]{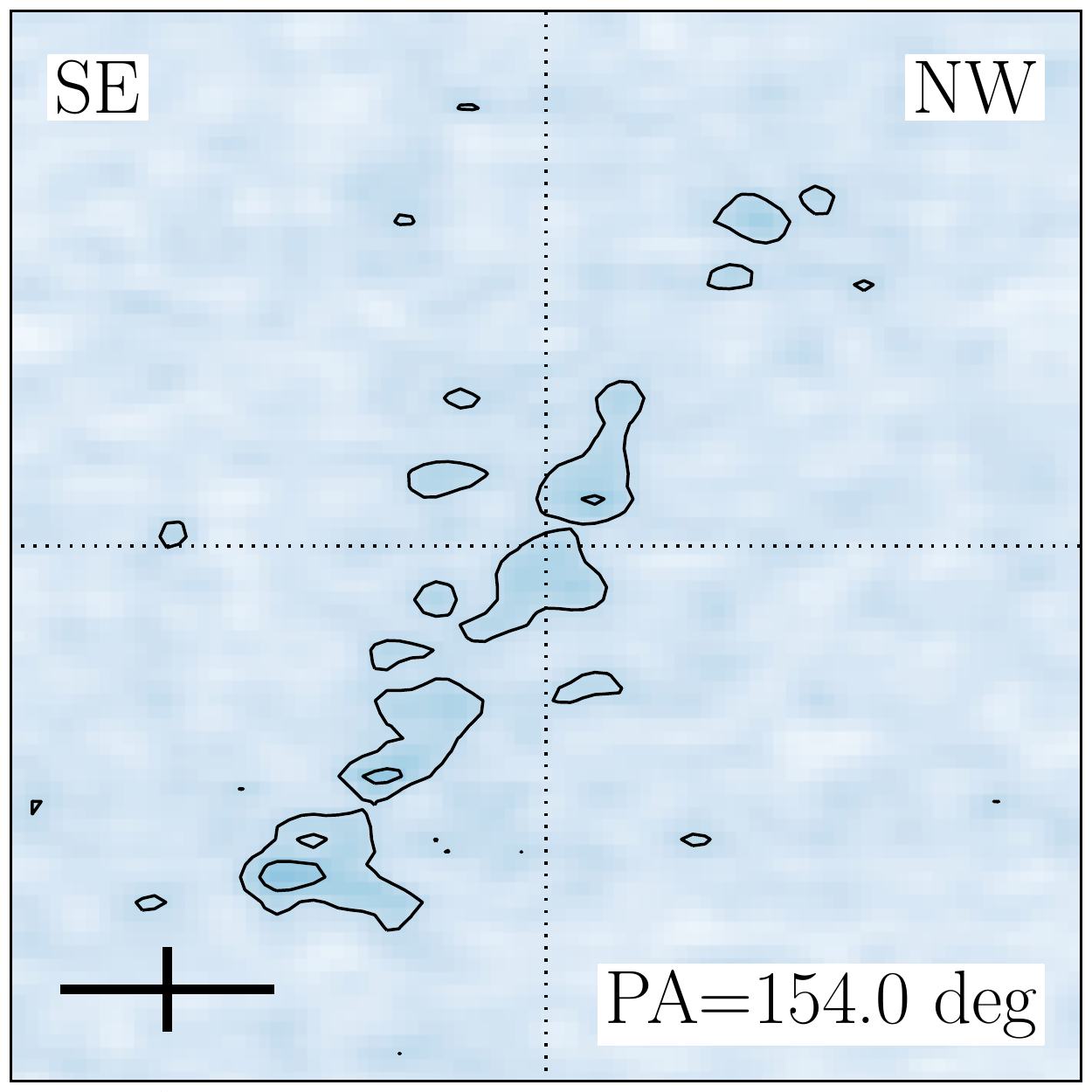}
\includegraphics[width=43mm]{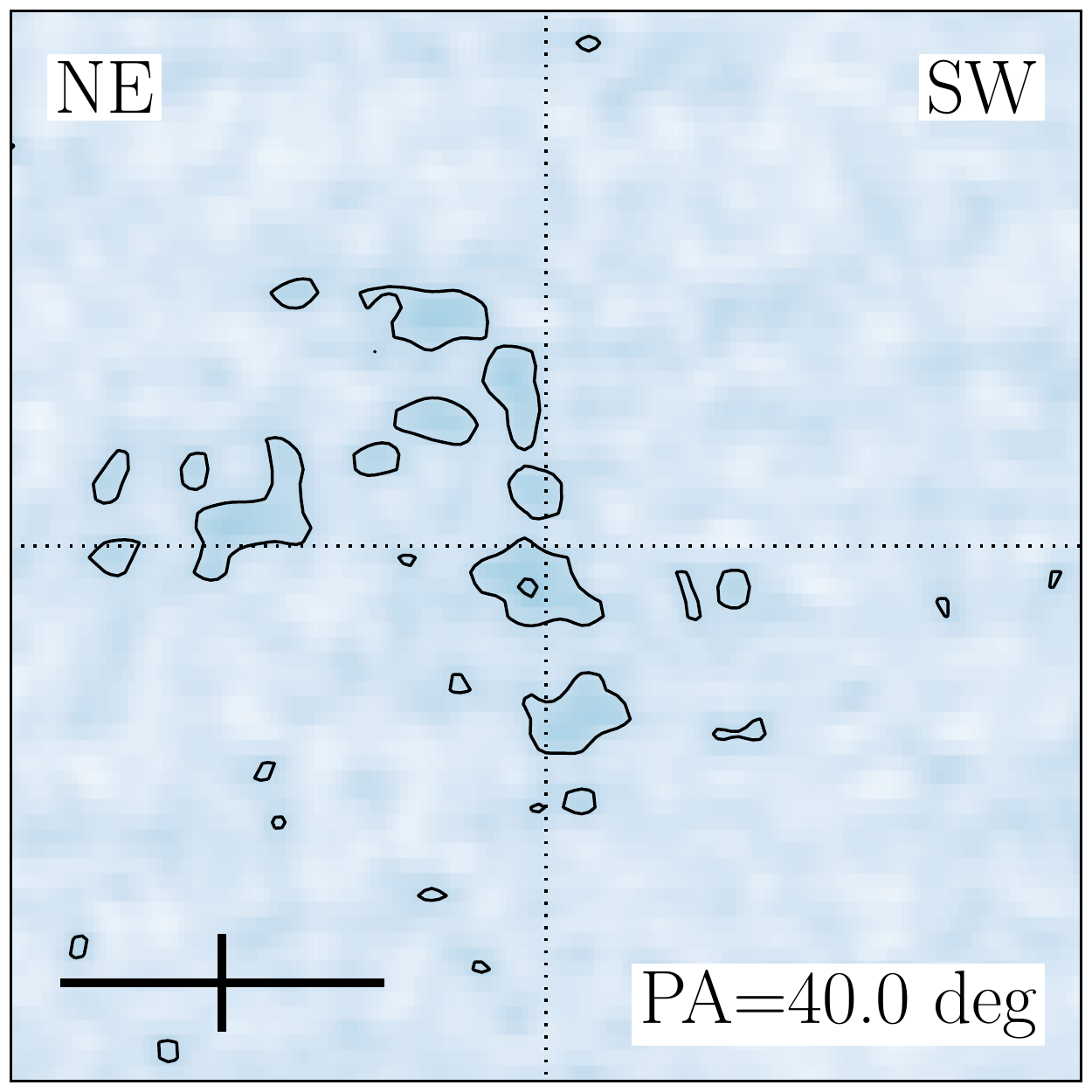}
\includegraphics[width=43mm]{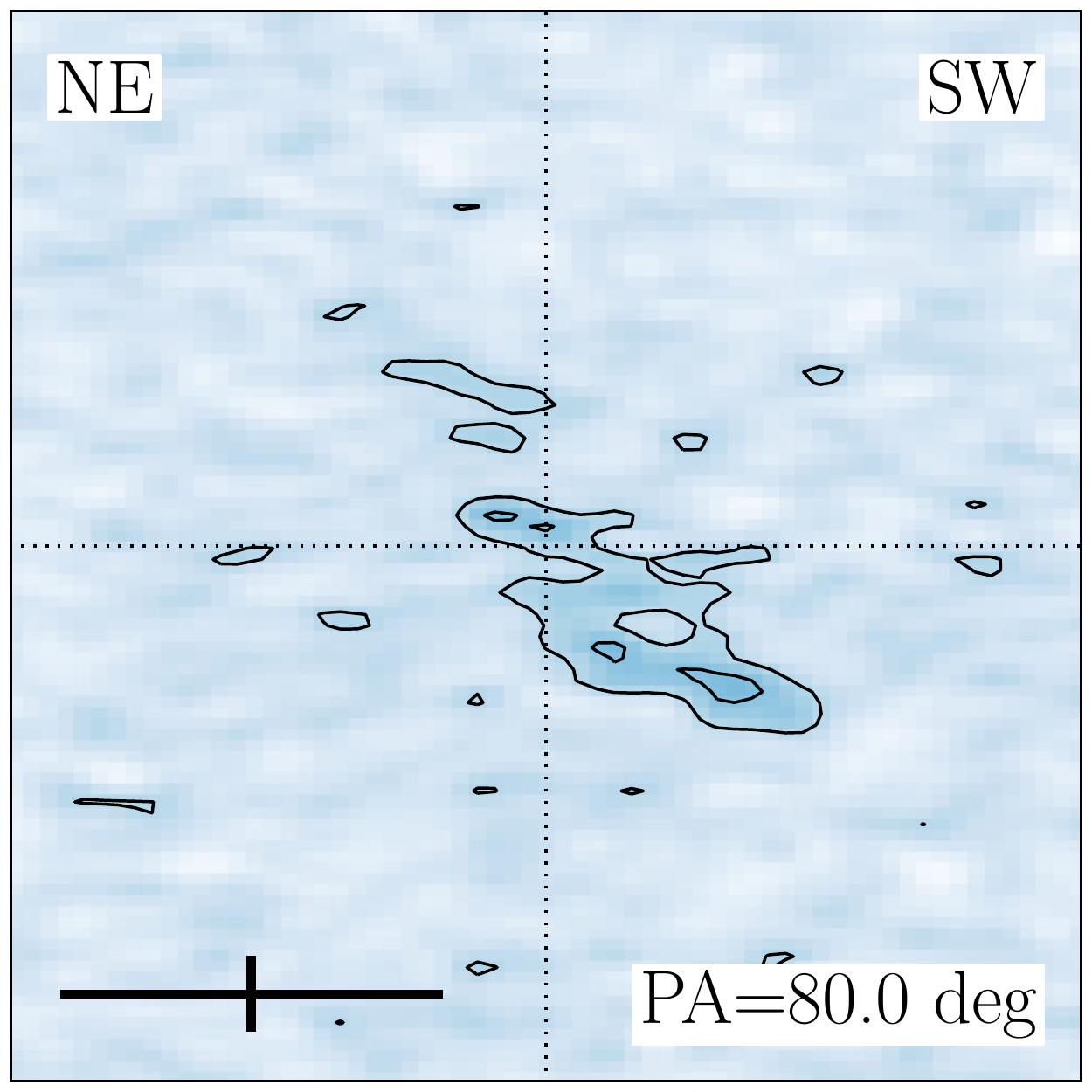}
\includegraphics[width=43mm]{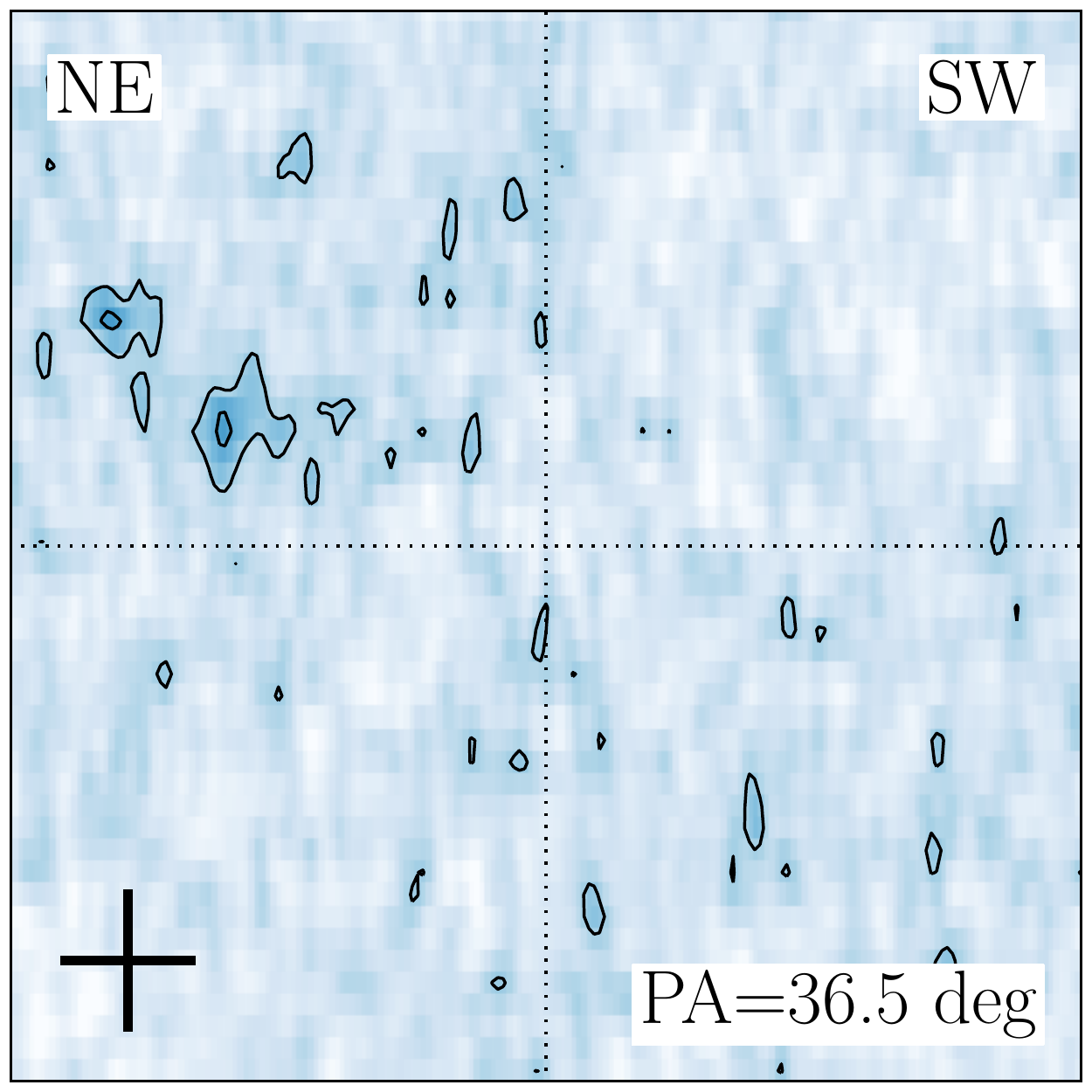}
\caption{A sequence of \hi-rich ETGs with increasingly less regular gas configurations (left to right). The sequence shows galaxies with \hi\ distributions of size similar to that of the stellar body. See the caption of Fig. \ref{fig:sample} for a description of the images. In this figure, position-velocity diagrams are drawn using the $R01$ cubes. Contours are drawn at $N_0\times2^n$ mJy beam$^{-1}$, $n=0,1,2,...$, where $N_0=0.5$ for NGC~2768, 0.75 for NGC~5422 and NGC~3489, 1.0 for NGC~7280.}
\label{fig:sequencee}
\end{figure*}

A similar sequence can be drawn for \hi-detected galaxies where the size of the gas distribution is comparable to that of the stellar body. Such a sequence, shown in Fig. \ref{fig:sequencee}, illustrates the continuity of \hi\ morphology going from the most regular $d$ objects (left) to the most irregular $u$'s (right). The sequence starts with a very regular small \hi\ disc. It continues with a very faint, small disc connected spatially and in velocity to a cloud outside the stellar body. This system appears fairly regular within the stellar body (given the low signal-to-noise ratio), but the outer cloud indicates that some accretion may be on-going. In fact, the \hi\ cloud suggests that small \hi\ discs may form by accreting gas from small companions or the surrounding medium (O10).

The third panel in Fig. \ref{fig:sequencee} shows a system which is difficult to classify. We detect \hi\ only on one side of the galaxy, and because of this we classify it as $u$. However, the position-velocity diagram shows a hint that gas may be present also on the other side of the galaxy at opposite velocity relative to systemic. This is the behaviour expected for a disc, so this galaxy may actually be a misclassified $d$. We conclude the sequence with a system where \hi\  is detected far from the stellar body but is clearly connected to it. The modest amount of gas and the size of the \hi\ distribution relative to the stellar body suggests that this system may evolve towards the $d$ class if left undisturbed.

\begin{figure*}
\includegraphics[width=43mm]{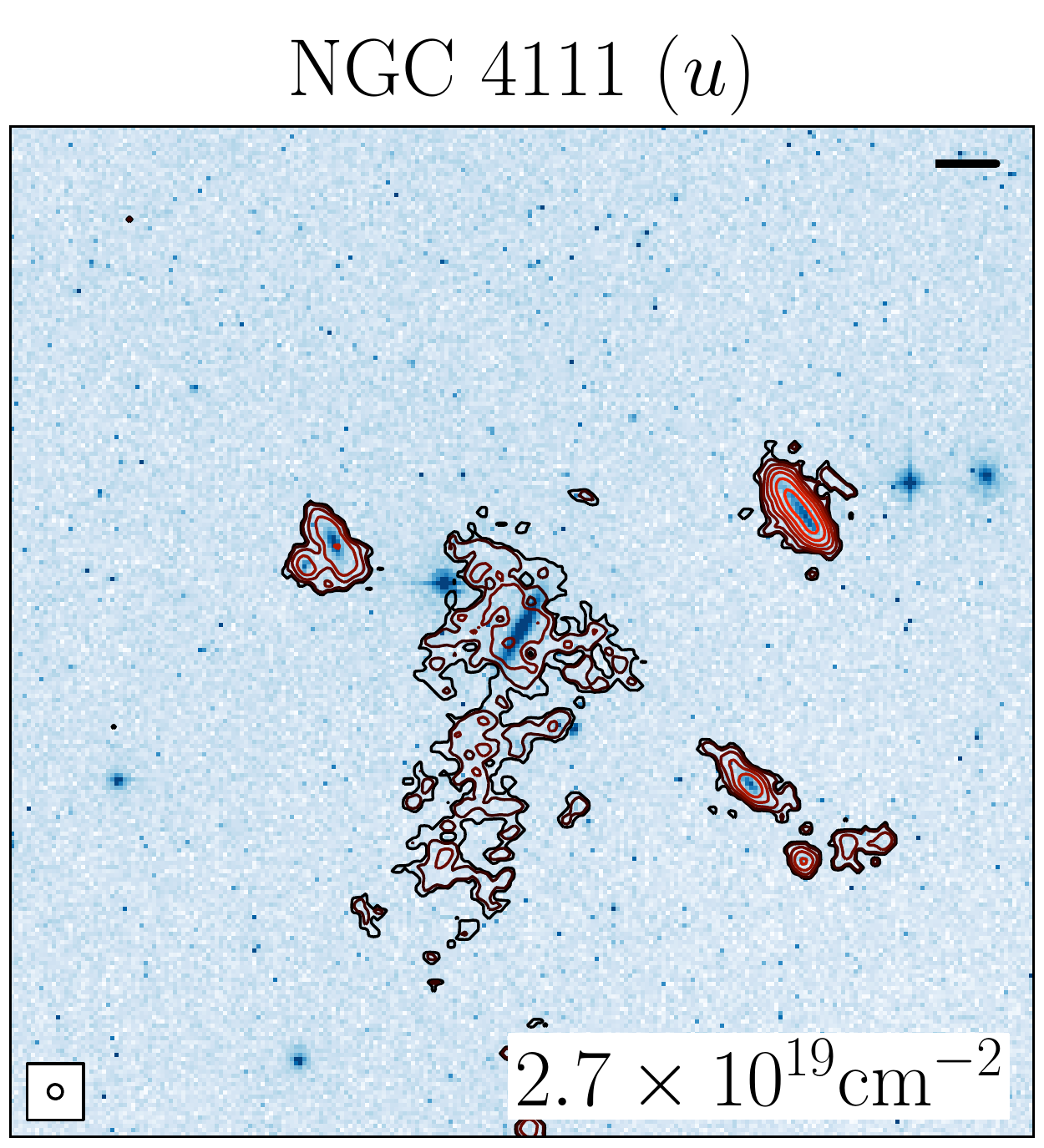}
\includegraphics[width=43mm]{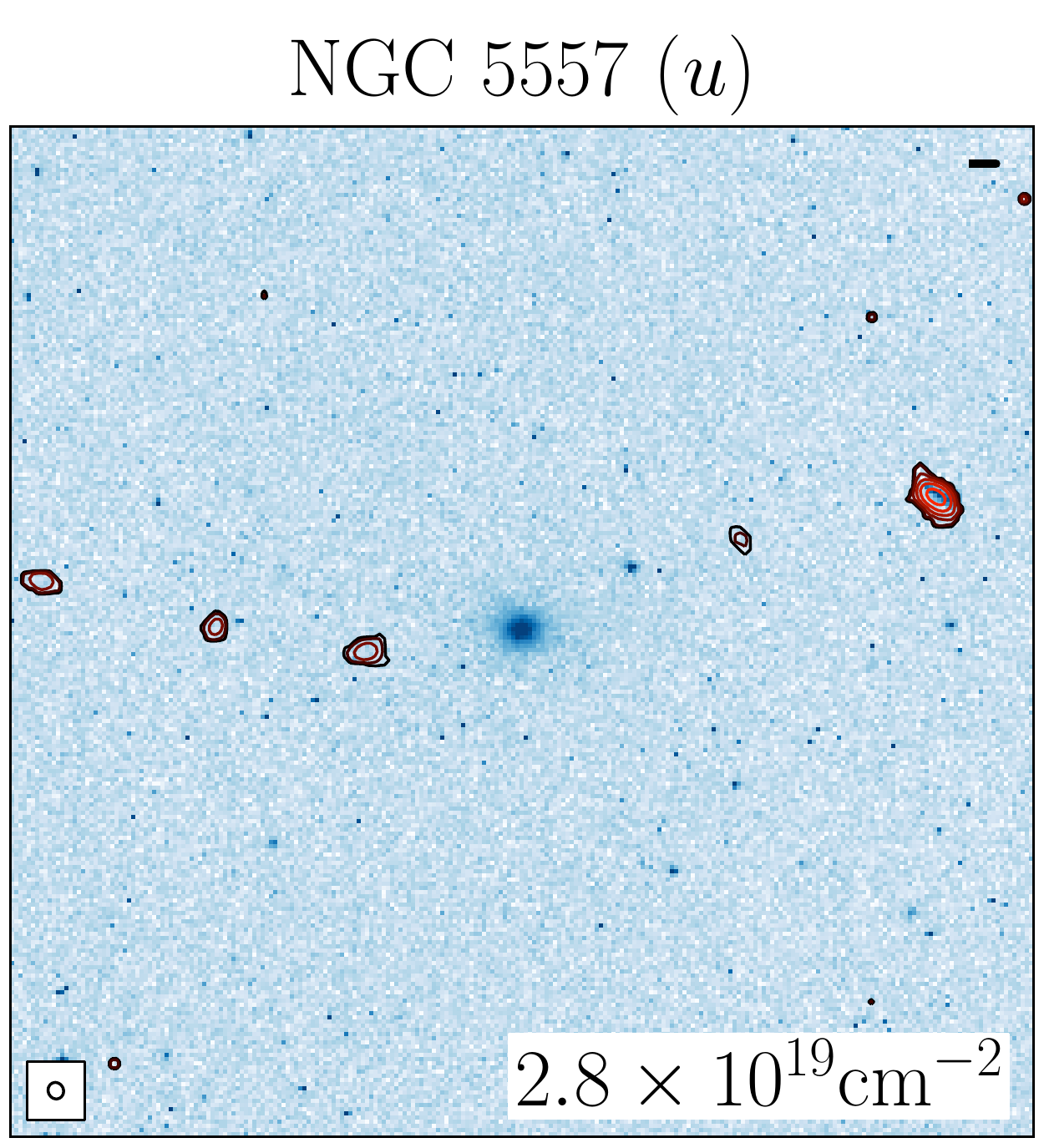}
\includegraphics[width=43mm]{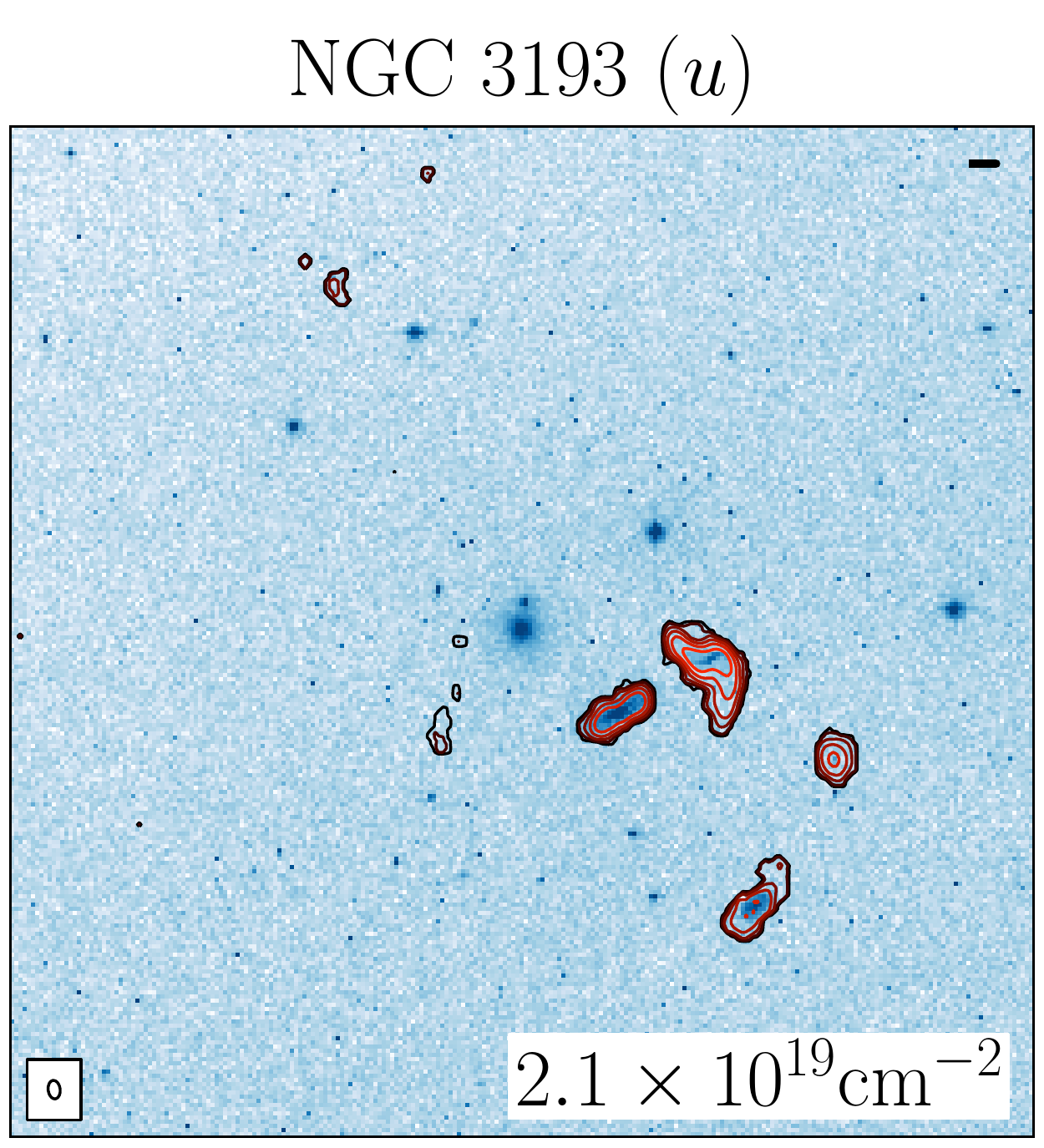}
\includegraphics[width=43mm]{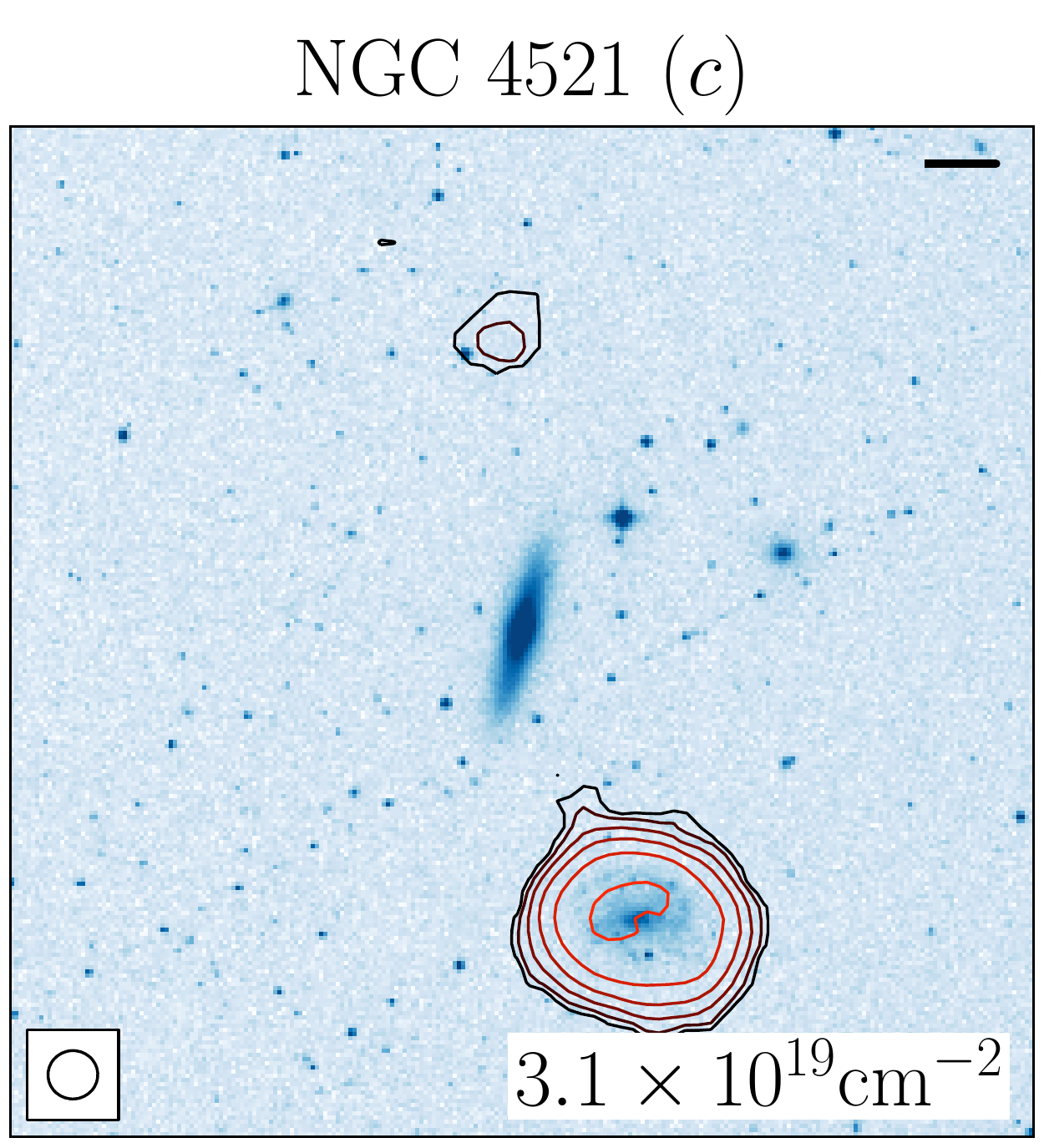}
\caption{A sequence of \hi-rich ETGs illustrating the continuity between class $u$ and class $c$. We refer to the caption of Fig. \ref{fig:gallery} for a description of the content of each image. In this figure the top-right scale bar indicates 10 kpc at the galaxy distance.}
\label{fig:sequenceb}
\end{figure*}

Another step in this continuum of \hi\ morphologies is illustrated in Fig. \ref{fig:sequenceb}. In this figure we show that some the unsettled systems appear as a bridge between class $u$ and class $c$. The sequence starts from an extended, unsettled \hi\ system. The large number of gas-rich neighbours suggests that galaxy interaction may be responsible of the disturbed \hi\ morphology \citep[see also][]{2001ASPC..240..867V}. The second and third galaxies on the sequence are surrounded by a few \hi\ clouds forming a coherent gaseous system -- in both cases the clouds are connected to each other and to the ETG in velocity and are likely to be the densest clumps of an \hi\ tail (for NGC~5557 see Paper IX). These systems represent a natural link to galaxies in group $c$ (represented by the fourth object in the sequence), where scattered clouds are detected around an ETG but their connection to it is much less obvious.

The final piece in this continuum of morphologies is provided by the continuity between $c$ and undetected objects. Many $c$ galaxies are difficult to classify as it is sometimes not clear whether we are seeing the densest gas clumps of a $u$ system, or gas is instead floating unbound in the inter-galactic medium. In fact, a number of ETGs which we classify as non detections are surrounded by \hi\ with properties similar to those of $c$ galaxies. We show these objects in Fig. \ref{fig:gallery_undet_a}, (the figure includes also galaxies close to or interacting with an \hi-rich neighbour, so that confusion prevents us from establishing whether they contain any \hi). This ambiguity has little effect on our discussion in the following sections as the \hi\ mass of $c$ objects is comparable to the detection limit of our observations (see Sec. \ref{sec:detlim}).

\subsection{Small and large \hi\ discs: relation to the kinematics of the host galaxy}
\label{sec:4.3}

\begin{figure*}
\includegraphics[width=33mm]{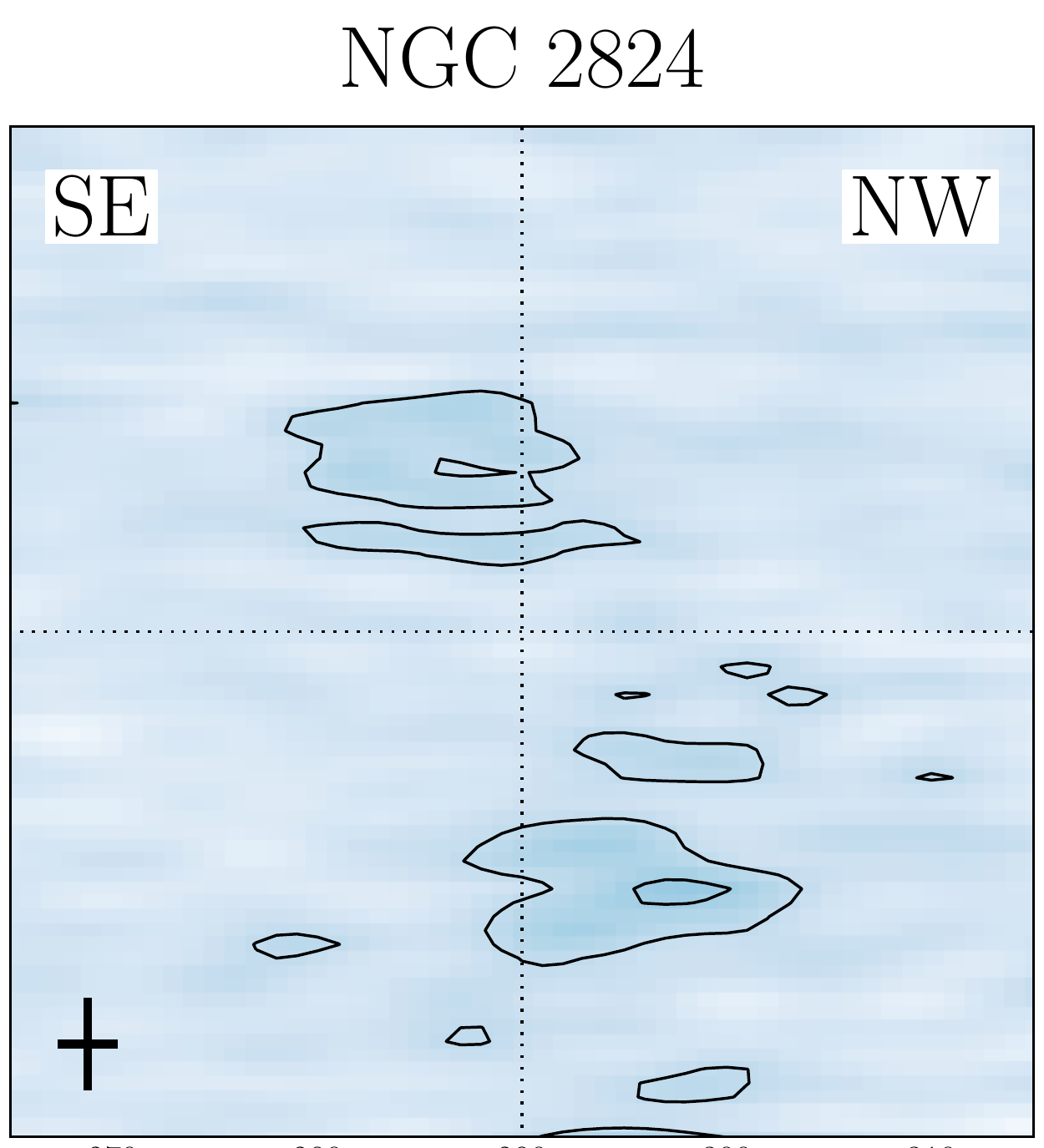}
\includegraphics[width=33mm]{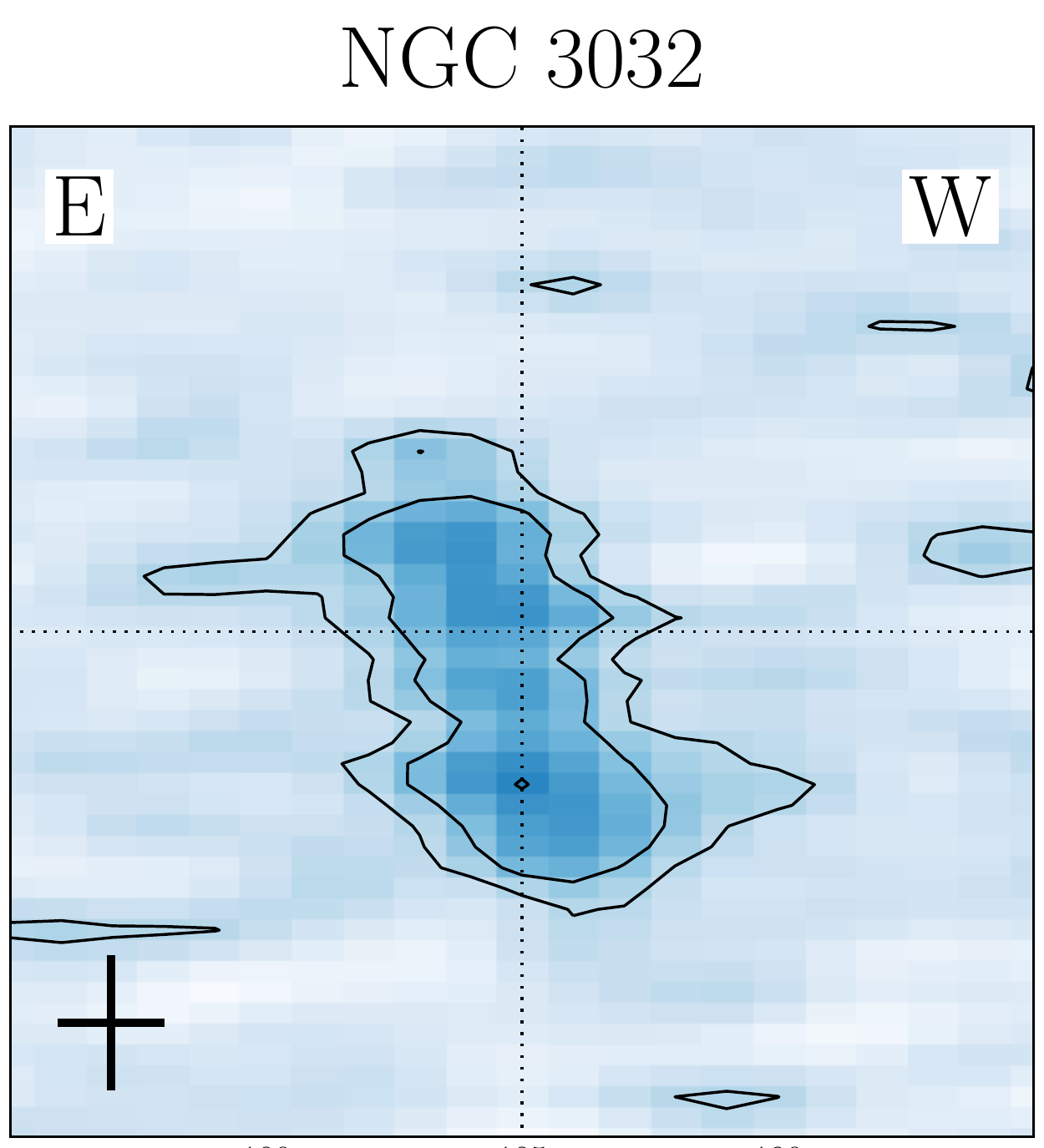}
\includegraphics[width=33mm]{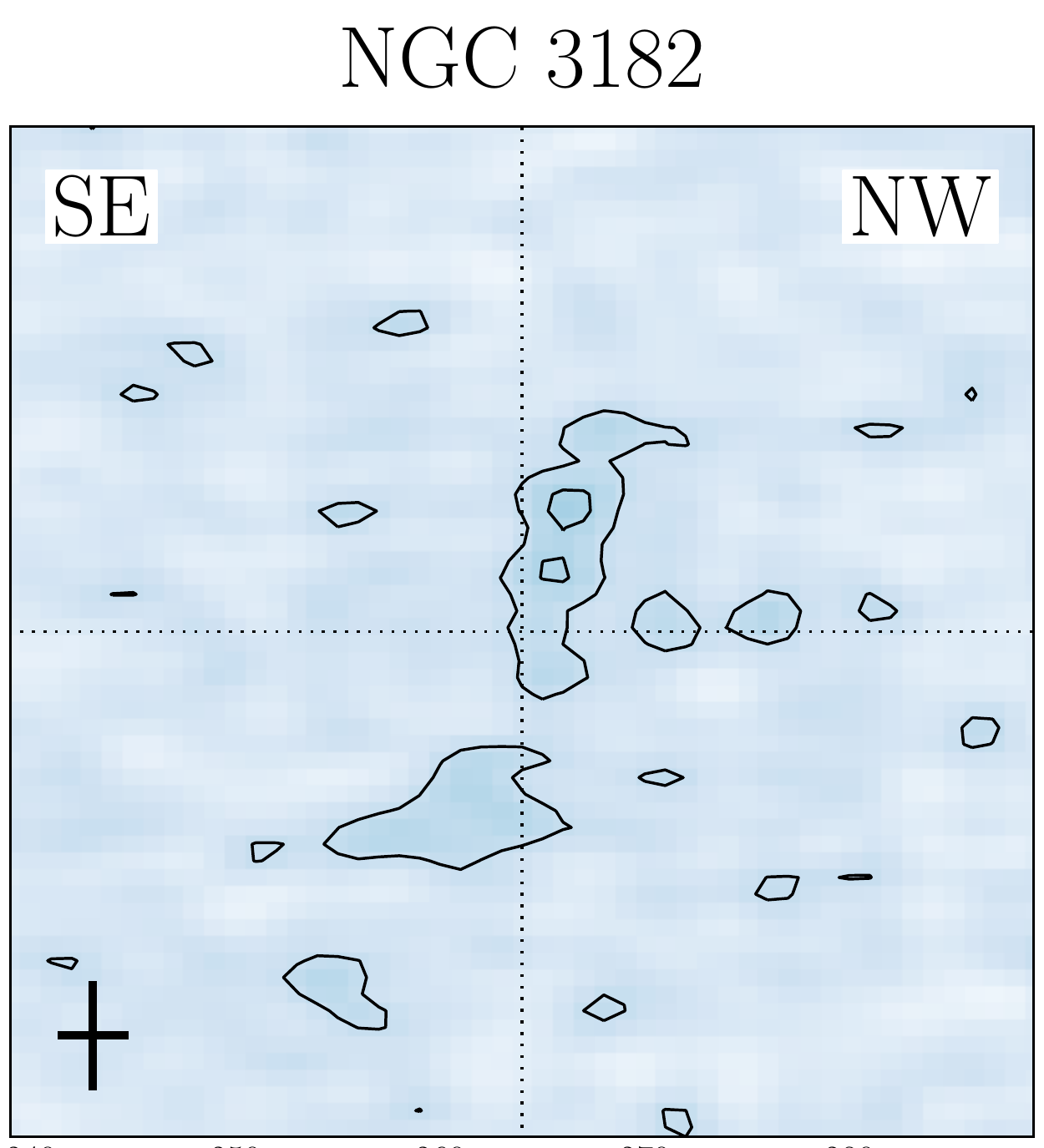}
\includegraphics[width=33mm]{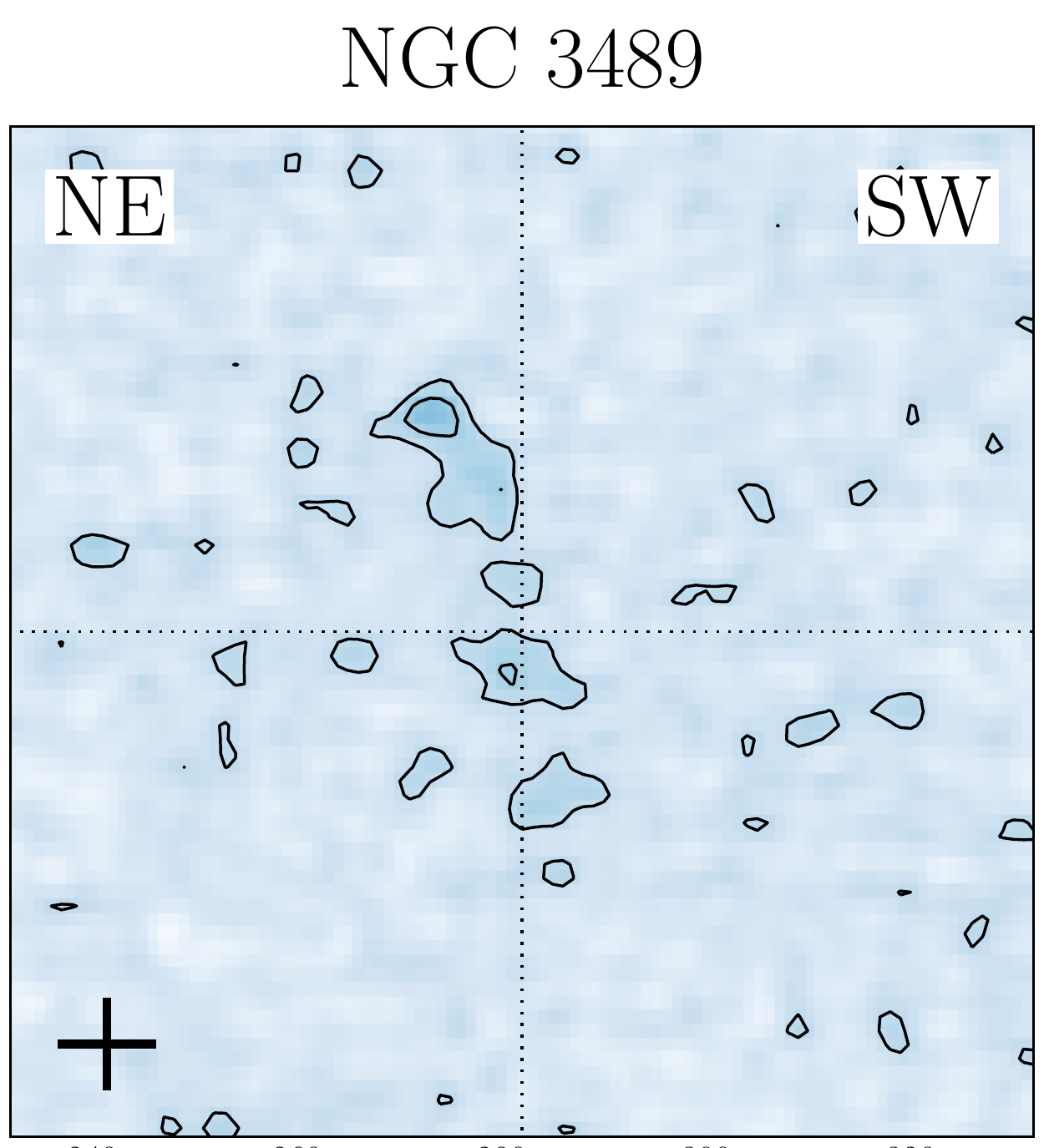}
\includegraphics[width=33mm]{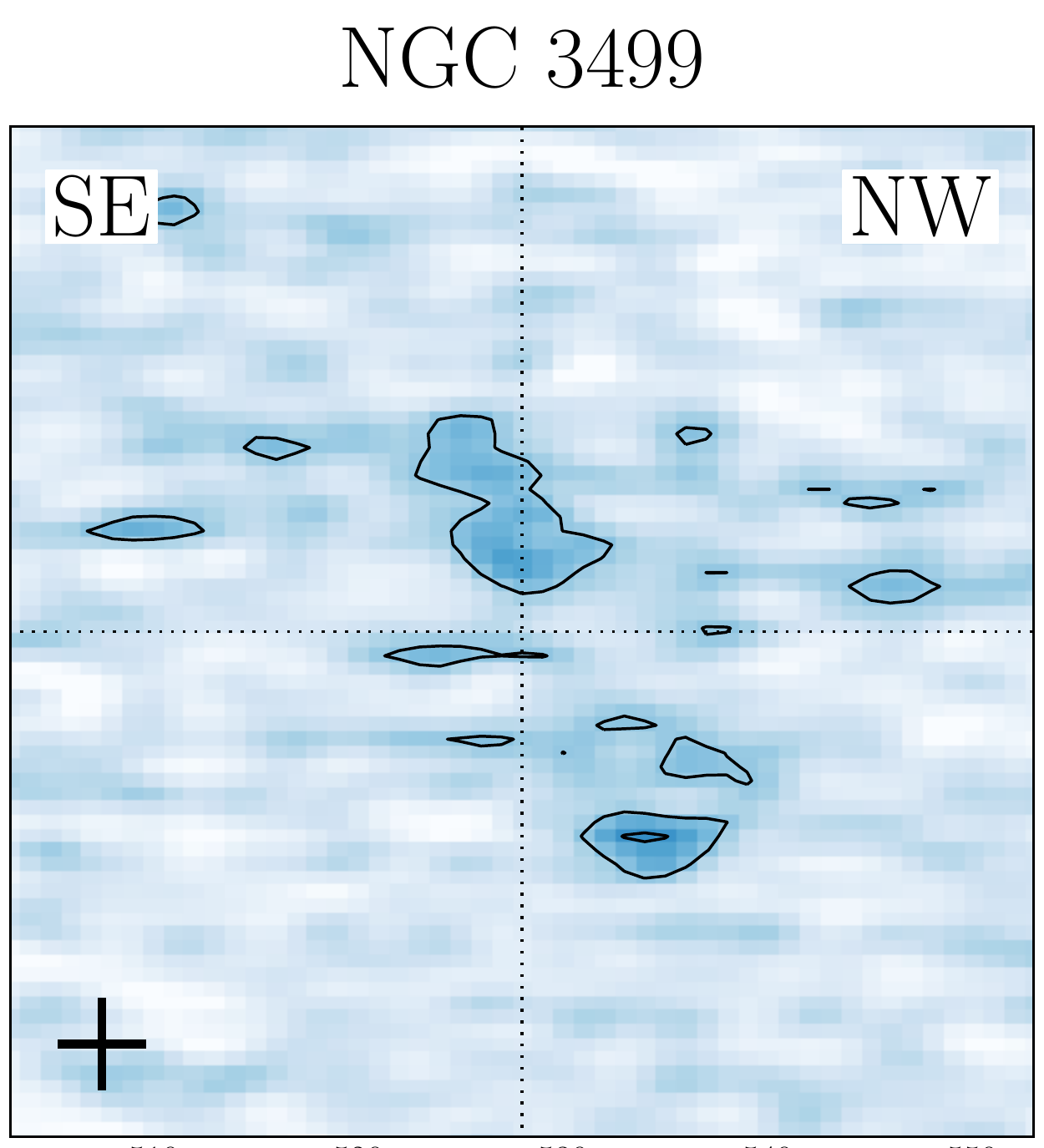}
\includegraphics[width=33mm]{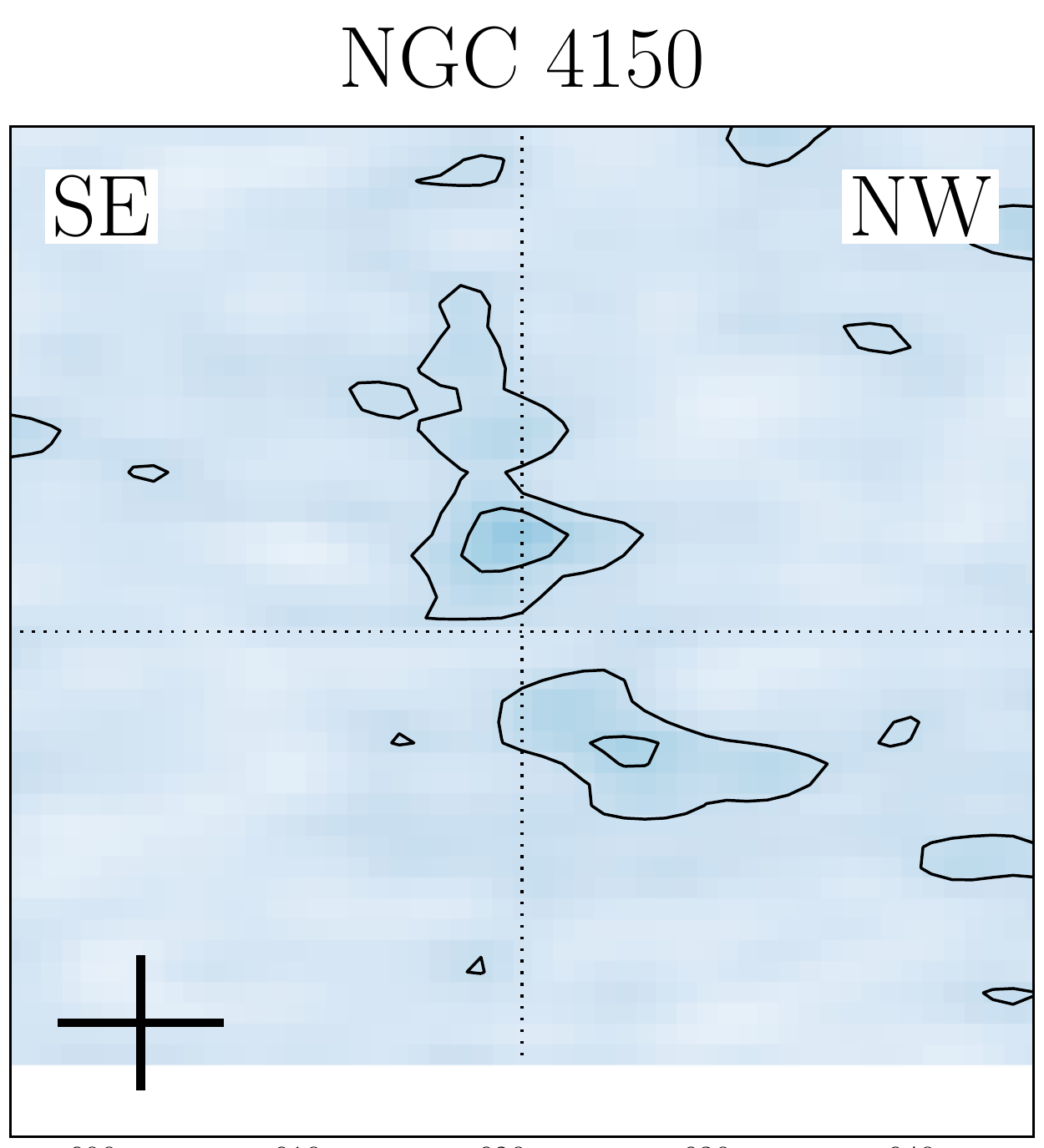}
\includegraphics[width=33mm]{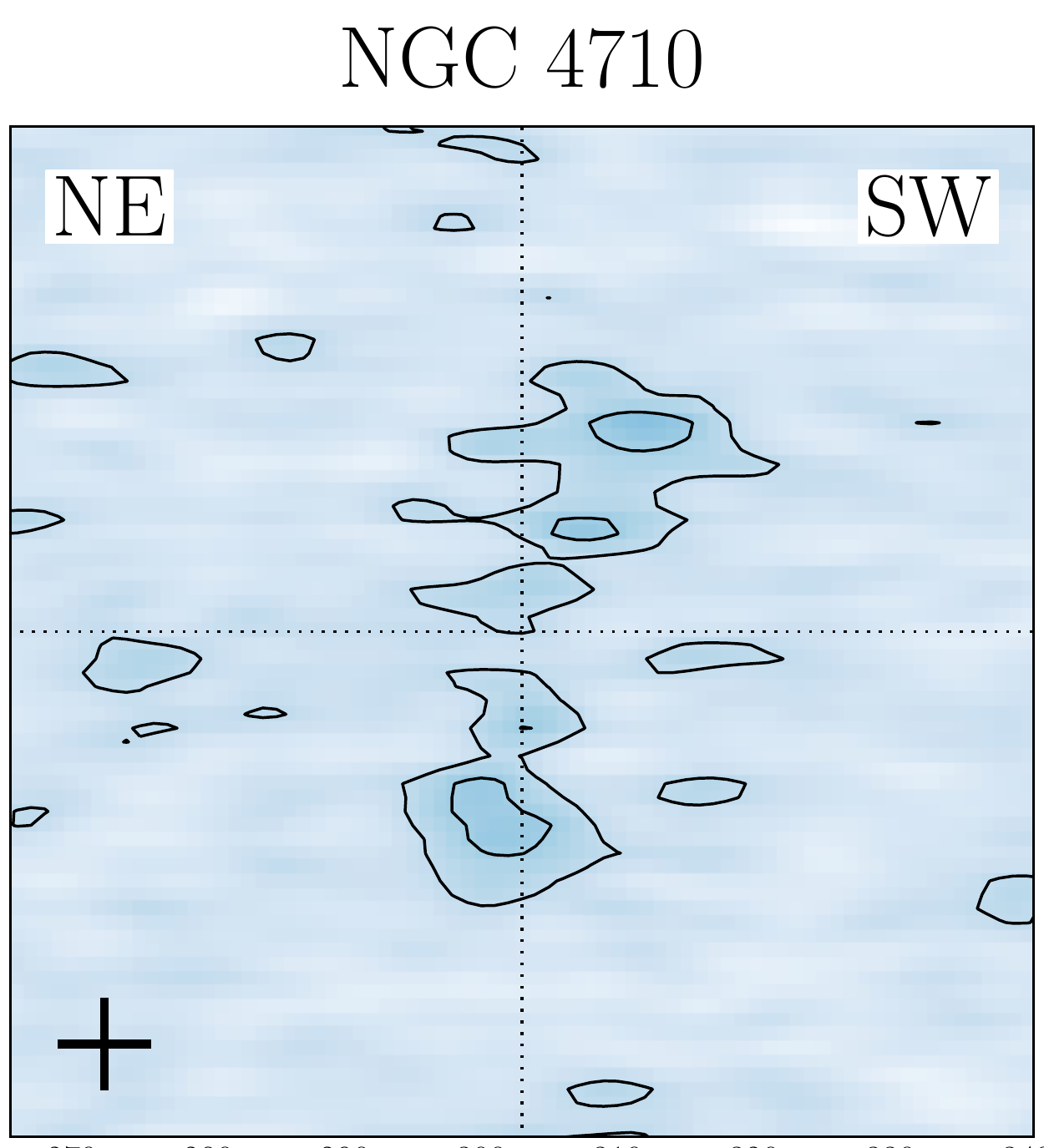}
\includegraphics[width=33mm]{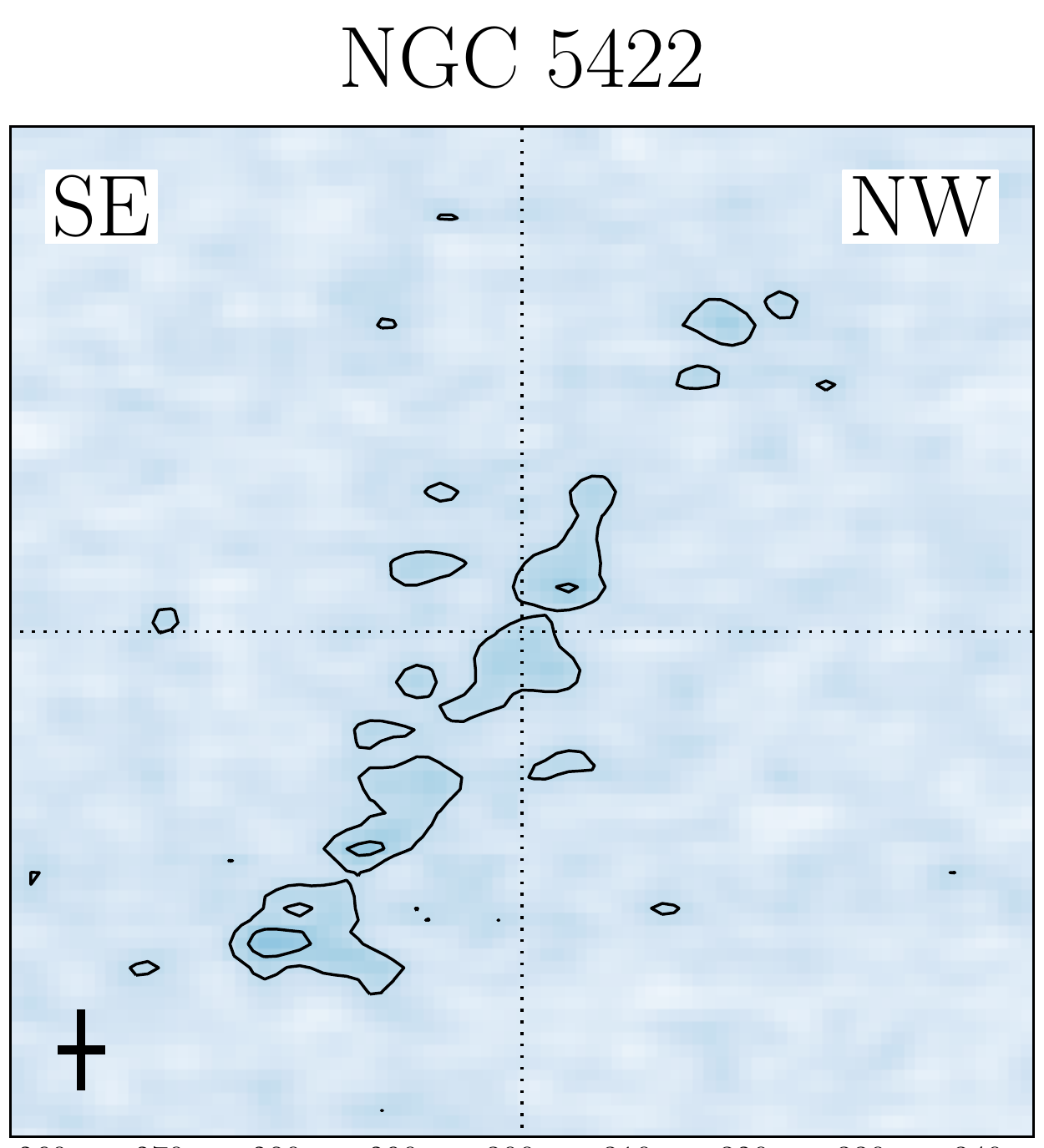}
\includegraphics[width=33mm]{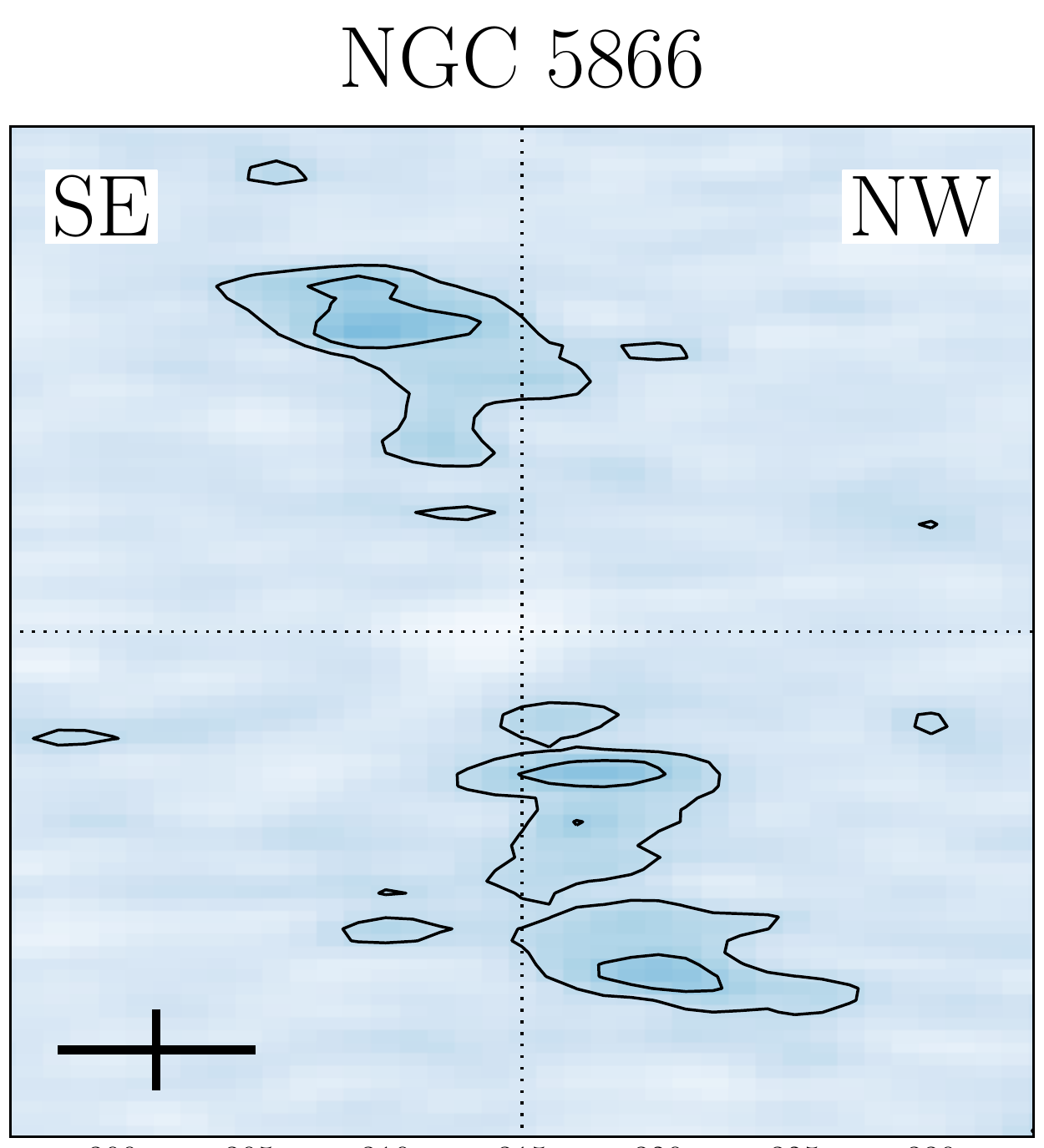}
\includegraphics[width=33mm]{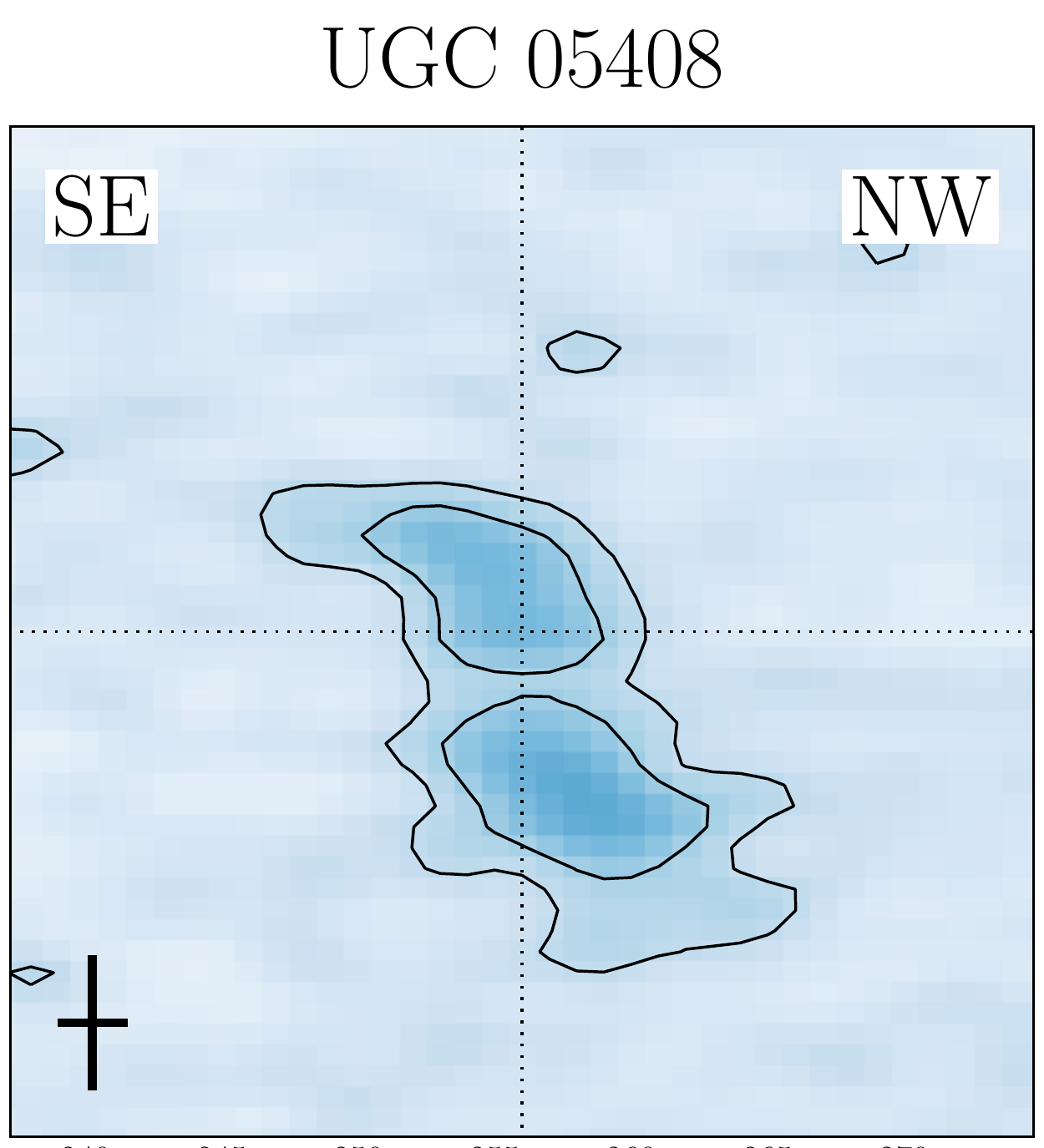}
\caption{Position-velocity diagrams of all $d$ galaxies along the stellar kinematical major axis given in Paper II (perpendicular to it for NGC~3499). In all diagrams the x axis represents angular offset and the y axis velocity offset. Dotted lines indicate the centre of the galaxy. The diagrams are drawn using the $R01$ cubes for all galaxies except NGC~3032. Contours start at 1.0 mJy beam$^{-1}$ for NGC~3032, at 0.6 mJy beam$^{-1}$ for NGC~3182, NGC~3499 and NGC~4150, and at 0.75 mJy beam$^{-1}$ for all other galaxies. Contours increase by a factor of two at each step. The cross on the bottom left indicates 2 kpc along the horizontal axis and 50 km s$^{-1}$ along the vertical axis.}
\label{fig:pv}
\end{figure*}

At the beginning of this section we introduced a criterion to distinguish between large and small rotating \hi\ systems (i.e., $D$ and $d$). We consider this distinction important as there are fundamental differences between these objects (besides their size). As mentioned, $D$'s contain between $10^8$ and $10^{10}$ \msun\ of \hi, while in $d$'s \mhi\ is typically below $10^8$ \msun. Furthermore,  the \hi\ kinematics is aligned with the stellar kinematics in most galaxies belonging to group $d$ -- 8/10 objects. Figure \ref{fig:pv} shows position-velocity diagrams of all $d$'s along the stellar kinematical major axis given in Paper II (perpendicular to it in the case of NGC~3499 -- see below). Comparing these diagrams with the stellar velocity fields published in Paper II we see that the \hi\ rotation is aligned with that of the stars in all galaxies except NGC~3032, where the \hi\ is counter-rotating, and NGC~3499, where the \hi\ is polar. Even in these two cases a connection between \hi\ and stellar body exists. In the former the gas disc is aligned with a kinematically-decoupled stellar component \citep[O10]{2006MNRAS.373..906M}. In the latter the \hi\ is aligned with a faint stellar disc and a dust lane visible in the SDSS image. 

In contrast with this result, visual inspection of the stellar velocity fields (Paper II) and the HI cubes shows that misaligned discs/rings are found in more than half (14/24) of all $D$ galaxies. This includes a number of polar rings and counter-rotating discs/rings (see Fig. \ref{fig:gallery} and column 7 in Table \ref{tab:sample}). From a kinematical point of view, it seems that $d$ \hi\ distributions are tightly coupled to their host galaxy, while in $D$'s the \hi\ kinematics is often different from the stellar one. A future paper will address the relation between \hi\ and stellar kinematics in more details, including the classification of ETGs in the \atlas\ sample in slow- and fast rotators (Paper III). \rm

\subsection{\hi, H$_2$ and star formation signatures}
\label{sec:h2}

As part of their study of 33 ETGs belonging to the SAURON sample, O10 show that all galaxies where \hi\ is detected within $\sim1R_\mathrm{e}$ contain molecular gas in their centre, and that H$_2$ dominates the ISM in this region. On the contrary, the CO detection rate is $\sim20$ percent for ETGs with no central \hi\ detection. Here we revisit their result by comparing the \hi\ properties of ETGs in the \atlashi\ sample to results of Paper II (which provides a catalogue of dust discs, dusty filaments and blue regions in all ETGs in the \atlas\ sample) and Paper IV (where we discuss the \atlas\ CO survey).

We detect dust/blue features and CO in 36 and 38 of the 166 ETGs in the \atlashi\ sample, respectively. The presence of dust/blue features and molecular gas are very tightly related to one another (Paper IV), so that most galaxies with CO also have dust/blue features and vice-versa. With reference to Table \ref{tab:summary}, dust/blue features and CO are present in $43\pm9$ percent of all \hi\ detections. In contrast the dust/blue and CO detection rate is $13\pm3$ percent for the \hi\ non-detections. Therefore, galaxies detected in \hi\ are $\sim3$ times  more likely to contain signatures of star formation than those with no detectable neutral hydrogen.

We make use of the classification presented above to investigate how \hi\ morphology is related to the occurrence of star formation. In the previous section we discuss the kinematical link between \hi\ and stellar body in $d$ ETGs. This result is complemented by the detection of star formation signatures in all these galaxies. As illustrated in Table \ref{tab:summary}, dust/blue features are detected in $10/10=100\pm32$ percent of them\footnote{Note that NGC~4150 is not listed as containing dust/blue features in Paper II because of the poor SDSS image quality. However HST observations by \cite{2011ApJ...727..115C} show that also this galaxy is characterised by a dusty and star-forming core.}. Furthermore, we detect CO in $8/10=80\pm28$ percent of them. The exceptions are NGC~3499 and NGC~5422, for which the upper limit on \mhtwo\ is $\sim5\times10^7$ \msun.

A further indication that stars are being formed efficiently in $d$ galaxies comes from the analysis of the ISM composition in their central region. As done in O10, we compare the H$_2$ mass given in Paper IV (measured within the $\sim$20 arcsec IRAM beam) to the \hi\ mass contained in one WSRT beam (whose minor axis is $\sim30$ arcsec) centred on the galaxy. This comparison shows that in $d$'s the \mhtwo/\mhi\ ratio is high -- always above $\sim1$, and $>10$ in 5 out of 8 galaxies detected in CO. Therefore, the ISM in these galaxies has similar properties as that in the central region of spirals \citep{2008AJ....136.2782L}, confirming that in many of them \hi\ is being turned into H$_2$ efficiently.

Signatures of star formation are less common in $D$ galaxies and depend on whether the \hi\ is distributed on a ring (10 objects -- see Table \ref{tab:summary}) or extends all the way to the centre of the galaxy (14 objects). We detect CO in 9 of the ETGs with central \hi\ ($64\pm21$ percent) and dust/blue features in 8 of them ($57\pm20$ percent). In contrast only 1 galaxy with an \hi\ ring is detected in CO ($10\pm10$ percent) and 2 (including the CO detection) exhibit dust/blue features ($20\pm14$ percent).

The physical state of the ISM in these objects may be different from that in $d$'s as the \mhtwo/\mhi\ ratio is somewhat lower. When CO is detected, we find \mhtwo/\mhi\ values in the range 0.5 - 3 for 80 percent of the $D$'s, and never larger than $\sim10$. When CO is not detected the upper limit on \mhtwo/\mhi\ is always above $\sim1$, so that these systems are not necessarily different from those with a CO detection (they may simply contain too little central \hi\ for molecular gas to be detected). Therefore,  also in $D$'s most of the central ISM is found in the form of molecular gas, but the conversion of \hi\ to H$_2$ may be less efficient than in $d$'s. One possibile explanation is that \hi\ in the centre of these galaxies has lower column density.

A similar result holds for $u$ galaxies. These systems are unsettled so that it is not always easy to judge whether \hi\ is present within the stellar body, or is simply projected onto it on the sky. However, \hi\ is certainly not present in the central region of NGC~3193, NGC~4026 (a ring with a large and massive \hi\ tail), NGC~5198 and NGC~5557. None of these systems hosts CO or dust/blue features. Of the remaining 10 $u$ galaxies, 3 host dust/blue features ($30\pm17$ percent) and 5 host molecular gas ($50\pm22$ percent). As in $D$'s, the typical (upper limit on) \mhtwo/\mhi\ is of the order of unity, and none of these objects has \mhtwo/\mhi$>10$.

\begin{figure*}
\includegraphics[width=18cm]{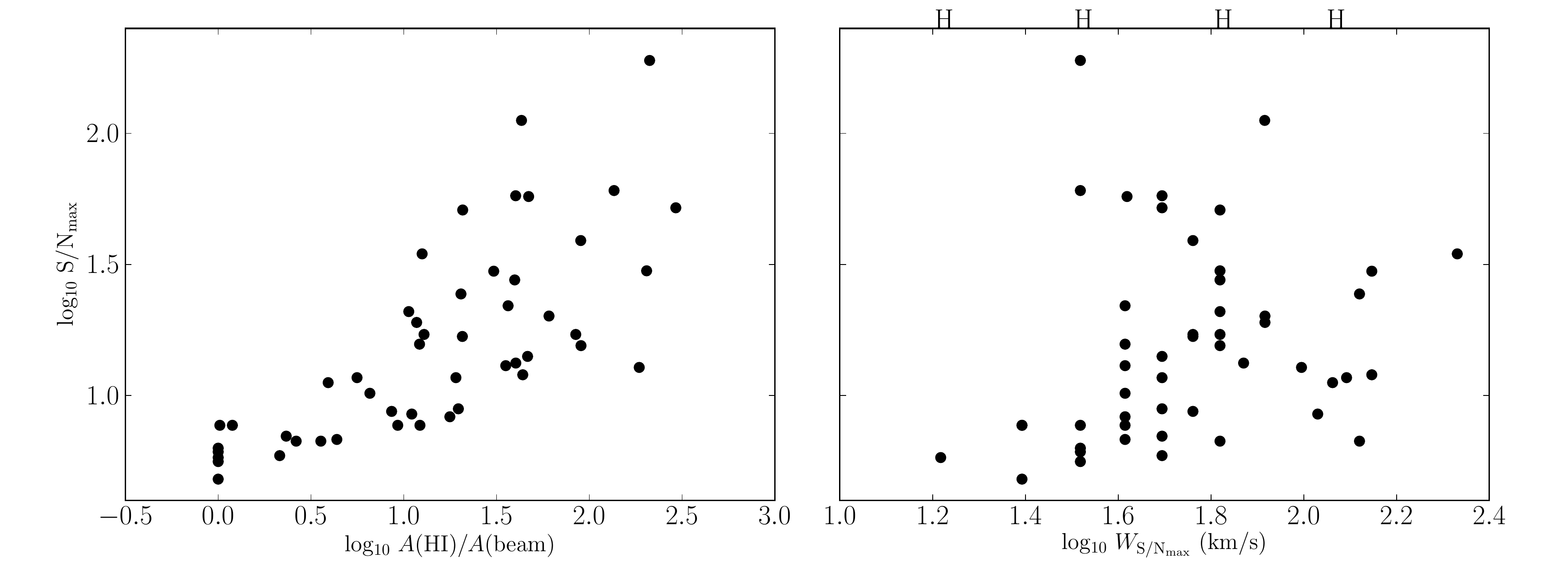}
\caption{Peak signal to noise \sn\ of all galaxies detected in the standard, 12-h \hi\ cubes (see text) plotted against the size of the \hi\ system in units of the beam (left), and the spectral line width at the position of \sn\ (right). The top $H$ symbols represent the $FWHM$ of the Hanning filters used to smooth the \hi\ cubes when searching for emission (see Sec. \ref{sec:data}).}
\label{fig:detlim}
\end{figure*}

Consistent with the above discussion, none of the 5 $c$'s exhibits dust/blue features or is detected in CO. We therefore confirm O10 result that the detection rate of star-formation signatures is high ($\sim2/3$) for ETGs hosting neutral hydrogen within $\sim1R_\mathrm{e}$, so that \hi\ seems to provide material for star formation. In particular, 100 percent of all $d$'s, $\sim60$ percent of all $D$'s with central \hi, and $\sim50$ percent of all $u$'s with central \hi\ exhibit such signatures. At the sensitivity of our data, even galaxies with central \hi\ but no star formation signatures are consistent with the picture that the central ISM of ETGs is dominated by molecular gas -- they may simply contain too little \hi\ for CO to be detected. In contrast with these results, galaxies surrounded by an \hi\ distribution with a central hole, and galaxies with no detectable \hi, show signs of recent star formation in a minority of the cases ($\sim15$ percent).

These trends indicate that \hi\ may play an important role in fuelling rejuvenation and the growth of a stellar disc in ETGs. In particular, in a number of $d$'s (the ETGs with the highest \hi-to-H$_2$ conversion rate in their centre) the \hi\ morphology suggests that recent accretion (possibly from small companions) is the source of the ISM (e.g., NGC~3032, NGC~3489, NGC~4150 -- see O10). Yet, 15/38 CO detections and 13/36 ETGs with dust blue/features are not detected in \hi. The H$_2$ mass of these systems is not particularly low, so they are galaxies with a very high \mhtwo/\mhi\ ratio (lower limits on this are above $\sim5$ in most of these cases). Hosting \hi\ at the present time (and at the sensitivity of our observations) is therefore not a necessary condition for an ETG to host star formation. It is possible that some ETGs have been stripped of most their \hi\ but not their molecular gas if they live in a dense environment with a hot medium (see Sec. \ref{sec:env} and Paper IV). However, more than half of the galaxies with signs of star formation but no detected \hi\ live outside Virgo, so this cannot be the only explanation. These results highlight the complex gas accretion history and interplay between different ISM phases of ETGs.

\subsection{\hi\ in absorption}

We conclude this section by noting that \hi\ gas can also be detected in absorption against a radio continuum central source (1.4 GHz continuum images are produced as part of the pipeline described in Sec. \ref{sec:data}).

For each object, we obtain continuum images using the line-free channels. The typical noise of the images is $\sim60$ $\mu$Jy beam$^{-1}$ and more than 40 percent of the galaxies were detected with radio power in the range $10^{18}$-10$^{22}$ W Hz$^{-1}$, consistent with previous studies of the continuum emission of this kind of objects \citep[e.g.,][]{1989MNRAS.240..591S,2007MNRAS.375..931M}. A complete discussion of these results will be presented in a future paper.

We look for \hi\ absorption in all objects with a continuum flux above 8 mJy. This results in three detections of objects where \hi\ is observed only in absorption (NGC~5322, NGC~5353 and PGC~029321), while in three more objects  absorption is detected in addition to emission (NGC~2824, NGC~3998 and NGC~5866; see Table \ref{tab:sample}). The continuum flux density of objects detected in \hi\ absorption ranges between 8 and $\sim30$ mJy. The \hi\ absorption is relatively narrow, 50 - 80 km/s, and in all cases it is centred on the systemic velocity. The absorbed flux ranges between 3 and 4 mJy. We derive optical depths of 7 - 10 percent and column densities of $4 - 7 \times 10^{20}$ cm$^{-2}$ (assuming T$_{\rm spin}$ = 100 K).

This absorption may be the result of small disc structures (i.e. discs that are not large enough to be seen in emission if a relatively strong continuum source is present). For example, two of the three galaxies detected only in absorption exhibit dust or blue features within the stellar body, consistent with the result described above for $d$ galaxies. Only one of them is detected in CO (PGC~029321).

It is worth noting that only 19 objects among the $\sim140$ observed with the WSRT have radio continuum  flux density  $S_{1.4GHz} > 8$ mJy. The detection rate of \hi\ in absorption is therefore $\sim30$ percent for galaxies with sufficient continuum emission. This underlines that the \hi\ detection rate of our study (in emission and absorption) is just a lower limit to the fraction of ETGs containing neutral hydrogen gas.

\section{\mhi\ detection limits}
\label{sec:detlim}

The analysis presented in the following sections is based on the images in Fig. \ref{fig:gallery}, the classification discussed above, the \hi\ mass measured from these images and listed in Table \ref{tab:sample}, as well as on \mhi\ upper limits  for the undetected galaxies. In this section we discuss how we calculate these upper limits.

The data cubes described in Sec. \ref{sec:data} give the \hi\ surface brightness distribution projected on the sky per unit velocity interval along the line of sight. The detection of \hi\ in a galaxy is therefore driven by its \it peak surface brightness \rm relative to the noise level at the angular and velocity resolution of the data, and not by the total mass of \hi\ hosted by the galaxy. This makes the calculation of upper limits \mhilim\ for undetected galaxies rather delicate.

The typical approach in the literature is to calculate \mhilim\ as the minimum detectable \hi\ mass within one angular-resolution element centred on the galaxy, and integrating over a velocity interval representative of the gas line-width. For example, if we calculate \mhilim\ as a 3$\sigma$ signal within one synthesised beam and within a 200 km/s velocity interval, we obtain \mhilim$=2.0 \cdot (d/\mathrm{10\ Mpc})^2 \times 10^6 $ \msun. However, a larger \hi\ mass can be hidden below the noise if it is spread over many beams. The first question we wish to answer is therefore whether we are missing a population of extended, massive \hi\ distributions with column density below the noise of our observations.

To find an answer we calculate the maximum signal-to-noise ratio \sn\ of the \hi\ emission for galaxies detected in the standard cubes\footnote{We therefore leave out NGC~3499, NGC~4150 and NGC~7332, which are detected only in the $R01$ cubes, and NGC~4406, for which only a total \hi\ image (and not the cube, necessary for this analysis) is available. We note that for some galaxies the available cubes are obtained after multiple 12-h integrations (see Table \ref{tab:sample}). For consistency, we re-analyse a single 12-h integration for these galaxies, and use the result in this section.}. This calculation is made including data-cubes at all angular and velocity resolutions listed in Sec 3 (i.e., all possible combinations of the velocity resolutions 16, 32, 64 and 112 km/s, and of the angular resolutions $\sim30$ and 60 arcsec). The value of \sn\ indicates how easily a galaxy is detected above the noise.

The left panel in Fig. \ref{fig:detlim} shows \sn\ plotted versus the area $A$(\hi) occupied by the detected gas in units of the beam area $A$(beam). This plot shows a trend of increasing \sn\ with increasing $A$(\hi)/$A$(beam). In particular, the majority of the detections have very high \sn\ (above 10), and this is true for nearly all detections larger than 10 beams. There is no hint of a population of very extended \hi\ systems close to the noise of our observations.

\begin{figure}
\includegraphics[width=8.5cm]{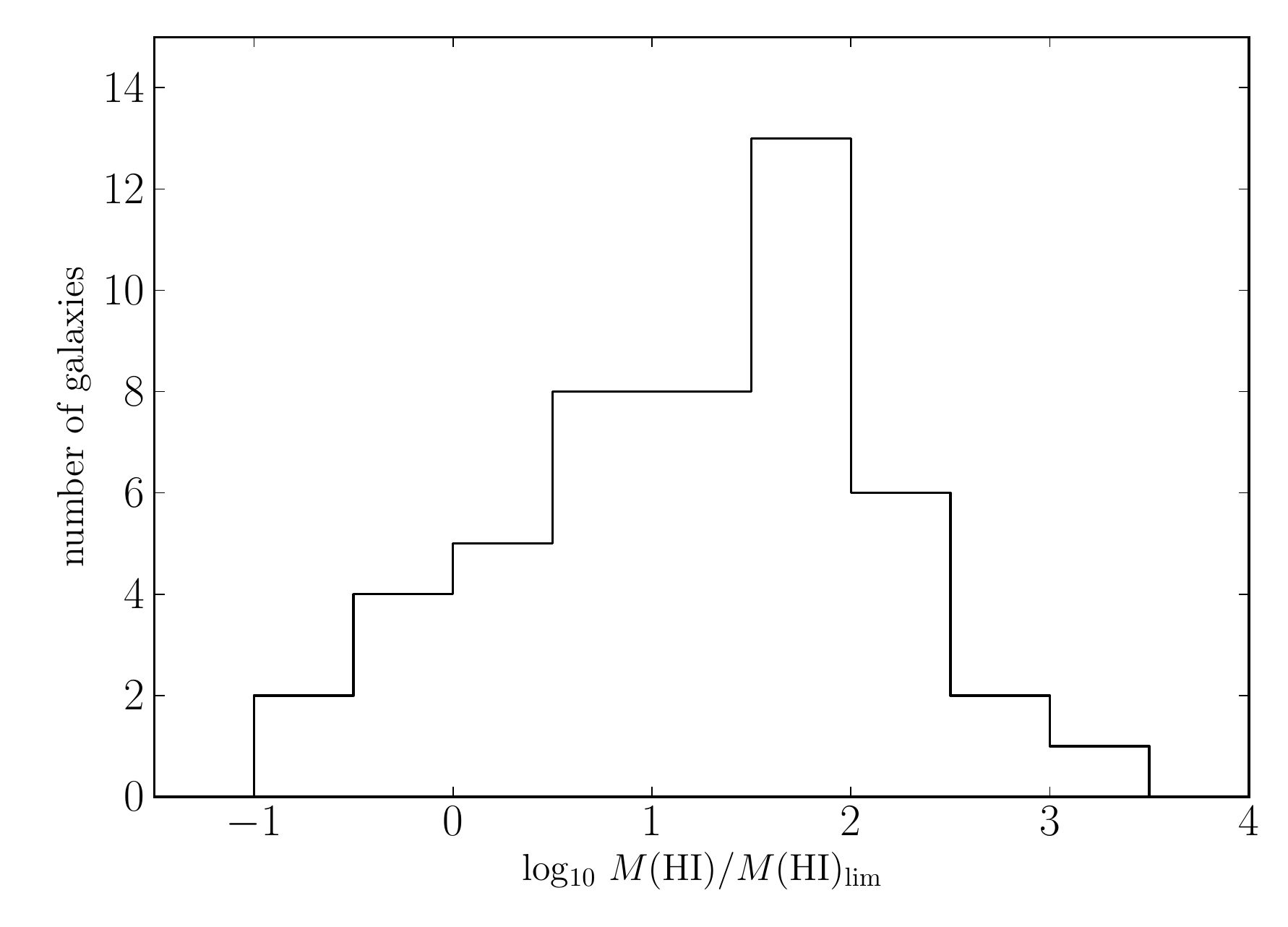}
\caption{Distribution of the \mhi/\mhilim\ ratio for galaxies detected in the standard, 12-h cubes (see text).}
\label{fig:detlim_fhi}
\end{figure}

The second question we wish to answer is what angular size and line-width we should assume to calculate \mhilim\ for the undetected galaxies. To find an answer we study the distribution of $A$(\hi)/$A$(beam) and of the line-width $W$ of low-\sn\ galaxies (\sn\ $<10$). The typical low-\sn\ detection is rather small on the sky. For these systems, the mode of the $A$(\hi)/$A$(beam) distribution is 1, and 2/3 of them have an area smaller than 6 beams, which we adopt as angular size for the calculation of upper limits.

We define $W$ as the line width including 80 percent of the detected emission at the sky coordinates of \sn. The right panel in Fig. \ref{fig:detlim} shows that \sn\ has little relation to $W$. The top $H$ symbols represent the \it FWHM \rm of the Hannning filters used to smooth the \hi\ cubes in the spectral direction. They demonstrate that we have properly sampled the $W$ range covered by the data. We find that median and mode of the overall $W$ distribution are $\sim 50$ km/s. Three quarters of all low-\sn\ galaxies have $W$ below this value, which we adopt as line-width for the calculation of upper limits.

Based on these results, we define \mhilim\ as a 3$\sigma$ signal obtained integrating over 6 beams and 50 km/s. For the typical noise of 0.5 mJy/beam at a 16 km/s velocity resolution, this corresponds to \mhilim$=2.4 \cdot (d/\mathrm{10\ Mpc})^2 \times 10^6 $ \msun. To verify that this is a reasonable upper limit we show in Fig. \ref{fig:detlim_fhi} the histogram of the \mhi/\mhilim\ ratio for the detected galaxies. For consistency, we make use of the \mhi\ value measured from the standard, 12 h-integration cube for all galaxies (this is lower than the \mhi\ value reported in Table \ref{tab:sample} for galaxies were additional \hi\ emission is detected in $R01$ or deep cubes).  Only 6 of the 49 galaxies used for this analysis have \mhi\ below \mhilim\ (NGC~2824, NGC~3182, NGC~3608, NGC~4710, NGC~5422, NGC~5866). This confirms the validity of our definition of upper limits.

As mentioned in Sec. \ref{sec:data}, we use ALFALFA spectra to determine \mhilim\ for about half of all Virgo ETGs. Upper limits from ALFALFA data are calculated on one resolution element. The ALFALFA beam has a FWHM of $\sim3.5$ arcmin. Therefore, its area is larger than the 6 WSRT beams over which we calculate upper limits for the WSRT data. As above, we assume a line-width of 50 km/s. ALFALFA spectra have a typical noise of $\sim3$ mJy/beam at 11 km/s resolution, so that \mhilim$=5.0 \cdot (d/\mathrm{10\ Mpc})^2 \times 10^6 $ \msun. The noise value of individual spectra was kindly provided by Riccardo Giovanelli.

\begin{figure}
\includegraphics[width=8.5cm]{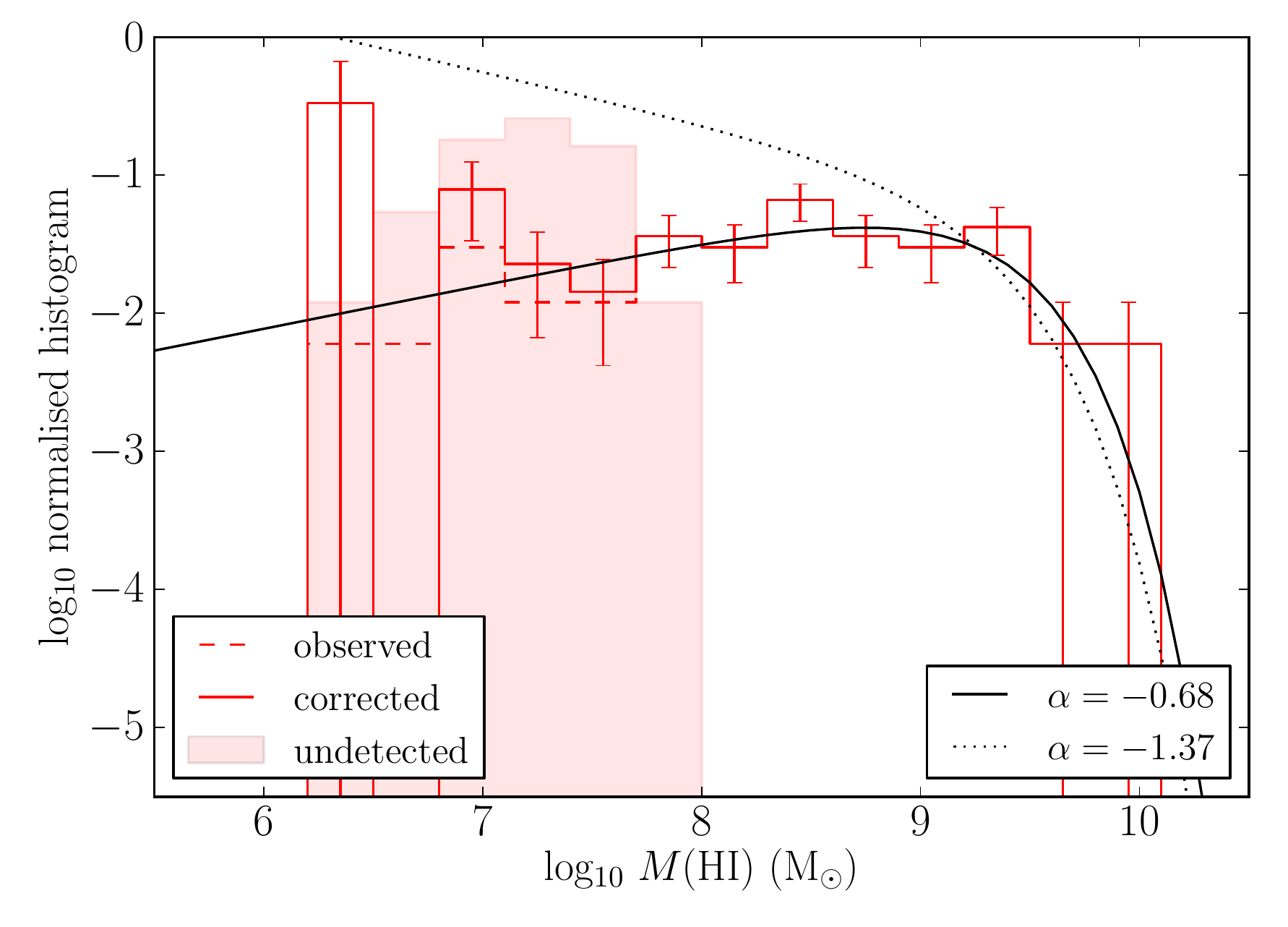}
\caption{Normalised histogram of \mhi\ for ETGs in the \atlas\ parent sample. The dashed red line refers to \hi-detected ETGs. The solid red line is the corrected histogram (see text). Error bars are obtained assuming a binomial distribution. The shaded histogram represents the distribution of upper limits for undetected galaxies. The solid  black line represents the best-fitting Schechter function. The dotted black line represents a Schechter function which is equal to the best-fitting one except for the slope $\alpha$, fixed to the value obtained by \protect \cite{2005MNRAS.359L..30Z}.}
\label{fig:schechter}
\end{figure}

\section{\hi\ mass function}
\label{sec:mf}

The large size of the \atlashi\ sample and the good sensitivity of our WSRT observations result in a relatively large number of \hi\ detections, making it possible to study the \hi\ mass function of ETGs. Figure \ref{fig:schechter} shows the distribution of \mhi\ for ETGs in the \atlashi\ sample. The red dashed line shows the observed distribution of \hi-detected ETGs. The shaded histogram shows the distribution of upper limits for undetected galaxies. Both distributions are normalised by the total number of galaxies in the sample.

We correct the histogram of detections for the incompleteness of our observations at low \mhi. We do this by normalising the number of galaxies detected in each bin by the number of galaxies $detectable$ in that bin (i.e., the number of galaxies with \mhilim\ below the upper end of the bin). This is similar to the $V/V_\mathrm{max}$ correction typically done in large flux-limited surveys \citep{1968ApJ...151..393S}. We show the corrected histogram with a red solid line in the figure. Error bars in the figure assume a binomial distribution.

The correction can be applied only to a sample of galaxies each with \mhi$\geq$\mhilim. Three \hi-detected ETGs do not meet this criterion (NGC~3182, NGC~3499 and NGC~7332). The corrected histogram in Fig. \ref{fig:schechter} are drawn ignoring these galaxies (i.e., considering them effectively as undetected).

The corrected \mhi\ distribution of ETGs is flat between a few times $10^8$ and $\sim3\times10^9$ \msun, and declines rapidly above the latter value. At lower masses it exhibits a slight decline all the way to \mhi$\sim10^7$ \msun, where it may start rising again (but uncertainties in this mass range are large). We quantify this behaviour by fitting a Schechter function to the corrected \mhi\ distribution \citep{1976ApJ...203..297S}:

\begin{equation}
dn \propto \left(M/M^*\right)^{1+\alpha} e^{-M/M^*} d \log_{10}M,
\end{equation}

\noindent where $dn$ is the number of galaxies within a logarithmic bin of width $d\log_{10}M$. The solid black line in the figure represents the best fit, which has parameters $M^*= \left( 1.8\pm 0.7 \right) \times 10^9$ \msun\ and $\alpha=-0.68\pm0.16$.

The above mass function is derived neglecting \hi\ non-detections. Figure \ref{fig:schechter} shows that this may lead us to underestimate the mass function (and overestimate $\alpha$) below \mhi\ $\sim5\times10^7$ \msun. We therefore repeat the fit using only the distribution above this mass. The result is $\alpha=-0.79\pm0.23$, consistent with the fit on the full \mhi\ range. We note that the shape of the ETG \hi\ mass function is not driven by environmental effects. The same shape is recovered when excluding the (mostly undetected) galaxies living inside Virgo.

The value of $M^*$ for ETGs is $\sim5$ times lower than the value typically found in large, blind \hi\ surveys \citep[e.g.,][]{2003AJ....125.2842Z,2005ApJ...621..215S,2005MNRAS.359L..30Z,2010ApJ...723.1359M}. Both our and these previous surveys are complete at such high \mhi\ levels, so that comparing the $M^*$ values is correct. This confirms that ETGs contain smaller amounts of \hi\ than other galaxies.

A comparison between the low-mass end slope $\alpha$ of the ETG \hi\ mass function and that found by previous authors for galaxies of different morphology is complicated because of the different way samples are selected. Our sample is magnitude-limited while previous studies analyse samples of \hi-detected galaxies with no selection on galaxy luminosity. This means that previous studies include fainter galaxies than we do. Such galaxies contribute to the low-mass end of the \hi\ mass function, causing $\alpha$ to decrease. Indeed, they find $\alpha$ values of $-1.2$ to $-1.4$, i.e., the mass function keeps rising with decreasing \mhi\ below $M^*$, unlike in our ETG sample (see dotted line in Fig. \ref{fig:schechter}). We present a comparison between ETGs and spiral galaxies with the same luminosity selection in the next section. \rm

\section{Comparison to spiral galaxies}
\label{sec:spir}

\begin{figure*}
\includegraphics[width=18cm]{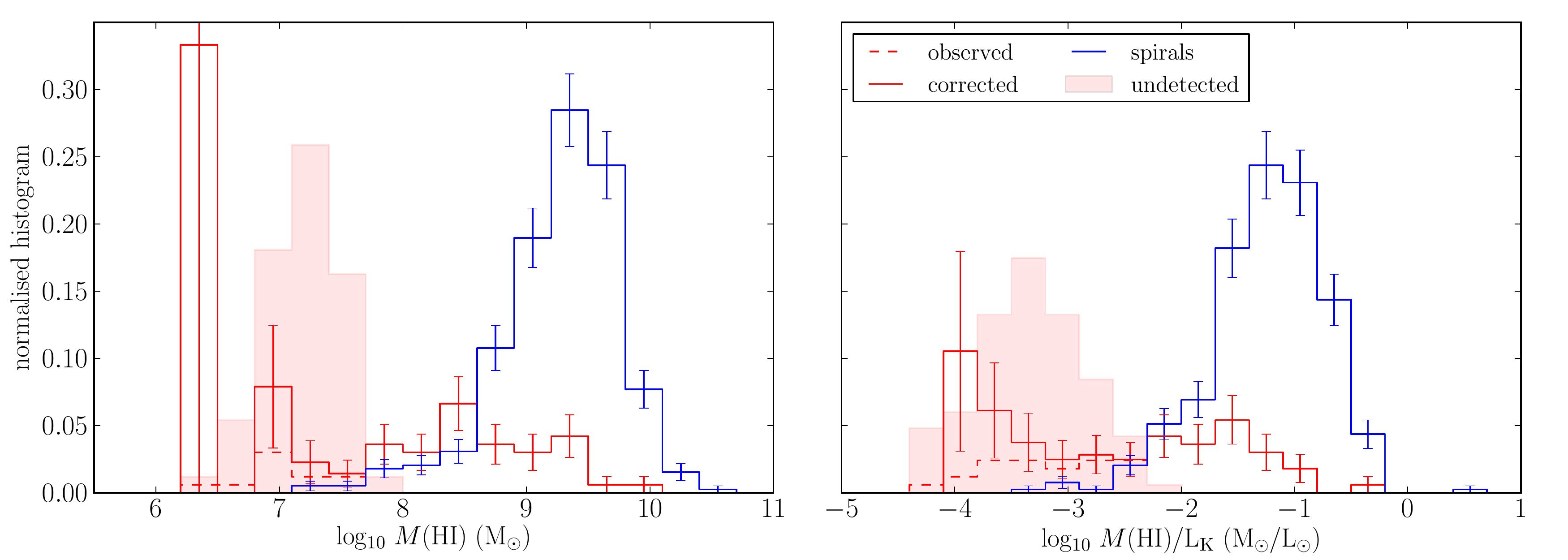}
\caption{Normalised histogram of \mhi\ (left) and \mhil\ (right) for ETGs and spirals in the \atlas\ parent sample. The dahsed red line refers to \hi-detected ETGs. The solid red line is the corrected histogram (see Sec. \ref{sec:mf}). The shaded histogram represents the distribution of upper limits for undetected galaxies. The solid blue line refers to spirals with available \hi\ flux in HyperLeda. Error bars are obtained assuming a binomial distribution.}
\label{fig:KMmhi0}
\end{figure*}

In Fig. \ref{fig:KMmhi0} we compare the \mhi\ and \mhil\ distribution of ETGs to those of spiral galaxies. Red histograms refer to ETGs in the \atlashi\ sample -- solid line, dashed line and shaded histogram have the same meaning as in Fig. \ref{fig:schechter}. The blue solid line represents the distribution of spirals belonging to the \atlas\ parent sample, with the additional selection criterion $\delta \geq 10$ deg as for the ETGs studied here (Sec. \ref{sec:sample}). Spirals in the \atlas\ parent sample have the same \mk\ selection as \atlas\ ETGs.

For spiral galaxies the value of \mhi\ is derived from the \hi\ flux available in HyperLeda \citep{2003A&A...412...57P}, assuming the distance given in Paper I. The \hi\ flux is available for $390/418=93$ percent of all spirals. From HyperLeda, it is not possible to know whether the remaining 28 spirals are undetected or unobserved in \hi. However, about 80 percent of them are fainter than \mk$=-22.5$, so that they are likely to contribute to the low-\mhi\ end of the spiral distribution. Furthermore, $\sim2/3$ of them have distance larger than $\sim30$ Mpc, which makes them difficult to detect depending on the depth of the data and may explain why they are missing from the HyperLeda database.  This small fraction of missing galaxies and the relatively high uncertainty on HyperLeda \hi\ fluxes are not particularly important for the type of comparison presented in this section.

As discussed in Sec. \ref{sec:mf}, the \mhi\ distribution of ETGs is very broad between (at least) a few times $10^7$ and a few times $10^9$ \msun, while at lower \hi\ masses it is uncertain because of many upper limits. Spirals in the \atlas\ parent sample look very different, with a much narrower distribution peaking at \mhi$\sim2\times10^9$ \msun\ and a tail towards \hi\ masses below a few times $10^8$ \msun. The number of galaxies in this tail might be slightly higher if most spiral galaxies without available \hi\ flux have small \mhi\ (as argued above). However, this will not be a very large effect, and Fig. \ref{fig:KMmhi0} shows that the relative number of galaxies with \mhi\ below a few times 10$^8$ \msun\ is larger in ETGs than in spirals when the same \mk\ selection is applied. Therefore, for \mk$<-21.5$ the slope $\alpha$ of the spiral \hi\ mass function might be much smaller than the value of $\sim-0.7$ derived for ETGs in the previous section. Overall, 86 percent of all ETGs and 87 percent of all spirals have \mhi\ below and above $5\times10^8$ \msun, respectively. \rm

The \mhil\ ratio follows a peaked distribution in spirals owing to the known correlation between \mhi\ and galaxy luminosity \citep{1978A&A....68..321S}. Such correlation does not hold for ETGs, so that their \mhil\ distribution is much broader. In fact, it is approximately flat between \mhil$=10^{-3}$ and $10^{-1}$ \mlsun\ and drops to zero above this value, where the distribution of spirals peaks. We find that 85 percent of all ETGs and 92 percent of all spirals have \mhil\ below and above $10^{-2}$ \mlsun, respectively.
 
Early-type galaxies and spirals are distributed very differently in Fig. \ref{fig:KMmhi0}, demonstrating that ETGs, as a family, contain much less neutral hydrogen gas than spirals (a similar indication comes from the low $M^*$ value of the best-fitting Schechter function described in the previous section). However, their distributions overlap significantly in the range  \mhi=$5\times10^7$ - $5\times10^{9}$ \msun\ and \mhil=$3\times10^{-3}$ - $10^{-1}$ \mlsun. \it A significant fraction of all ETGs contain as much \hi\ as spiral galaxies\rm.

This is in agreement with results of \cite{2010MNRAS.403..683C} as part of the GASS survey of massive galaxies. They show a weak relation between \mhi/$M_\mathrm{star}$ and the concentration index measured from galaxies' optical images (their Fig. 8). This index is proxy for the bulge-to-disc ratio of a galaxy \citep[e.g.,][]{2009MNRAS.394.1213W}, so that it is often used to select samples of early- or late-type systems. Adopting standard criteria for this selection, \cite{2010MNRAS.403..683C} results show that spiral galaxies are on average \hi-richer than ETGs, and that most \hi\ non-detections are ETGs for \mhilim/$M_\mathrm{star}$ of a few percent. However, consistent with our result, they also show that there is substantial overlap between the gas content of early- and late-type galaxies.

Another element of similarity between spirals and ETGs with \hi\ is that the majority of all \hi-rich ETGs show disc-like \hi\ morphology (64 percent over the entire sample and 80 percent above \mhi=$5\times10^8$ \msun). Therefore, a significant population of ETGs exists which has similar \mhi, \mhil\ and \hi\ morphology as spiral galaxies. What is then the difference between the \hi\ properties of these two types of galaxies?

\begin{figure}
\includegraphics[width=8.5cm]{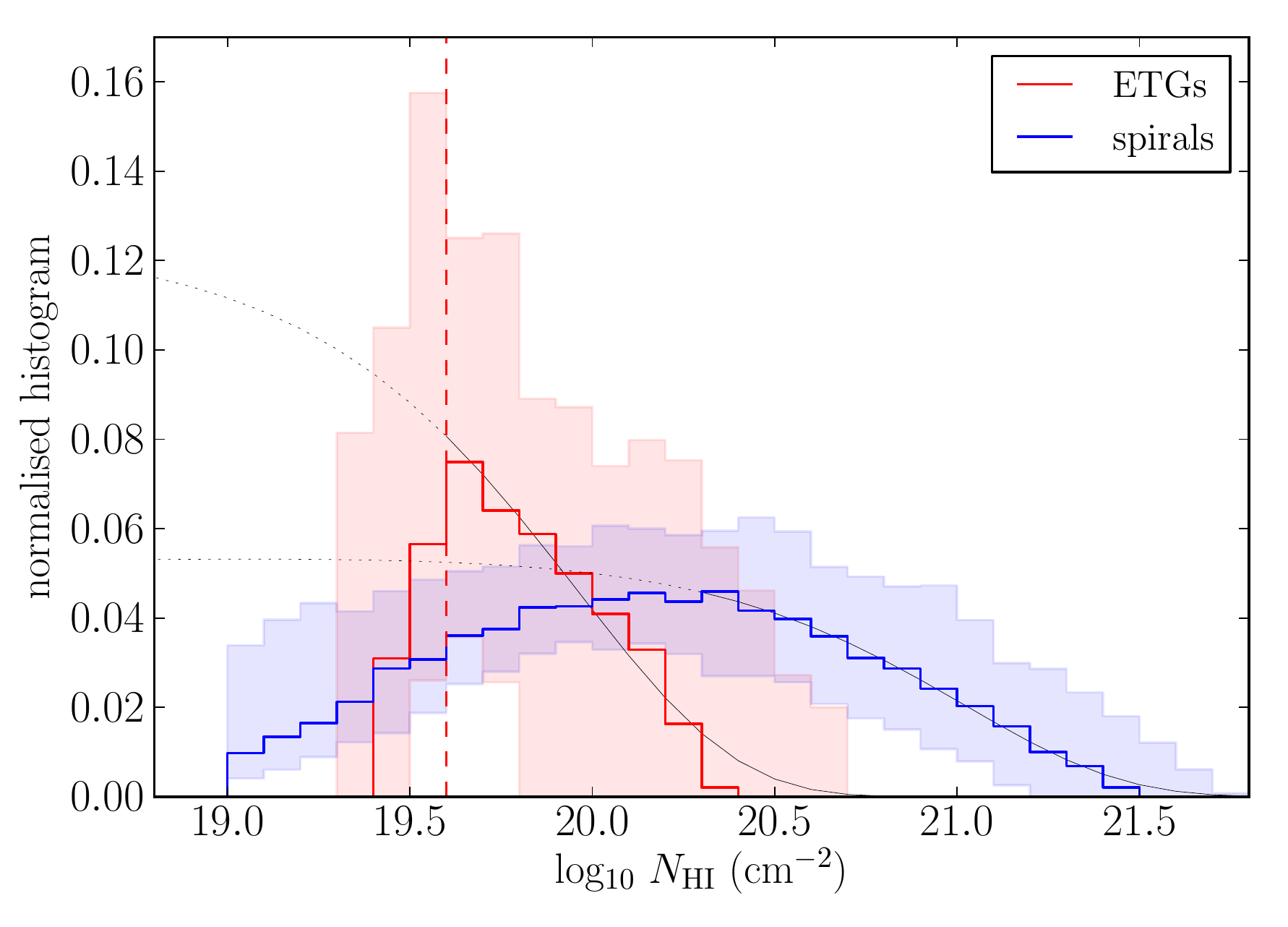}
\caption{Distribution of \hi\ column density \Nhi. The solid red line shows the median distribution for detected ETGs. The solid blue line shows the median distribution for spirals in the WHISP sample (see text). The shaded area corresponds to the 16th and 84th percentile (equivalent to $\pm1\sigma$ for a Gaussian distribution). Thin black lines represent the best-fitting Schechter function for ETGs and spirals, with a dotted line showing the extrapolation of the best fit to column density below the sensitivity of the observations. The dashed red line represents the column density sensitivity of the shallowest total \hi\ images presented in Fig. \ref{fig:gallery}.}
\label{fig:nhi}
\end{figure}

To investigate this we compare the distribution of \hi\ column density \Nhi\ in detected ETGs and spirals. For the latter we use total-\hi\ images constructed from WSRT datacubes as part of the WHISP survey \citep{2002ASPC..276...84V}. We make this comparison as fair as possible by: \it(i)\rm\ using WHISP images at a 30 arcsec angular resolution, similar to that of our ETG \hi\ images, and \it(ii)\rm\ studying only WHISP galaxies which either belong to the \atlas\ parent sample (54 galaxies), or fall within the same recessional velocity and $M_\mathrm{B}$ range as spirals in the \atlas\ parent sample but are outside the sky area covered by it (29 galaxies).

We build the normalised \Nhi\ histogram for each detected ETG and spiral separately. To do so we use for each galaxy all pixels with detected \hi\ emission in the total \hi\ image. Figure \ref{fig:nhi} shows the median of all ETG histograms (red line) and of all spiral histograms (blue line). Both distributions increase with decreasing \Nhi\ down to the sensitivity limit of the observations. WHISP data cover a wide range in sensitivity and are on average a factor of $\sim5$ shallower than our observations of ETGs. The relative normalisation of ETG and spiral \Nhi\ distributions is therefore uncertain.

We choose to normalise the histograms in the figure to the integral of the respective best-fitting Schechter function above \Nhi$=10^{19}$ cm$^{-2}$. These best fits are shown by solid black lines in the \Nhi\ range where the fit was performed, and their extrapolation to lower column density is represented by a dotted black line. For spiral galaxies we find $\alpha=-0.99\pm0.04$ and $N^*=\left( 1.03\pm0.07\right)\times10^{21}$ cm$^{-2}$. For ETGs $N^*$ is an order of magnitude lower, $\left( 9.2\pm0.6\right) \times10^{19}$ cm$^{-2}$. This value is close to our sensitivity threshold (red dashed line in the figure) so that $\alpha$ is not well constrained for ETGs. Therefore, we assume the same flat slope found for spiral galaxies ($\alpha\sim-1$)\footnote{The best-fitting Schechter function to the \Nhi\ distribution of spirals has a flat slope at low column densities not fully probed by the WHISP data. We verify the prediction of this fit at low column density by analysing a very deep WSRT \hi\ image of a prototypical nearly face-on spiral galaxy, NGC~6946 \citep{2008A&A...490..555B}. We find that the \Nhi\ distribution in this galaxy is flat and in excellent agreement with our best-fitting spiral Schechter function between a few times $10^{18}$ and $\sim10^{20}$ cm$^{-2}$, confirming the validity of our fit.}. We note that, although in different contexts, several authors have successfully parameterised  the distribution of \Nhi\ in galaxies using a Schechter function \citep[e.g.,][]{2005MNRAS.364.1467Z}.

Figure \ref{fig:nhi} and the above Schechter function parameters show a significant difference between ETGs and spirals. The \Nhi\ distribution of spirals is very broad and stretches up to a few times $10^{21}$ cm$^{-2}$. In contrast, the distribution of \Nhi\ in ETGs drops very quickly and, on average, stops at $\sim3\times10^{20}$ cm$^{-2}$. Most ETGs never reach column densities above $5\times10^{20}$ cm$^{-2}$ at the resolution of our data. This value is the mean column density of \hi\ within the bright disc of spiral galaxies, derived from the tight relation between \hi\ mass and \hi\ radius within the 1 \msun/pc$^2=1.25\times 10^{20}$ cm$^{-2}$ isophote \citep[e.g.,][]{1997A&A...324..877B,2005A&A...442..137N}. Therefore, ETGs rarely host neutral atomic gas as dense as even the \it average \rm column density in the disc of spirals. It is interesting to note that the few ETGs with \hi\ at about 10$^{21}$ cm$^{-2}$ -- NGC~2685, NGC~2764, NGC~3619 and NGC~7465 -- are all galaxies with large amounts of molecular gas (Paper IV), complex and prominent dust distributions and star-forming regions (Paper II).

So we find that, as far as \hi\ properties are concerned, the main difference between ETGs with large amounts of \hi\ and spirals is that the former miss the high-column-density \hi\ typical of the bright stellar disc of the latter. This is reasonable as the star-formation rate per unit area is much larger in spirals than in ETGs. Instead, the \hi\ found in ETGs has column densities similar to those observed in the outer regions of spirals. In spirals, gas in these regions often exhibits warps and unsettled morphology and kinematics, and is taken as a signature of the on-going accretion of gas on the host galaxy \citep{2008A&ARv..15..189S}. This is similar to what we find in ETGs. As mentioned in Sec. \ref{sec:morph}, most of the \hi-rich ETGs have a disc- or ring-like morphology, but in many cases part of the gas has not settled yet (or has been disturbed recently). This suggests that despite appearing as extremely different objects in their central regions, \hi-rich ETGs and spirals may look very similar in their outskirts.

Based on numerical simulations, several authors have argued that it is possible to form an ETG with a disc by merging two gas-rich galaxies \citep[e.g.,][]{2001ApJ...555L..91N,2002MNRAS.333..481B,2005ApJ...622L...9S,2006MNRAS.372..839N,2006ApJ...641...21R}. The distribution of the gas around the simulated merger remnants resembles in many aspects the \hi\ systems discussed here. The size and mass of the distribution varies greatly depending on the properties of the merging galaxies and the merger geometry, and evolves with time since the merging. Both gas discs and rings can form as a result of these events, and gaseous tidal tails can extend to large distance from the remnant and be re-accreted at a later time. Our results suggest that, in order for the remnant to be an ETG, gas must settle on these discs in such a way to keep its column density low. Future comparison to simulations could explore this aspect in more details.

\section{Relation between \hi, galaxy luminosity and environment}
\label{sec:env}

\begin{figure*}
\includegraphics[width=18cm]{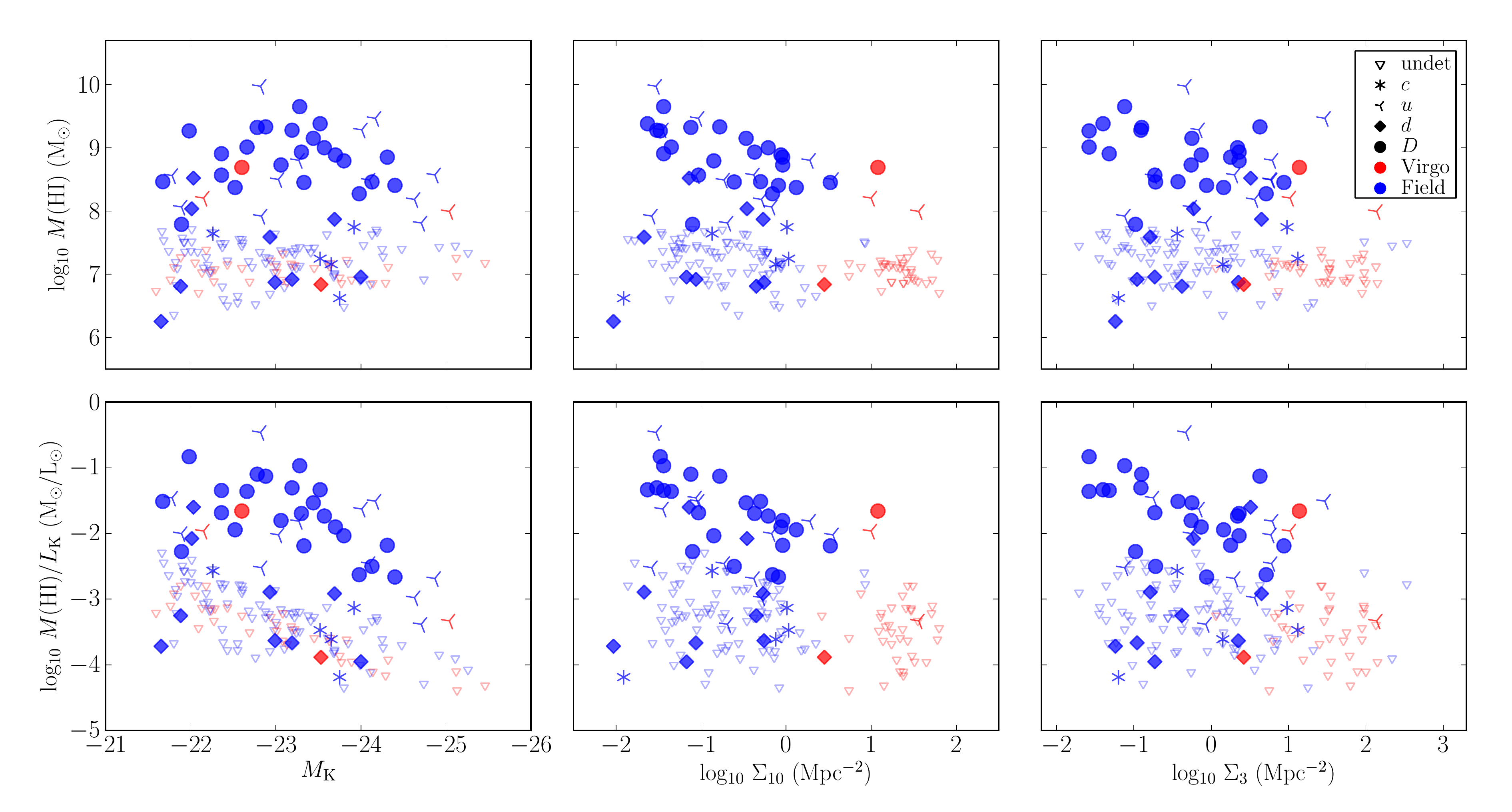}
\caption{\mhi\ (top) and \mhil\ (bottom) plotted against galaxy \mk\ (left), $\Sigma_{10}$ (middle), and $\Sigma_3$ (right). Different symbols represent different \hi\ morphologies as explained in the legend. Red and blue markers represent galaxies inside and outside the Virgo cluster, respectively.}
\label{fig:mhi}
\end{figure*}

In previous sections we describe the distribution of ETG \hi\ morphology, \mhi\ and \mhil. In this section we investigate whether these properties depend on galaxy luminosity and environment.

Figure \ref{fig:mhi} shows \mhi\ (top) and \mhil\ (bottom) of all galaxies in the \atlashi\ sample as a function of galaxy magnitude $M_\mathrm{K}$ (left), large-scale environment density $\Sigma_{10}$ (middle) and local environment density $\Sigma_3$ (right). Values of $M_\mathrm{K}$  are listed in Paper I. Values of $\Sigma_{10}$ and $\Sigma_3$ are listed in Paper VII (see Sec. \ref{sec:sample} for the definition of these parameters).

In the figure, triangles indicate upper limits on \mhi. Other markers represent different \hi\ morphologies as explained in the legend on the top-right panel (see Sec. \ref{sec:morph} for a definition of \hi\ morphological classes). Red and blue markers represent galaxies inside and outside the Virgo cluster, respectively.

We note that the \atlashi\ sample shows no relation between \mk\ and $\Sigma_{10}$ or $\Sigma_{3}$. Therefore, trends with environment can be separated from trends with galaxy luminosity.

\subsection{\hi\ and galaxy luminosity}

There is no strong trend of \hi\ properties with galaxy luminosity. Figure \ref{fig:mhi} hints to a lack of \hi\ at bright \mk. This is in agreement with results by \cite{2010MNRAS.403..683C}, although a comparison of our results is not straightforward because of the different sample selections. In our case however, the small number of bright galaxies makes this trend relatively weak from a statistical point of view. For example, we consider the two sub-samples of galaxies above and below \mhi=$5\times10^8$ \msun, respectively. According to a two-sample KS test, the hypothesis that their \mk\ distributions are drawn from a same parent distribution is rejected just at the $\sim 93$ percent confidence level.

Massive galaxies appear to exhibit a disturbed \hi\ morphology more frequently than fainter systems. The fraction of $u$'s is $7/33=21\pm8$ percent for the brightest 20 percent of the \atlashi\ sample, and $7/133=5\pm2$ percent for the remaining galaxies. This seems to indicate that recent (major or minor) interaction with other gas-rich galaxies is relatively common for massive ETGs. For example, Paper IX discuss evidence of the major-merger origin of 2 of these massive $u$ galaxies (NGC~0680 and NGC~5557) based on the \hi\ data presented here and on deep optical imaging, and argue that the formation event must have occurred after $z\sim0.5$. However, while \hi\ observations can reveal these past interactions, in a large fraction of these systems the \hi\ is distributed at large distance from the galaxy so that it is not clear whether in all cases gas is or will be accreted on the central region of the ETG (e.g., NGC~3193 and NGC~5557 in Fig. \ref{fig:sequenceb}).

Figure 8 suggests also that the fraction of $d$'s is higher in fainter galaxies: $4/33=12\pm6$ percent for the faintest 20 percent of the \atlashi\ sample, against $6/133=5\pm2$ percent for the rest of the galaxies. However, within the error bars this difference is only marginal.

\subsection{\hi\ mass and environment}
\label{sec:envelope}

Several previous authors reported a strong difference in \hi\ detection rate between ETGs living inside and outside the Virgo cluster. Within the \atlashi\ sample we detect $4/39=10\pm5$ percent  of all ETGs in Virgo and $49/127=39\pm6$ percent of all ETGs outside Virgo. This is in good agreement with results from \citeauthor{2007A&A...474..851D} (\citeyear{2007A&A...474..851D}; $2/32=6\pm4$ percent inside Virgo for $M_\mathrm{B}\leq-18$, comparable to our \mk\ selection), \citeauthor{2009A&A...498..407G} (\citeyear{2009A&A...498..407G}; $5/14=36\pm16$ percent outside Virgo with the same $M_\mathrm{B}$ selection) and O10 on the Sauron sample ($1/13=8\pm8$ percent inside and $14/20=70\pm19$ percent outside Virgo; the latter is slightly higher than our result but the typical distance of galaxies in the Sauron sample is lower and, unilke O10, our detection rate does not include a number of cases where it is unclear whether \hi\ belongs to the observed ETG or not -- see Fig. \ref{fig:gallery_undet_a}). Our result represents a significant improvement on the statistical significance of these detection rates.

While we confirm the Virgo vs. non-Virgo dichotomy, we note that environmental effects are more subtle than this. Firstly, \mhi\ and \mhil\ vary smoothly with environment density rather than showing clear breaks and sharp transitions. Figure \ref{fig:mhi} suggests the presence of an envelope of decreasing \mhi\ and \mhil\ with increasing environment density over the entire density range covered by our sample.

One cause of this envelope is that the \hi-richest ETGs live in the poorest environment. To investigate the significance of this result we compare the $\Sigma_{10}$ distributions of galaxies with \mhil\ above and below $3\times10^{-2}$ \mlsun. A two-sample KS test rejects the hypothesis that the two distributions are the same at the 99.98 percent confidence level (99.74 percent if we exclude galaxies in Virgo from this analysis). Similarly, the $\Sigma_{3}$ distributions of the same two samples are different at the 99.6 percent confidence level.

We find a similar result in our study of the CO content of ETGs (Paper IV) -- the most CO-rich ETGs live in the poorest environments. Therefore, the observation of both \hi\ and H$_2$ suggests that ETGs living in low-density environment accrete more cold gas (possibly from the surrounding medium) and/or can retain it for a longer period of time. This can be caused by the lack of both large neighbouring galaxies and a hot medium in such poor environments.

Another clue to the different gas-accretion histories of ETGs living in different environments might be the way star formation activity changes as a function of environment density. We divide the full \atlas\ sample in four $\Sigma_{10}$ bins with bin edges at $\mathrm{log_{10}} \ \Sigma_{10}/$Mpc$^2=-2,-1,0,1,2$. Moving from the poorest to the richest environment we find dust/blue features (indicative of star formation; see Sec. \ref{sec:h2}), in $21/65$, $18/105$, $6/47$ and $6/43$ of all galaxies. We conclude that star formation signatures are  more frequent at the poorest environment ($32\pm7$ percent) than in all the other $\Sigma_{10}$ bins ($\sim15\pm5$ percent in each of them separately).

Variations in the physical properties of the medium (e.g., density and temperature) and in the rate of galaxy interaction might be causing the gradual decrease of \mhi\ and \mhil\ with environment density. Such transition in \hi\ properties is visible also within the Virgo cluster, suggesting that galaxies living at the outskirts of Virgo are different from those closer to its centre.

\subsection{Centre and outskirts of the Virgo cluster}

Galaxies detected in Virgo are NGC~4262, NGC~4406, NGC~4694 and NGC~4710. Their projected distance from M87 is 1.0, 0.5, 1.3 and 1.6 Mpc, respectively. Surface-brightness fluctuation distances are available for NGC~4262 and NGC~4406, and place them at a distance of 1.8 and 0.4 Mpc from M87 along the line of sight, respectively (Paper I). NGC~4406 is the only galaxy detected in \hi\ close to the cluster centre, so that the \hi\ detection rate in this region is just $1/26=4\pm4$ percent. In contrast, the detection rate outside 1 Mpc from M87 is $3/13=23\pm13$ percent. So we find that \hi\ detection rate, \mhi\ and \mhil\ (see Fig. \ref{fig:mhi})  in the centre of Virgo are significantly lower than outside the cluster, while Virgo's outskirts appear as a transition region.

In Paper VII we find that the variation of galaxy morphological mix with environment density also continues within Virgo. In particular, the morphology-density relation changes slope at the Virgo cluster centre, where all Virgo slow rotators are found. This is in agreement with the above \hi\ result and strengthens the conclusion that different processes for the formation of ETGs must be at work inside the cluster core. These processes have the effect of generating a population of ETGs which is poorer of \hi\ and significantly richer of slow rotators than galaxies in less dense environments, including the cluster outskirts.

The above results are in good agreement with conclusions from the VIVA \hi\ survey of spirals in Virgo by \cite{2009AJ....138.1741C}. They find very \hi-rich spirals with a gas disc larger than the optical disc (typical of field spirals) only further than 1 Mpc from M87 in projection. Closer to the cluster centre they find only galaxies whose \hi\ disc is truncated to the size of the stellar disc (or even smaller for systems very close to M87), and galaxies with gas tails caused by a combination of ram-pressure stripping and tidal interaction with neighbouring galaxies. The 1 Mpc cluster-centric radius around which \cite{2009AJ....138.1741C} see a transition in the \hi\ properties of spirals matches our result on the variation of ETG \hi\ properties within Virgo.

\begin{figure*}
\includegraphics[width=34mm]{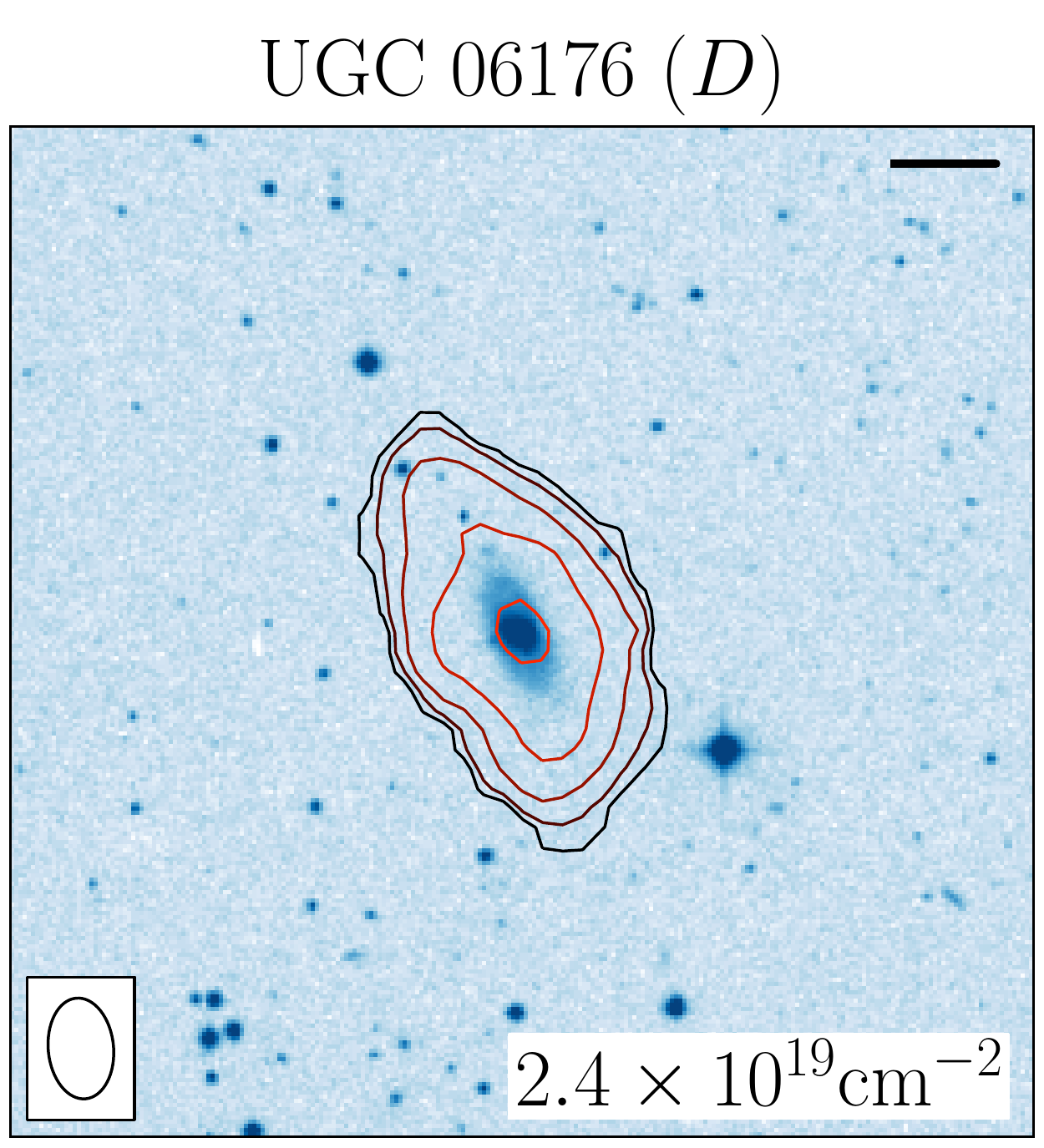} 
\includegraphics[width=34mm]{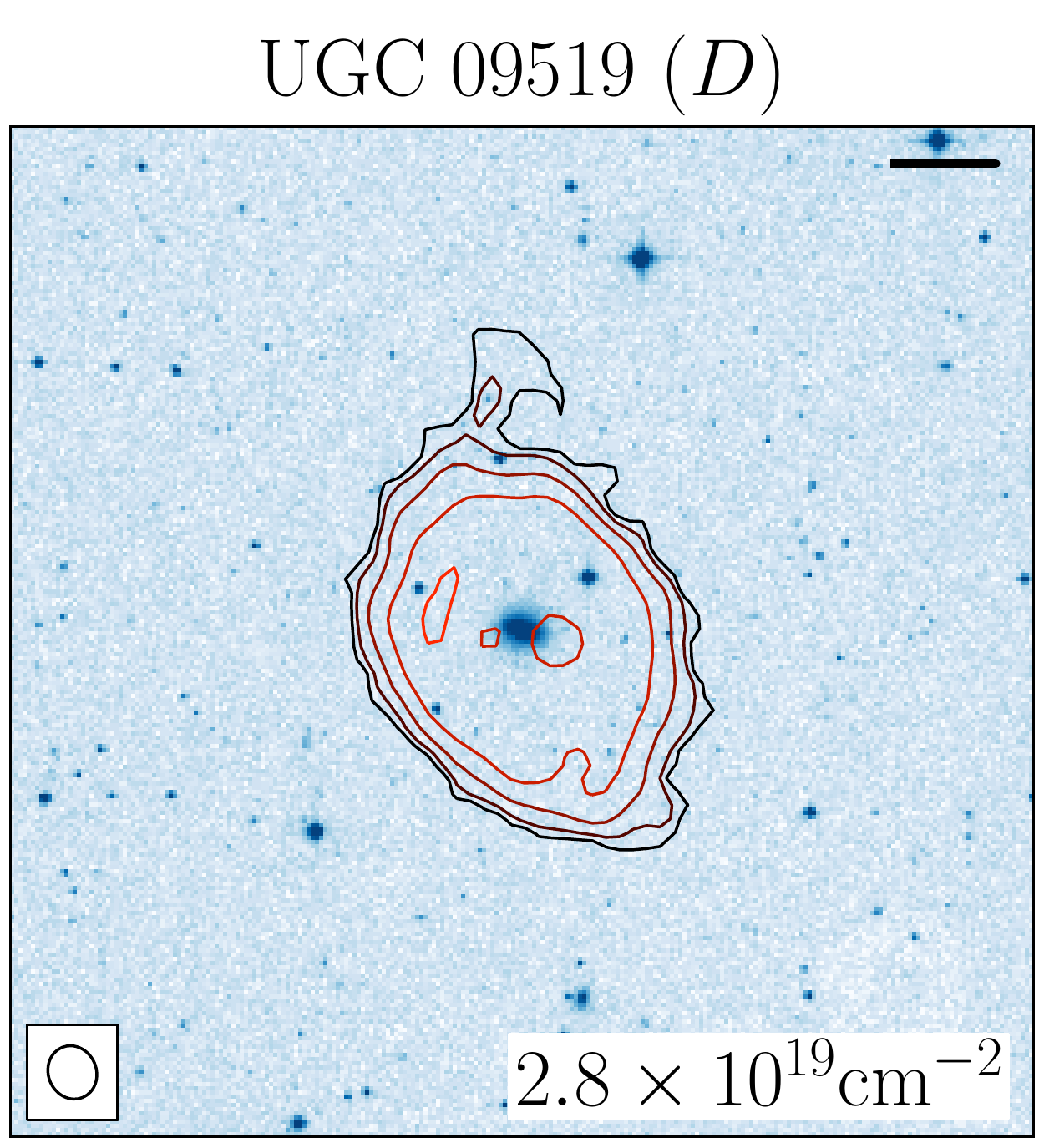} 
\includegraphics[width=34mm]{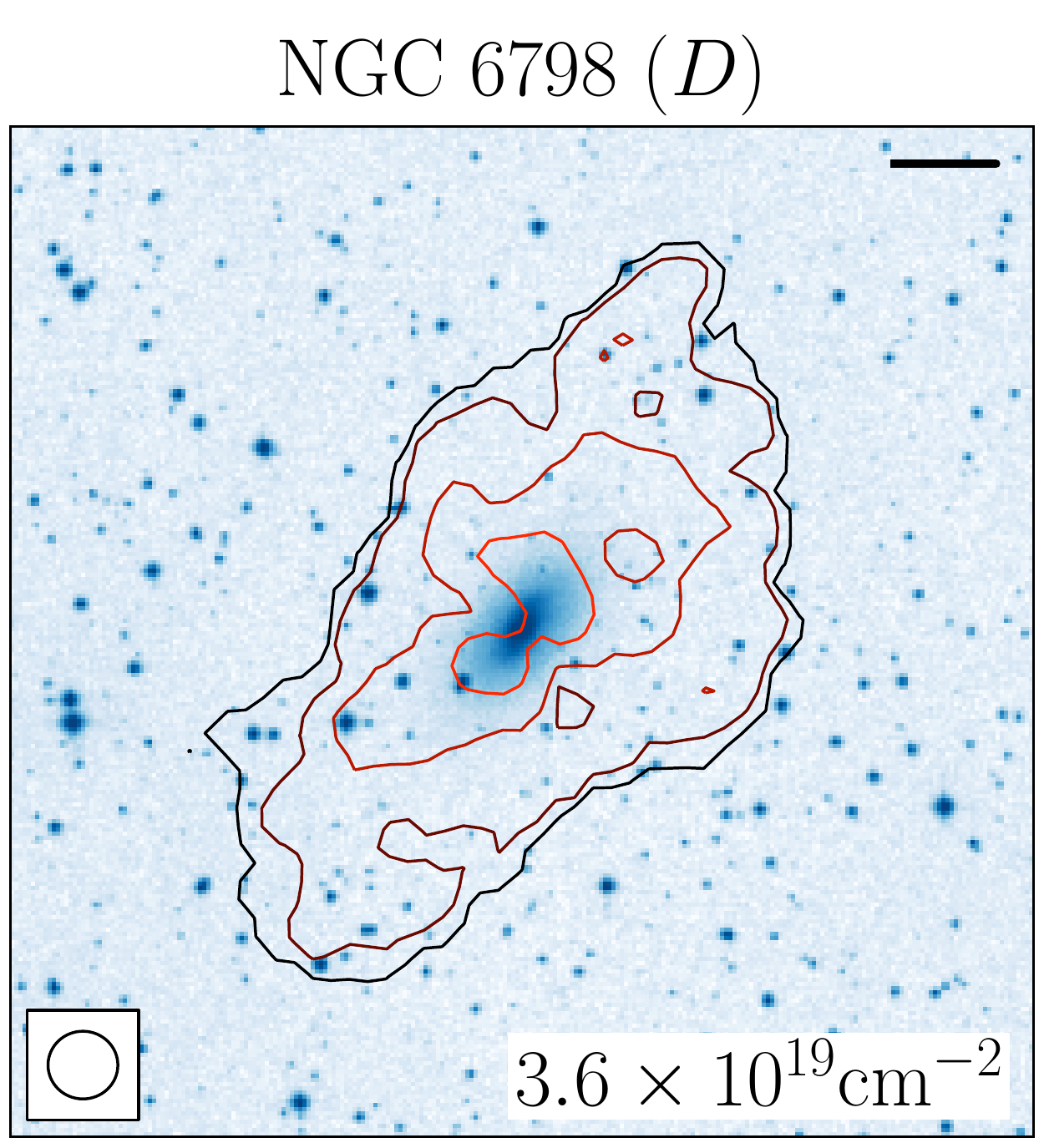} 
\includegraphics[width=34mm]{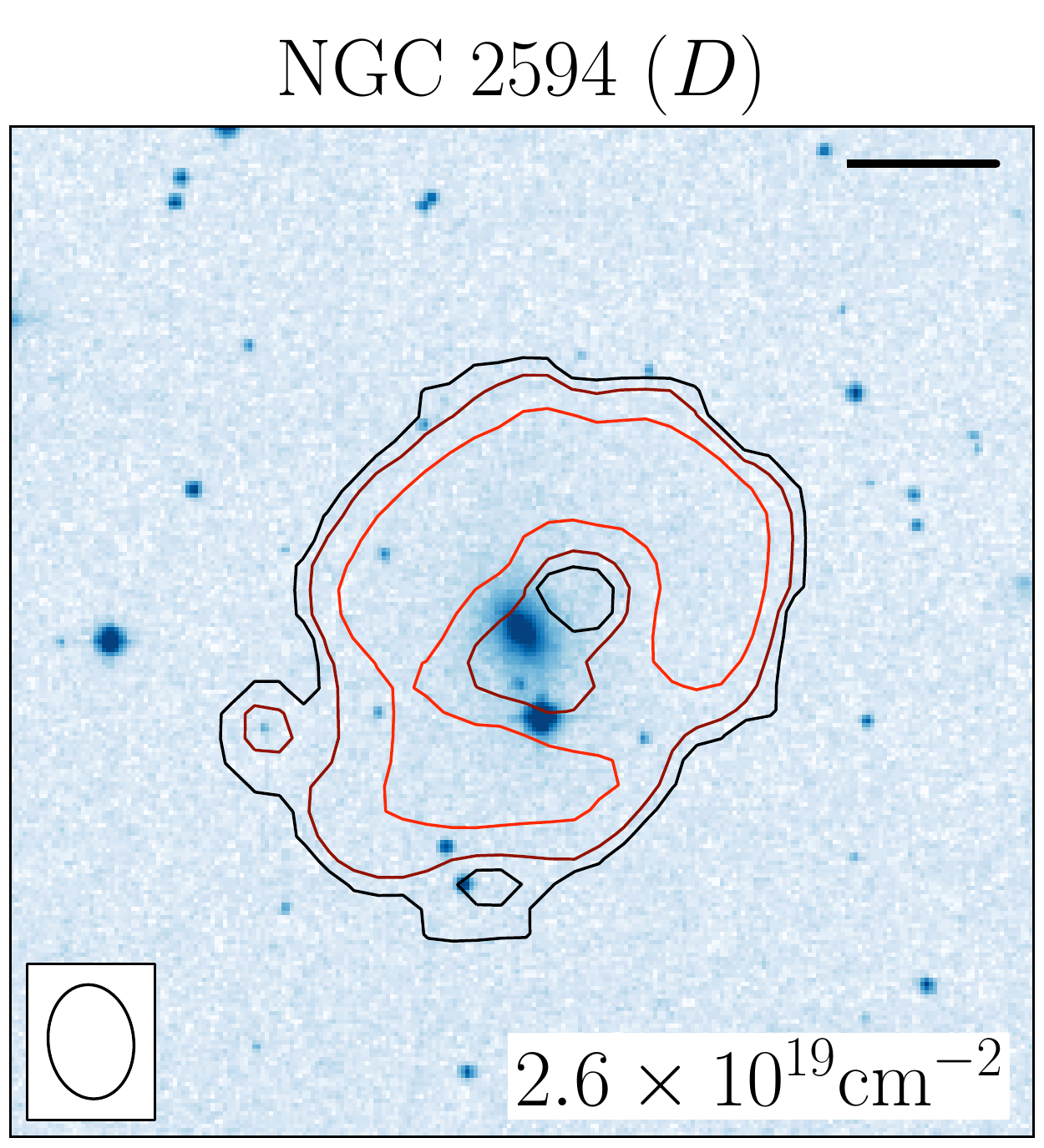} 
\includegraphics[width=34mm]{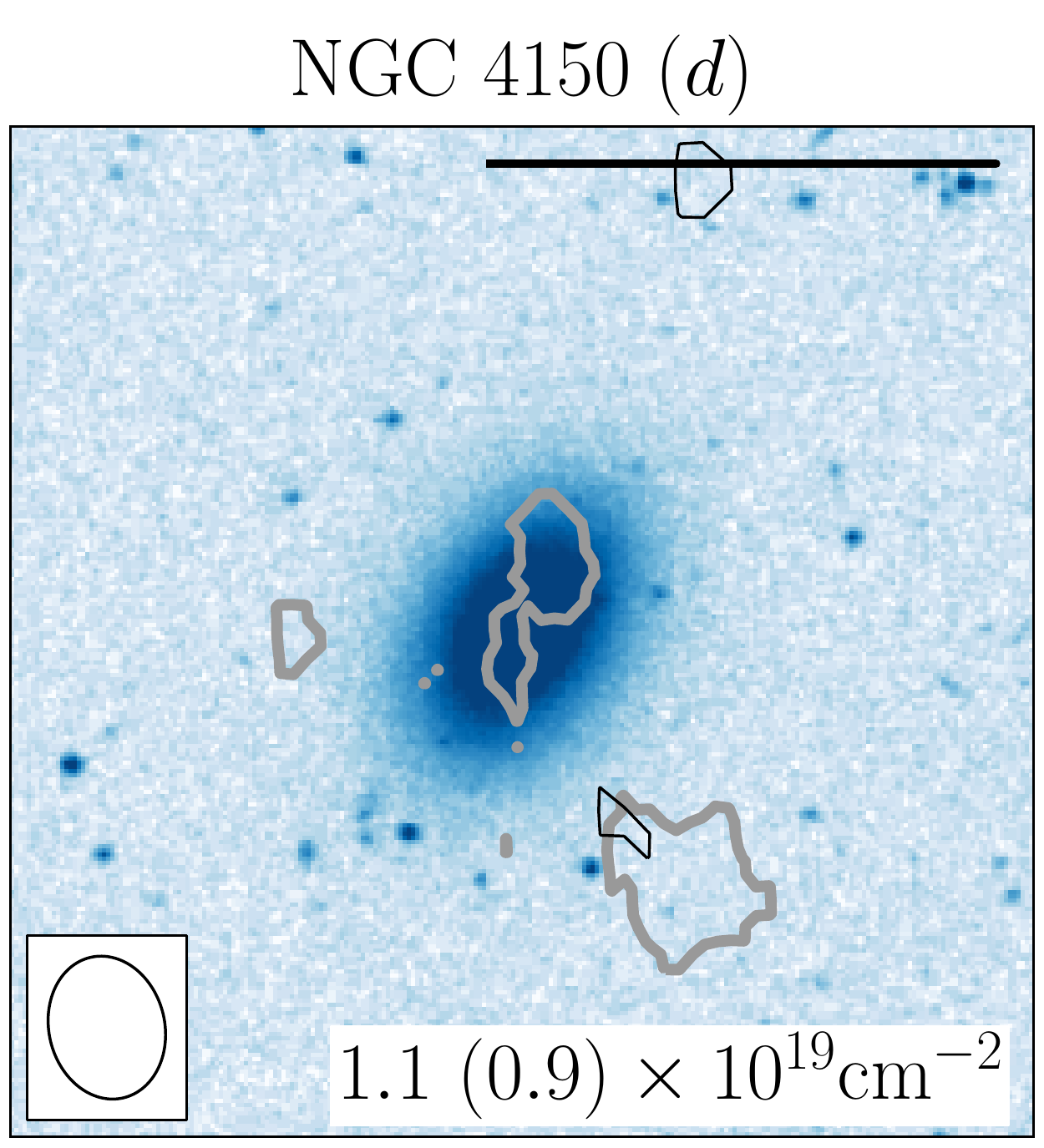} 
\includegraphics[width=34mm]{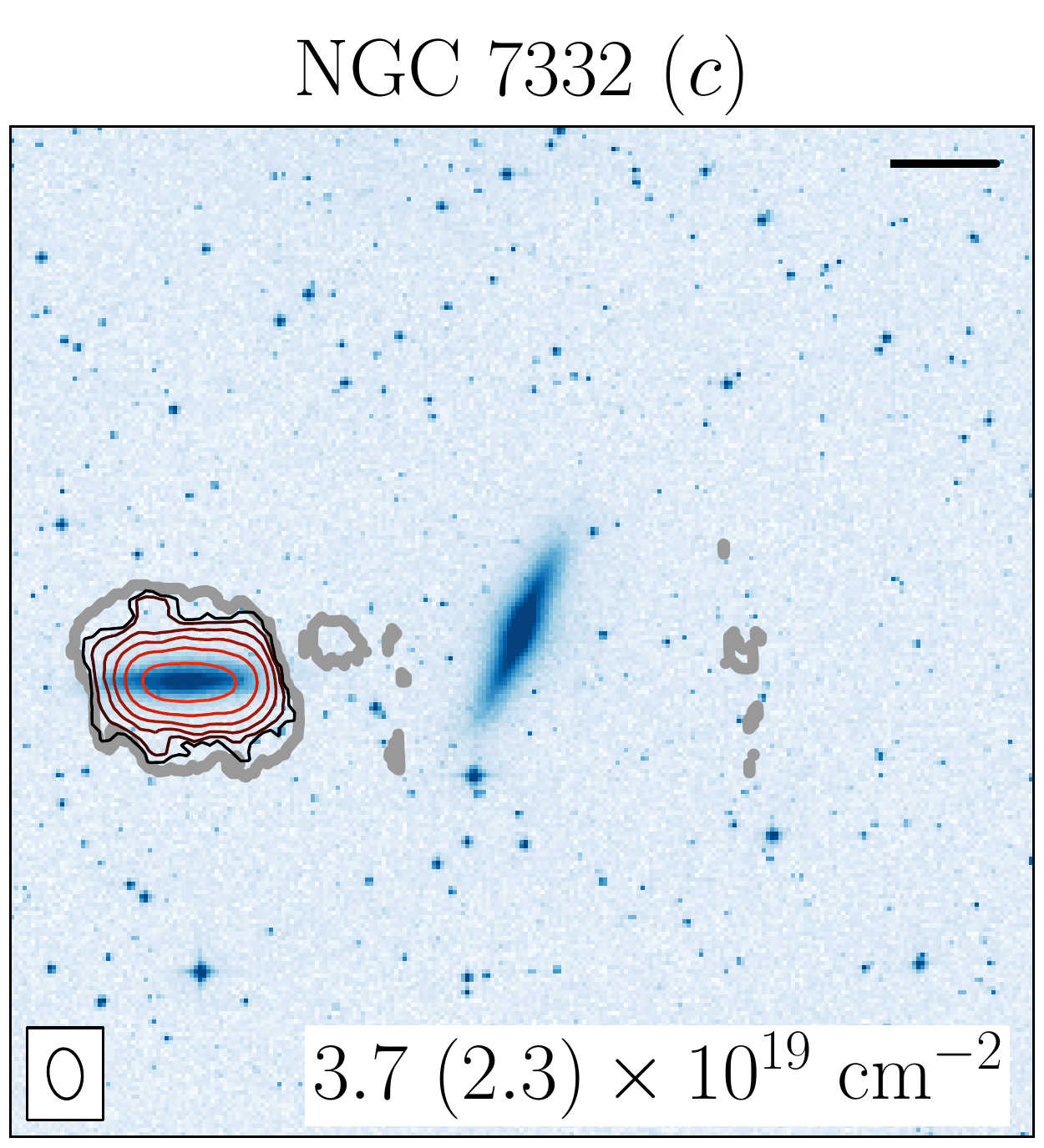} 
\includegraphics[width=34mm]{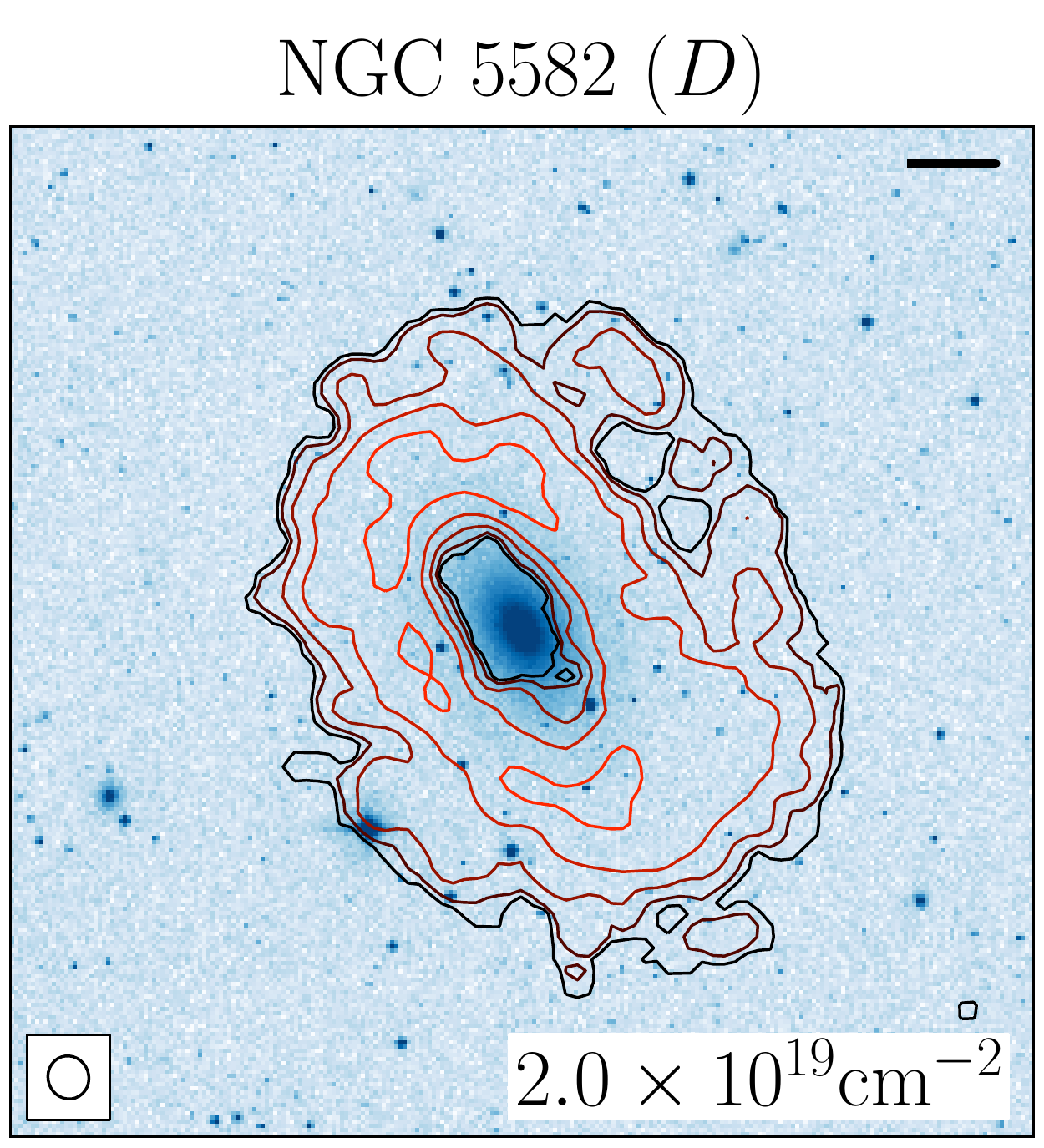} 
\includegraphics[width=34mm]{figures/HISPECIALfigures/NGC7280.pdf} 
\includegraphics[width=34mm]{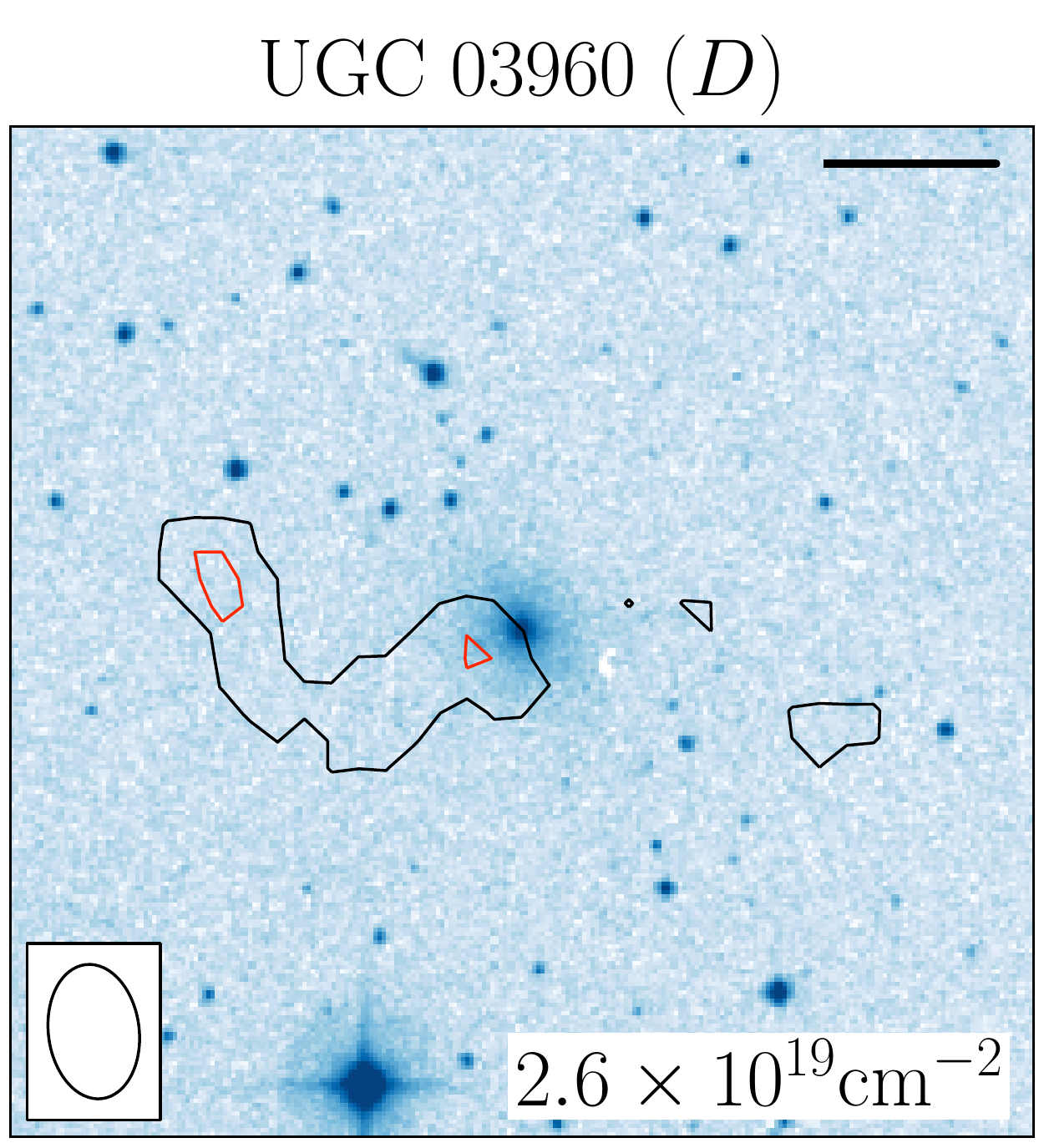} 
\includegraphics[width=34mm]{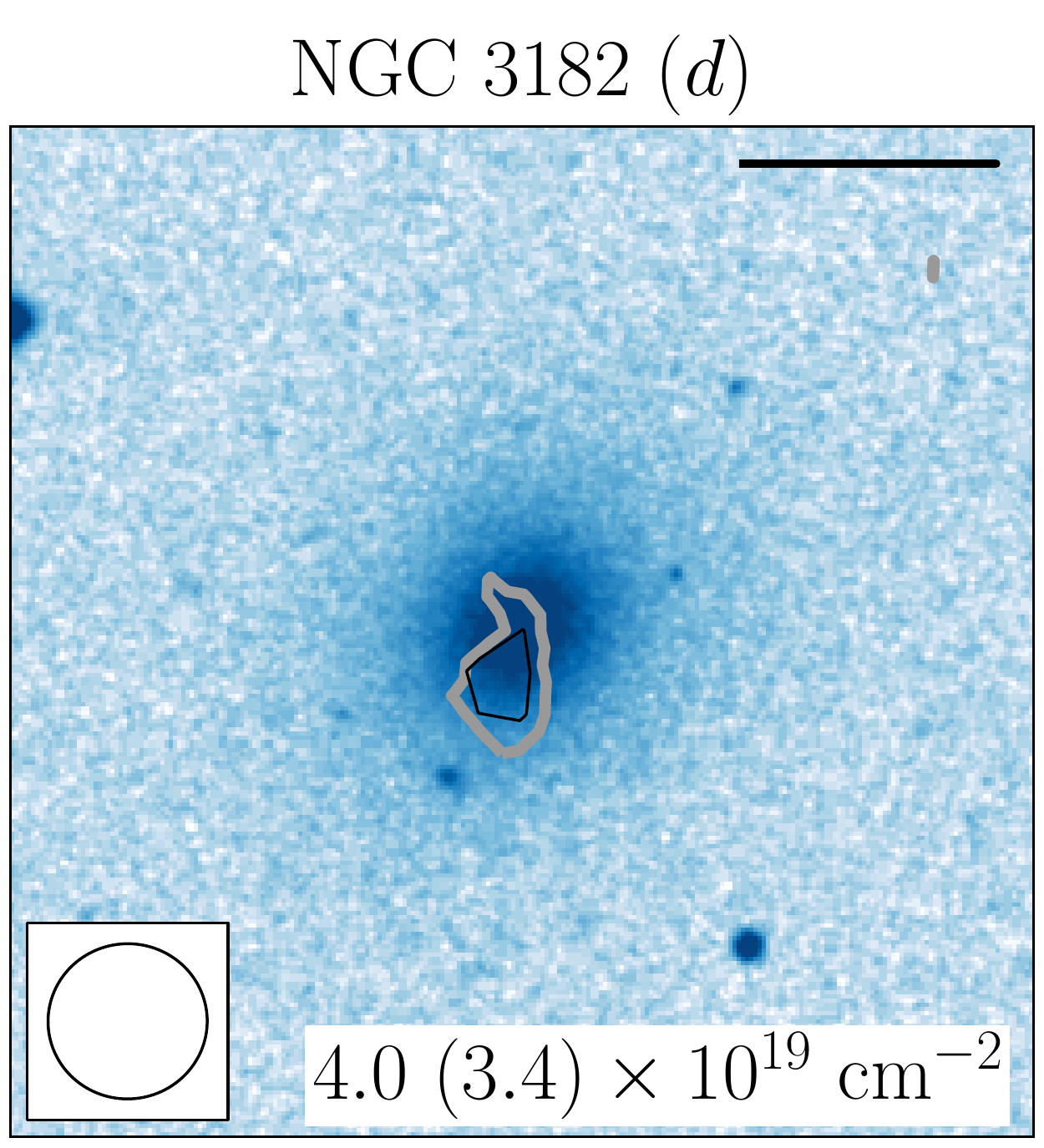} 
\caption{The ten \hi\ detections with the lowest $\Sigma_{3}$. Images are sorted according to increasing $\Sigma_3$, left to right, top to bottom. We refer to the caption of Fig. \ref{fig:gallery} for a description of the content of each image. In this figure the top-right scale bar indicates 10 kpc at the galaxy distance}
\label{fig:poorenv}
\end{figure*}

\begin{figure*}
\includegraphics[width=34mm]{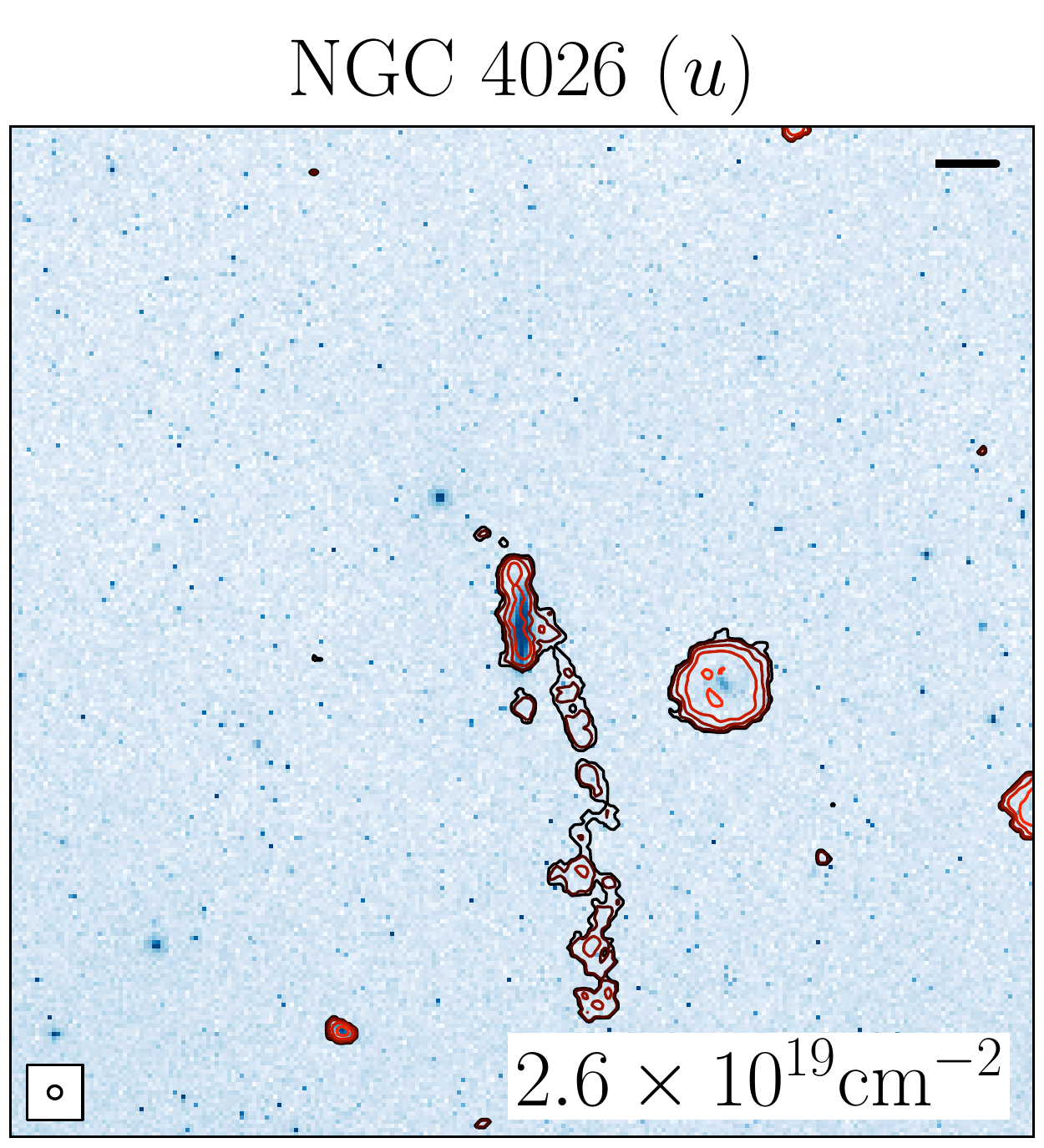} 
\includegraphics[width=34mm]{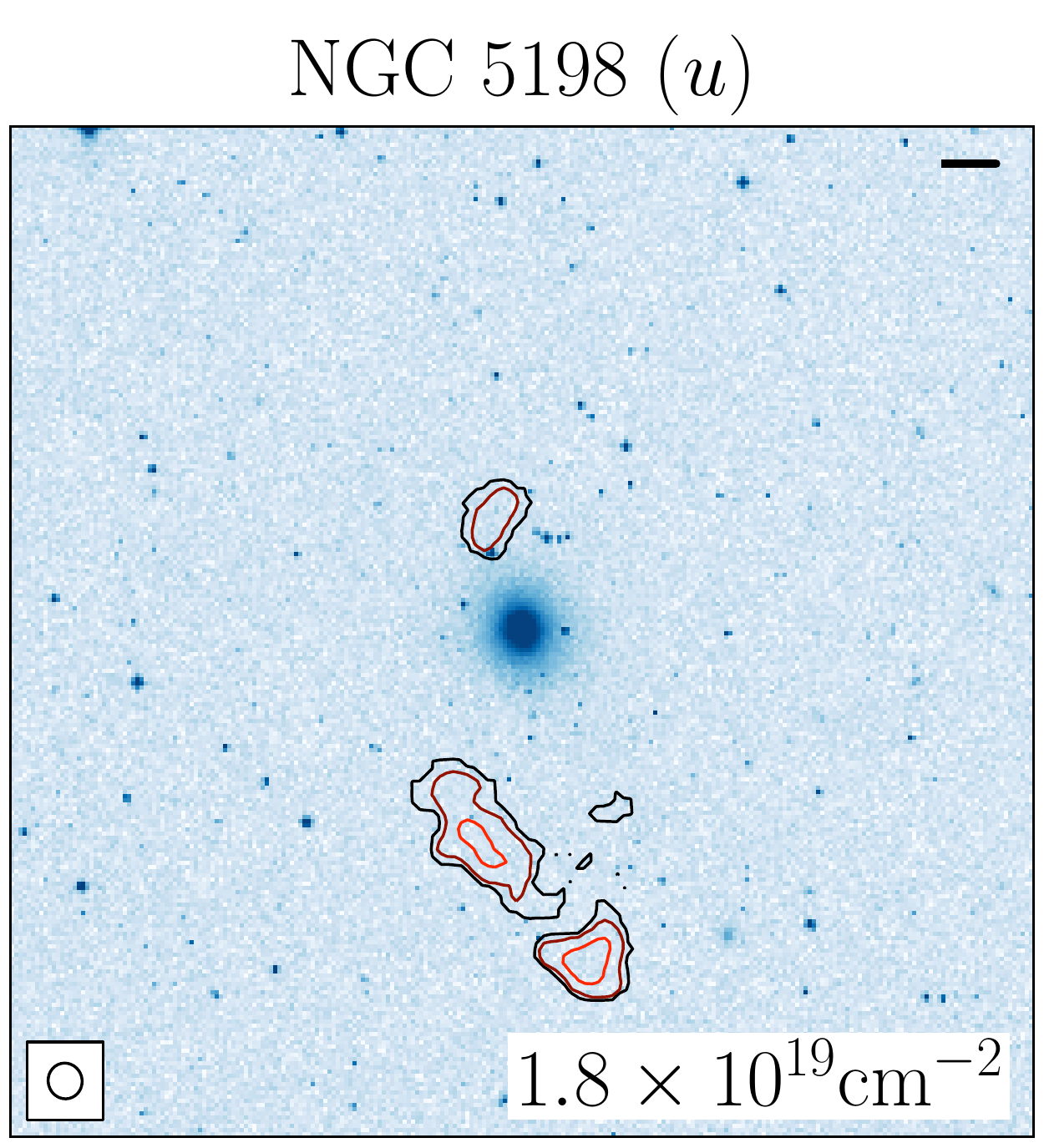} 
\includegraphics[width=34mm]{figures/HISPECIALfigures/NGC4111.pdf} 
\includegraphics[width=34mm]{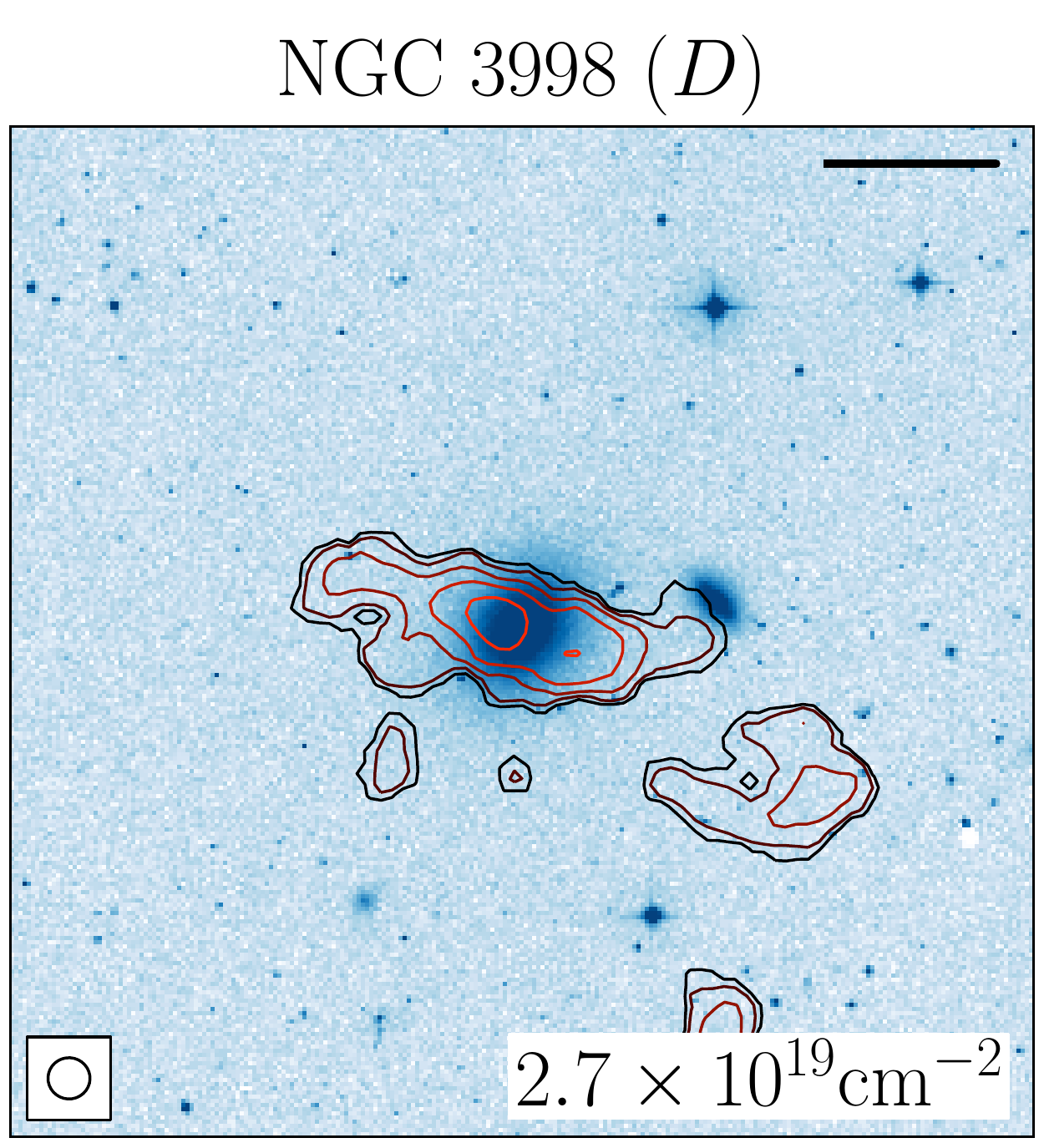} 
\includegraphics[width=34mm]{figures/HISPECIALfigures/NGC4521.pdf} 
\includegraphics[width=34mm]{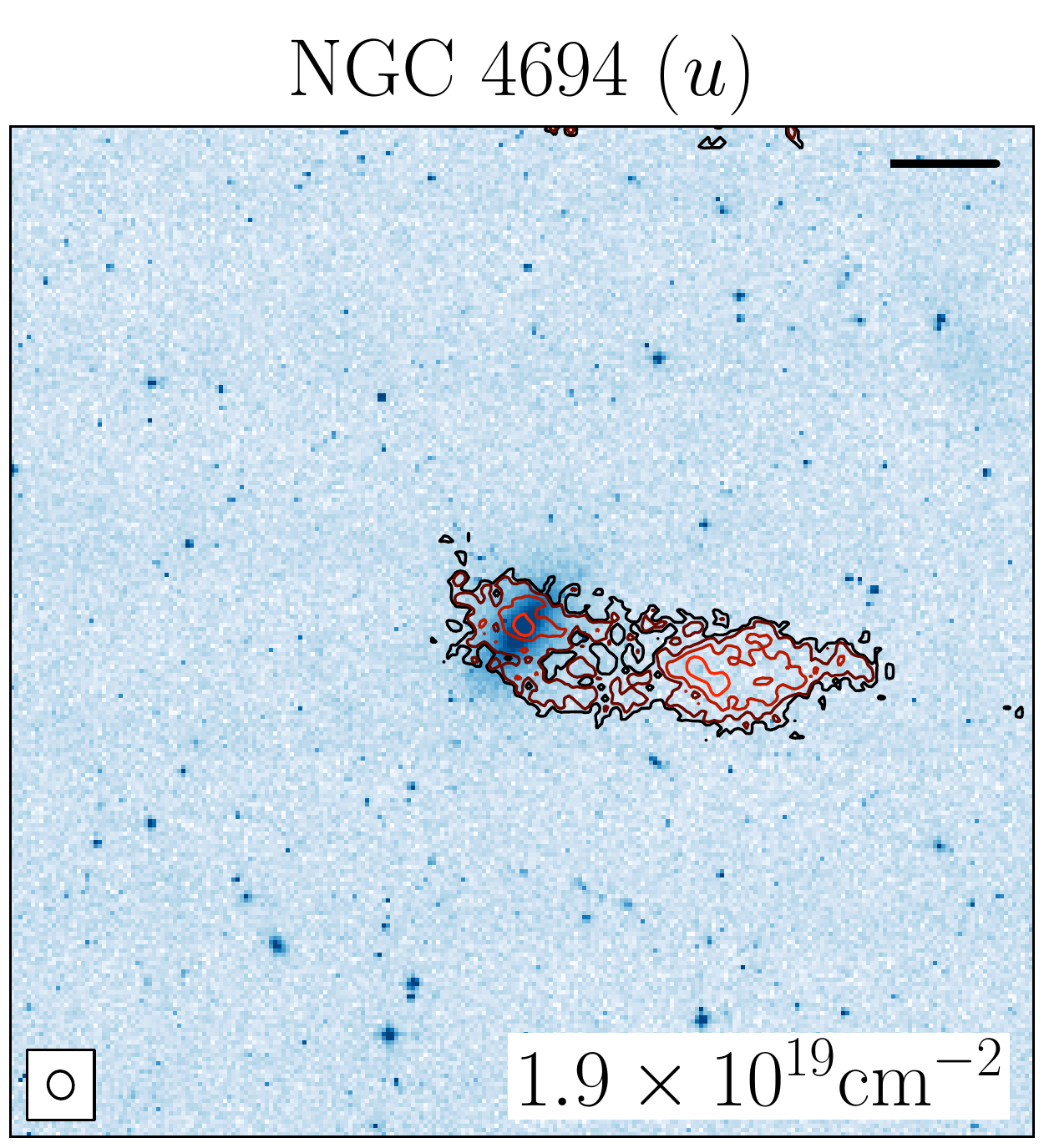} 
\includegraphics[width=34mm]{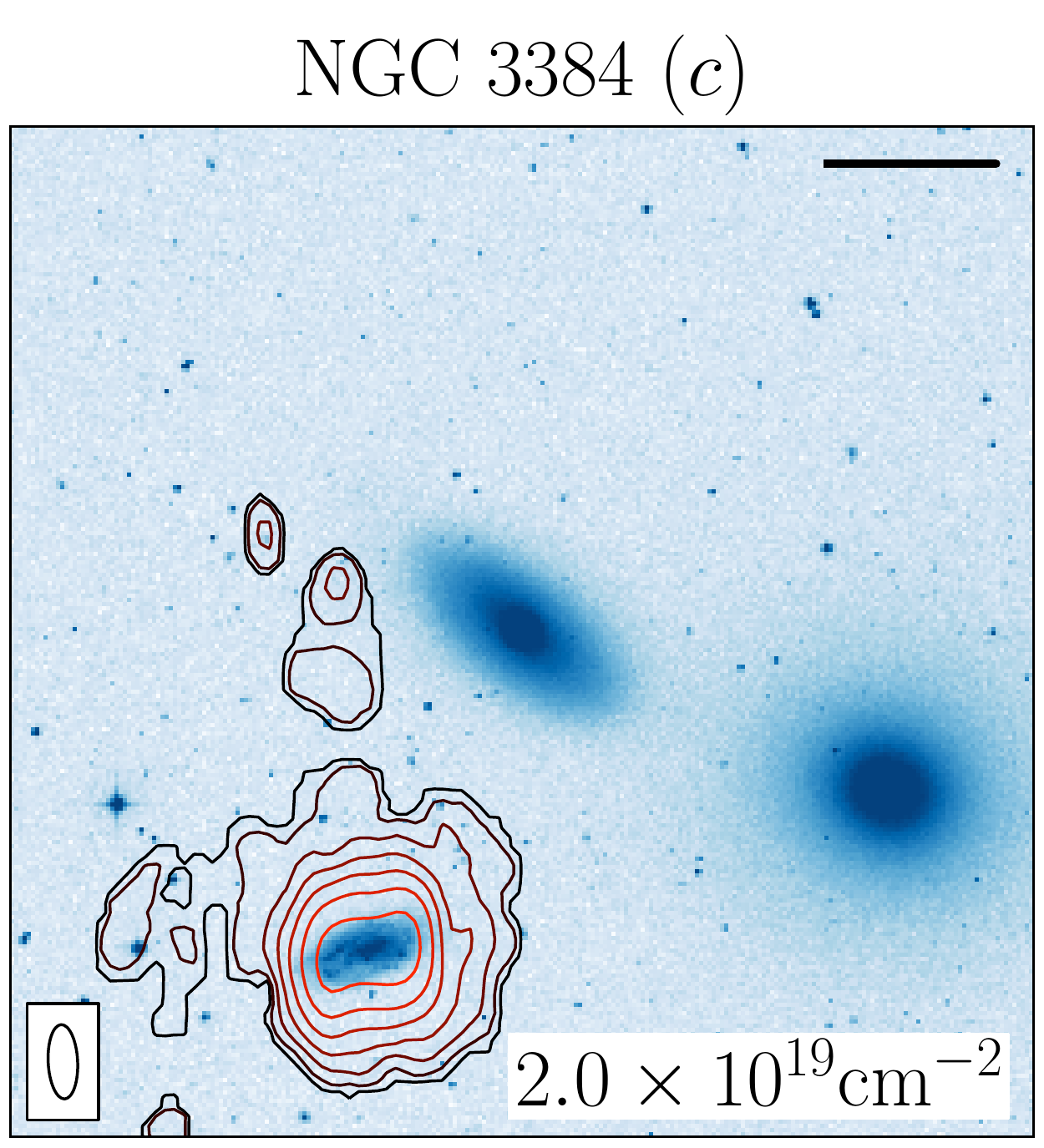} 
\includegraphics[width=34mm]{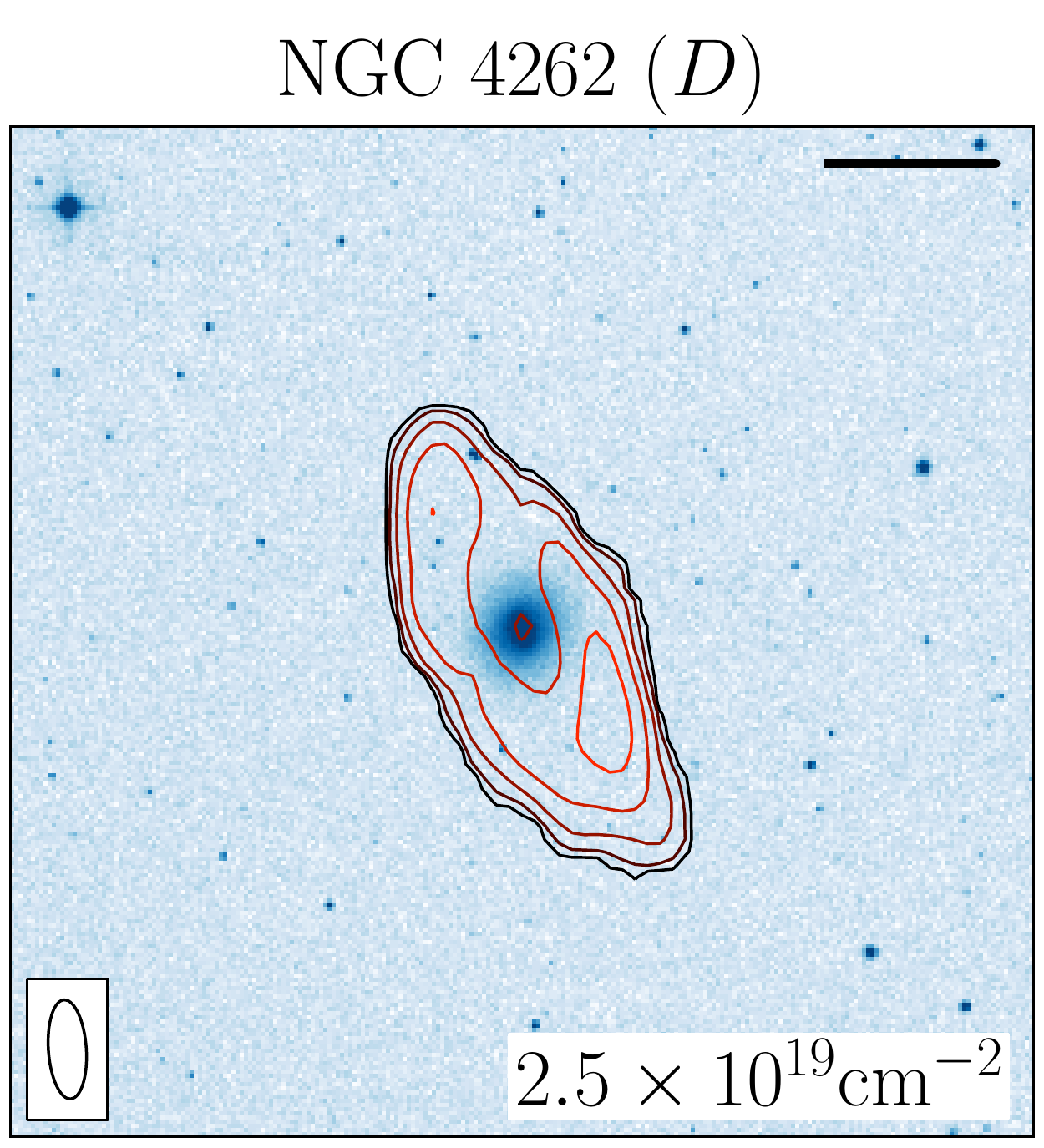} 
\includegraphics[width=34mm]{figures/HISPECIALfigures/NGC0680.pdf} 
\includegraphics[width=34mm]{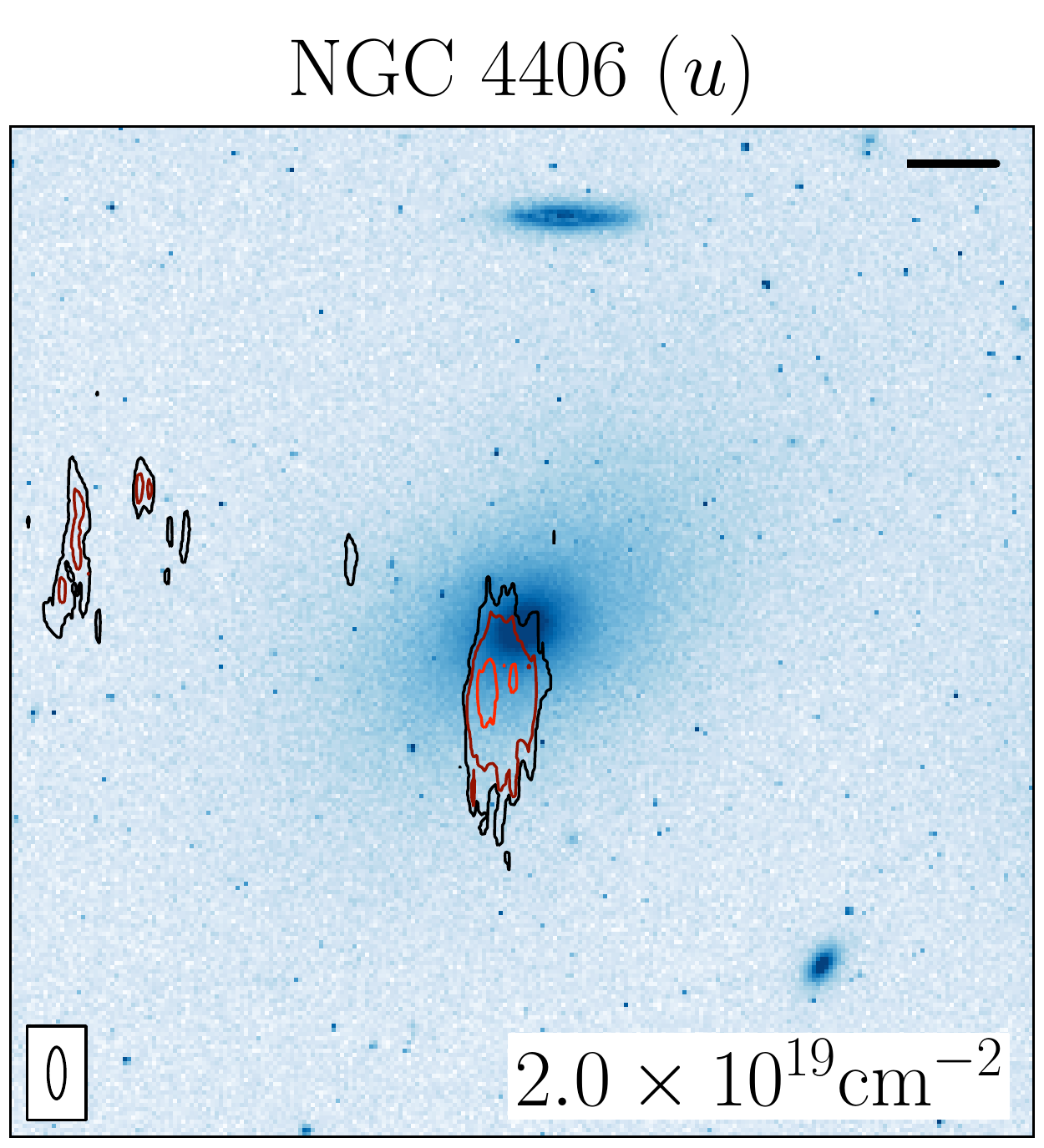} 
\caption{The ten \hi\ detections with the highest $\Sigma_{3}$. Images are sorted according to increasing $\Sigma_3$, left to right, top to bottom. We refer to the caption of Fig. \ref{fig:gallery} for a description of the content of each image. In this figure the top-right scale bar indicates 10 kpc at the galaxy distance}
\label{fig:richenv}
\end{figure*}

Neutral hydrogen is known to be quickly removed from galaxies as they move within the hot intra-cluster medium \citep[e.g.,][]{1972ApJ...176....1G}. This is indeed the common interpretation of the fact that Virgo ETGs, which are a virialised population and have therefore completed many orbits within the cluster, are so \hi-poor \citep[e.g.,][O10]{2007A&A...474..851D}. Because of these considerations, the relatively high \mhi\ and \hi\ detection rate of ETGs in the outskirts of Virgo suggests that at least a fraction of them is falling now for the first time in the cluster (a similar argument is commonly used to claim that Virgo spirals are making their first passage through the cluster). This implies that at least a fraction of all Virgo galaxies may have joined the cluster already with an early-type morphology, rather than having become ETGs within the cluster. Their current morphology may result from pre-processing inside galaxy groups, as suggested also by our result that the main driver of galaxies' morphology-density relation are processes occurring at the galaxy-group scale (Paper VII).

The \hi\ properties of ETGs in Virgo are in contrast with results of our CO observations (Paper IV). Unlike \hi-rich objects,  ETGs with molecular gas do not avoid the cluster centre. Their detection rate is $5/33=15\pm7$ percent within 1 Mpc from M87 and $6/25=24\pm10$ percent farther than 1 Mpc from M87 (for comparison, it is $47/202=23\pm3$ percent outside Virgo). Given the errorbars, it is still possible that the CO detection rate is lower in the centre of Virgo than at its outskirts or outside the cluster. However, it is clear that the difference between ETGs living in different environments is much more obvious in \hi\ than in H$_2$ (see \citealt{1989ApJ...344..171K} for a similar result on spiral galaxies).

The H$_2$ mass of ETGs inside the cluster is also consistent with that of galaxies outside it (with the aforementioned exception of the very H$_2$-rich galaxies in the poorest environment; see Sec. \ref{sec:envelope}). This implies that the lack of \hi\ in Virgo is related to a high \mhtwo/\mhi\ ratio rather than an overall gas deficiency. A possible explanation is that \hi\ is more easily stripped than H$_2$ when a galaxy moves through a hot medium. This is in general true over the full extent of a galaxy because \hi\ is typically distributed out to larger radii than CO and is therefore less bound to the galaxy \citep{2009ApJ...697.1811F}. This situation is particularly frequent in ETGs, where the \hi\ often extends to tens of kpc from the galaxy (see Fig. \ref{fig:gallery}). However, many ETGs outside the cluster contain \hi\ in their central regions, so the question is why we do not detect such central \hi\ (but do detect central CO) in Virgo.

Neutral hydrogen in the centre of ETGs might be more easily stripped than CO because of its lower surface density \citep{2011arXiv1106.2554Y}. Another important indication is that ETGs with CO appear as a virialised population within Virgo (Paper IV), so that they may have completed many orbits within the cluster. A prolonged residence within the cluster might be the reason why these galaxies are so \hi\ poor. They are not only stripped of their gas, but the lack of other \hi-rich companions means that no further \hi\ accretion can occur (see O10). While this seems a plausible picture, we note that at least in some cases CO is stripped together with the \hi\ from the centre of galaxies \citep{2008A&A...491..455V}, so that we do not have a definitive answer to this problem yet.

\subsection{\hi\ morphology-density relation}

In Sec. \ref{sec:envelope} we show that galaxies in poor environment can reach higher \hi\ masses than those found at higher density. These objects also tend to have a more regular \hi\ morphology than gas-rich ETGs in denser environments. We illustrate the relation between \hi\ morphological mix and environment density in Figs. \ref{fig:poorenv} and \ref{fig:richenv}. These show total \hi\ images of the ten detections with the lowest and highest $\Sigma_3$, respectively, sorted according to increasing $\Sigma_3$ (left to right, top to bottom). Figure \ref{fig:richenv} includes 3/4 of all Virgo detections, all 3 Ursa Major detections (NGC~3998, NGC~4026 and NGC~4111; \hi\ in these galaxies is discussed also by \citealt{2001ASPC..240..867V}) and 4 galaxies living in relatively rich groups. The fraction of settled \hi\ systems (i.e., $D$ and $d$) is high at the lowest environment densities: $8/10=80\pm28$ percent, with the only $u$ system possibly being a misclassified $d$ (see Sec. \ref{sec:morph}). In contrast, it is very low at the highest densities: $2/10=20\pm14$ percent. At high $\Sigma_3$  most \hi-rich systems look disturbed (including some classified as $D$, like NGC~3998).

The trend of \hi\ morphology with environment density is stronger using $\Sigma_3$ than $\Sigma_{10}$. For example, galaxies with a very disturbed \hi\ morphology figure among the ten objects with the lowest $\Sigma_{10}$ (NGC~1023, NGC~7465); and the fraction of settled \hi\ systems at the lowest $\Sigma_{10}$ is the same as that at the highest $\Sigma_{10}$ ($5/10$ against $4/10$). This indicates that processes influencing the \hi\ morphology occur on a galaxy-group scale, and that a disturbed \hi\ morphology is more clearly related to the presence of close neighbours than to an overall, large-scale environment over-density.

It is interesting that a similar conclusion was reached in Paper VII based on the morphology-density relation of galaxies within the \atlas\ volume. There we found that processes occurring on a galaxy-group scale must be the main driver of the relation, which is tighter and steeper using $\Sigma_3$ rather than $\Sigma_{10}$. The variation of \hi\ morphology as a function of local environment density confirms the importance of such processes for the evolution of ETGs.

\section{Summary and Conclusions}
\label{sec:summ}

We study the \hi\ properties of all 166 ETGs in the volume-limited \atlas\ sample above $\delta=10$ deg and further than 15 arcmin from Virgo A. The sample includes galaxies within 42 Mpc and brighter than \mk=$-21.5$. This is the largest sample of ETGs with deep, interferometric \hi\ observations to date.

We confirm earlier findings that the \hi\ detection rate of ETGs depends strongly on environment density \citep[][O10]{2007A&A...474..851D,2009A&A...498..407G}. We detect $4/39=10\pm5$ percent of all ETGs inside the Virgo galaxy cluster, and $49/127=39\pm6$ percent of all ETGs outside it. This is consistent with results from previous authors, although on a stronger statistical basis. The lower detection rate of ETGs in \hi\ surveys of larger volumes is easily explained by their much poorer sensitivity. For example the \mhi\ detection limit in the study of \cite{2011ApJ...732...92T} is above $10^9$ \msun\ in most of the surveyed volume. Figure \ref{fig:KMmhi0} shows that, at such sensitivity, their survey is expected to detect only $\sim10$ percent of all ETGs.

We classify galaxies according to their \hi\ morphology. We find that 1/5 of all ETGs (and 1/4 of all ETGs outside Virgo) host \hi\ distributed on a regularly rotating disc or ring. These systems represent the majority of the \hi\ detections, 64 percent, so that if an ETG is detected in \hi\ it will most likely host a gas disc or ring. Another 8 percent of all ETGs (26 percent of all detections) host \hi\ in an unsettled configuration, while the remaining detected galaxies are surrounded by \hi\ clouds scattered around the stellar body and not obviously associated to it.

Although most detections can be easily classified within this scheme, we do find a number of galaxies intermediate (or in transition) between classes. Indeed, we claim that the \hi\ morphology of ETGs varies in a continuous way, going from very regular discs and rings to unsettled gas distributions, and from these to systems of scattered clouds which may or may not be associated to the host galaxy. This continuity may be related to the time passed since the last major episode of gas accretion or stripping relative to the \hi\ orbital time in each galaxy.

We divide regular, rotating \hi\ systems in two classes based on their size relative to the stellar body of the host galaxy: $D$ (large discs/rings -- 24 galaxies) and $d$ (small discs -- 10 galaxies). The former contain between $10^8$ and $10^{10}$ \msun\ of gas distributed out to tens of kpc from the stellar body. In half of these systems the \hi\ is morphologically or kinematically misaligned with respect to the stellar body. In contrast, $d$'s contain typically less than $10^8$ \msun\ of \hi\ confined within the stellar body and morphologically and/or kinematically aligned to it in nearly all cases.

We investigate the role of \hi\ in fuelling star formation within the host ETG. Confirming results by O10, we find that the detection rate of signatures of star formation (molecular gas, dust discs/filaments and blue features) is high ($\sim70$ percent) for ETGs with \hi\ within the stellar body. Namely, such features are found in all $d$'s, $\sim60$ percent of al $D$'s with central \hi\ and $\sim50$ percent of all unsettled \hi\ systems with central \hi. On the contrary, they are found in just $\sim15$ percent of all ETGs with no central \hi\ (or no \hi\ at all at the sensitivity of our observations), highlighting the role of \hi\ in enriching ETGs with material for star formation.

The ISM in the centre of ETGs is dominated by molecular gas, whose mass is a factor of a few up to $\sim100$ larger than the \hi\ mass. Galaxies hosting a small \hi\ disc ($d$) reach particularly high \mhi/\mhtwo\ values (larger than 10 in half of the cases), suggesting that they are forming molecular gas at efficiencies comparable to those of spirals.

We parameterise the \hi\ mass function of ETGs with a Schechter function with $M^*= \left( 1.8\pm 0.7 \right) \times 10^9$ \msun\ and $\alpha=-0.68\pm0.16$. The value of $M^*$ is $\sim5$ times lower than typical values found for samples dominated by spiral galaxies, confirming that ETGs are, as a a family, poorer of \hi. The value of the faint-end slope $\alpha$ means that the mass function for a \mk-limited sample decreases with decreasing \mhi\ below $M^*$.

We compare the \hi\ properties of ETGs to those of spirals. The \mhi\ and \mhil\ distributions of ETGs are very broad, reflecting the large variety of \hi\ content of these galaxies, and confirming the lack of correlation between ETG \hi\ mass and luminosity. The distributions of spirals are much narrower and shifted towards larger \mhi\ and \mhil. Yet, we find a substantial overlap in the \mhi\ and \mhil\ distributions of ETGs and spirals (consistent with \citealt{2010MNRAS.403..683C}). A significant fraction of all ETGs can have as much \hi\ as spiral galaxies, and in the majority of the cases this is distributed on a large, rotating disc or ring.

We investigate the difference between such \hi-rich ETGs and spiral galaxies. We find that in the former the \hi\ column density is typically very low. Early-type galaxies rarely reach the gas density typical of the bright stellar disc of spiral galaxies, consistent with their lower star formation rate per unit area. The few galaxies that do, show clear signs of star formation and prominent dust features. Gas column density appears therefeore as the decisive factor in determining whether a galaxy is of early or late morphological type.

We find a minor dependence of ETG \hi\ properties on galaxy luminosity. Very luminous galaxies seem to contain less \hi, but the only clear result is that \hi\ in these systems is typically found in an unsettled configuration. This may reflect a higher rate of interaction with gas-rich companions for massive ETGs compared to fainter objects, although it is not clear how much of this gas is eventually accreted on the centre of the host galaxy.

We do find clear trends of \hi\ properties with environment. These go well beyond the known Virgo-vs.-non Virgo dichotomy mentioned at the beginning of this section. We find a smooth envelope of decreasing \mhi\ and \mhil\ with environment density. Consistent with results from the \atlas\ CO survey presented in Paper IV, we find that the gas-richest galaxies live in poor environments. This indicates that the cold-gas accretion rate is higher at these densities, and/or that these galaxies can retain their gas reservoir for longer periods of time. This effect is likely driven by a combination of a relatively quiet merging history and by the  lack of a hot medium, and is accompanied by a higher fraction of objects exhibiting signs of recent star-formation in this environment.

The gradual decrease of \hi\ mass with environment density continues within the Virgo cluster. Consistent with results from the study of spiral galaxies \citep{2009AJ....138.1741C}, we find that the \hi\ properties of ETGs living at the outskirts of Virgo  are intermediate between those of galaxies outside Virgo and those of the very \hi-poor objects in its central region. Being \hi-rich, galaxies at the Virgo outskirts are likely falling in the cluster for the first time, and they do so already with an early-type morphology. This suggests that pre-processing  on a galaxy-group scale is fundamental for the evolution of ETGs, as concluded also in Paper VII.

Galaxies inside $\sim1$ Mpc from the centre of the cluster are significantly poorer of \hi\ than galaxies outside Virgo, with a drop in the \hi\ detection rate to just a few percent. This agrees with the observation of truncated \hi\ discs in spiral galaxies in this region, and resonates with our finding that the fraction of slow rotating ETGs, which is very low at all environment densities, increases in the core of Virgo. This highlights that different processes must drive the evolution of galaxies deep inside the cluster, among which the presence of a hot medium (and therefore lack of \hi).

Unlike \hi, molecular gas can be found in ETGs living deep inside Virgo. These systems have \mhtwo\ similar to that of galaxies outside Virgo, so that the lack of \hi\ seems to imply a high molecular-to-atomic mass ratio. A possible interpretation is that molecular gas is more difficult to remove via ram-pressure stripping than \hi, so that ETGs that have long been in the cluster (and therefore long been stripped of their \hi) can still host molecular gas deep in the centre of their gravitational potential (Paper IV). However, it is not clear why \hi\ is not detected in the centre of these galaxies, as observations show that CO can be stripped together with \hi\ in at least some cases \citep{2008A&A...491..455V}.

We find that the \hi\ morphology also depends on environment density. In poor environment the typical \hi\ detection exhibits a large, settled \hi\ disc or ring. These systems must have been in place for many Gyr and indicate that very recent accretion/merging events have been minor. On the contrary, disturbed and unsettled \hi\ morphologies are very frequent at higher environment density. Such \hi\ morphology-density relation is clearer when measuring environment density on a local scale, indicating that processes occurring on a galaxy-group scale are driving factors for the evolution of ETGs. This agrees with the study of the kinematical morphology-density relation presented in Paper VII, where we conclude that processes occurring on this scale determine the increase of the ETG fraction (and decrease of the spiral fraction) with environment density.

As a concluding remark, we have demonstrated that a large fraction of ETGs contain significant amounts of \hi. It is quite normal for these galaxies to host neutral hydrogen gas, and \hi-rich ETGs should not  be regarded as odd or rare systems. On the contrary, \hi\ seems to be playing an important role in replenishing ETGs with cold gas and fuelling the formation of new stars and a stellar disc in these systems. The amount and morphology of this \hi\ depend weakly on  galaxy luminosity, one of the main parameters determining some of the most important galaxy properties such as their structure and stellar populations. Even trends with environment density, another fundamental parameter for galaxy evolution, are characterised by a very large scatter. This scatter suggests that the gas accretion history of ETGs varies widely from galaxy to galaxy even at fixed galaxy mass and environment density. A logical consequence of this large scatter is that ETGs are not such because they lack \hi\ -- many of them contain plenty of it. Instead we have shown that most ETGs must accrete gas in such a way that it remains at low column density -- too low to support the higher levels of star formation seen in spiral galaxies.

\section*{Acknowledgments}

PS would like to thank Mike Sipior for his help setting up the WSRT data reduction pipeline at ASTRON, Riccardo Giovanelli for providing ALFALFA spectra, and Gyula J{\'o}zsa for scheduling the WSRT observations.

MC acknowledges support from a Royal Society University Research Fellowship.

This work was supported by the rolling grants `Astrophysics at Oxford' PP/E001114/1 and ST/H002456/1 and visitors grants PPA/V/S/2002/00553, PP/E001564/1 and ST/H504862/1 from the UK Research Councils. RLD acknowledges travel and computer grants from Christ Church, Oxford and support from the Royal Society in the form of a Wolfson Merit Award 502011.K502/jd. RLD also acknowledges the support of the ESO Visitor Programme which funded a 3 month stay in 2010.

SK acknowledges support from the the Royal Society Joint Projects Grant JP0869822.

RMcD is supported by the Gemini Observatory, which is operated by the Association of Universities for Research in Astronomy, Inc., on behalf of the international Gemini partnership of Argentina, Australia, Brazil, Canada, Chile, the United Kingdom, and the United States of America.

TN and MBois acknowledge support from the DFG Cluster of Excellence `Origin and Structure of the Universe'.

MS acknowledges support from a STFC Advanced Fellowship ST/F009186/1.

NS and TD acknowledge support from an STFC studentship.

MBois has received, during this research, funding from the European Research Council under the Advanced Grant Program Num 267399-Momentum.

The authors acknowledge financial support from ESO. 

\bibliographystyle{mn2e}
\bibliography{a3dhi}

\appendix

\section{\hi\ gallery}

\begin{figure*}
\includegraphics[width=50mm]{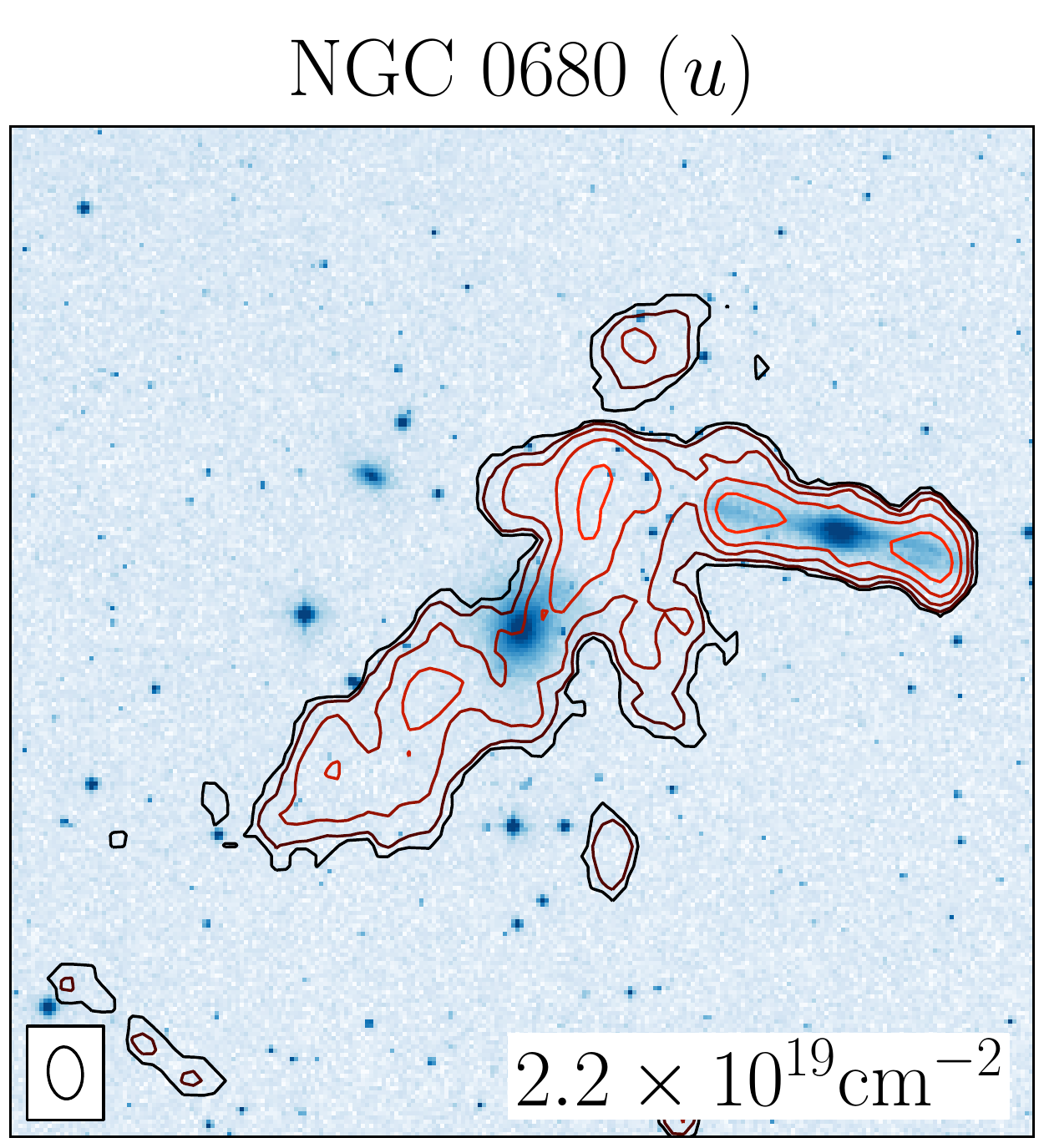} 
\includegraphics[width=50mm]{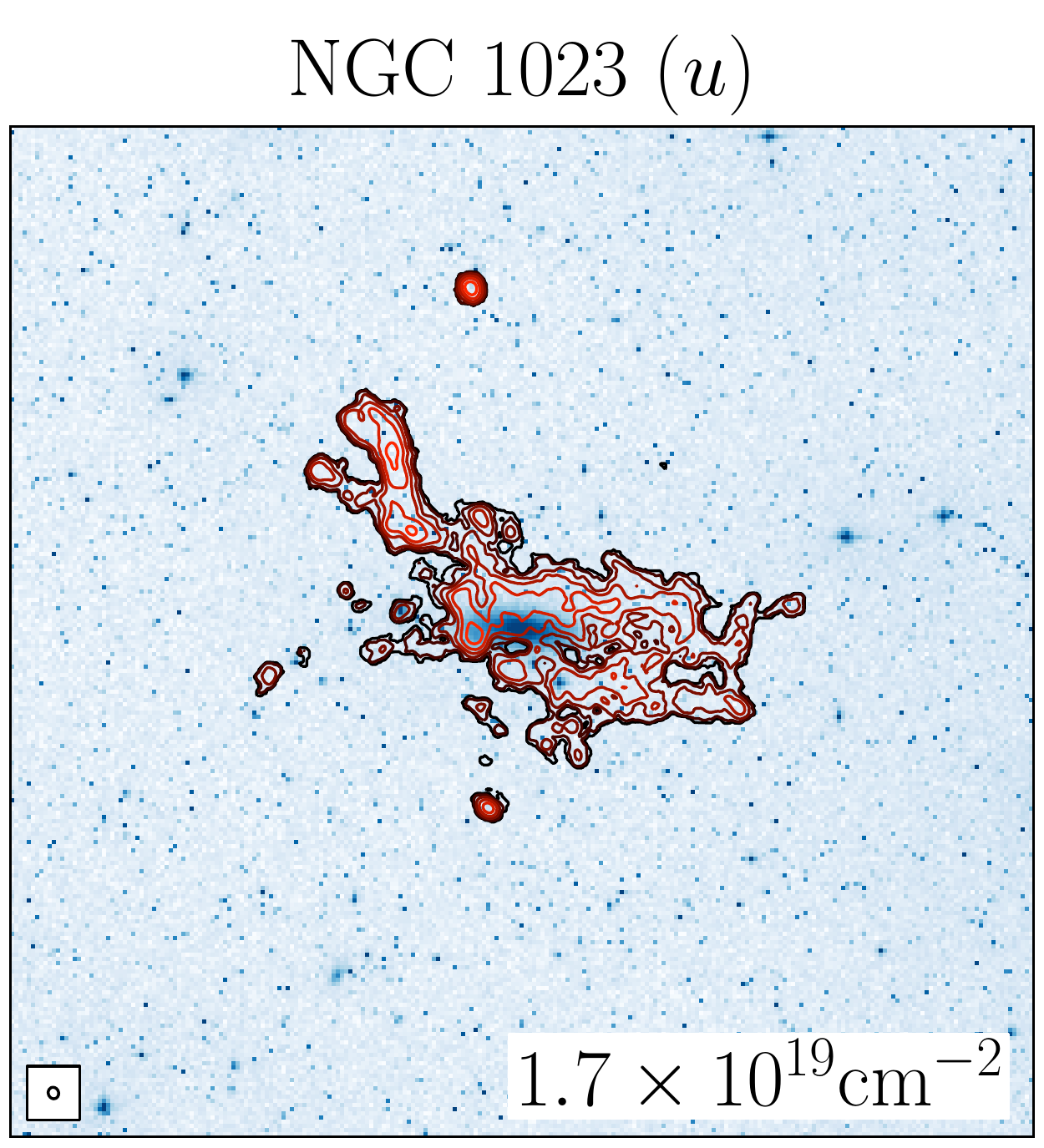} 
\includegraphics[width=50mm]{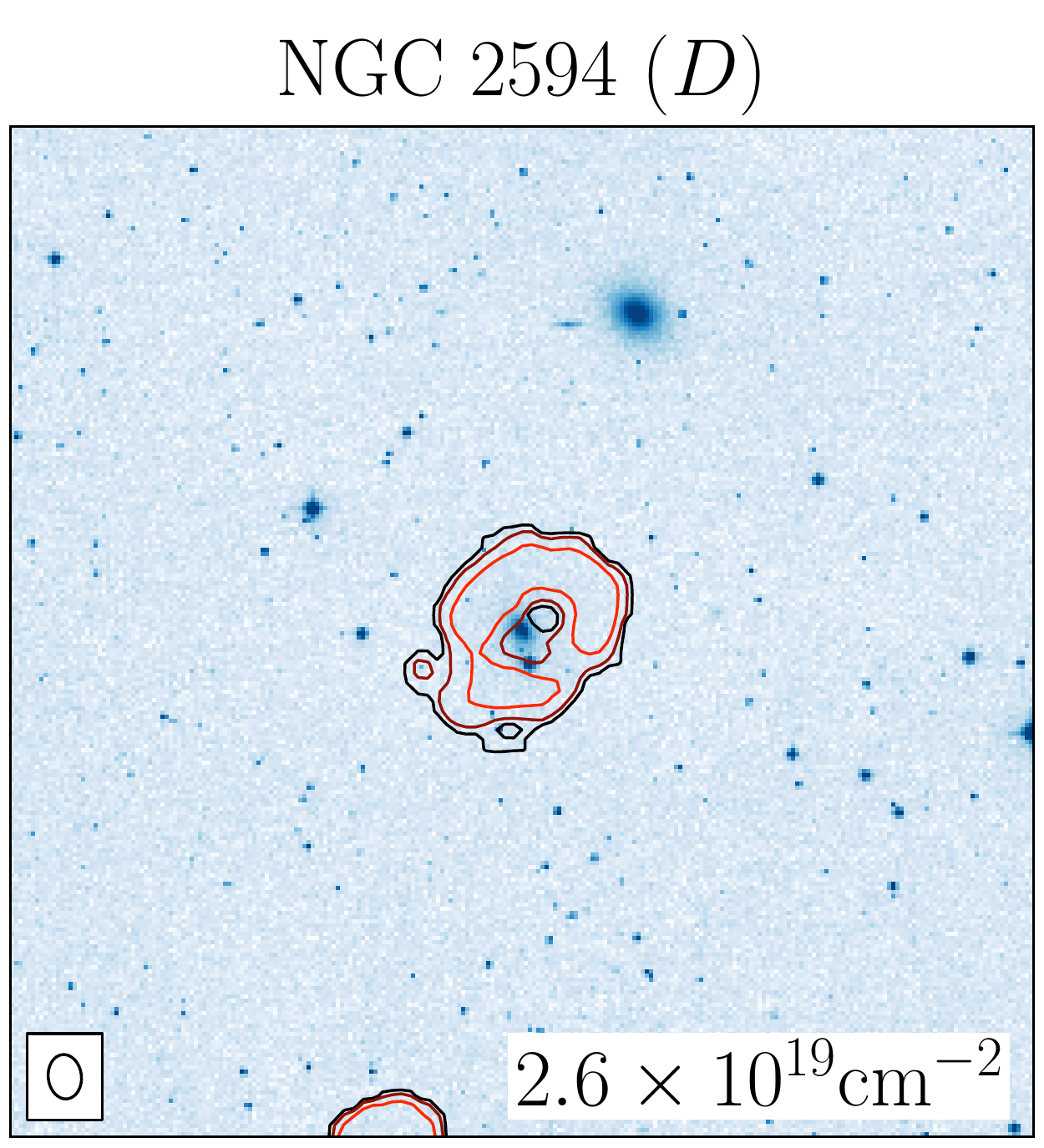} 
\includegraphics[width=50mm]{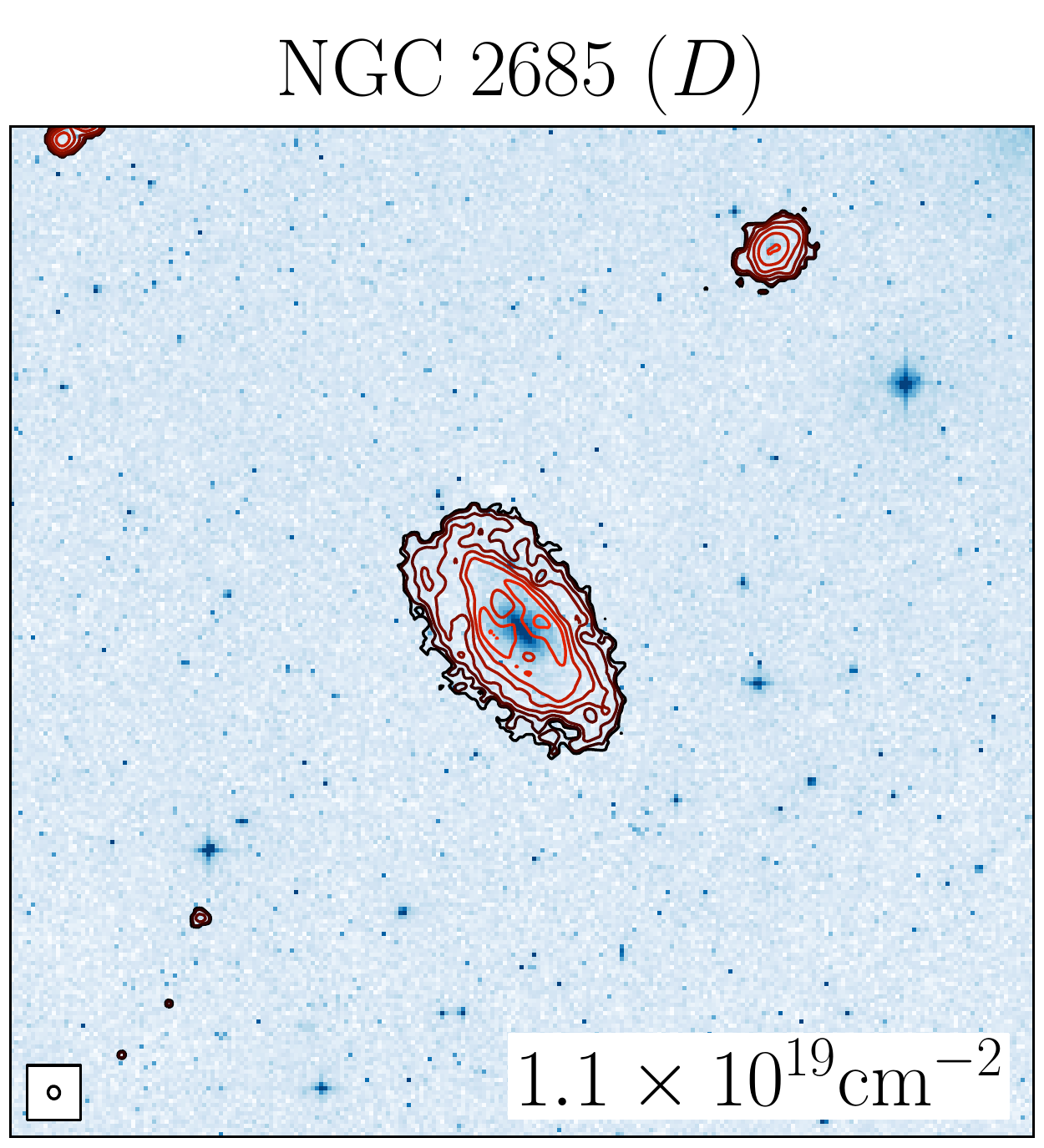} 
\includegraphics[width=50mm]{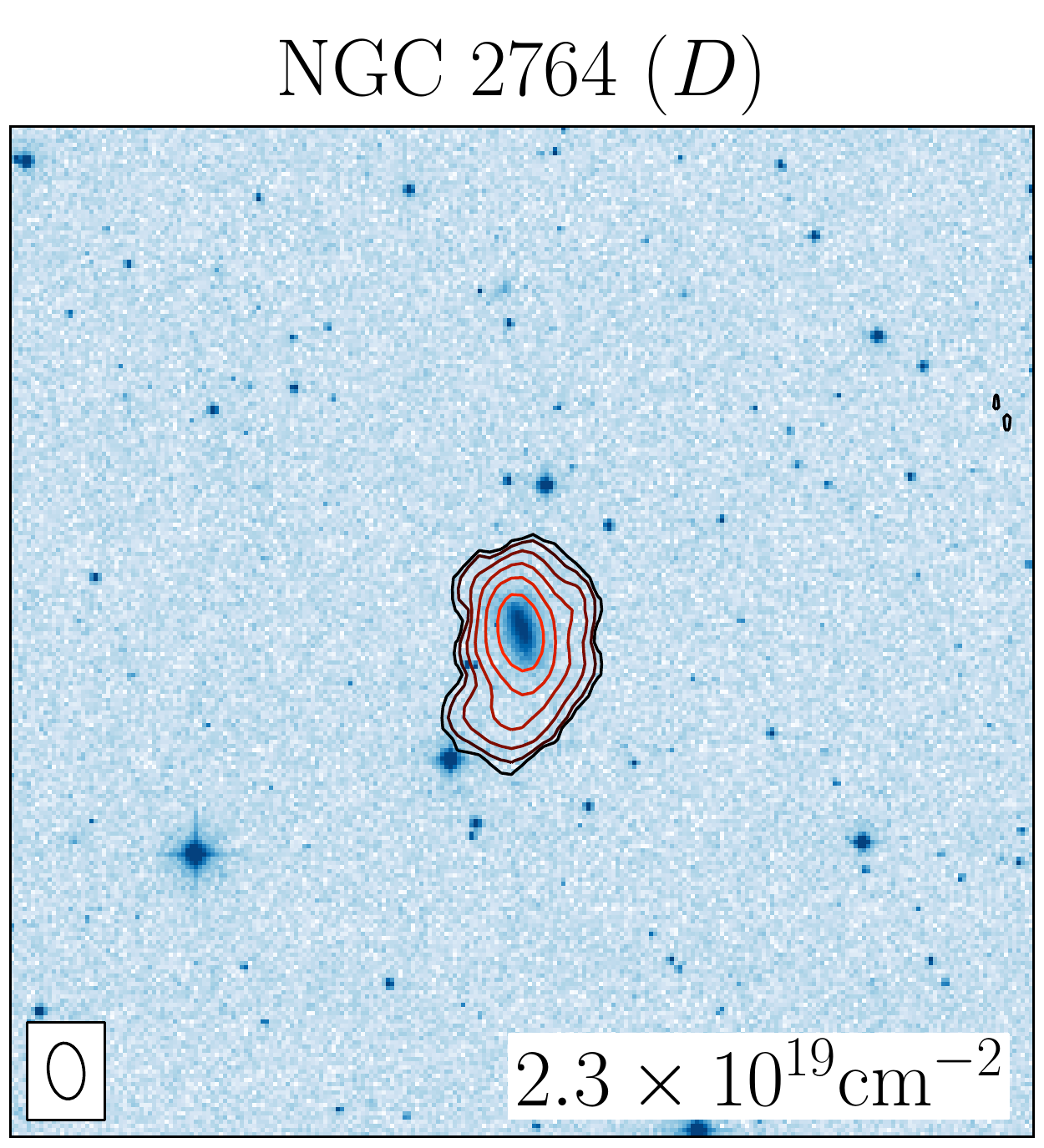} 
\includegraphics[width=50mm]{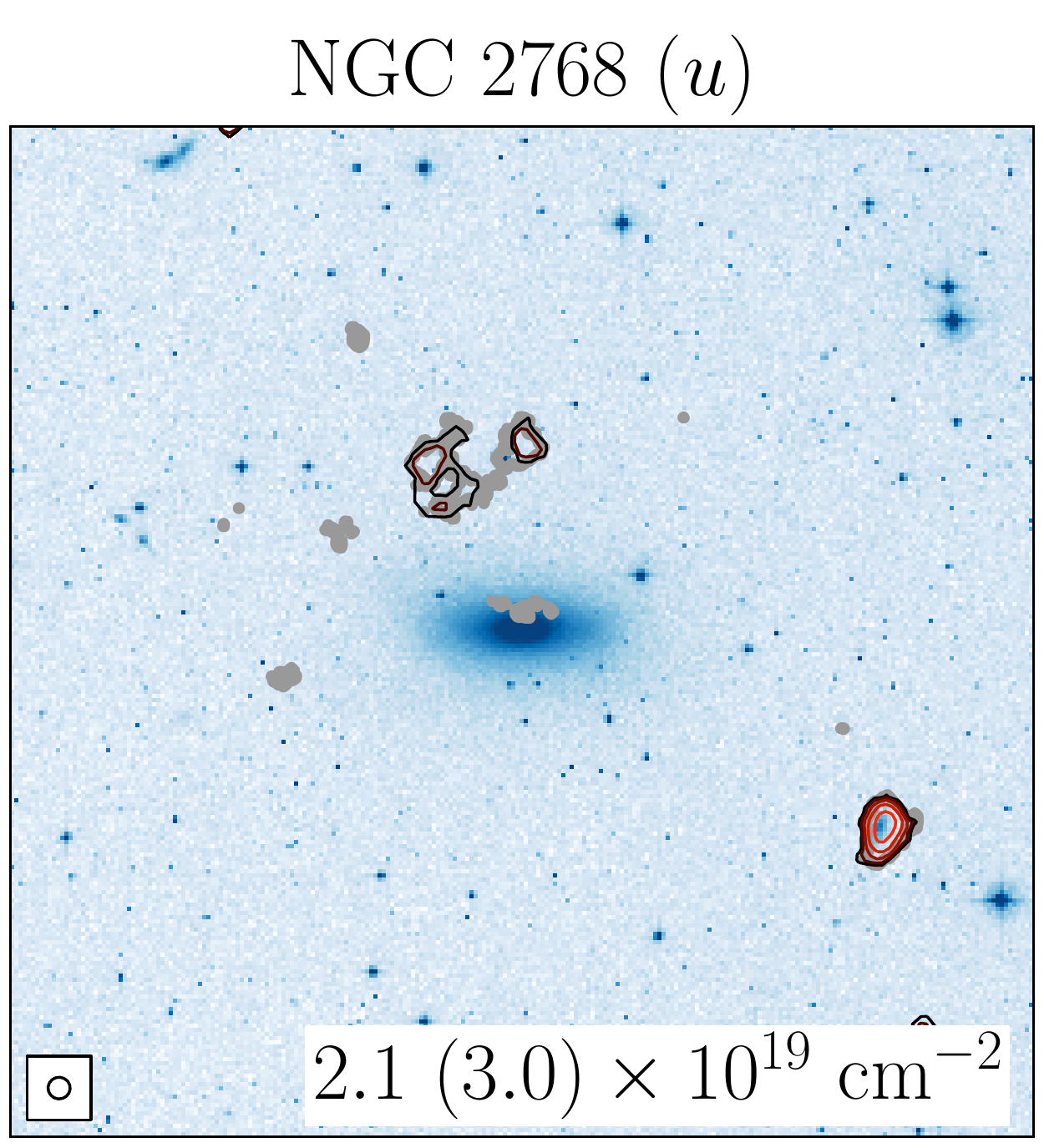} 
\includegraphics[width=50mm]{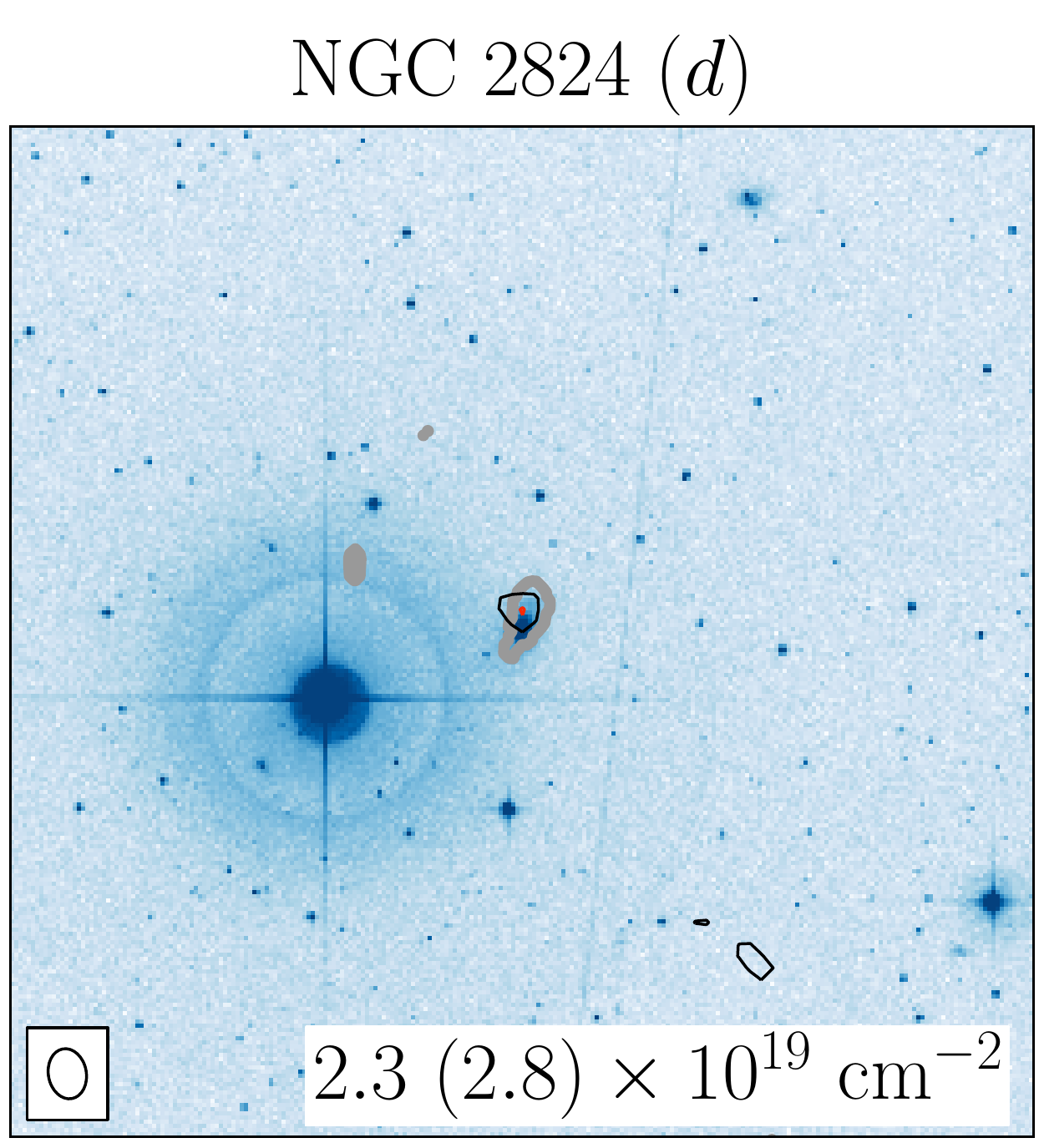} 
\includegraphics[width=50mm]{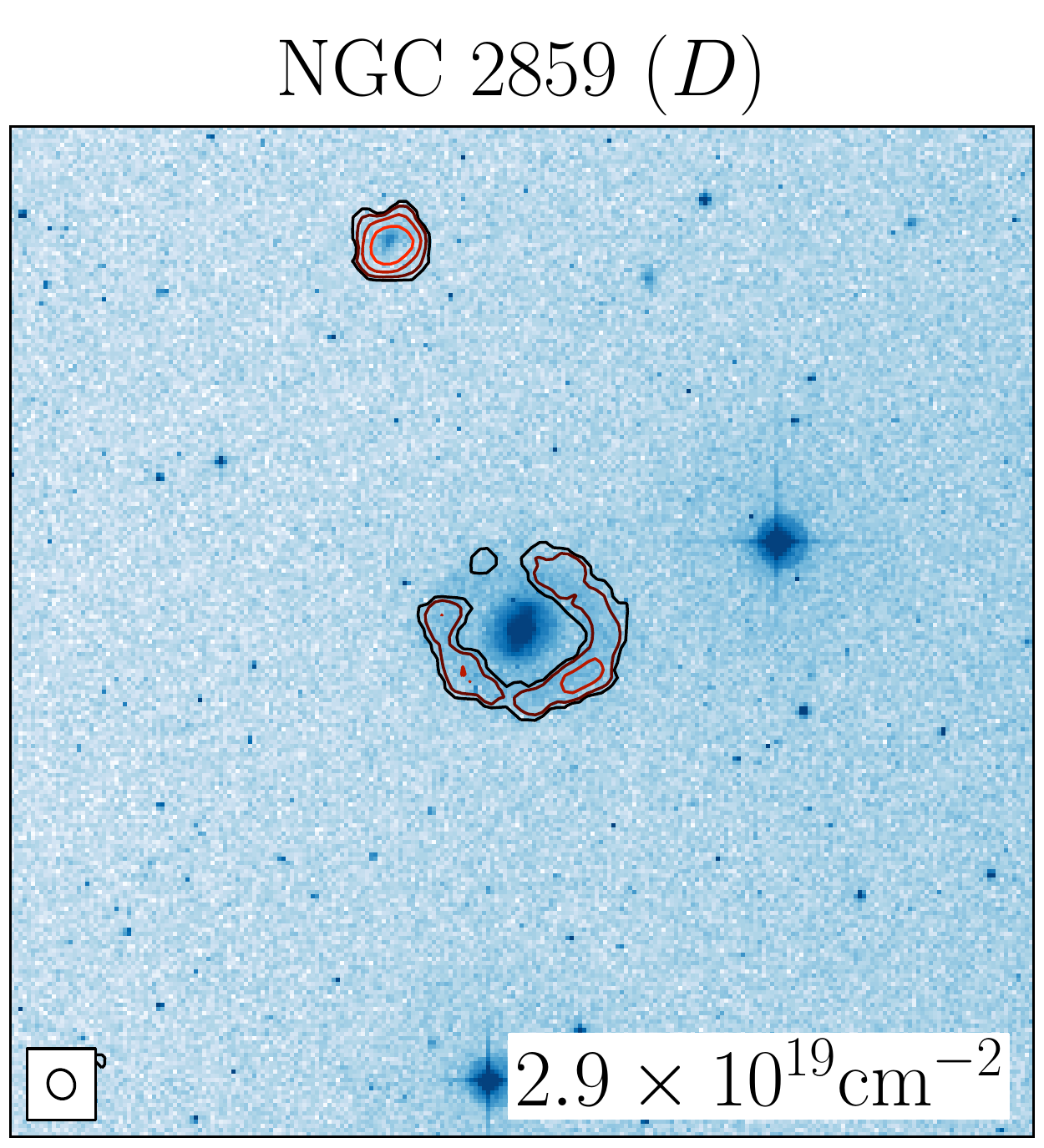} 
\includegraphics[width=50mm]{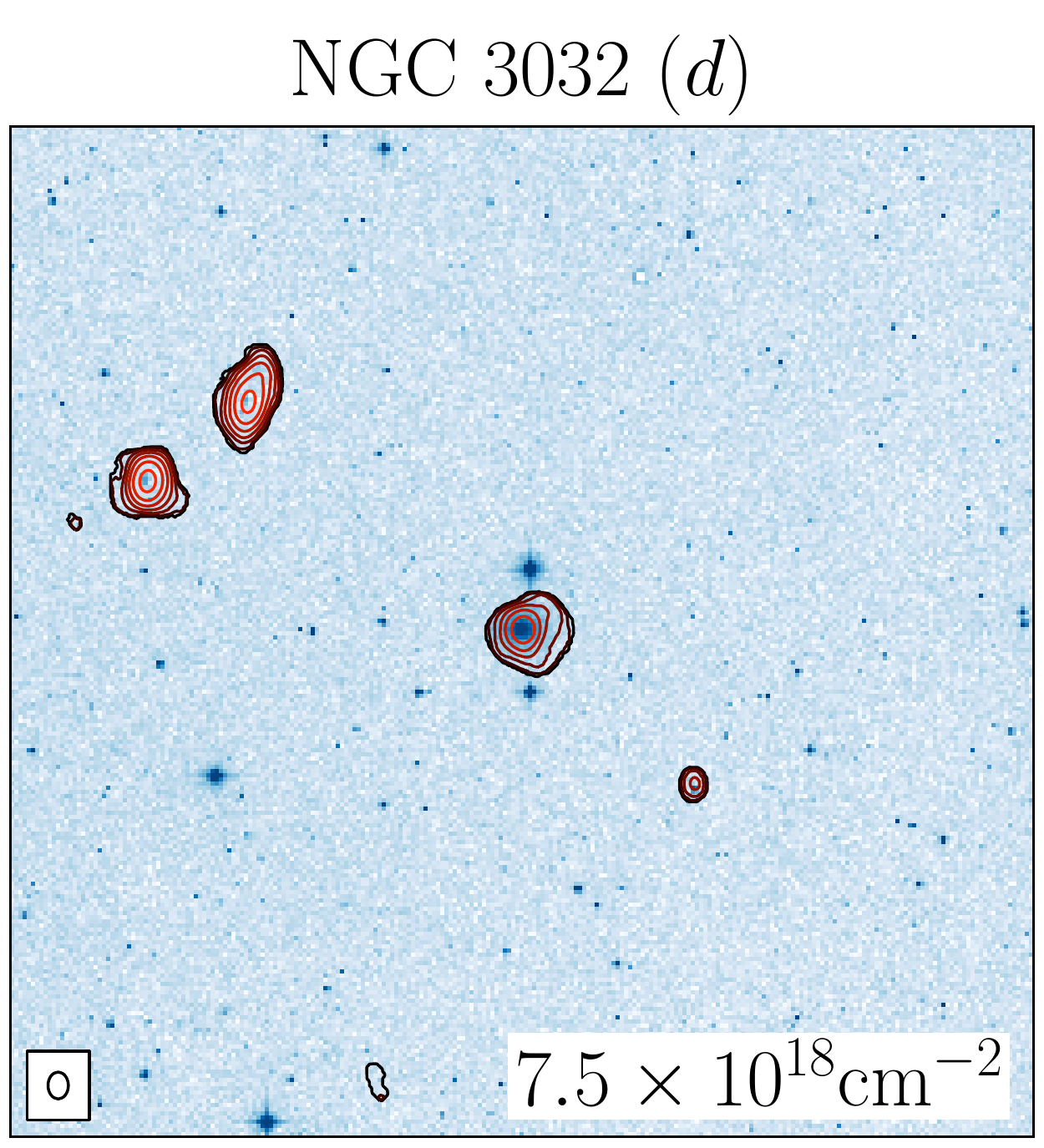} 
\includegraphics[width=50mm]{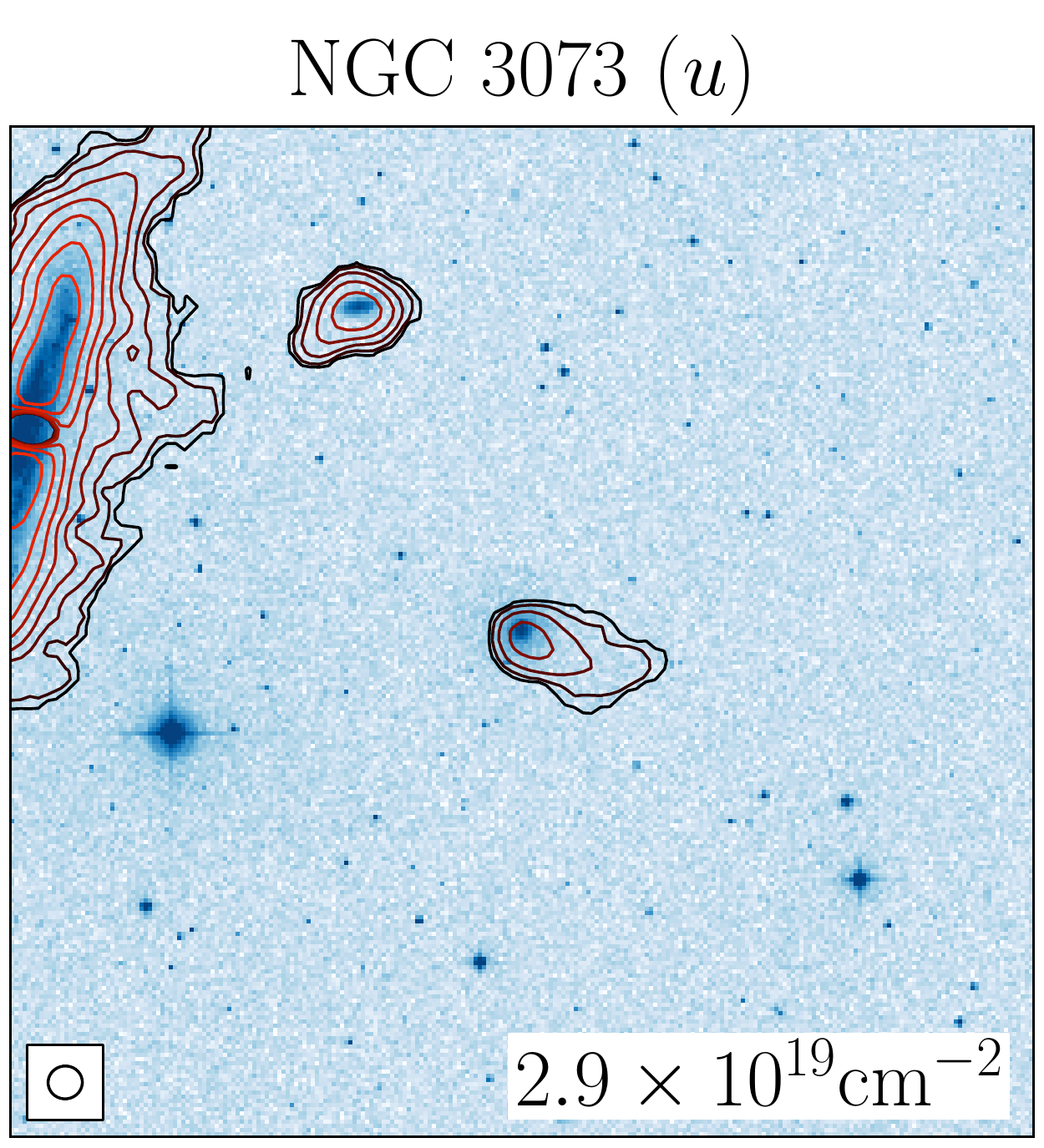} 
\includegraphics[width=50mm]{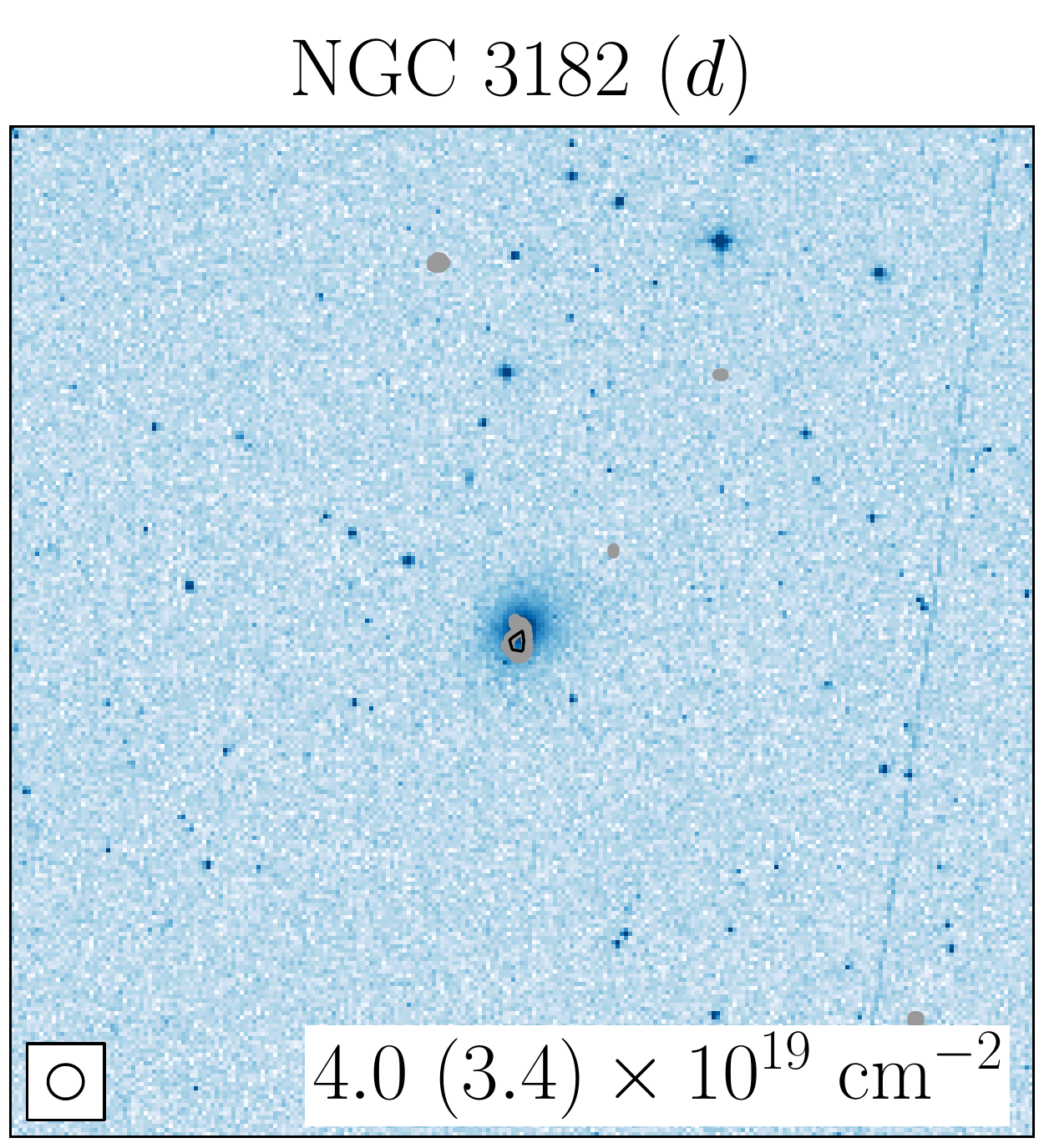} 
\includegraphics[width=50mm]{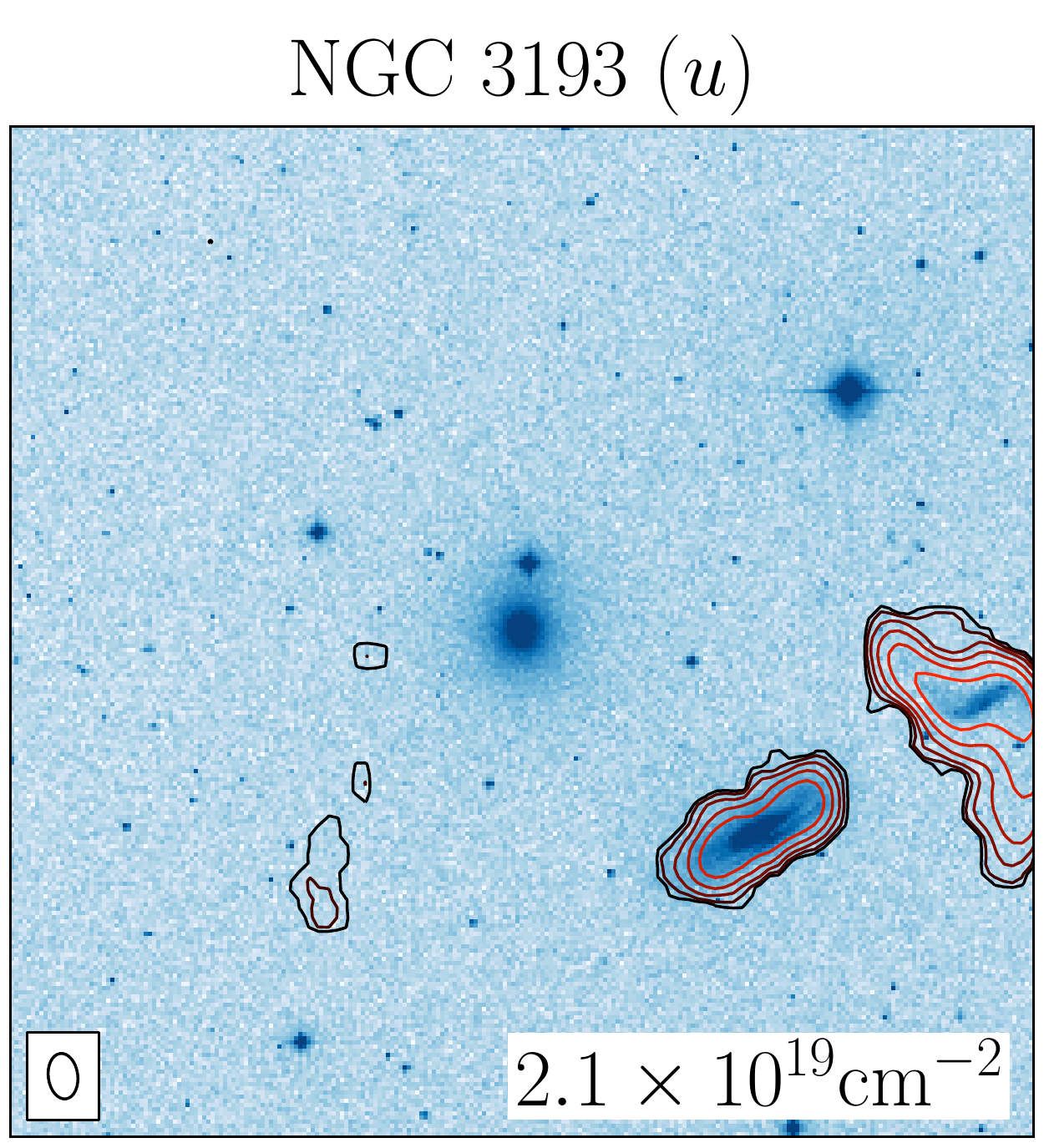} 
\caption{Total-\hi\ contours on top of DSS images of all \hi-detected galaxies. Images are centred on the target galaxy and show a fixed area of $180 \times 180$ kpc$^2$ at the galaxy distance. Contour levels are \Nhi=$N_0 \times 2^n$ with $N_0$ given at the bottom of each image and $n=0,1,2,...$ Contours are coloured black to red, faint to bright. The beam is shown on the bottom left. In the cases where the $R01$ image reveals \hi\ missing from the standard image we show its lowest contour with a thick grey line (contour level in parenthesis).}
\label{fig:gallery}
\end{figure*}

\addtocounter{figure}{-1}
\begin{figure*}
\includegraphics[width=50mm]{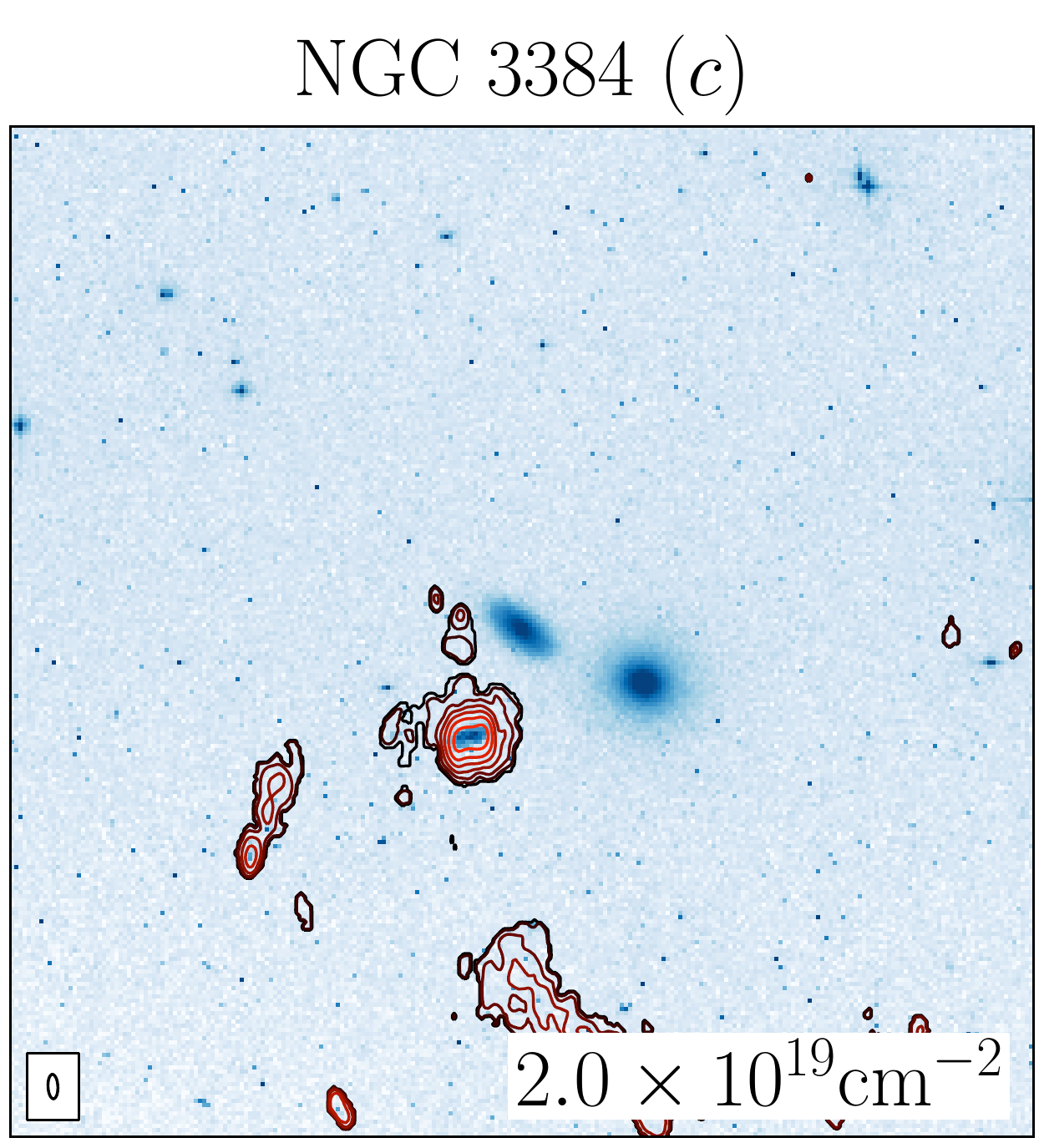} 
\includegraphics[width=50mm]{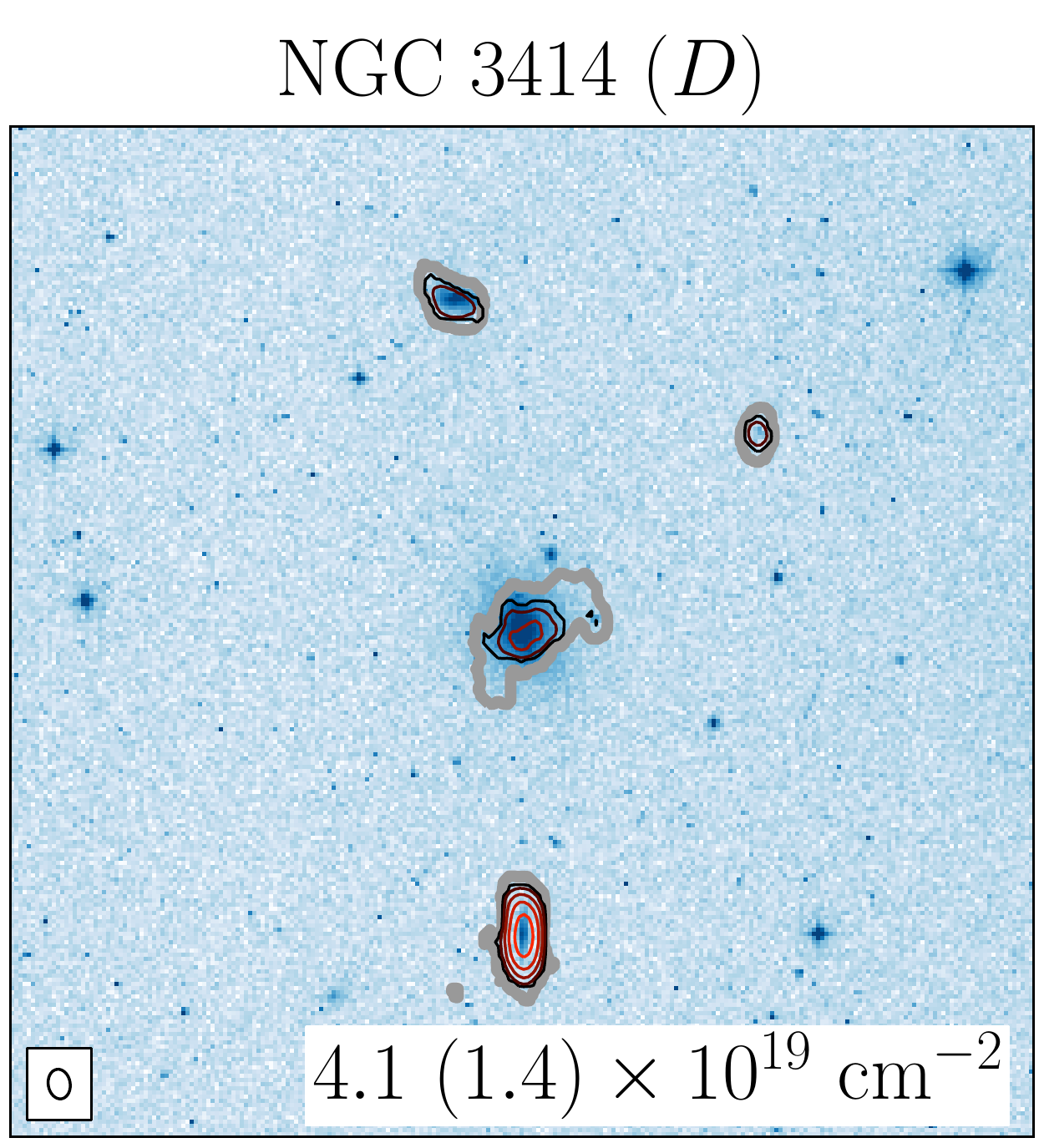} 
\includegraphics[width=50mm]{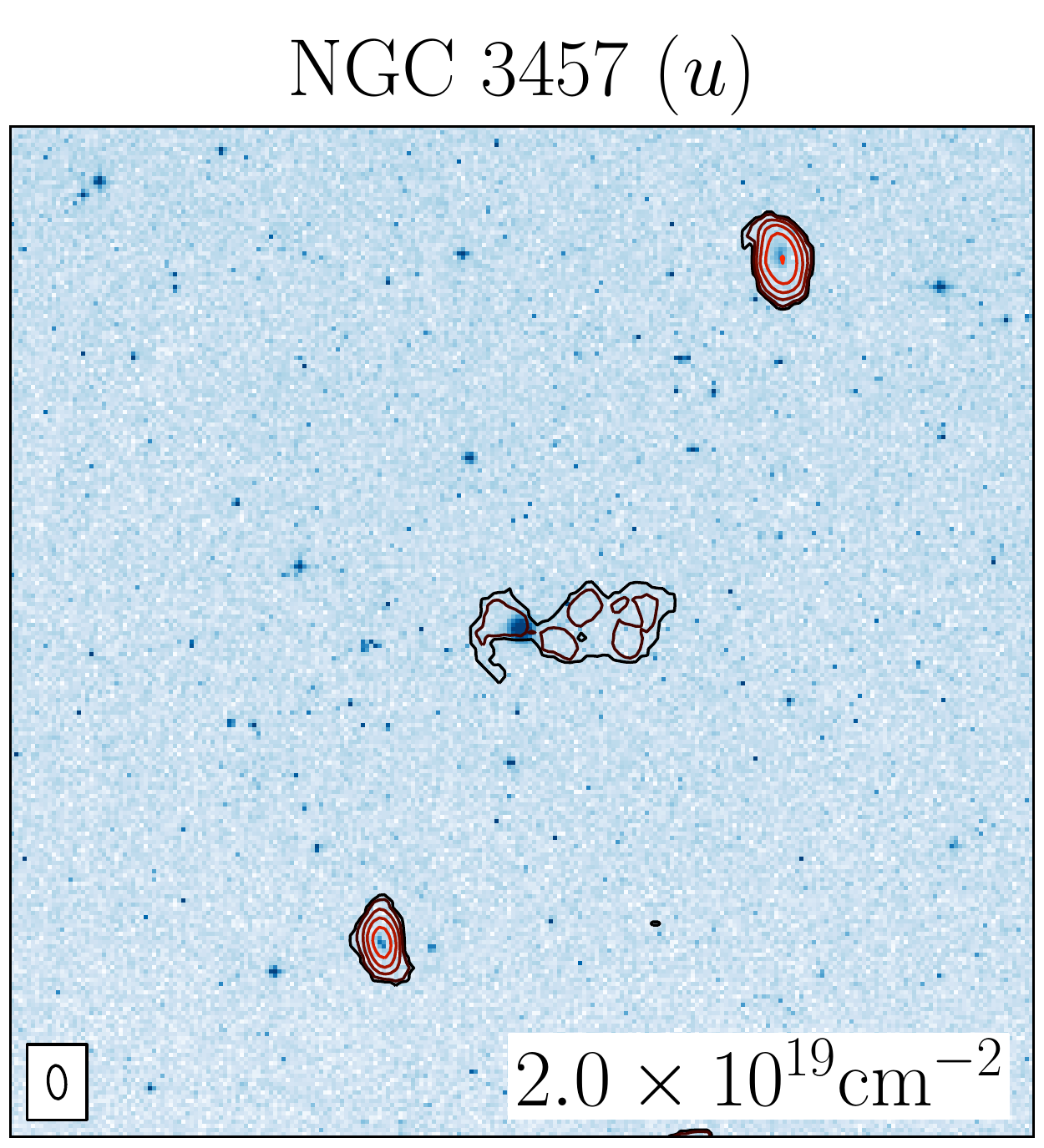} 
\includegraphics[width=50mm]{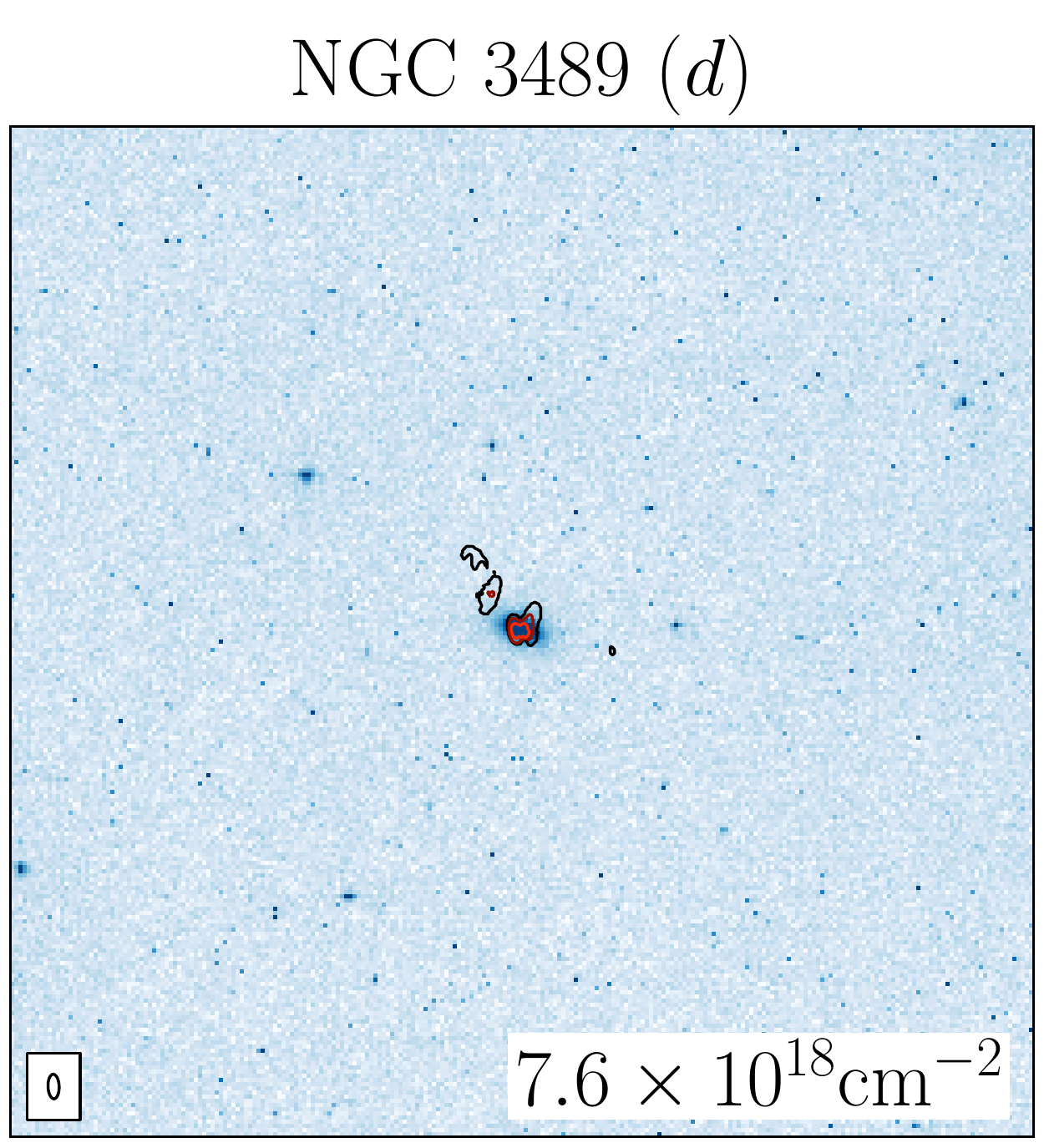} 
\includegraphics[width=50mm]{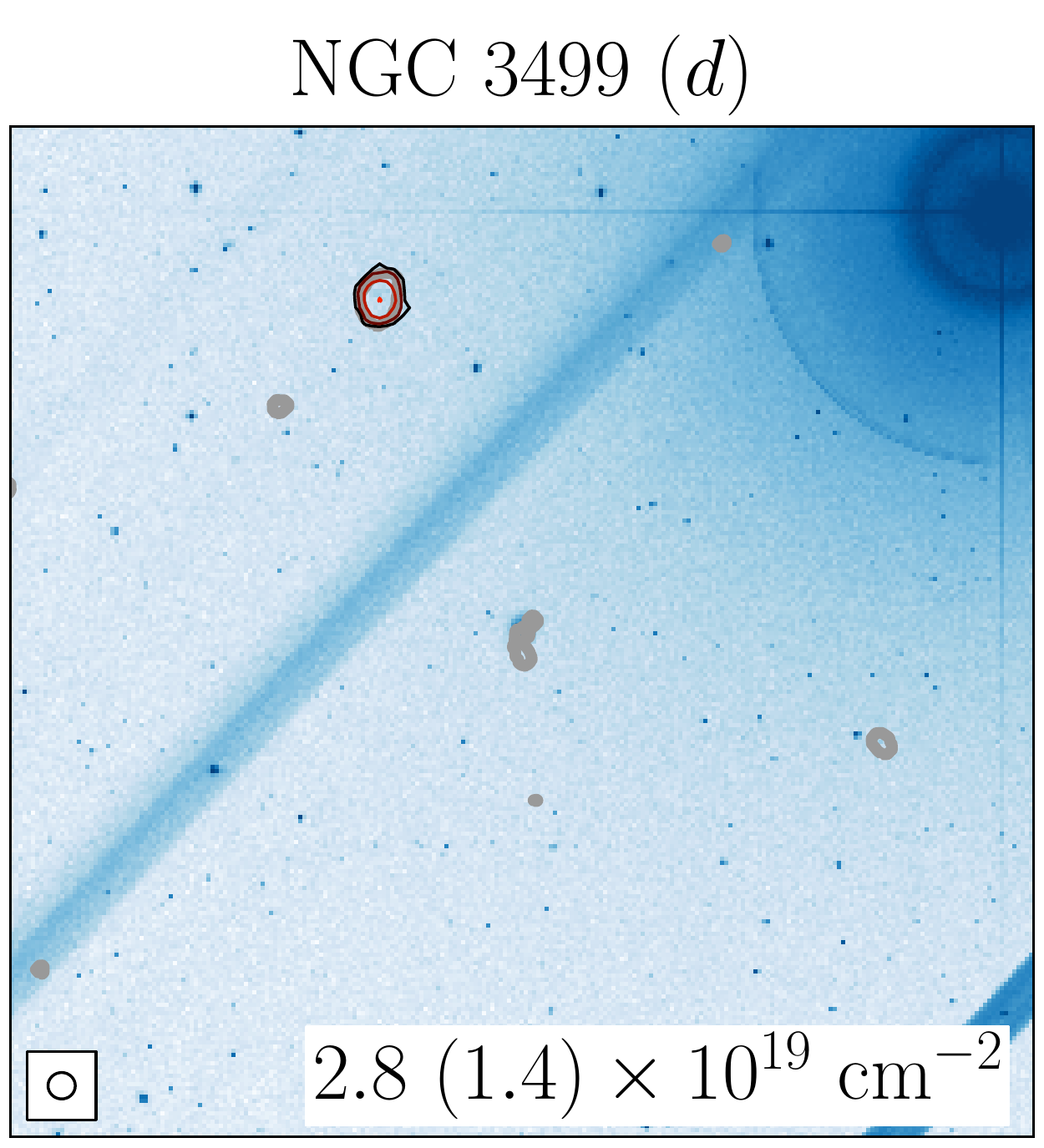} 
\includegraphics[width=50mm]{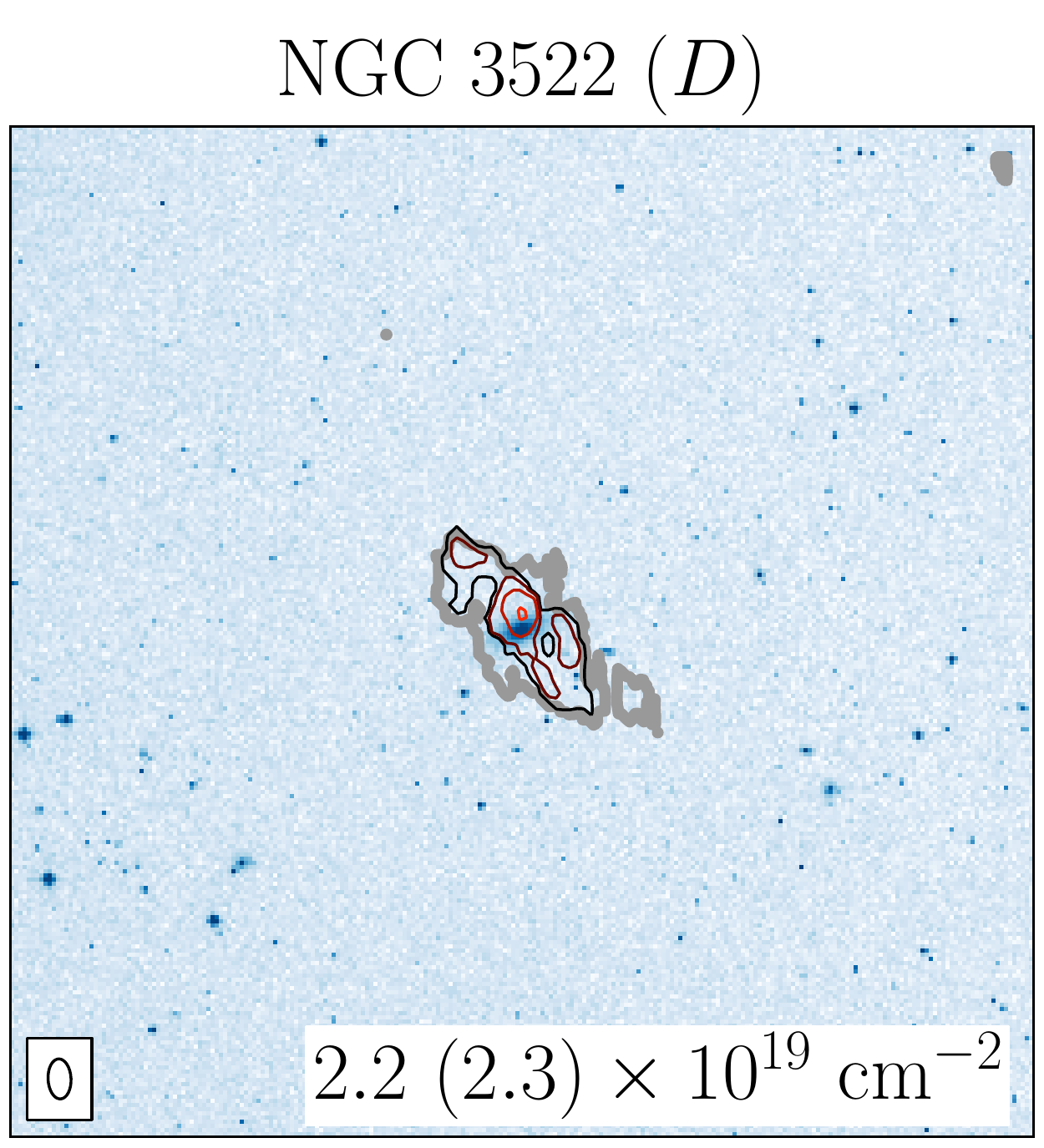} 
\includegraphics[width=50mm]{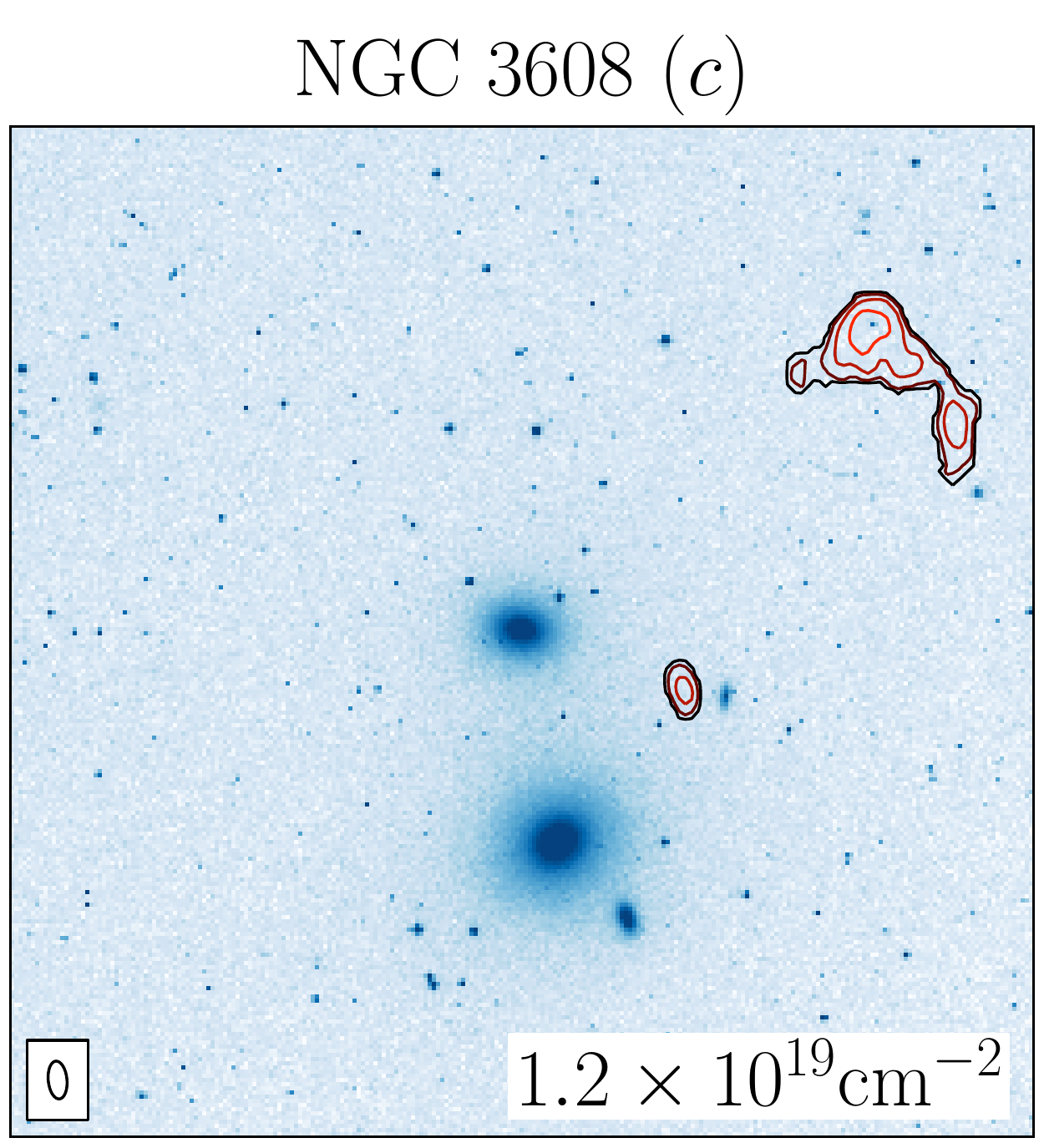} 
\includegraphics[width=50mm]{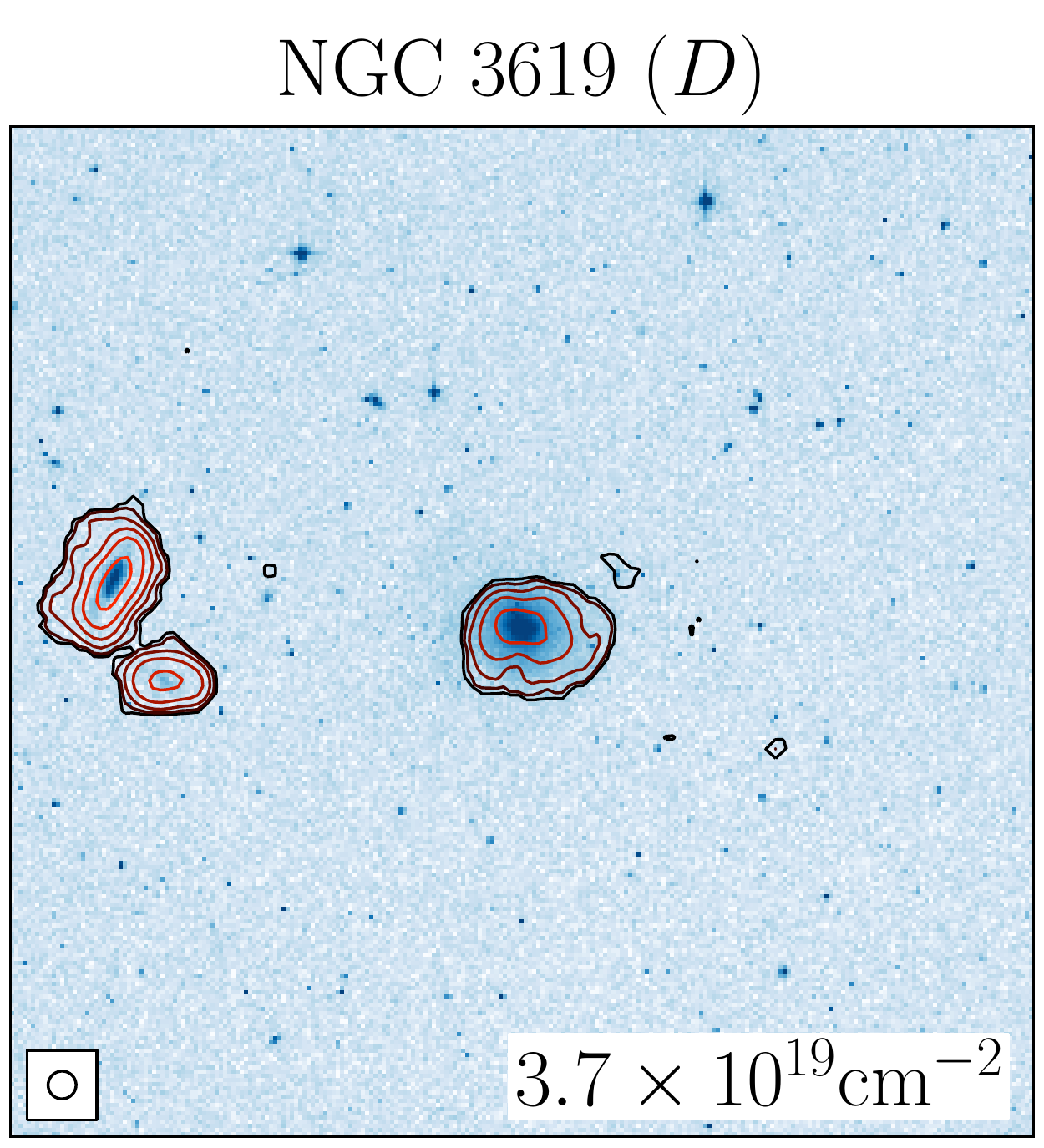} 
\includegraphics[width=50mm]{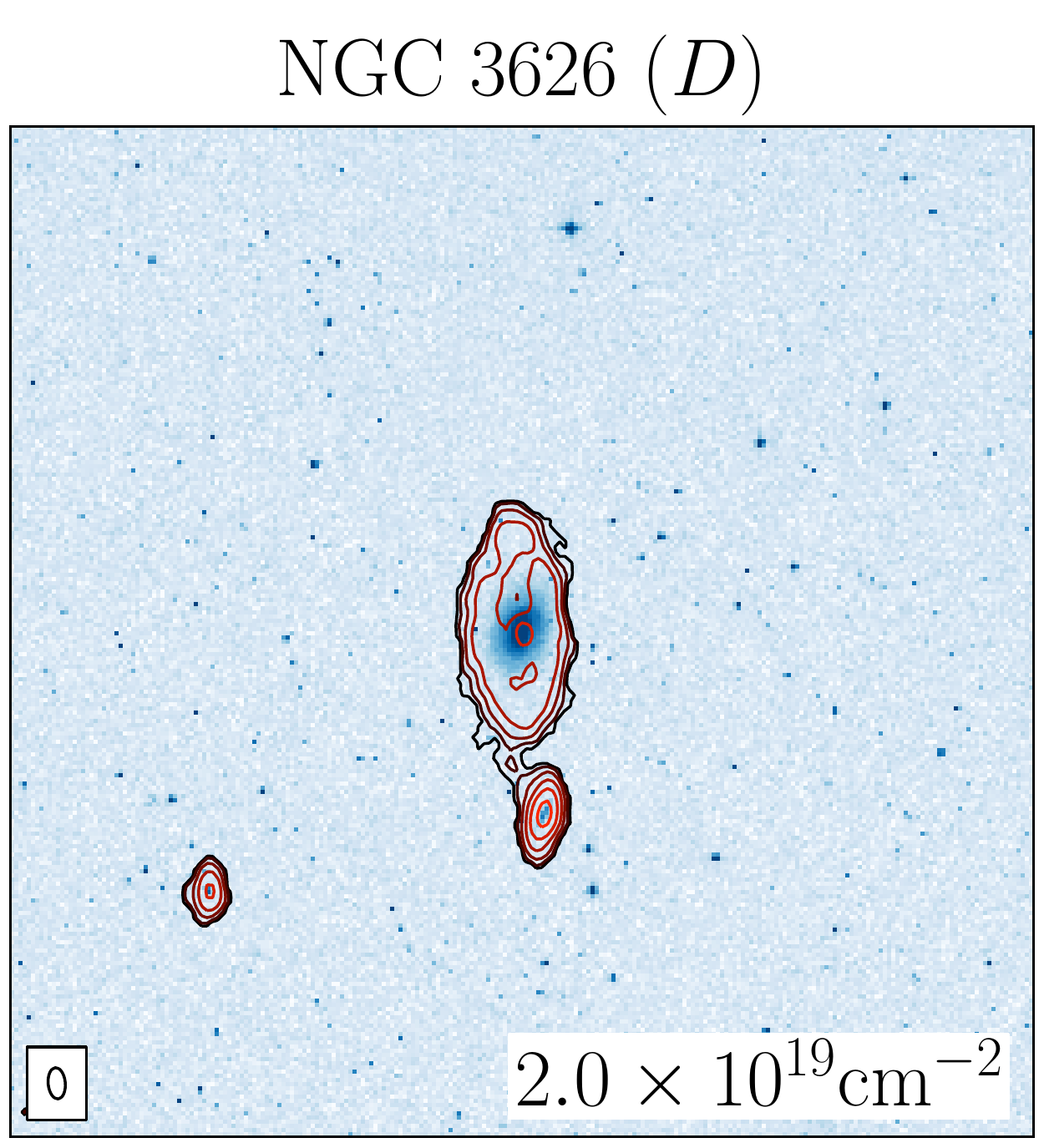} 
\includegraphics[width=50mm]{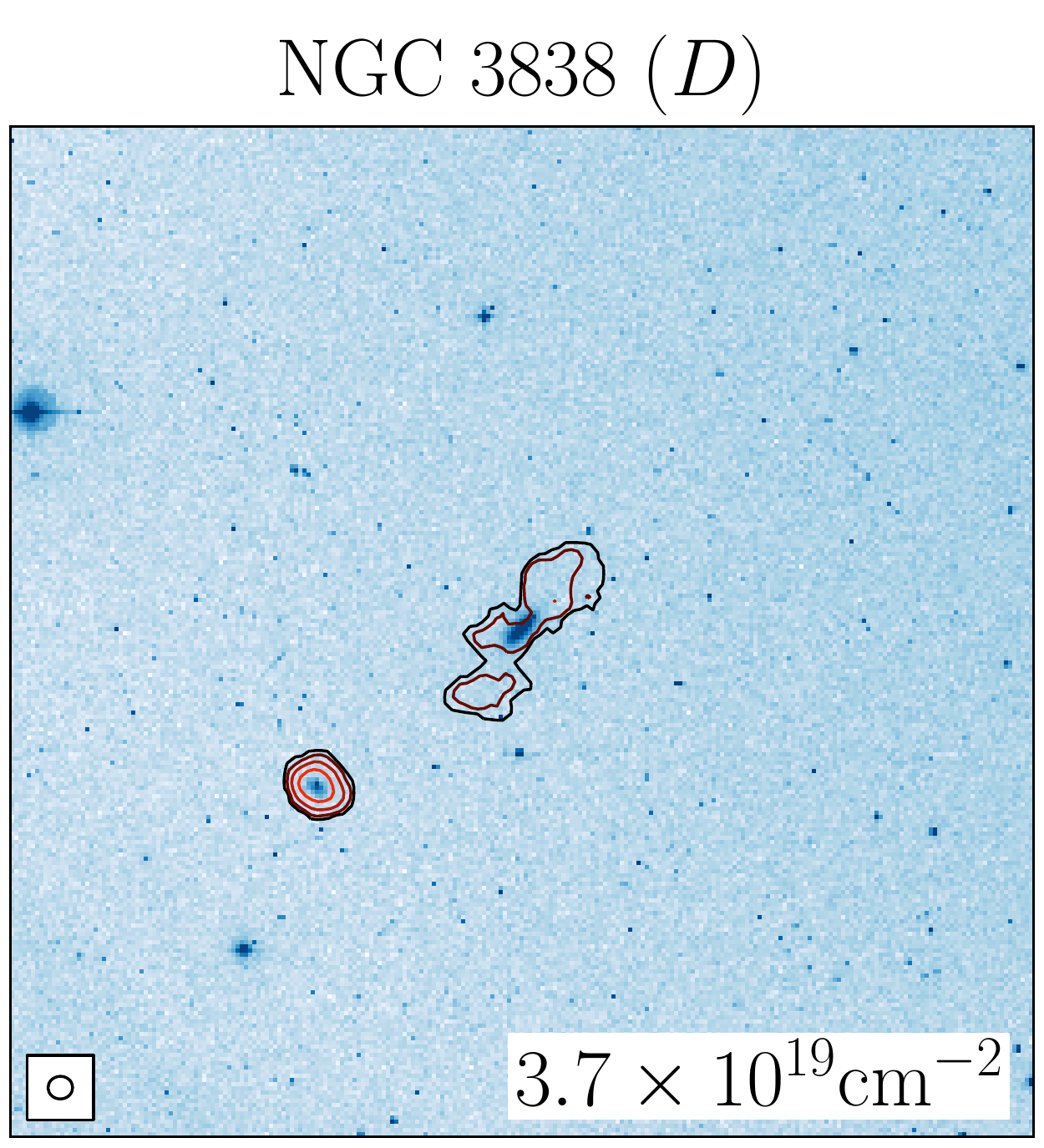} 
\includegraphics[width=50mm]{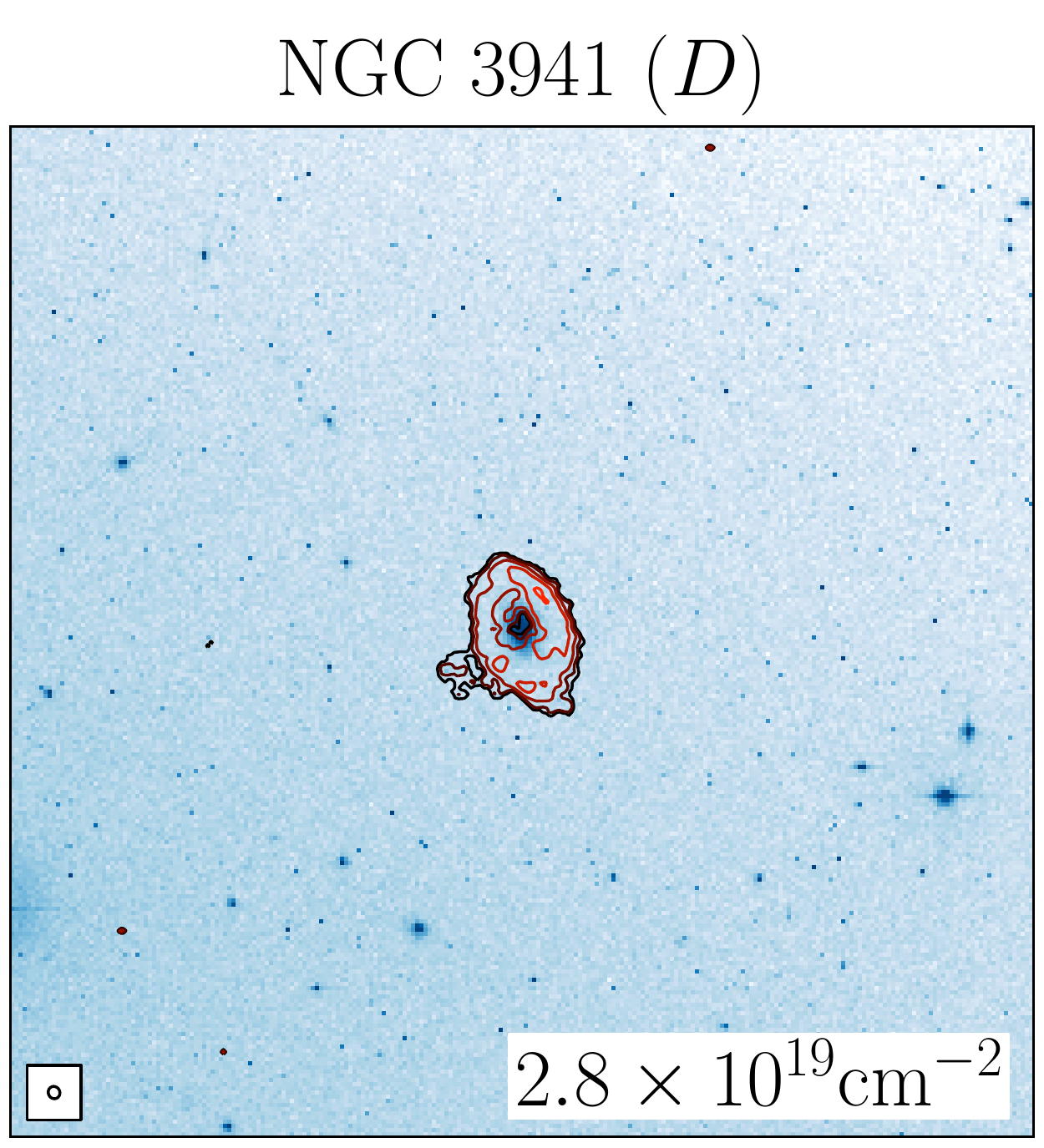} 
\includegraphics[width=50mm]{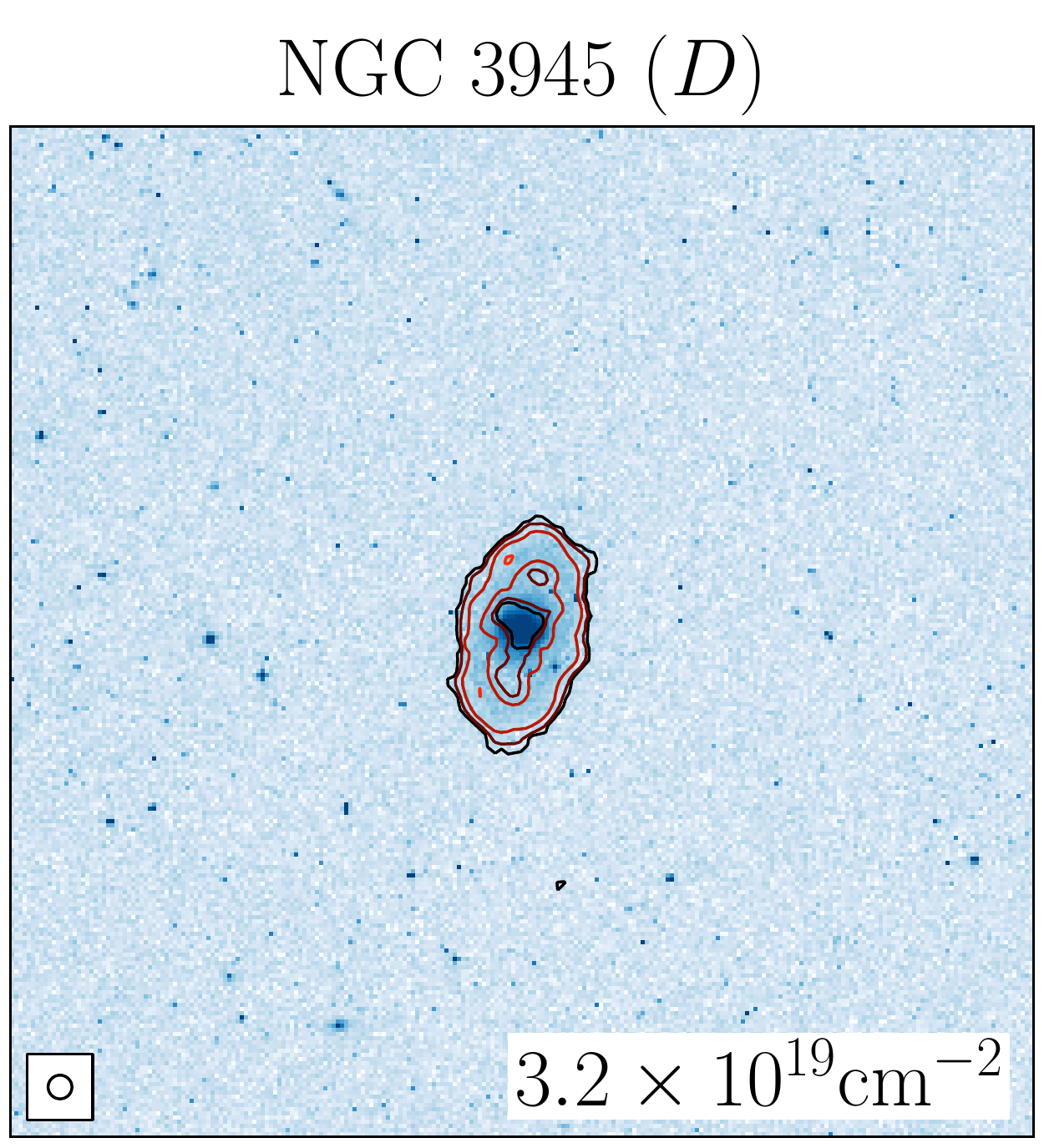} 
\caption{\it Continued \rm}
\end{figure*}

\addtocounter{figure}{-1}
\begin{figure*}
\includegraphics[width=50mm]{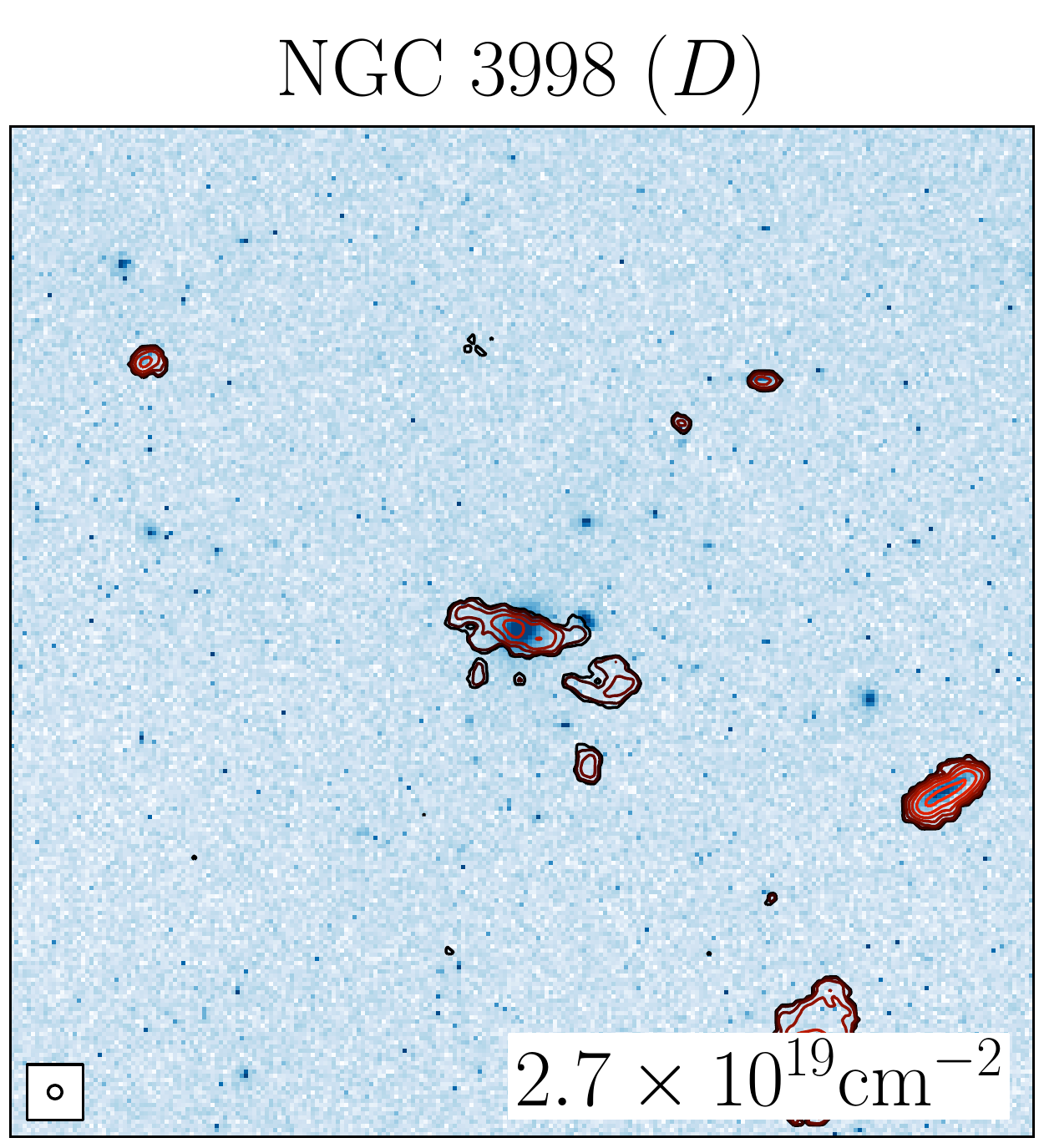} 
\includegraphics[width=50mm]{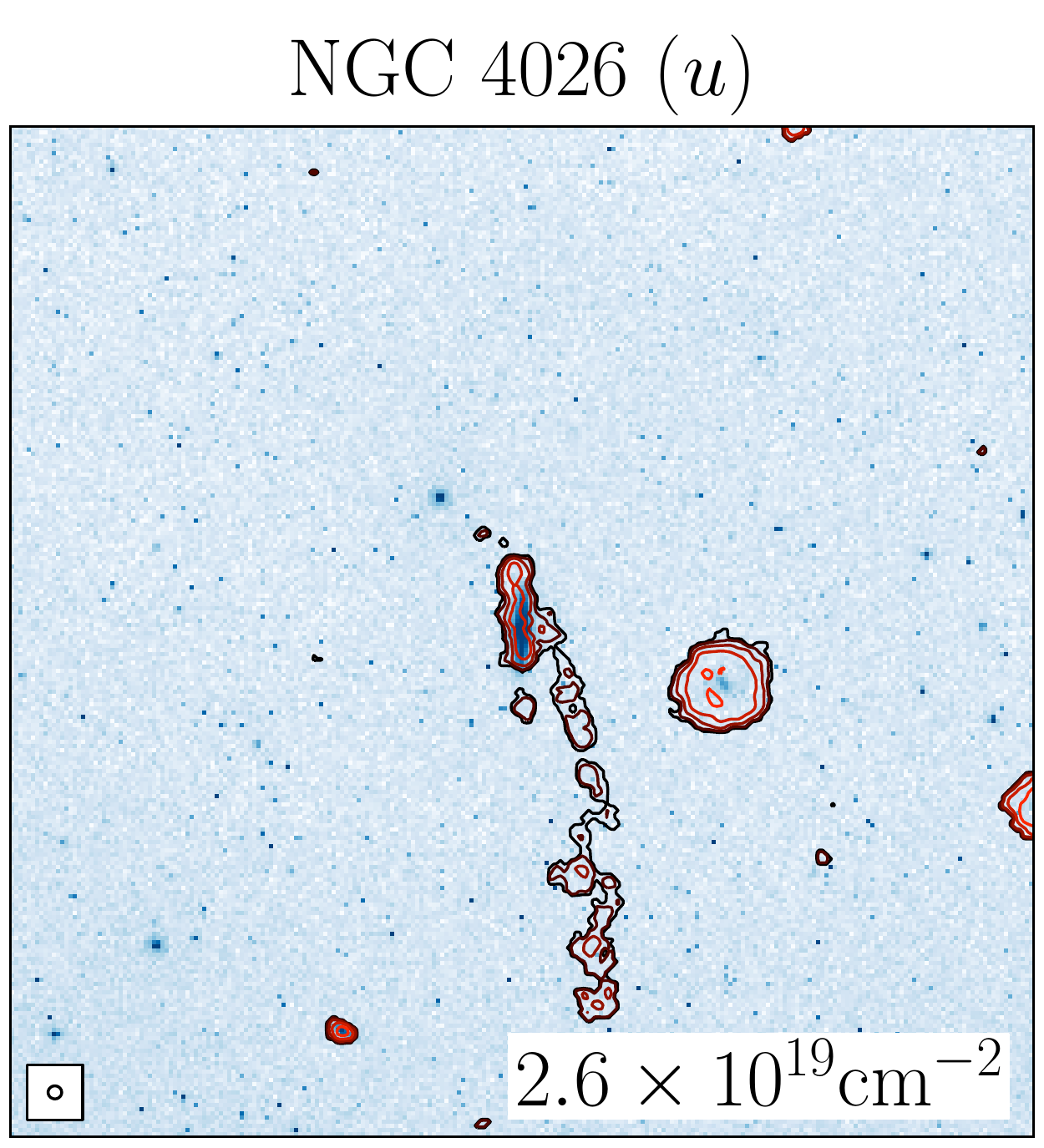} 
\includegraphics[width=50mm]{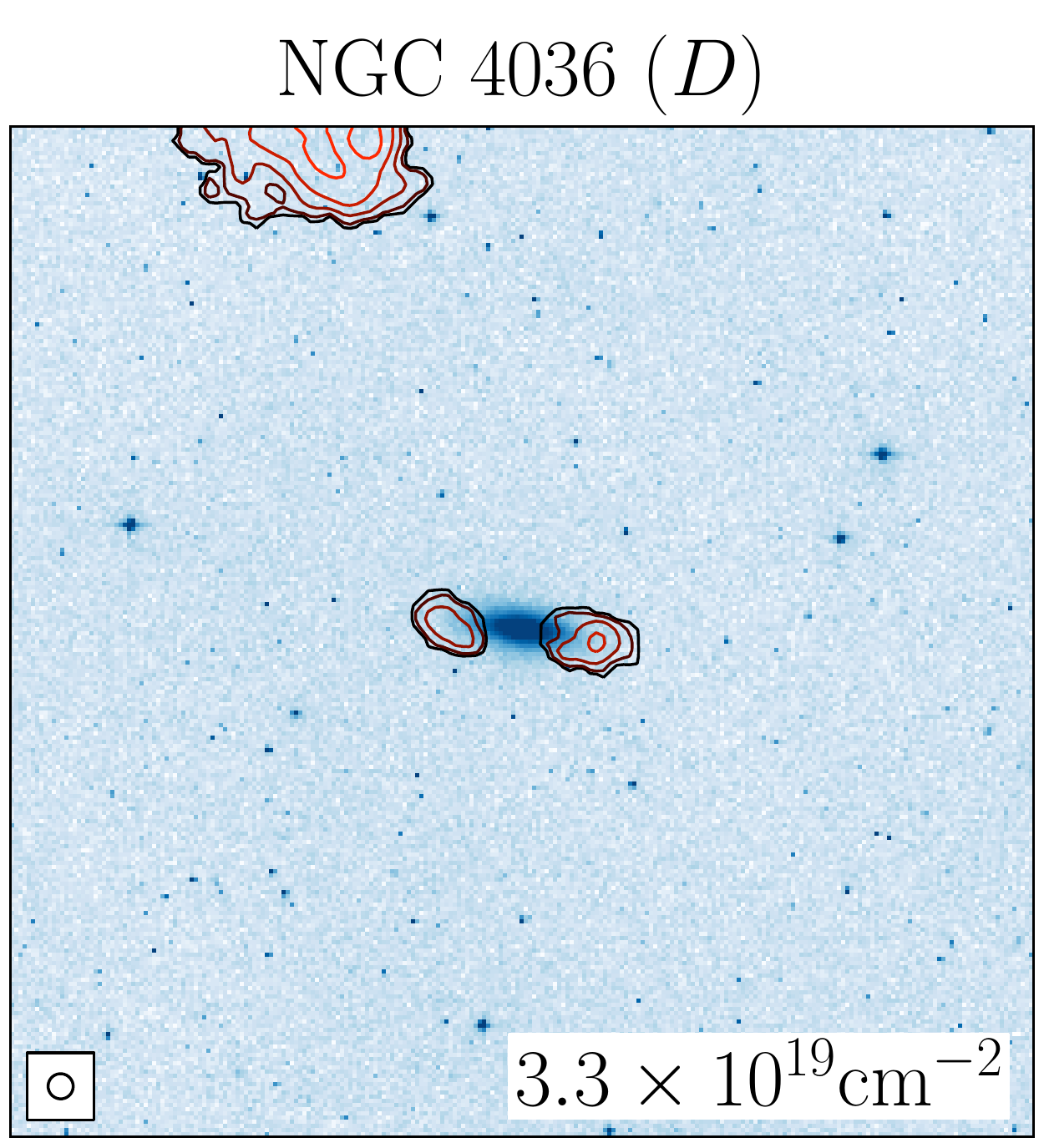} 
\includegraphics[width=50mm]{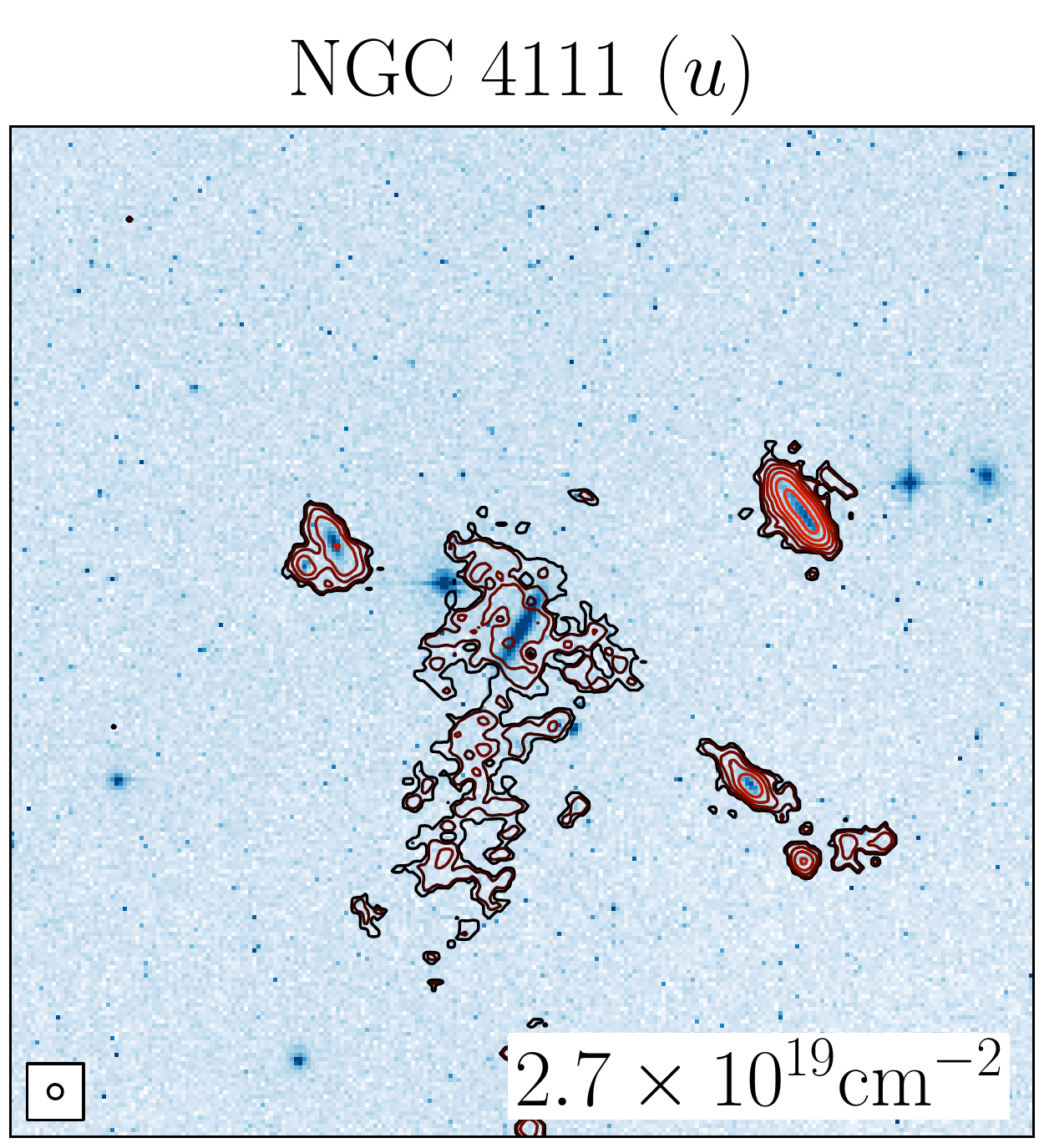} 
\includegraphics[width=50mm]{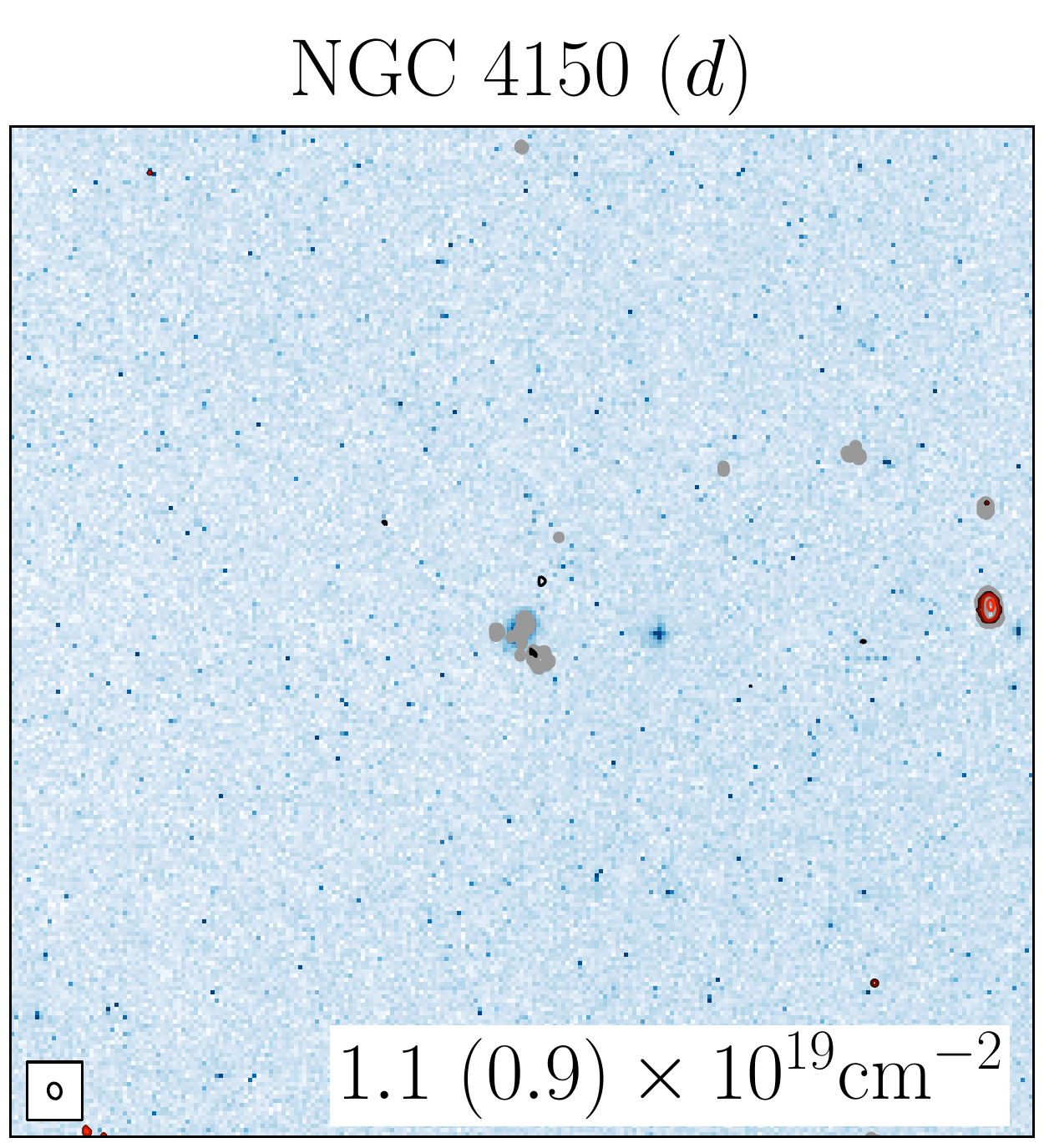} 
\includegraphics[width=50mm]{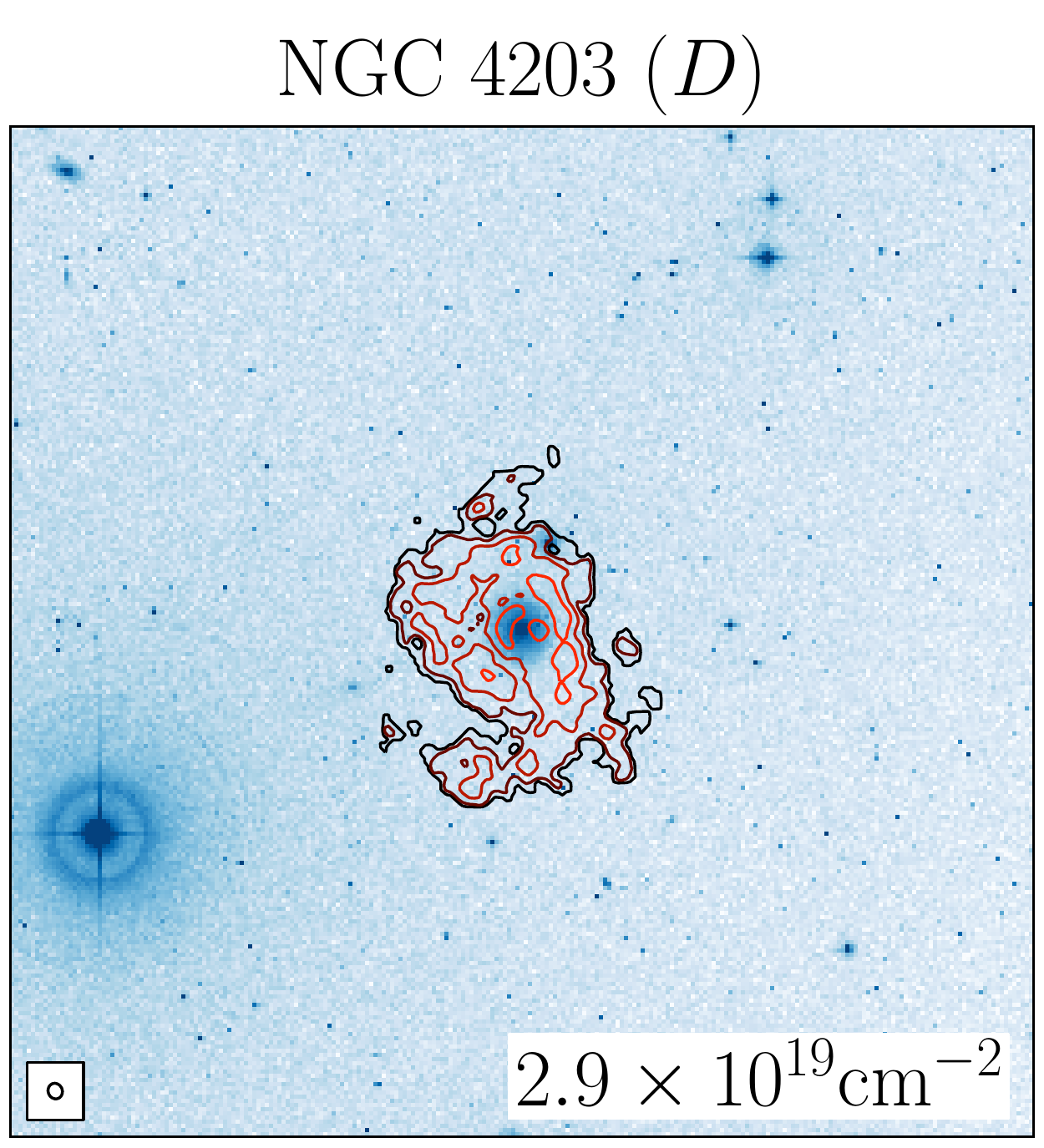} 
\includegraphics[width=50mm]{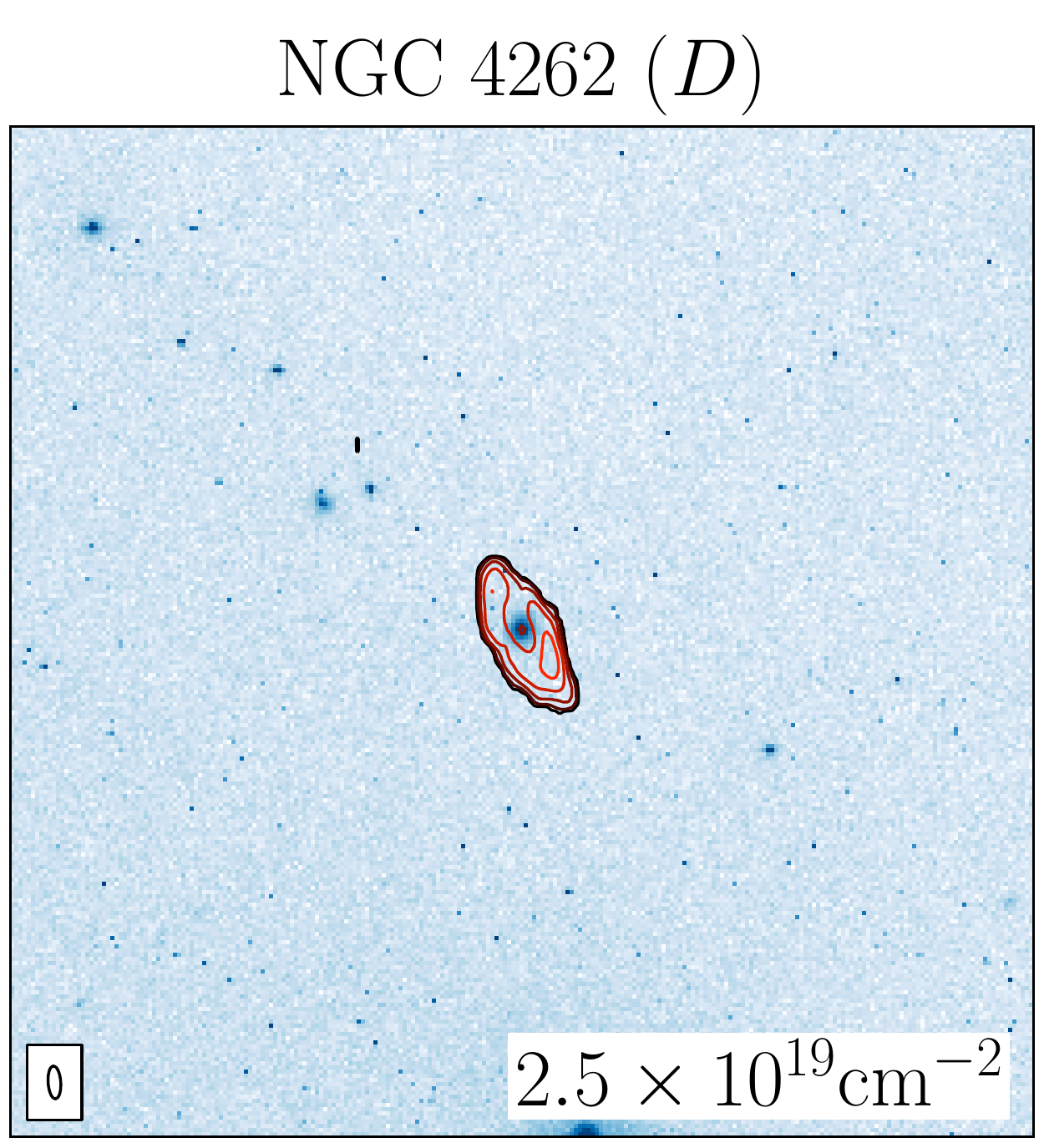} 
\includegraphics[width=50mm]{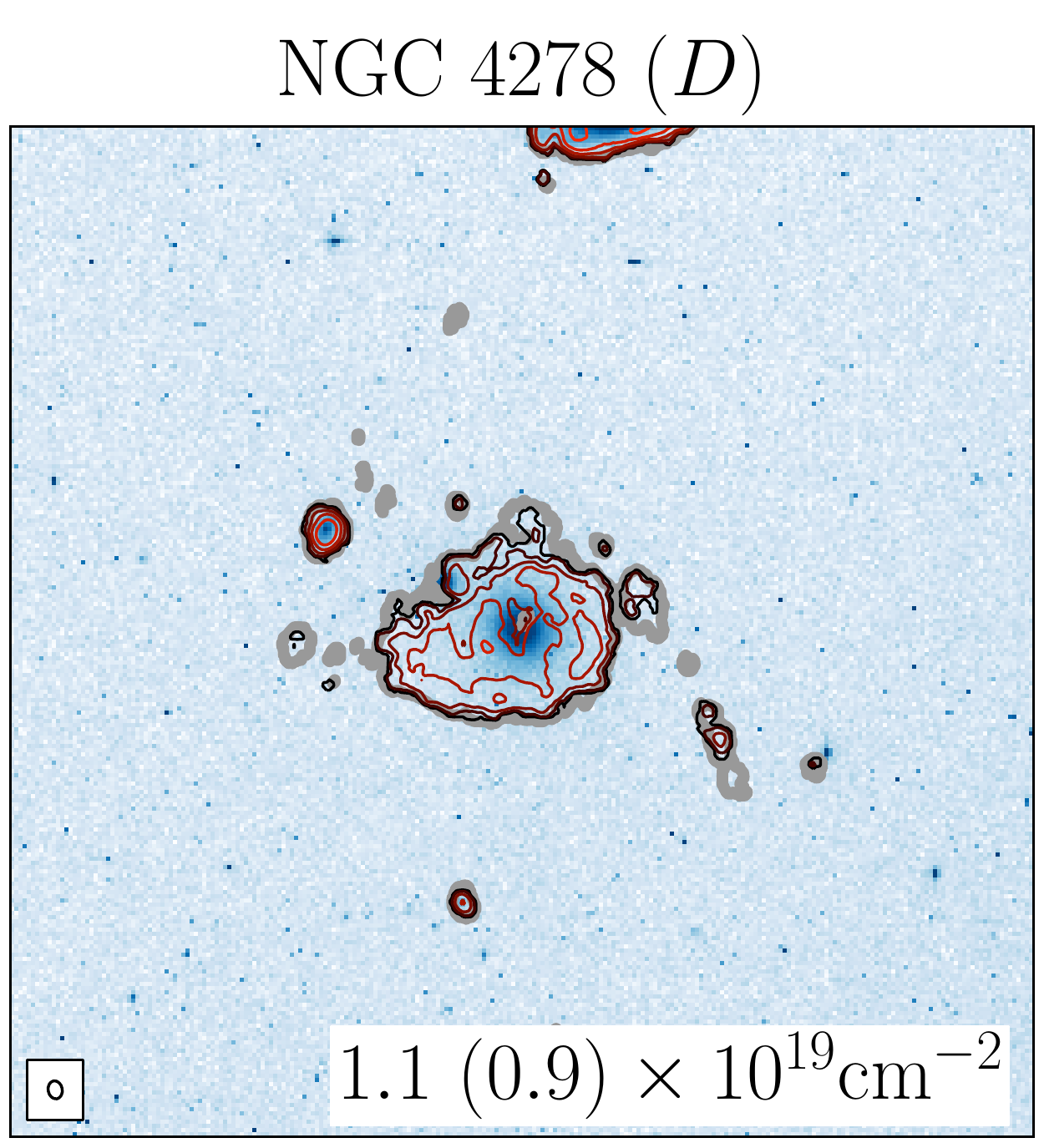} 
\includegraphics[width=50mm]{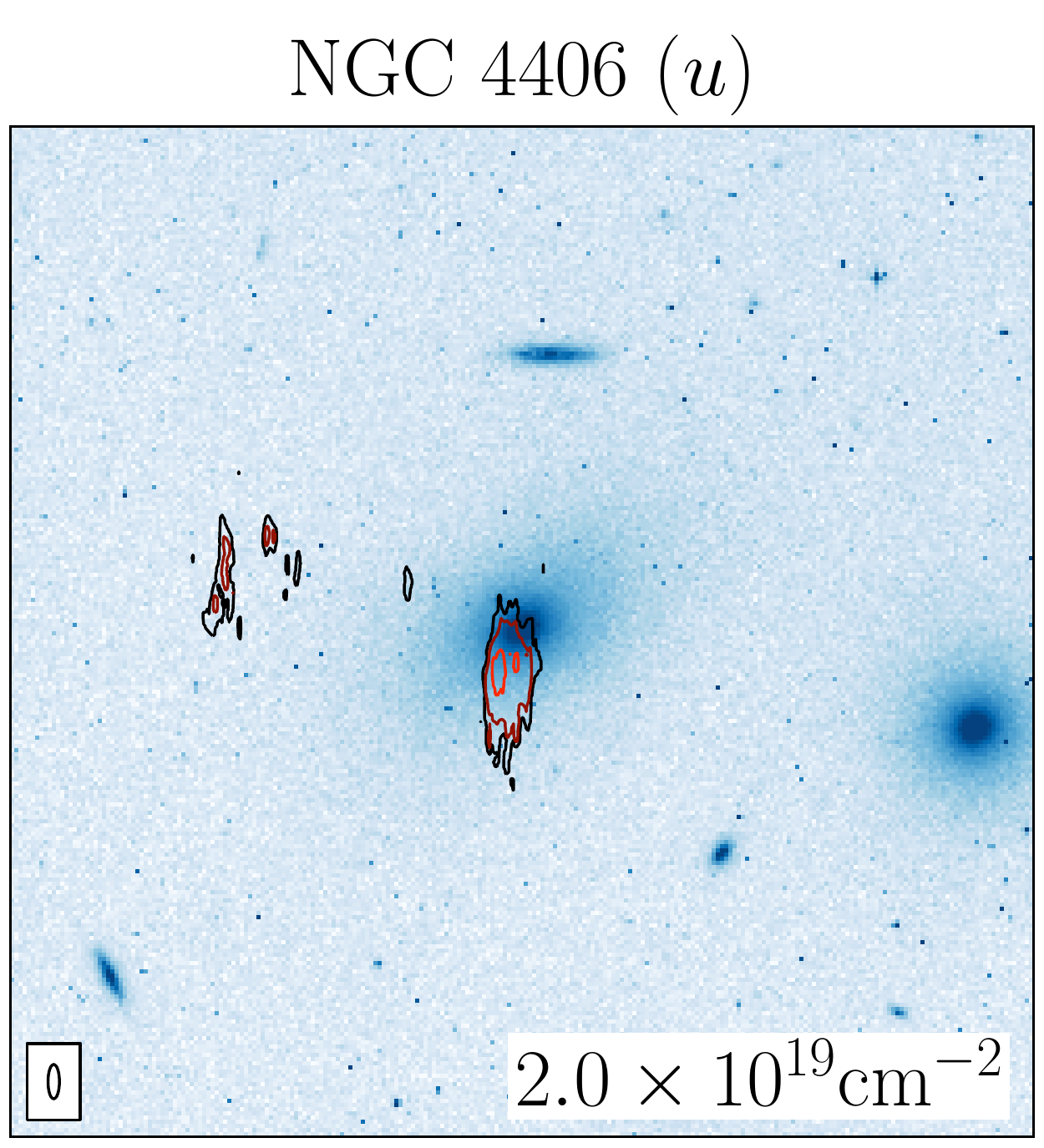} 
\includegraphics[width=50mm]{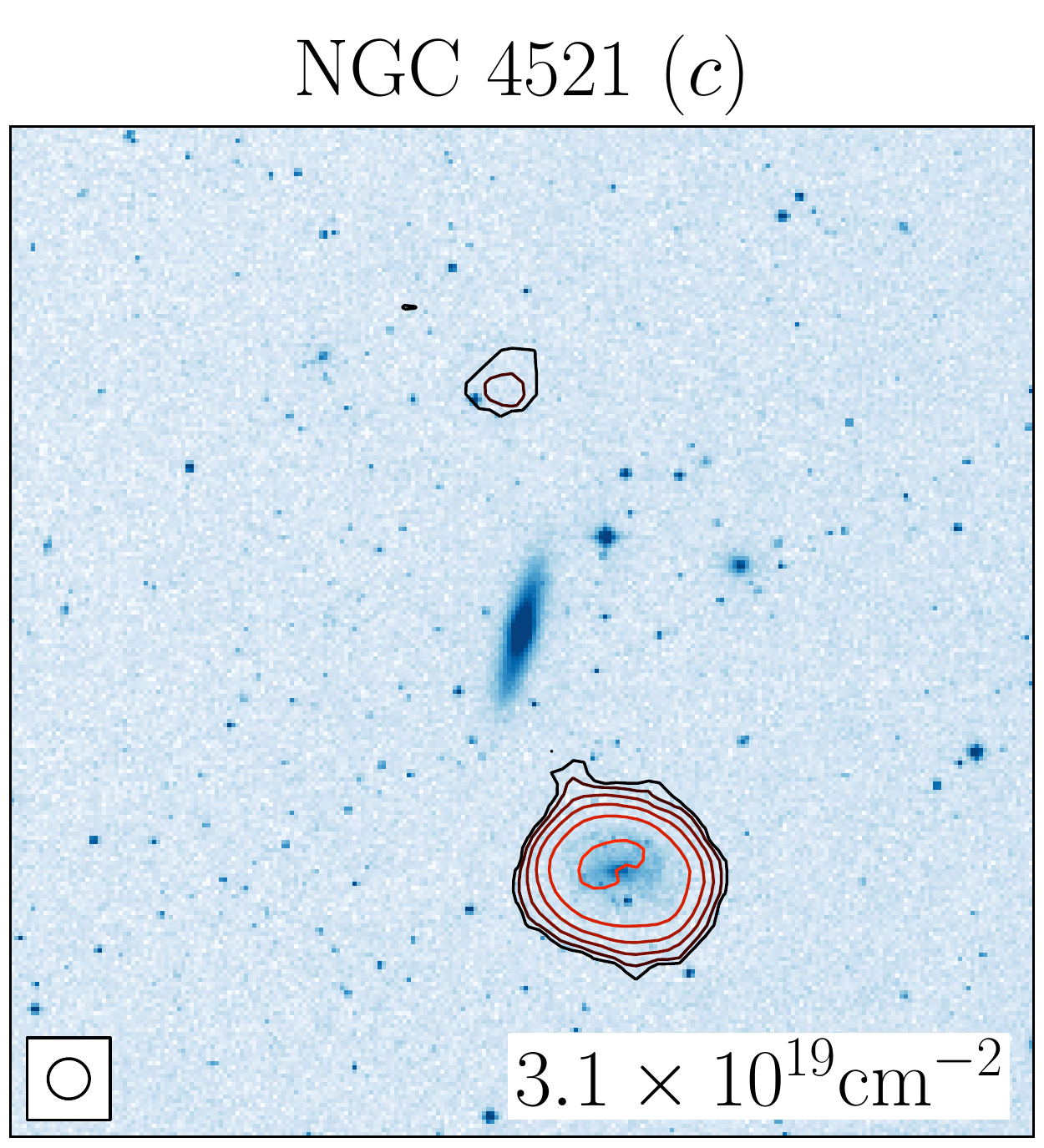} 
\includegraphics[width=50mm]{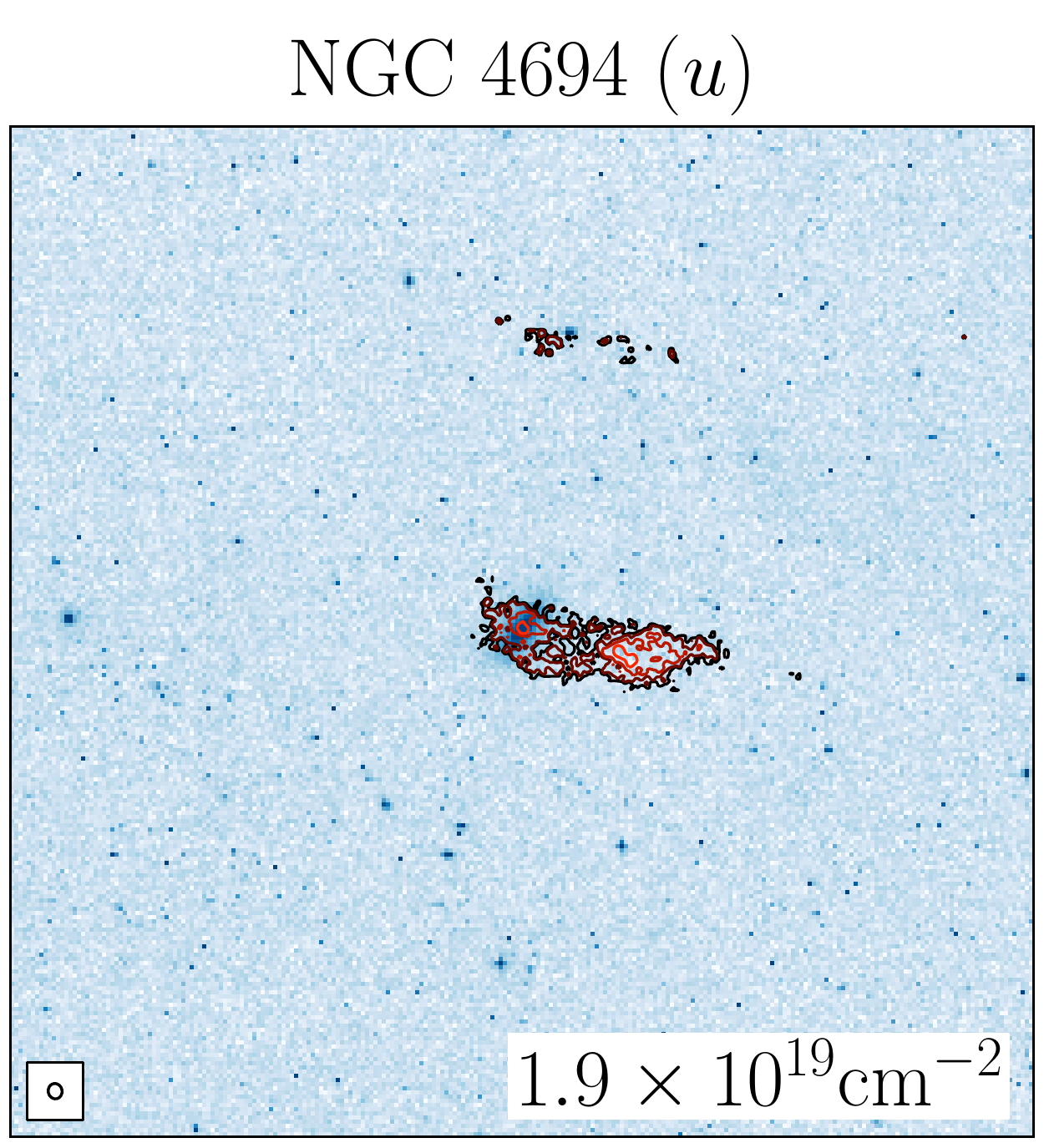} 
\includegraphics[width=50mm]{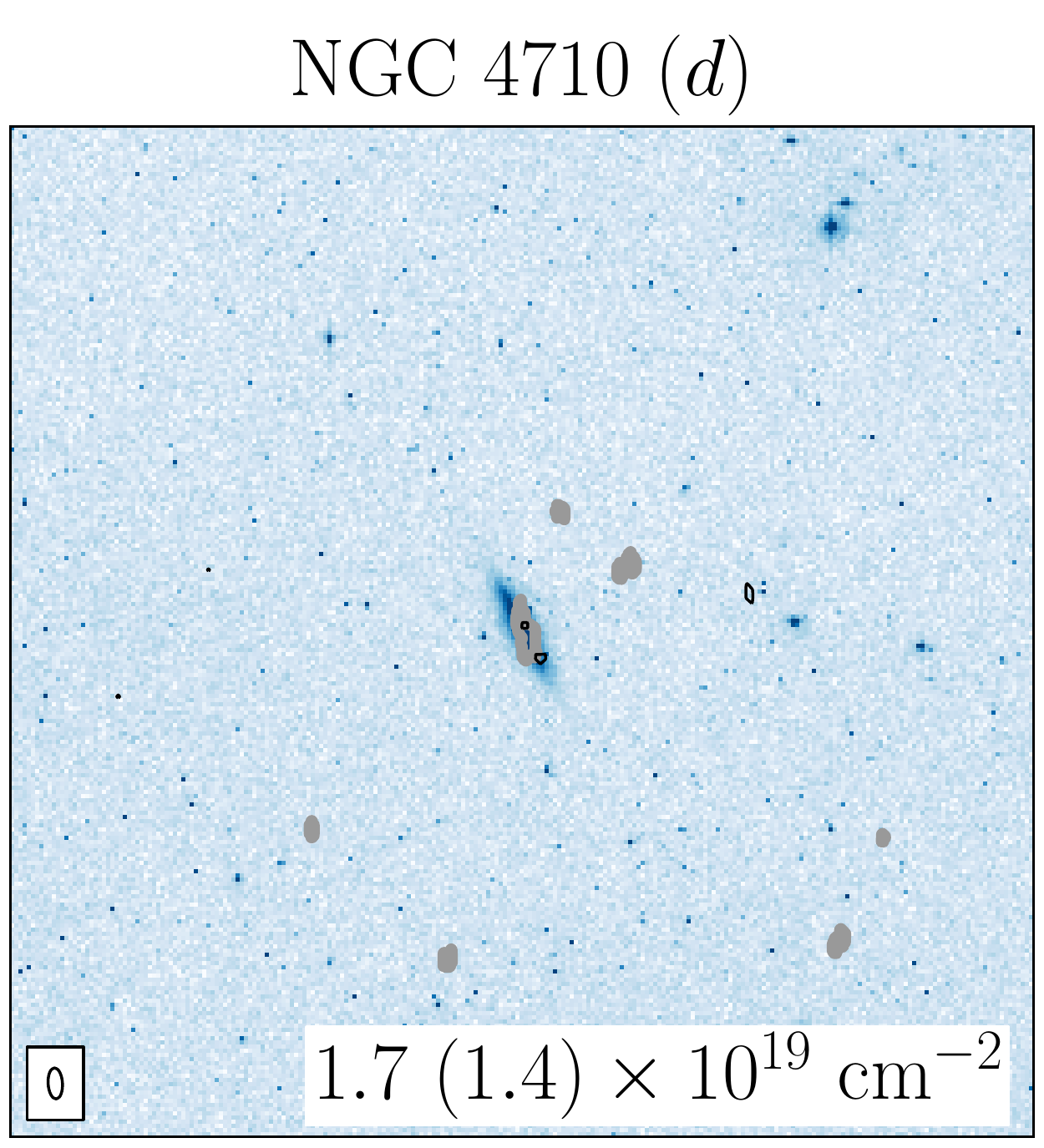} 
\caption{\it Continued \rm}
\end{figure*}

\addtocounter{figure}{-1}
\begin{figure*}
\includegraphics[width=50mm]{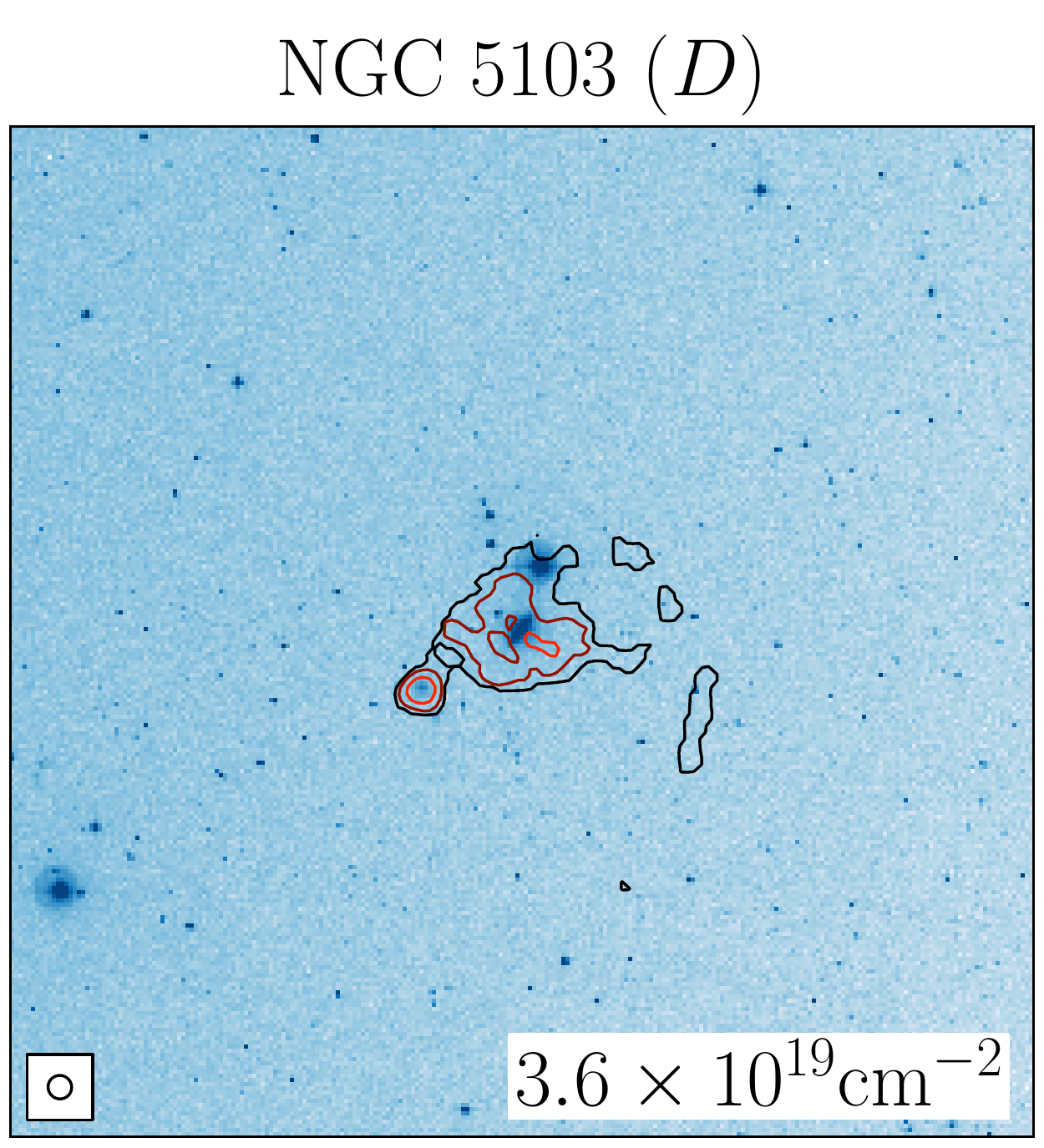} 
\includegraphics[width=50mm]{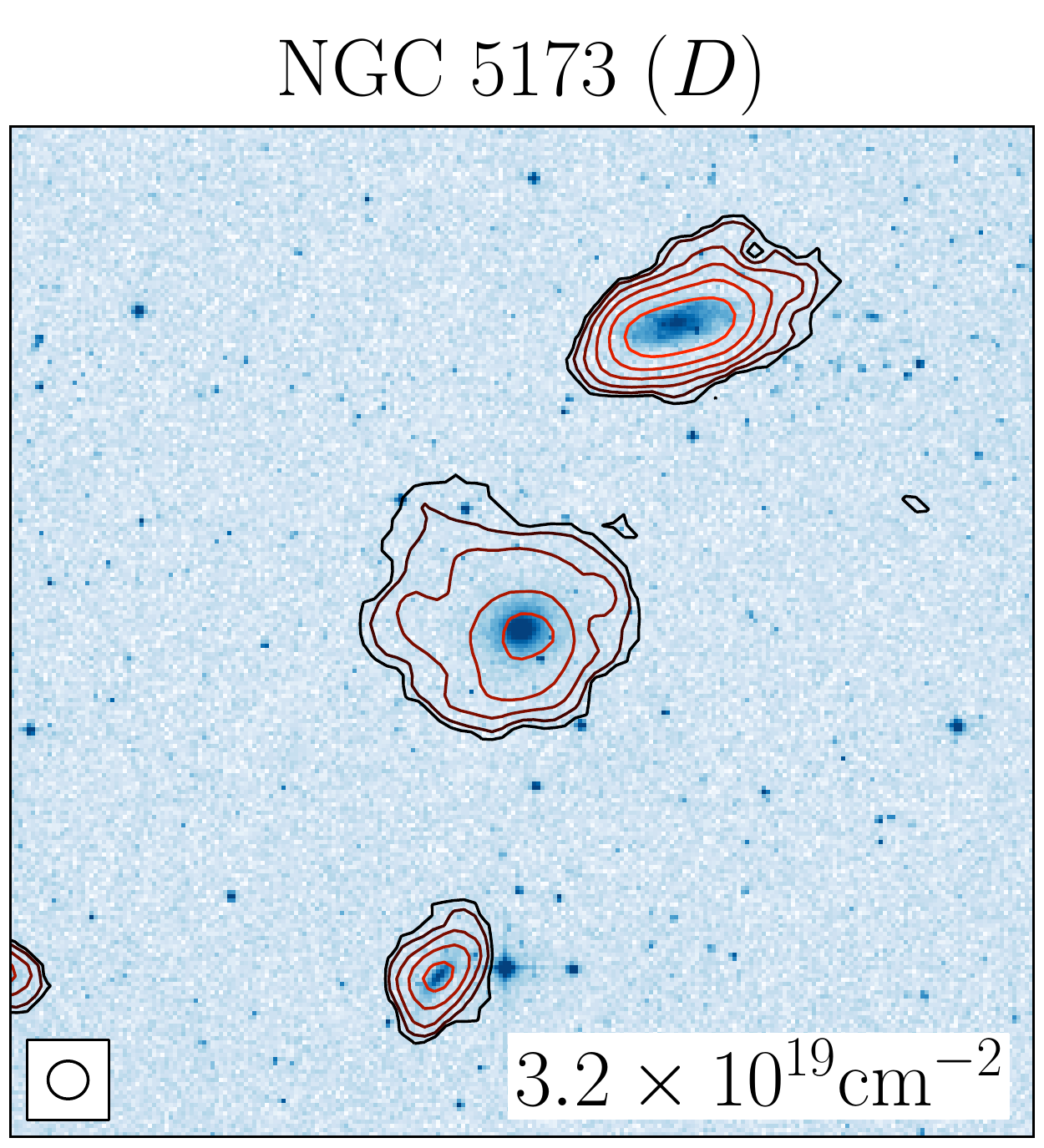} 
\includegraphics[width=50mm]{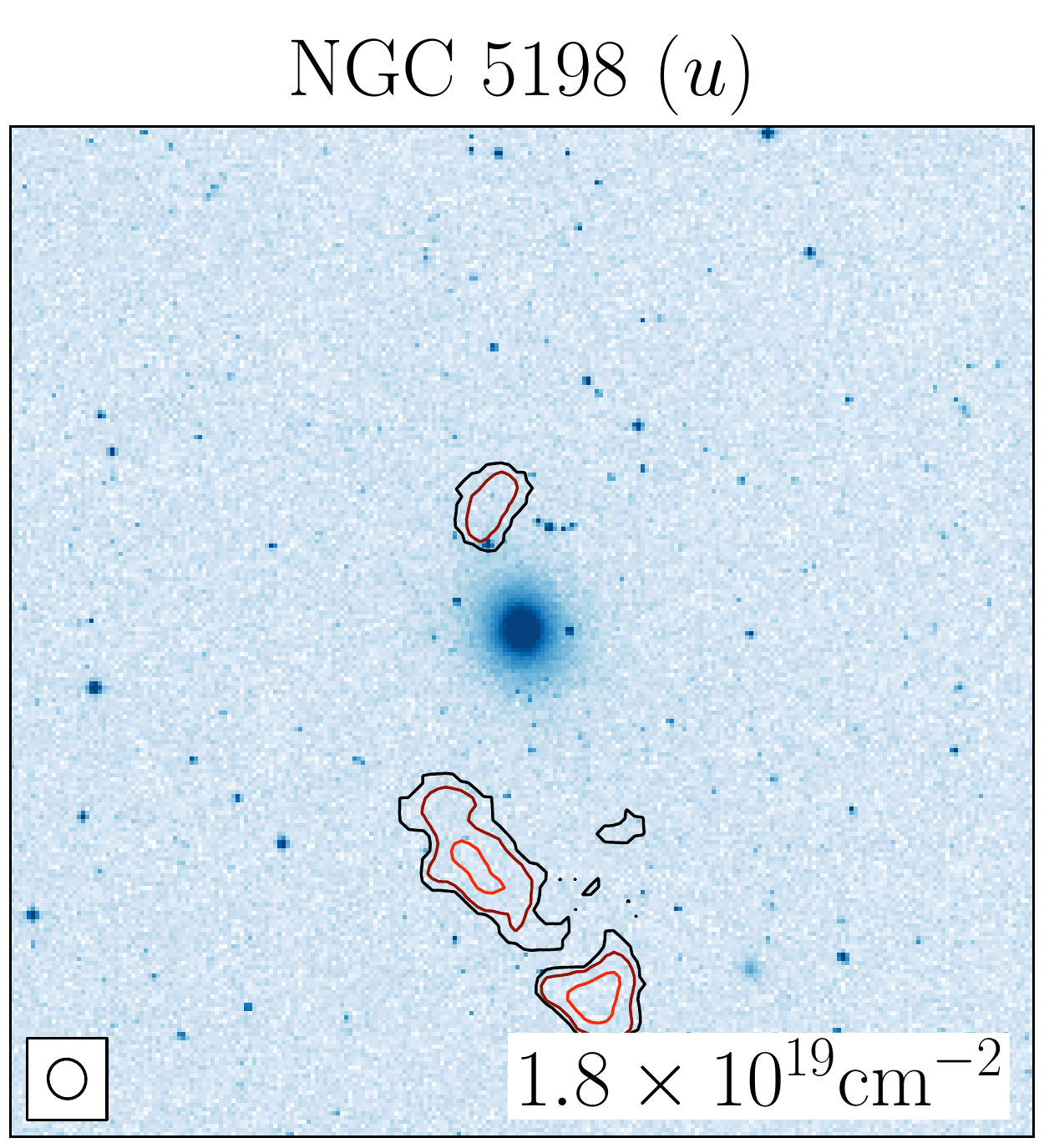} 
\includegraphics[width=50mm]{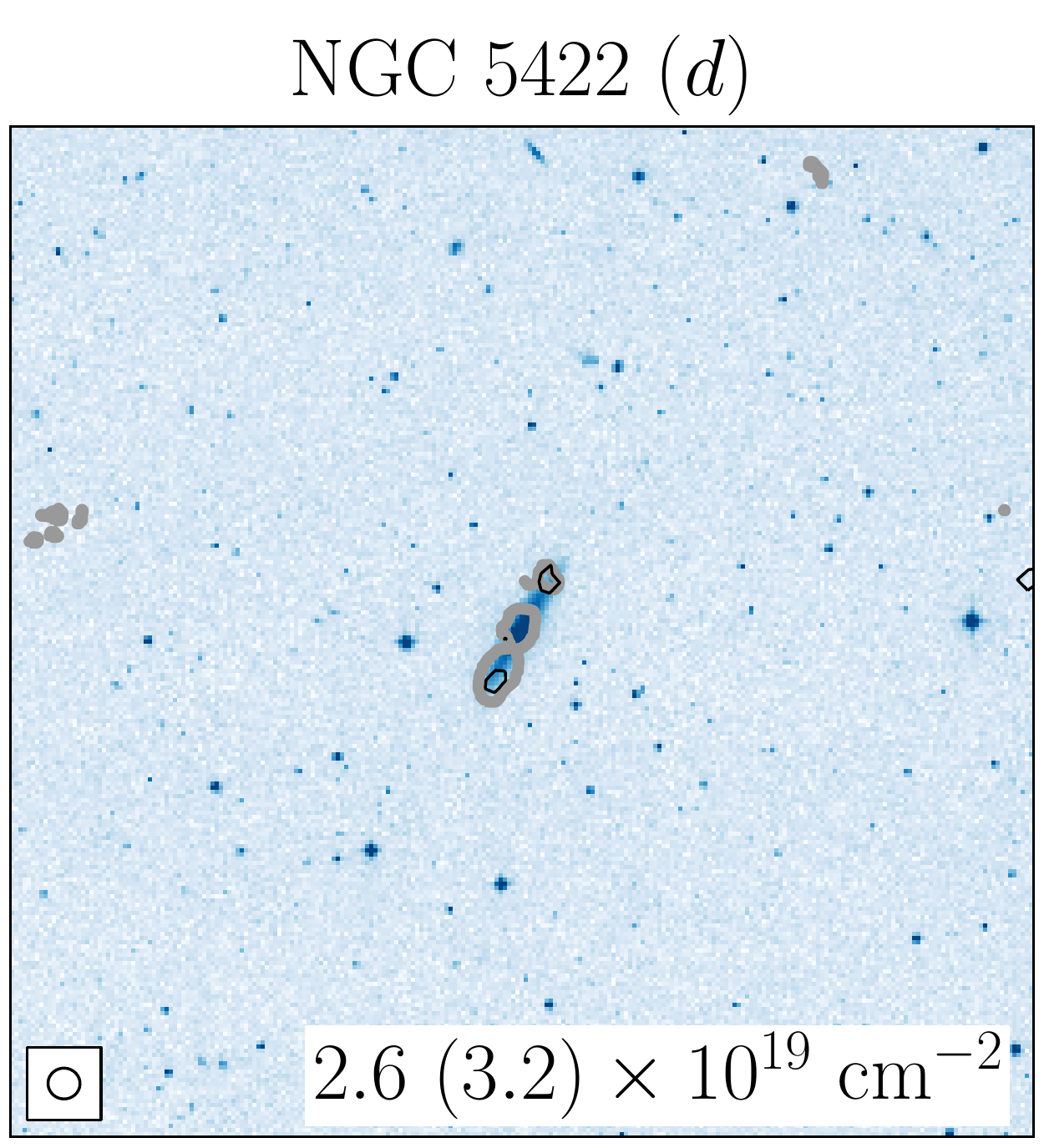} 
\includegraphics[width=50mm]{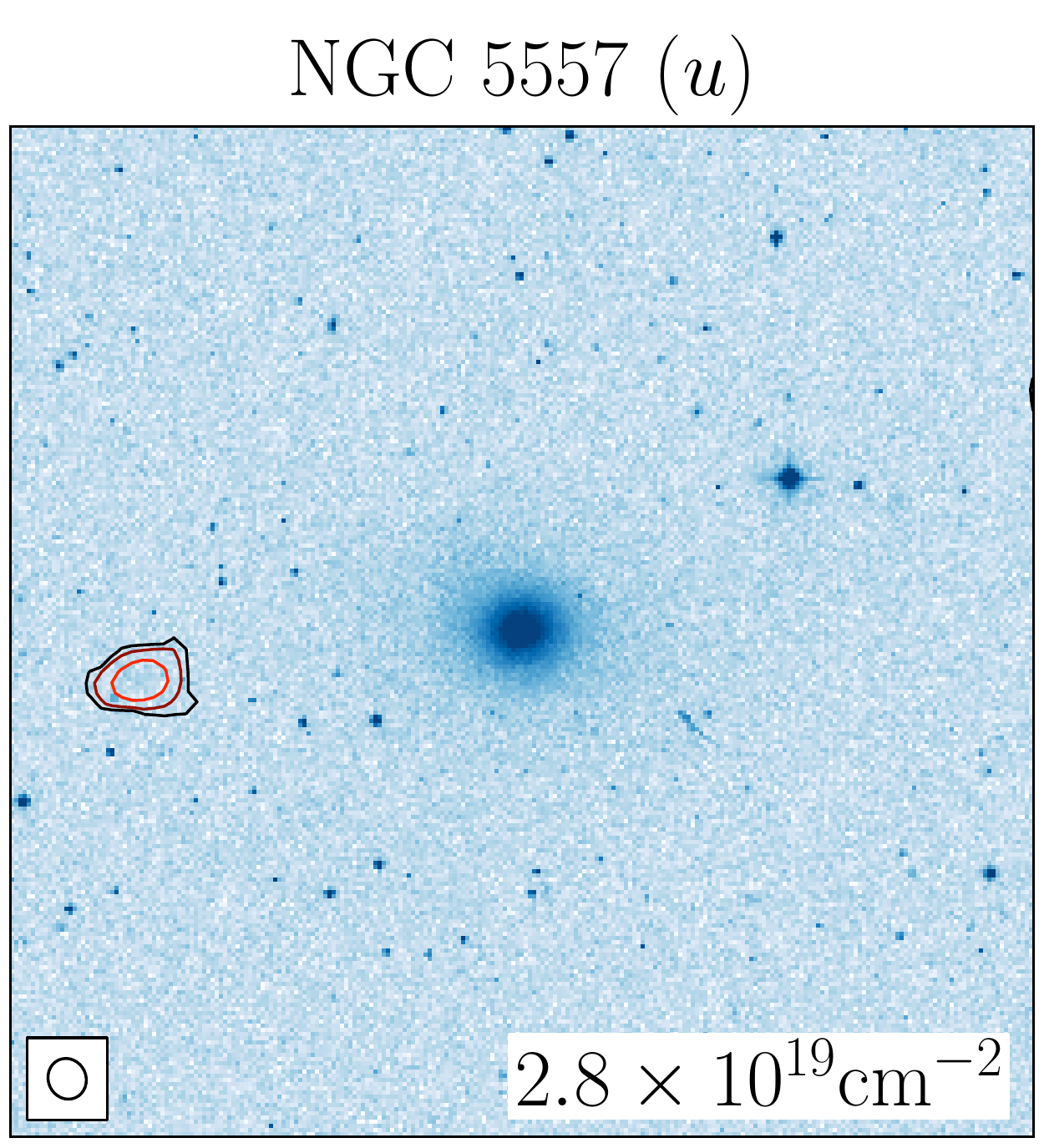} 
\includegraphics[width=50mm]{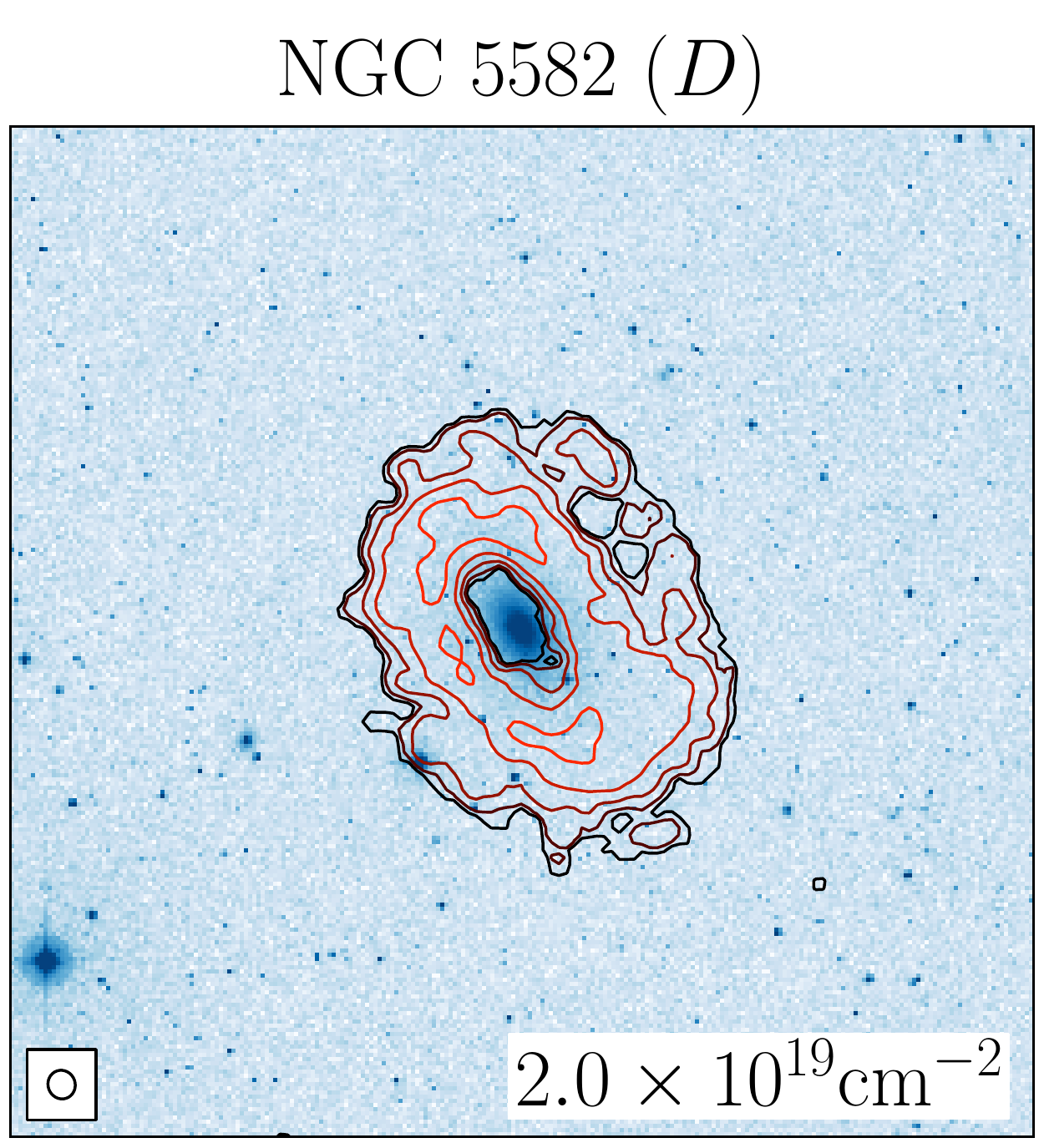} 
\includegraphics[width=50mm]{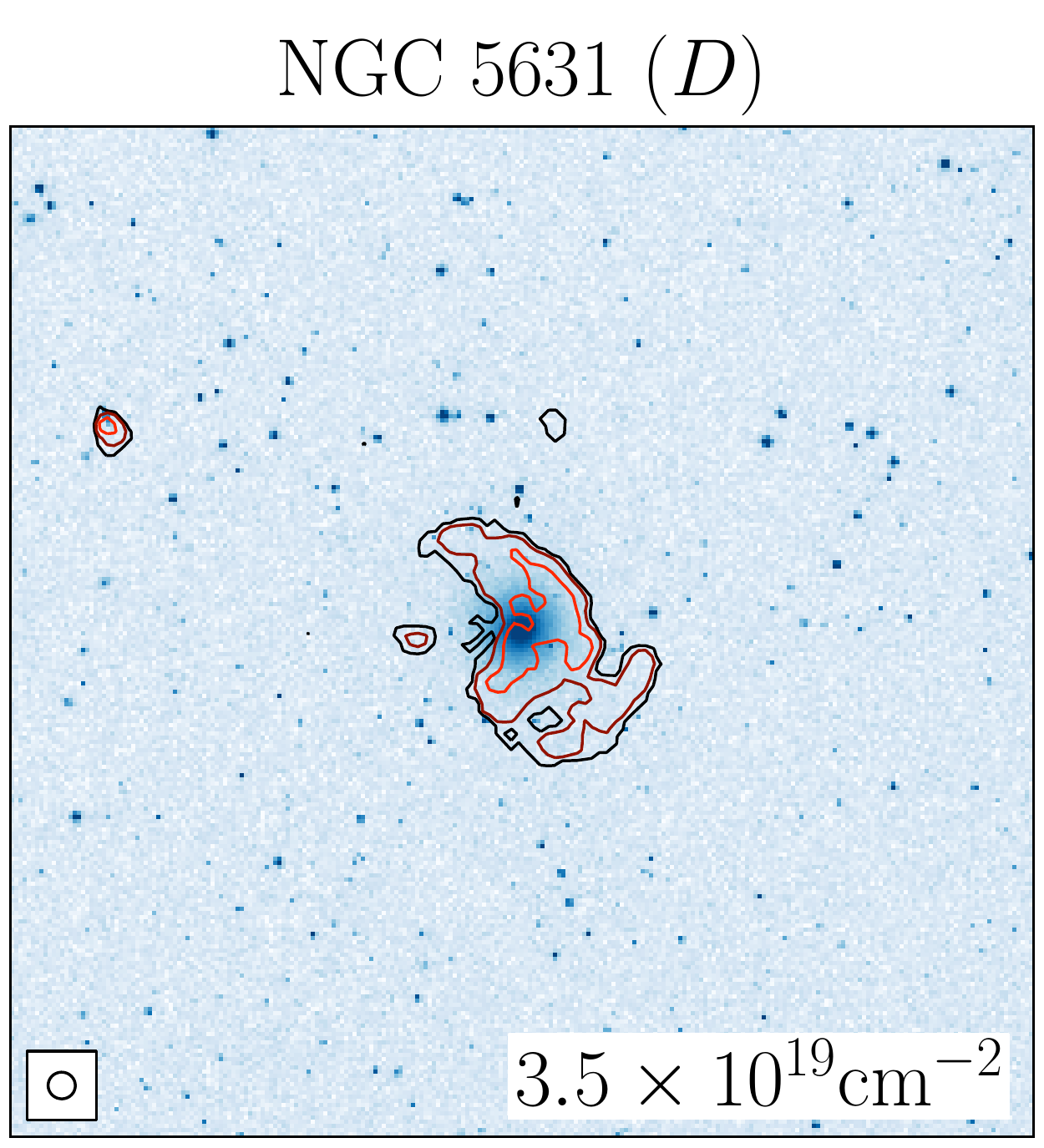} 
\includegraphics[width=50mm]{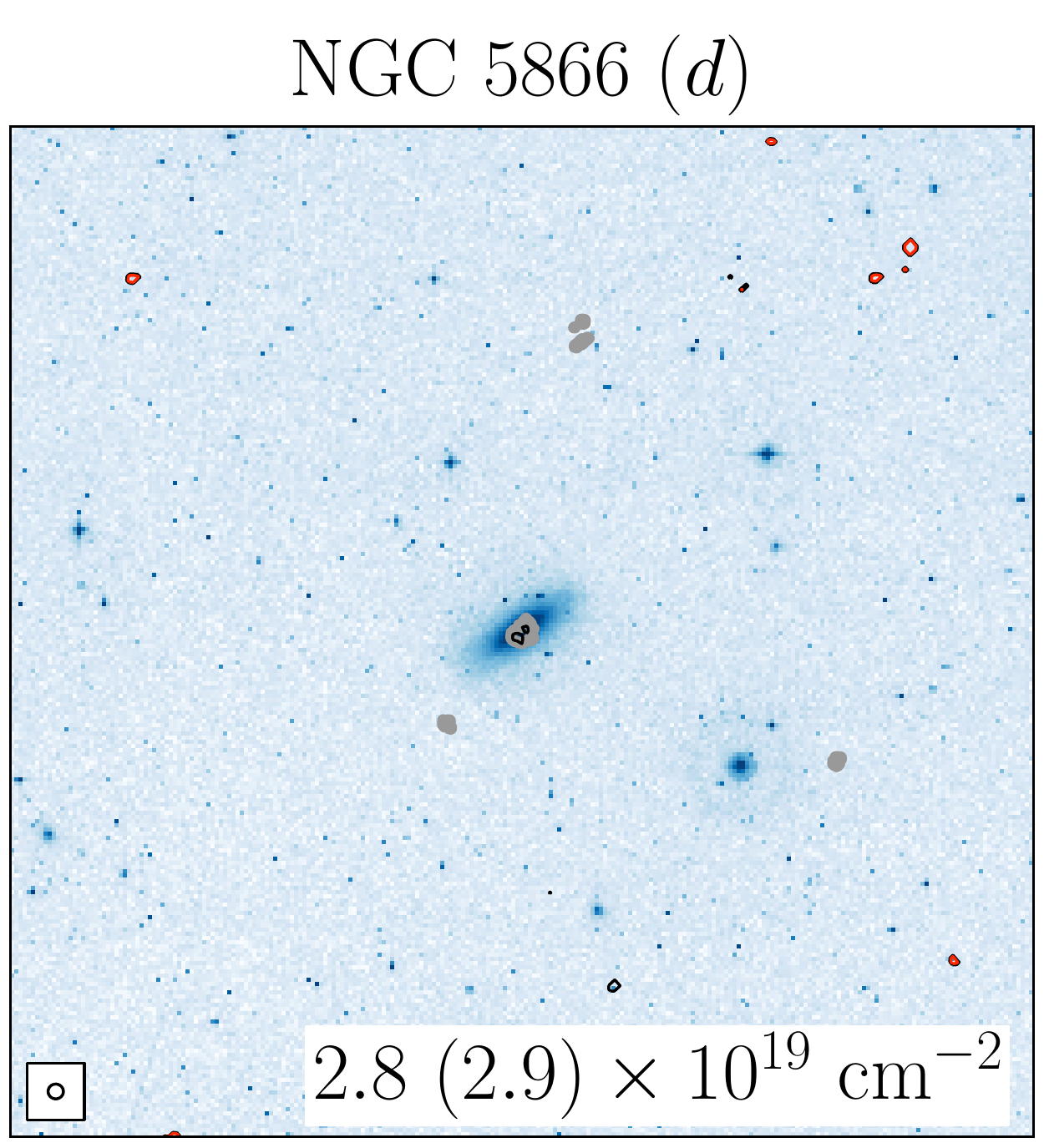} 
\includegraphics[width=50mm]{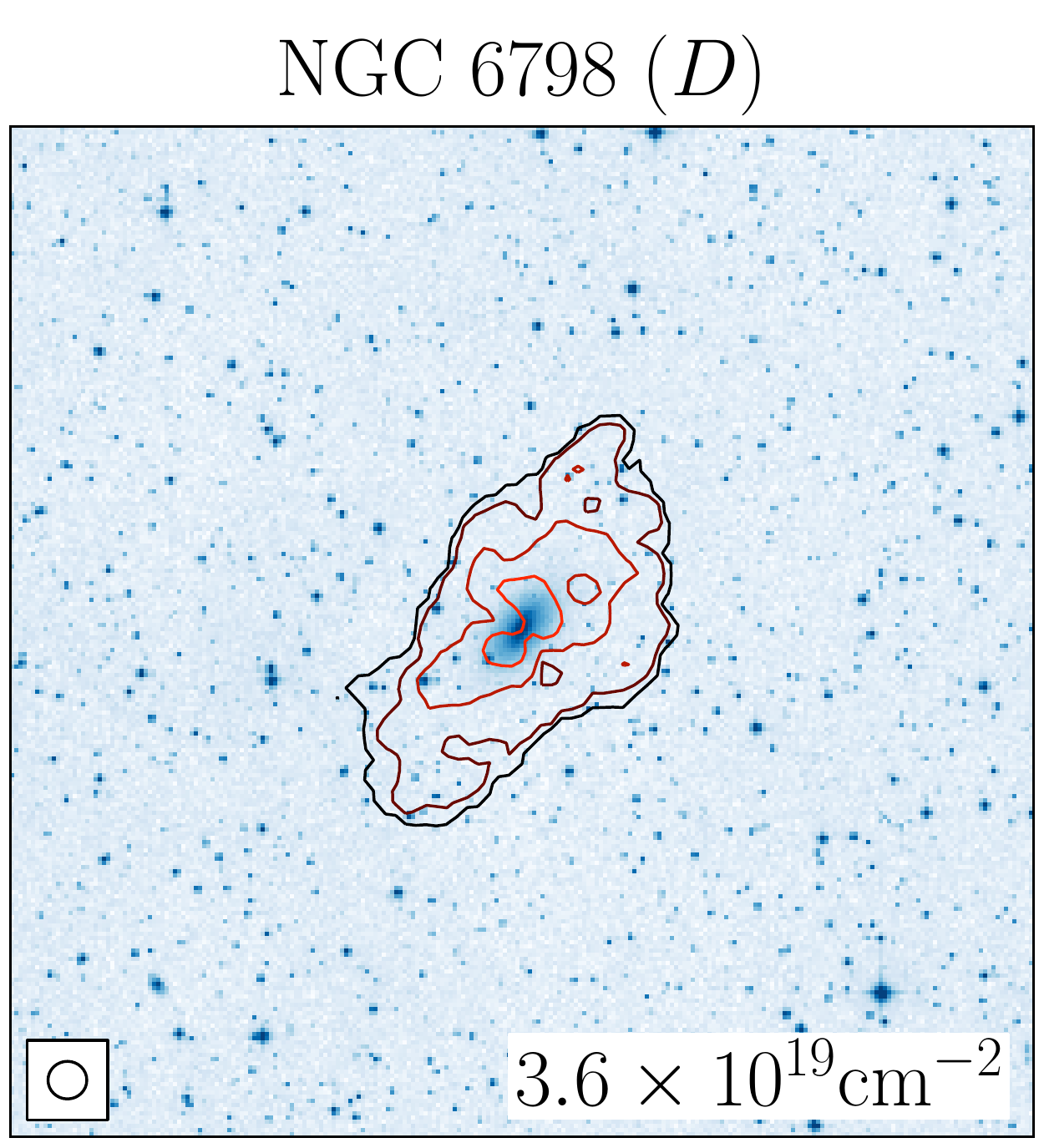} 
\includegraphics[width=50mm]{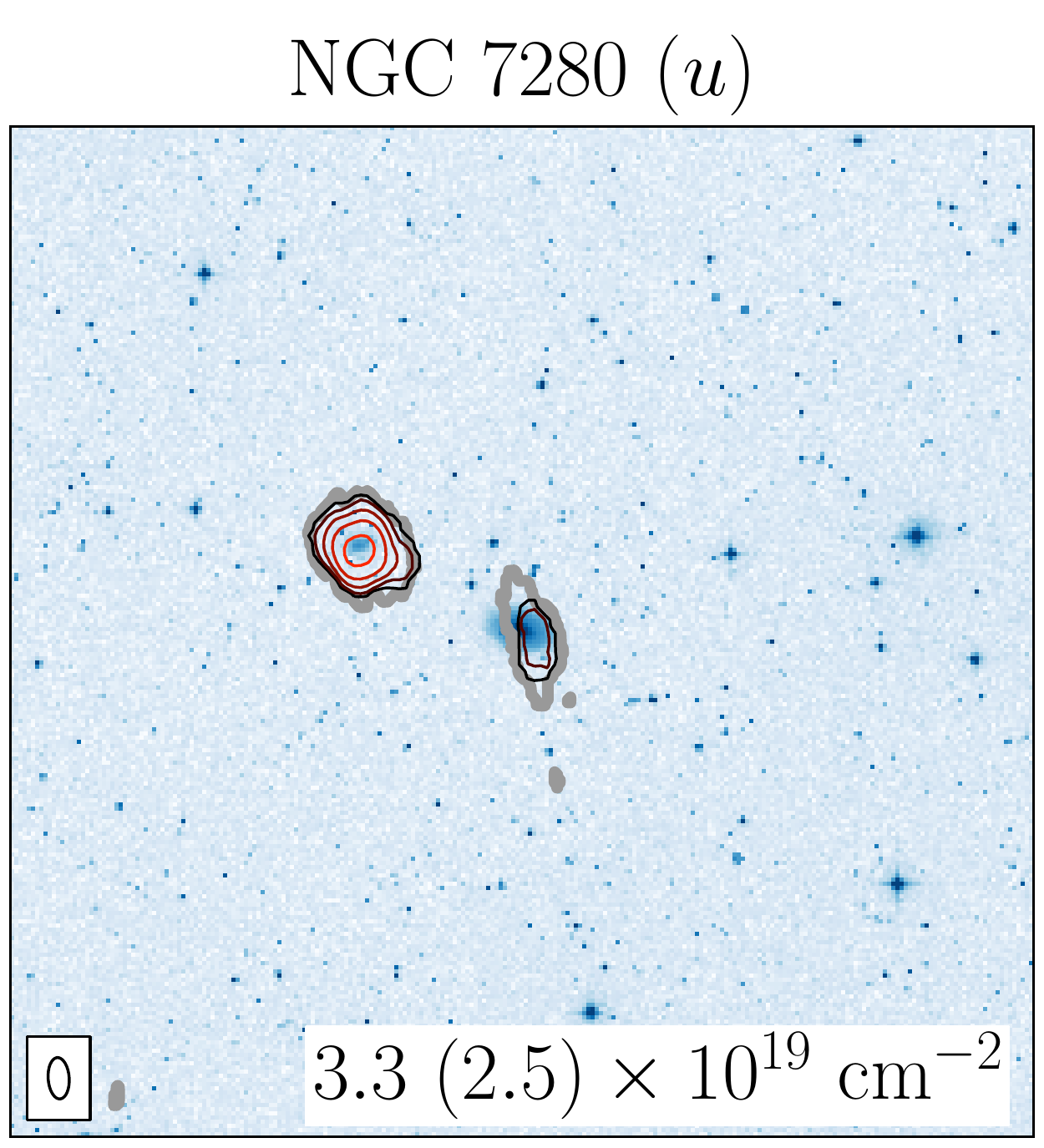} 
\includegraphics[width=50mm]{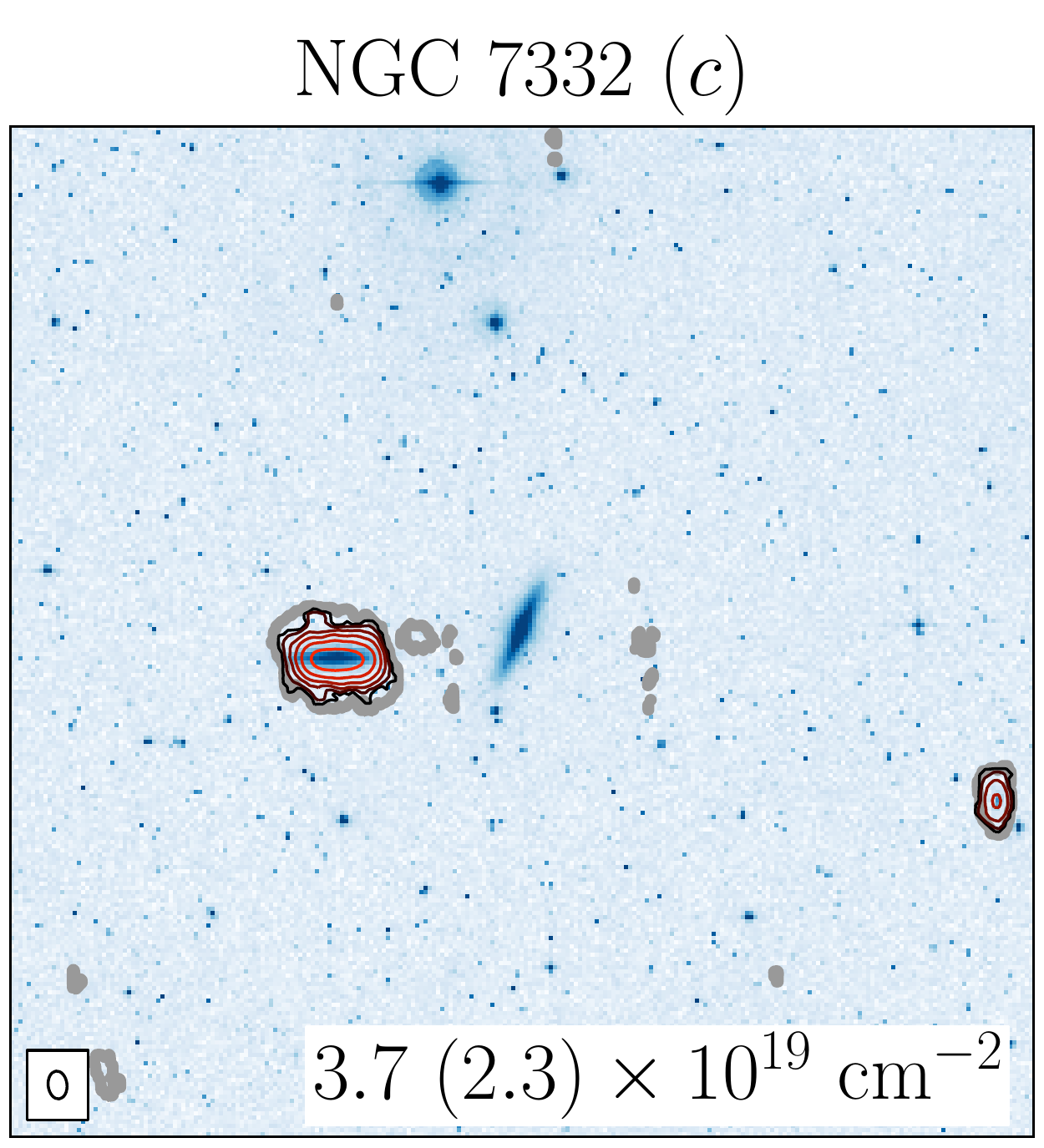} 
\includegraphics[width=50mm]{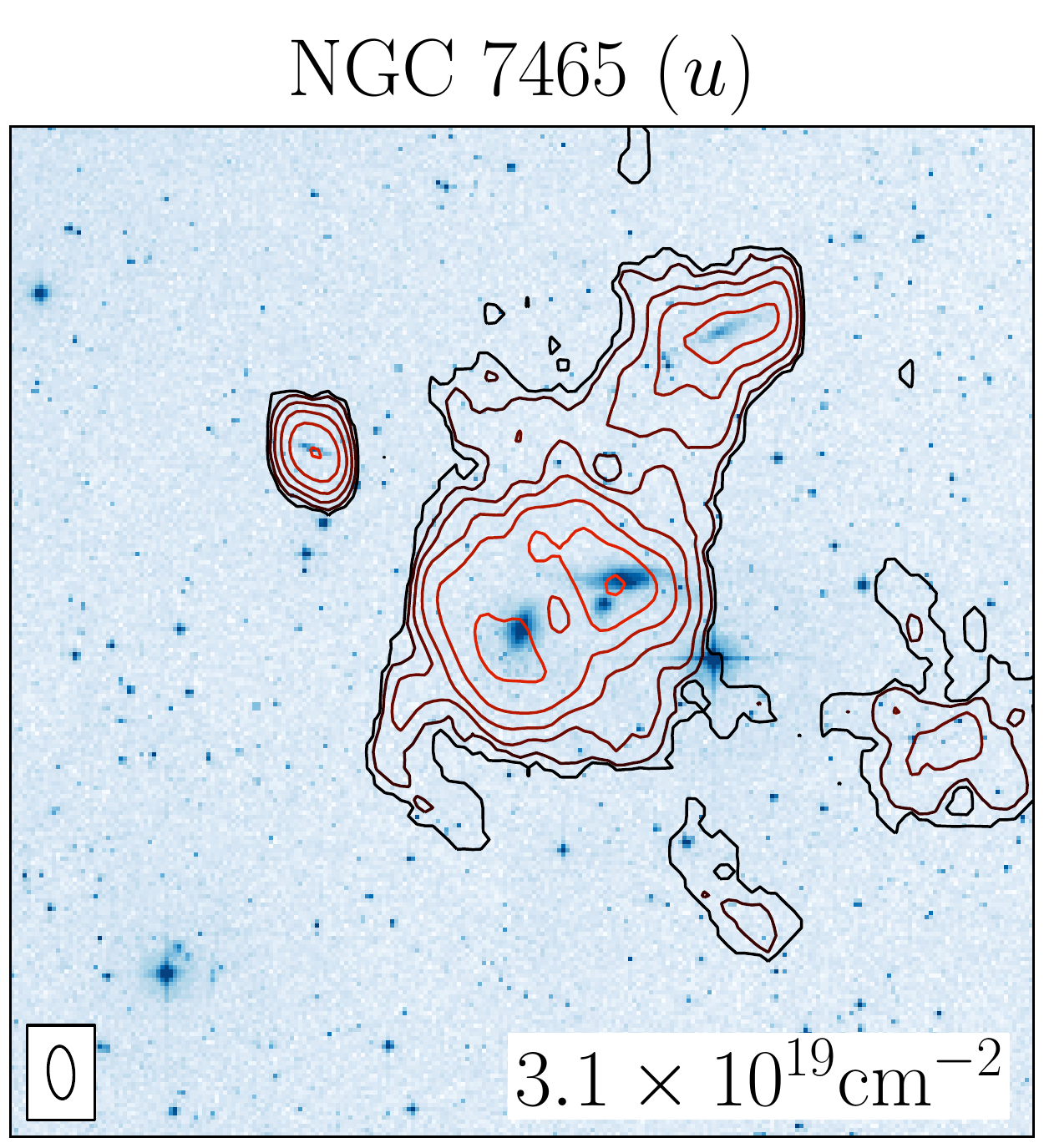} 
\caption{\it Continued \rm}
\end{figure*}

\addtocounter{figure}{-1}
\begin{figure*}
\includegraphics[width=50mm]{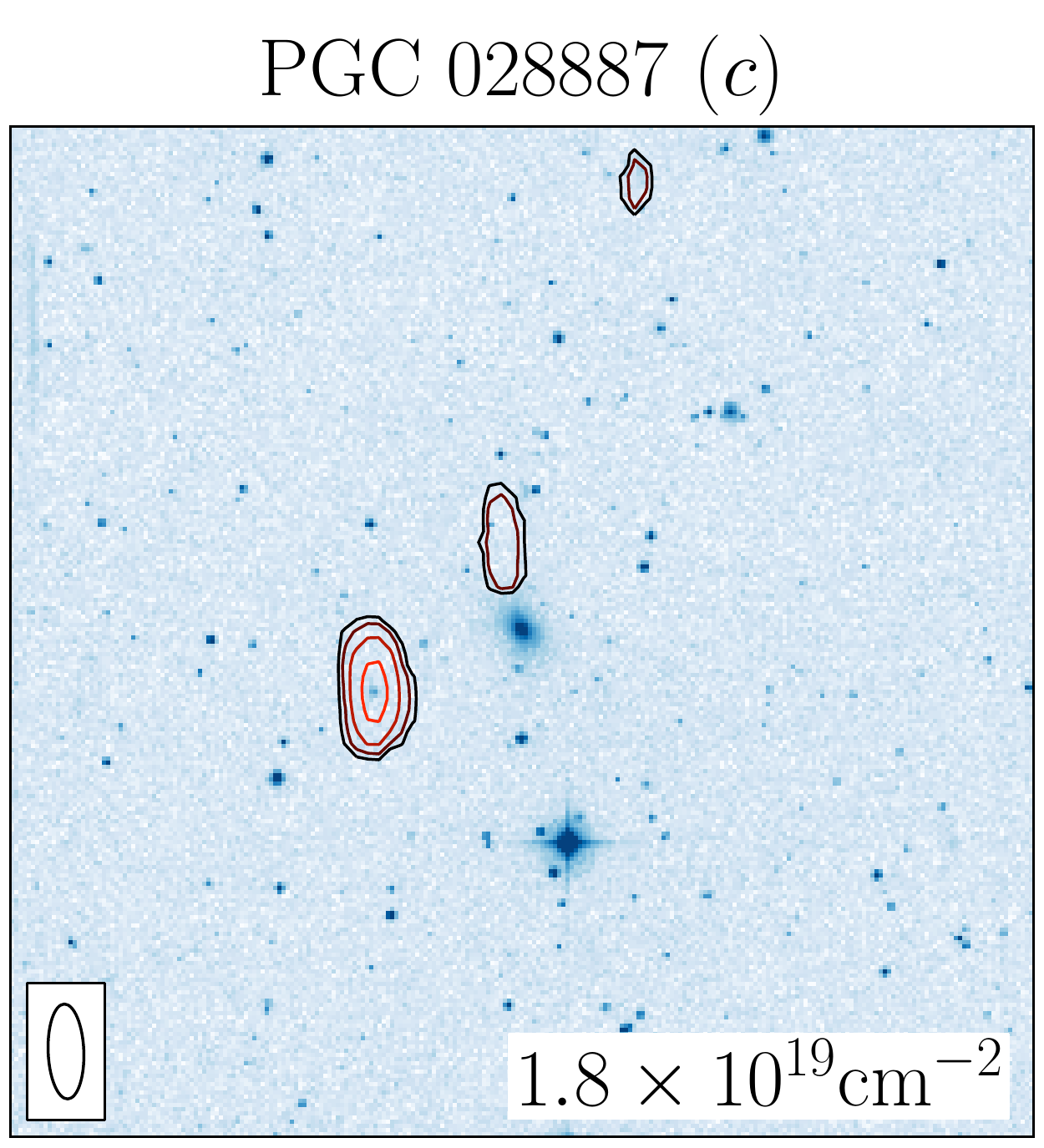} 
\includegraphics[width=50mm]{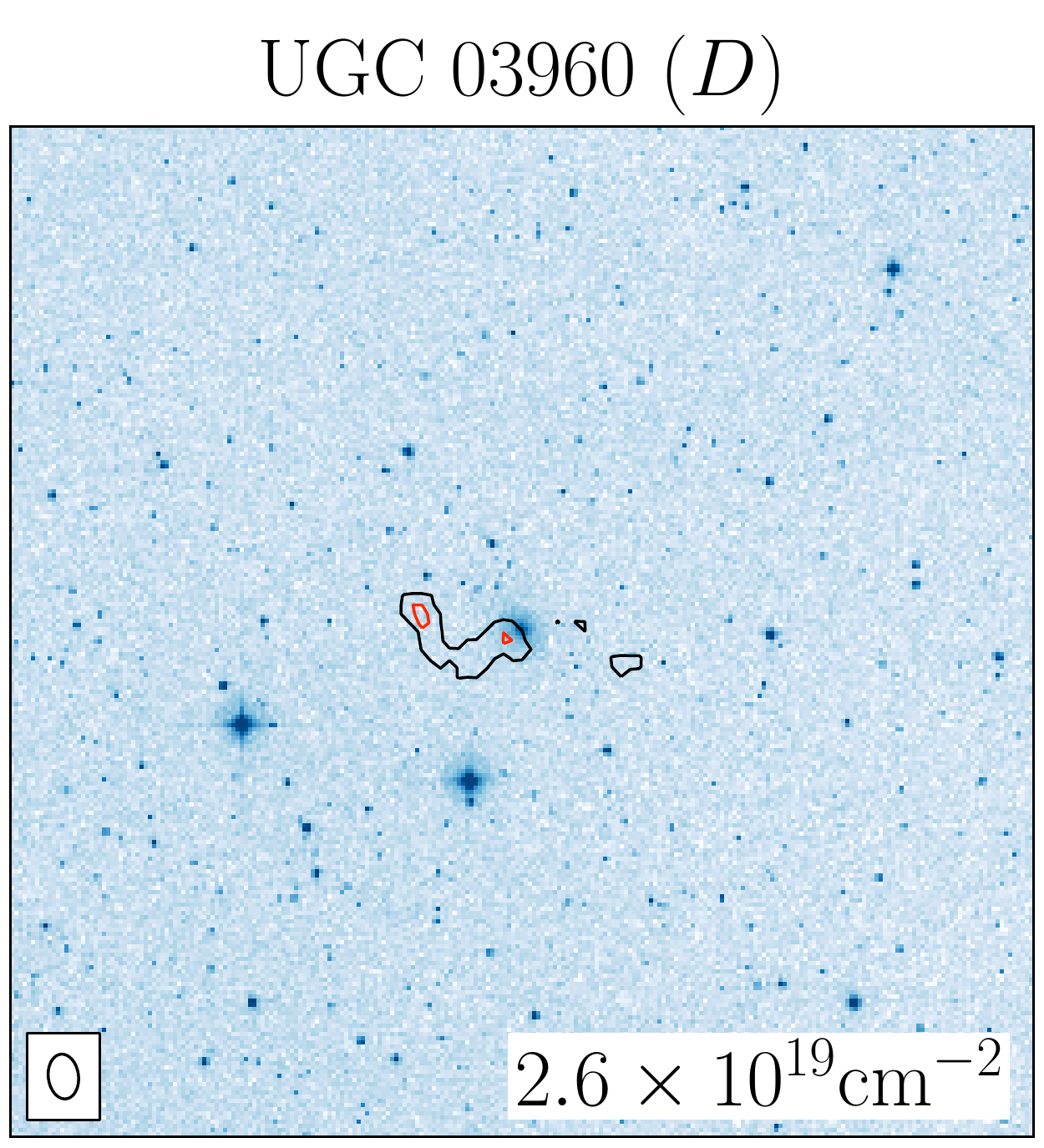} 
\includegraphics[width=50mm]{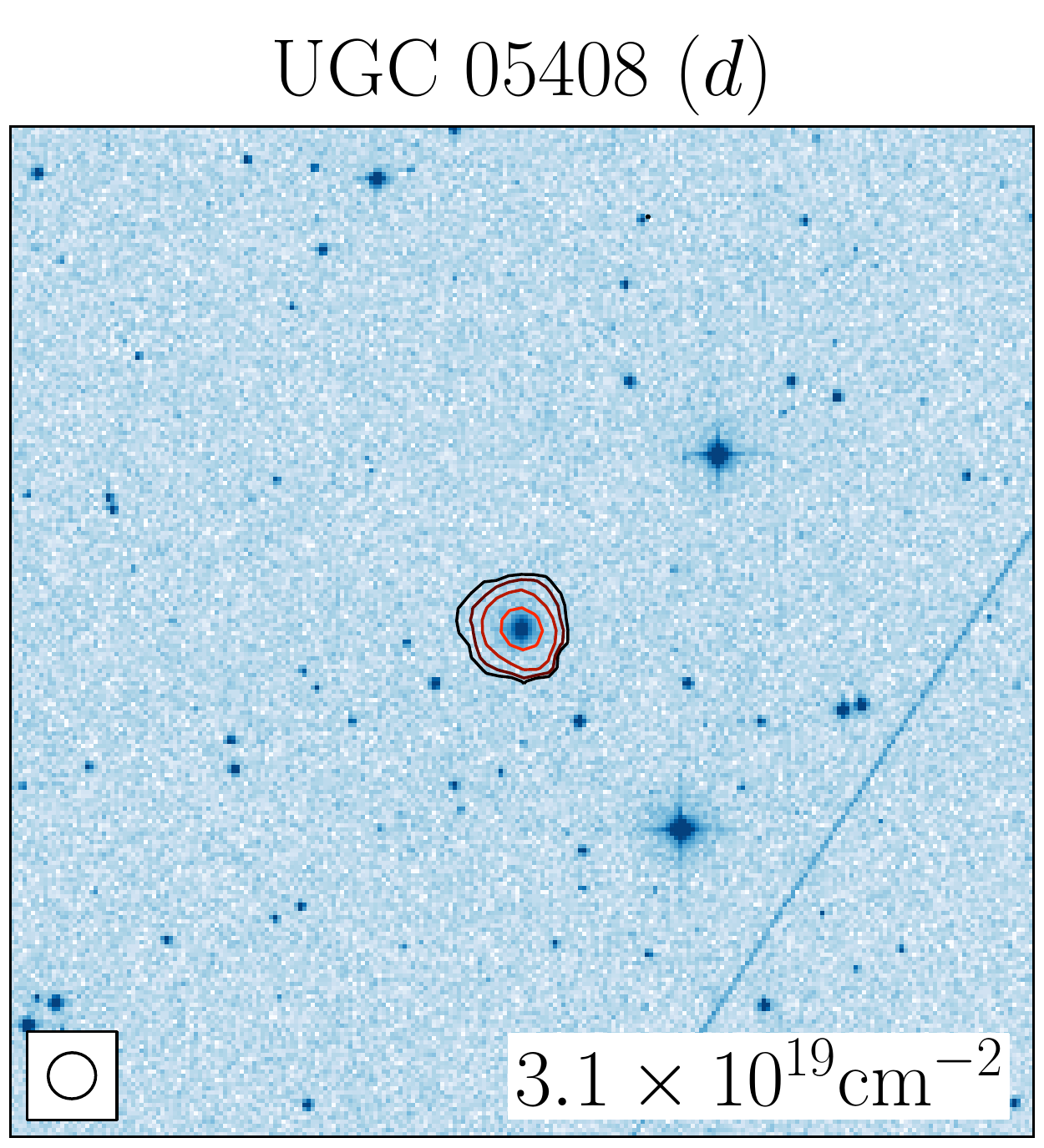} 
\includegraphics[width=50mm]{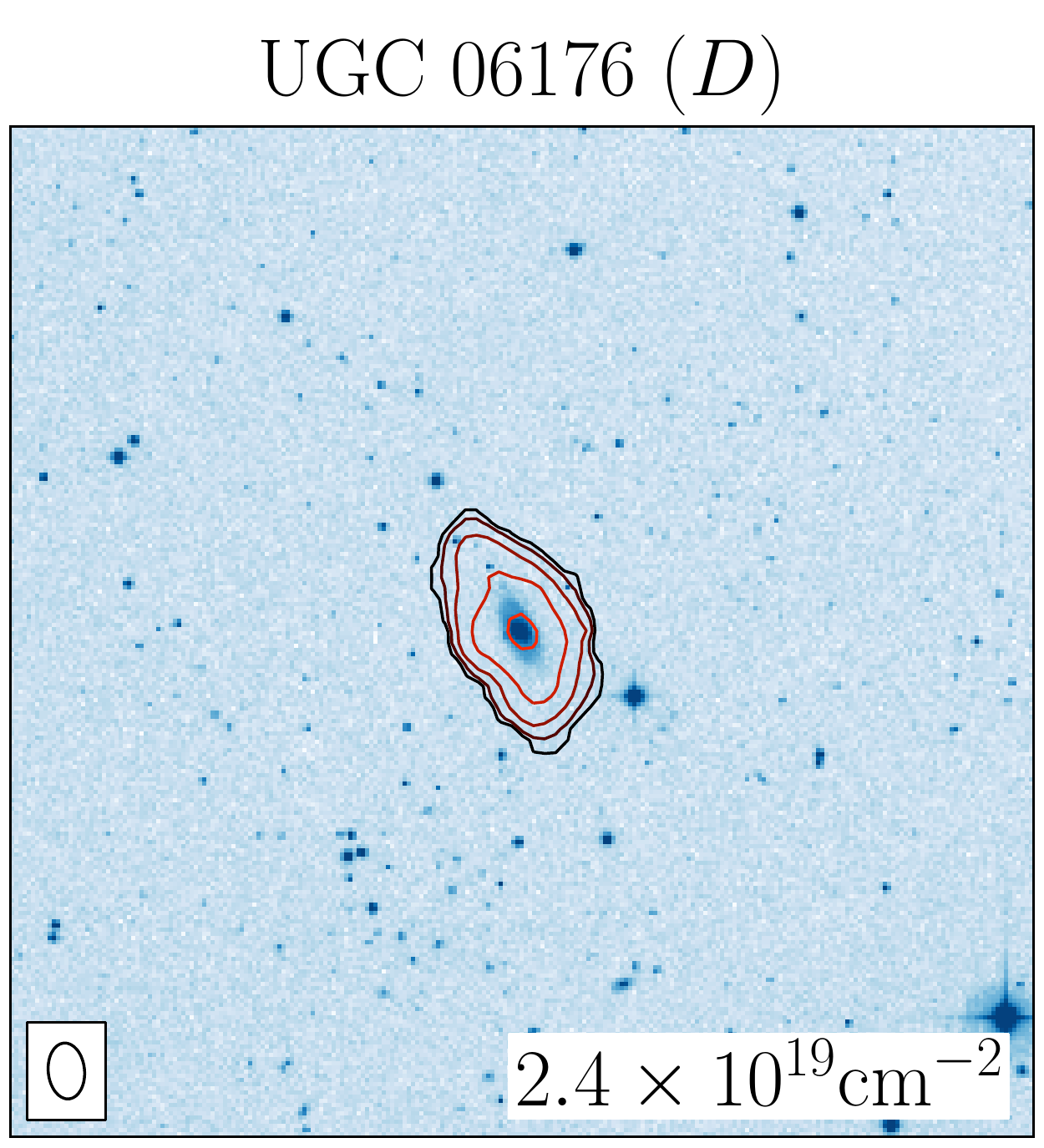} 
\includegraphics[width=50mm]{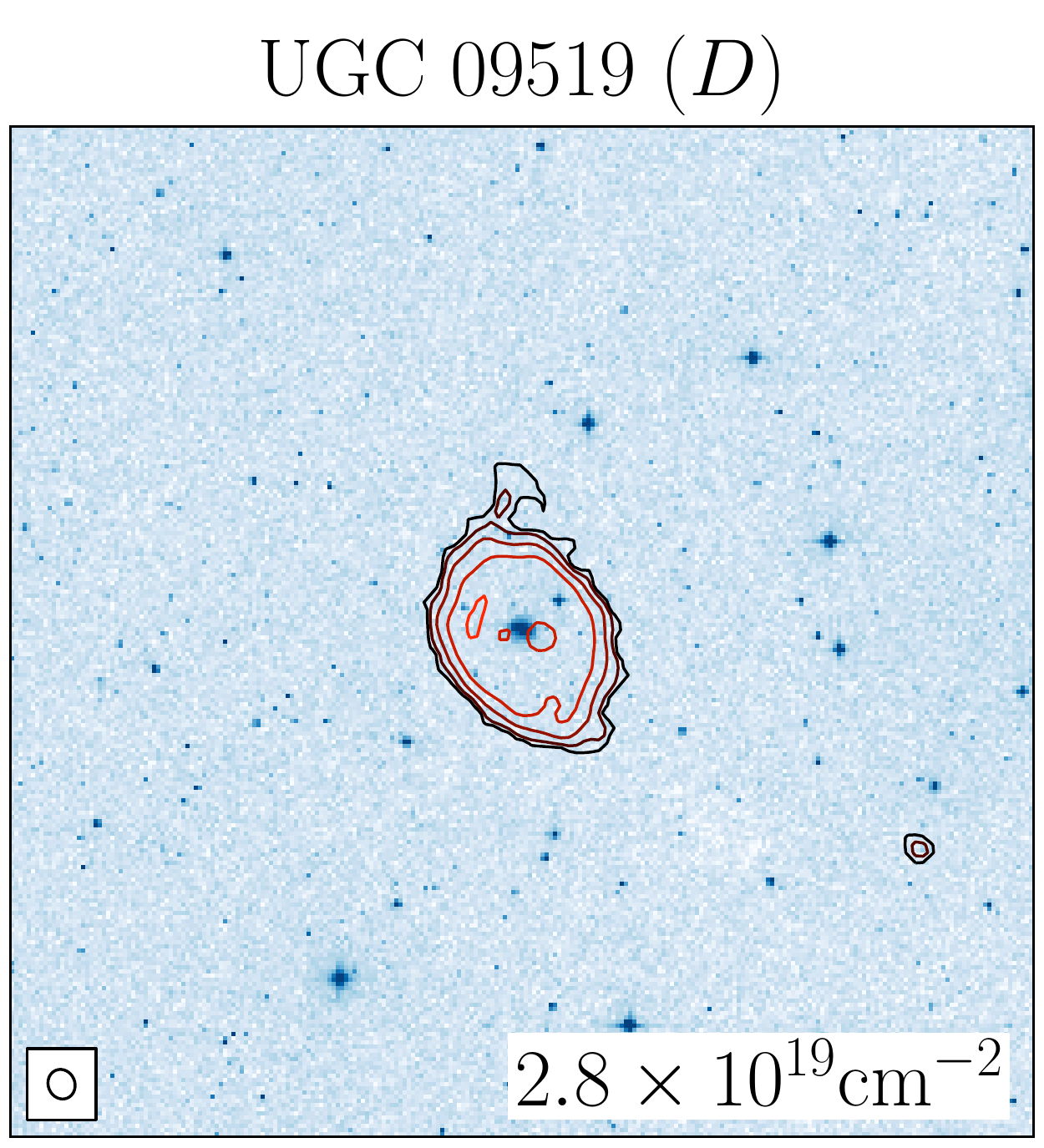} 
\caption{\it Continued \rm}
\end{figure*}

\begin{figure*}
\includegraphics[width=50mm]{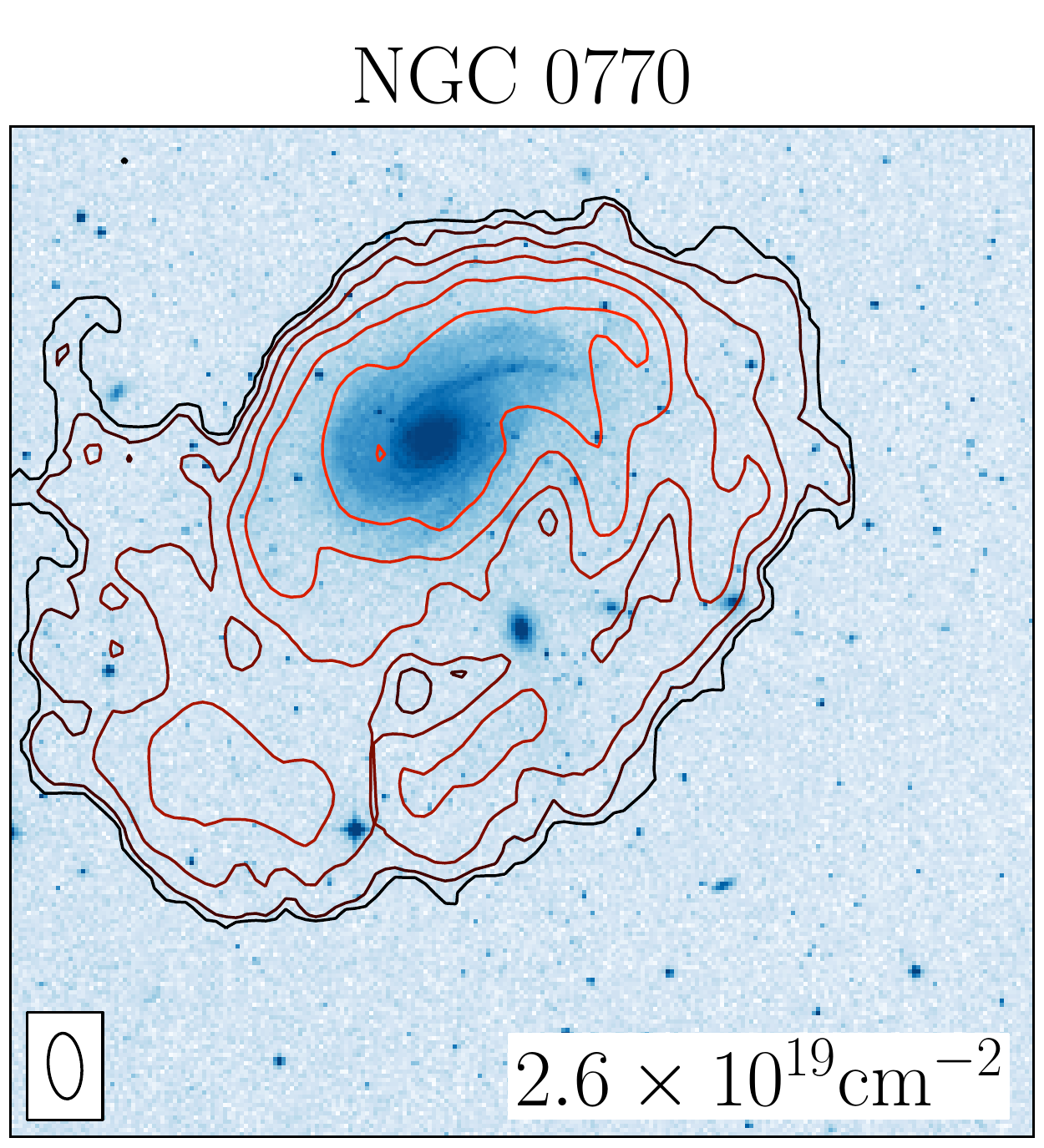}
\includegraphics[width=50mm]{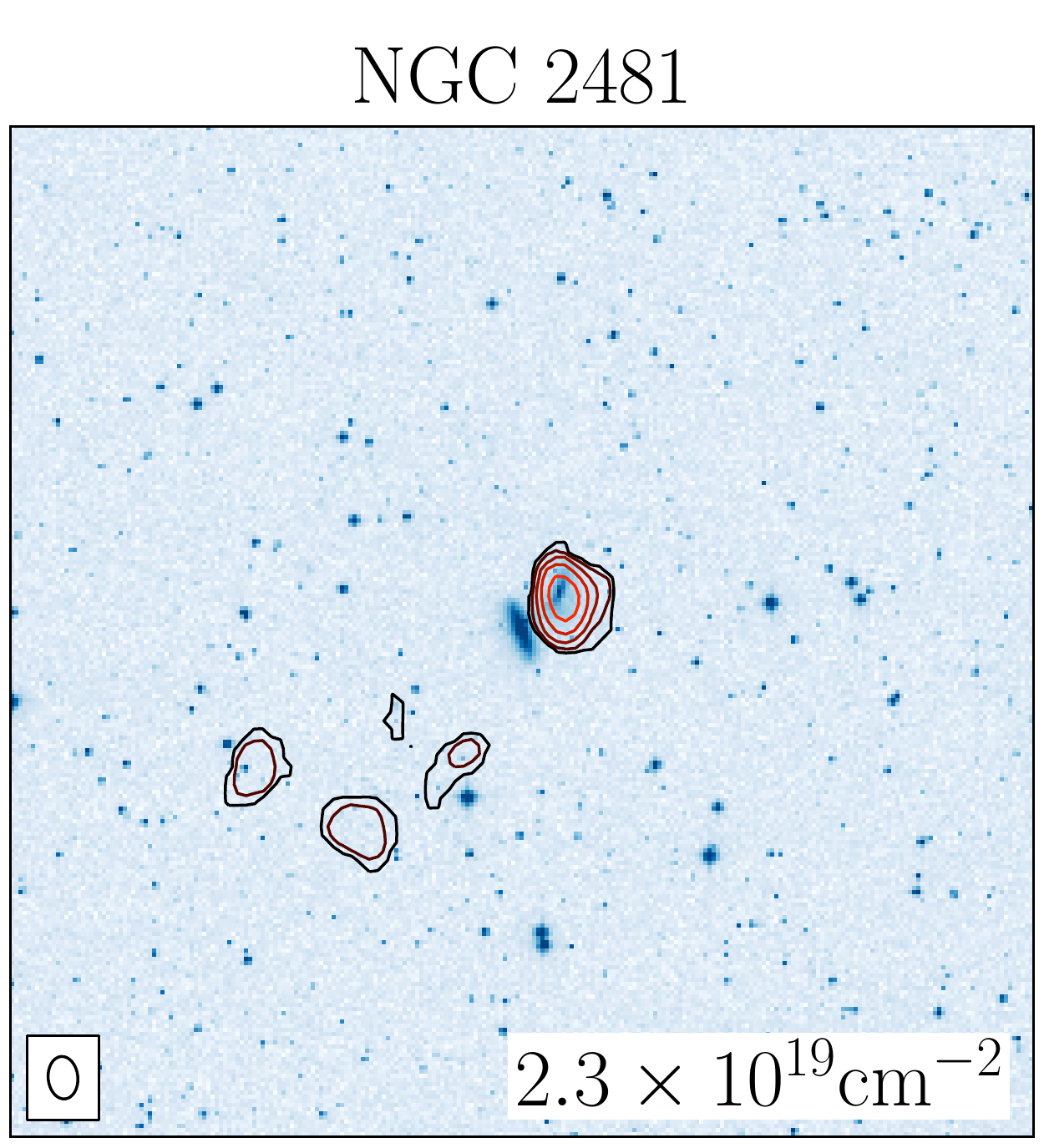}
\includegraphics[width=50mm]{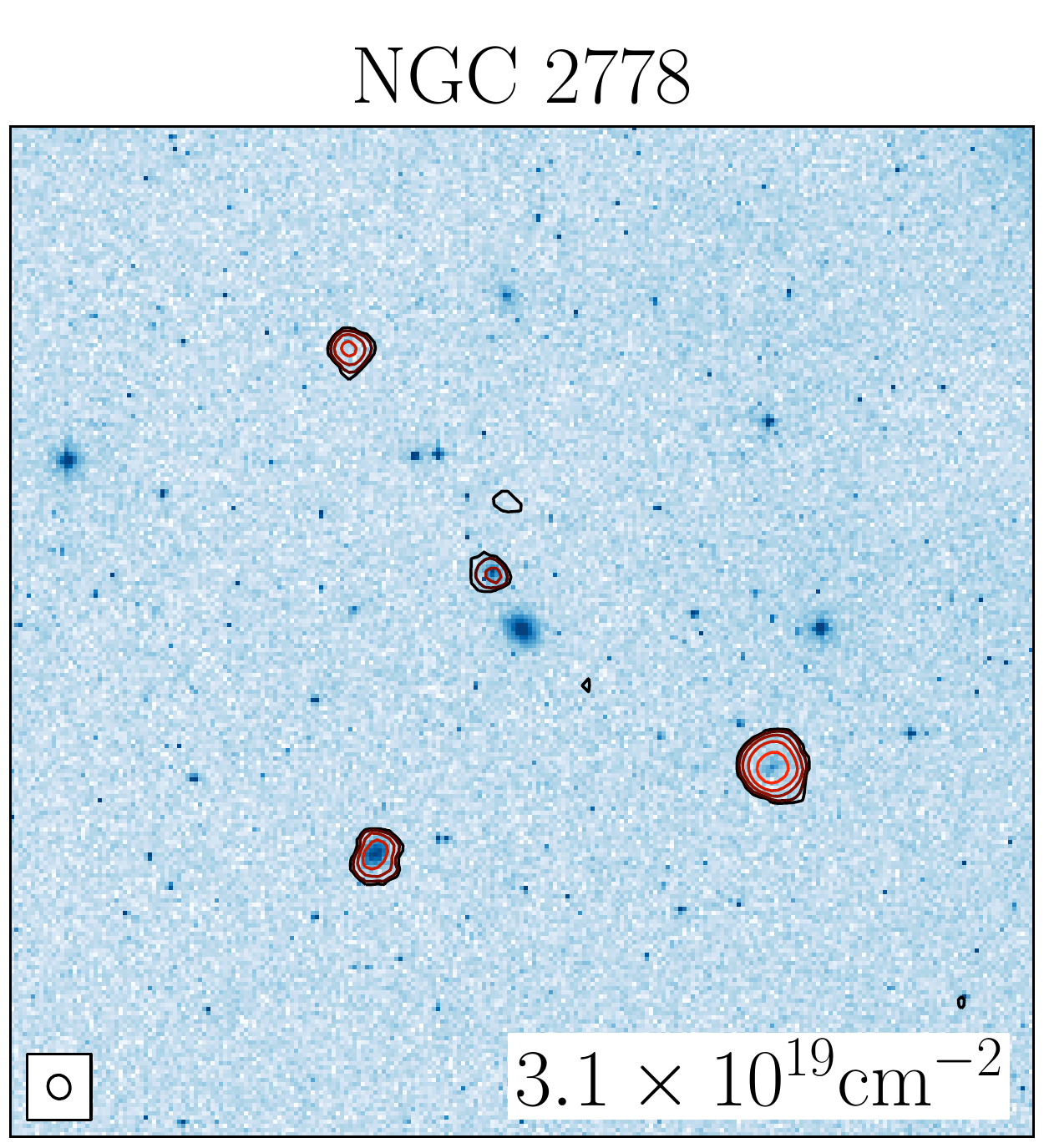}
\includegraphics[width=50mm]{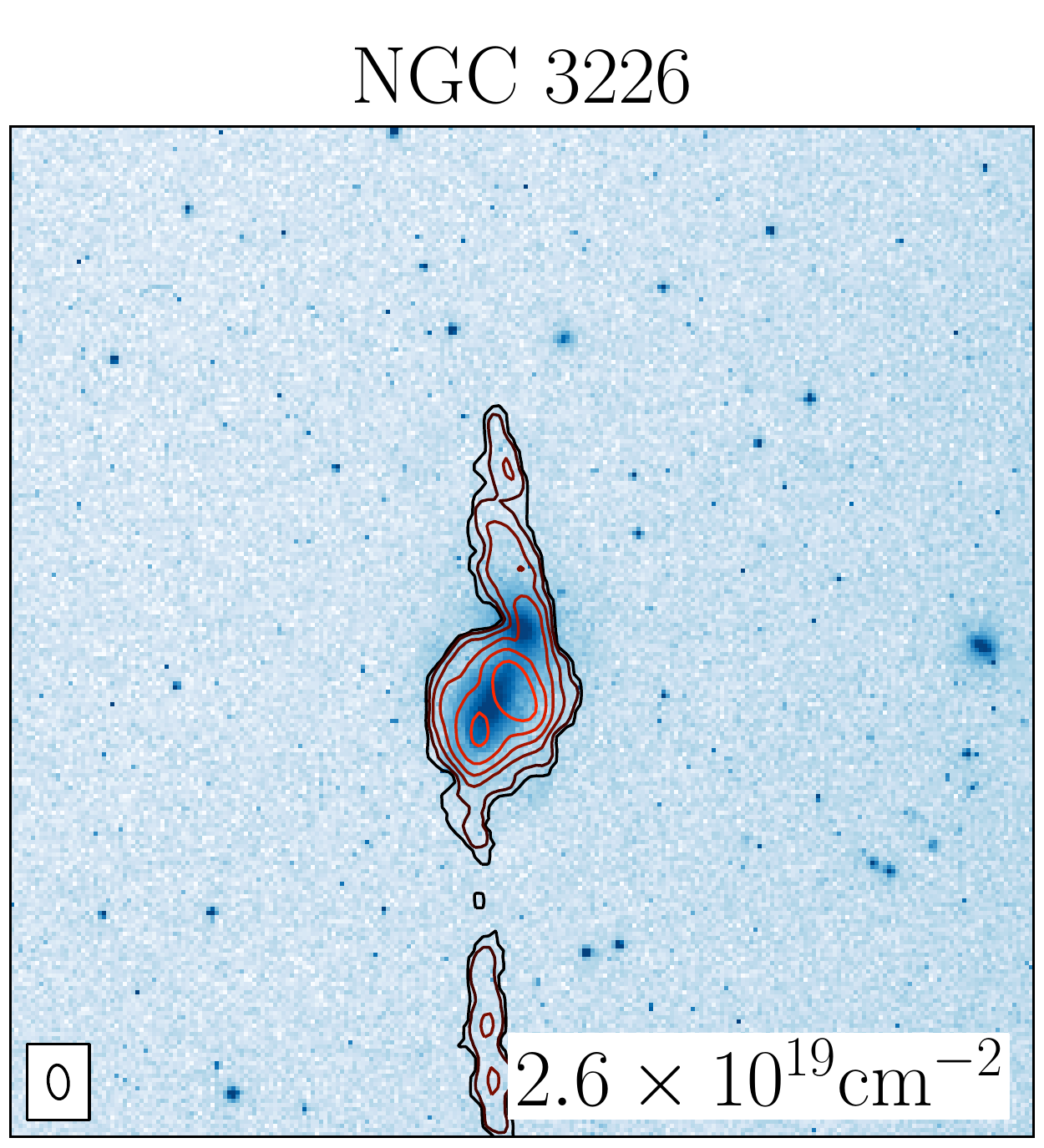}
\includegraphics[width=50mm]{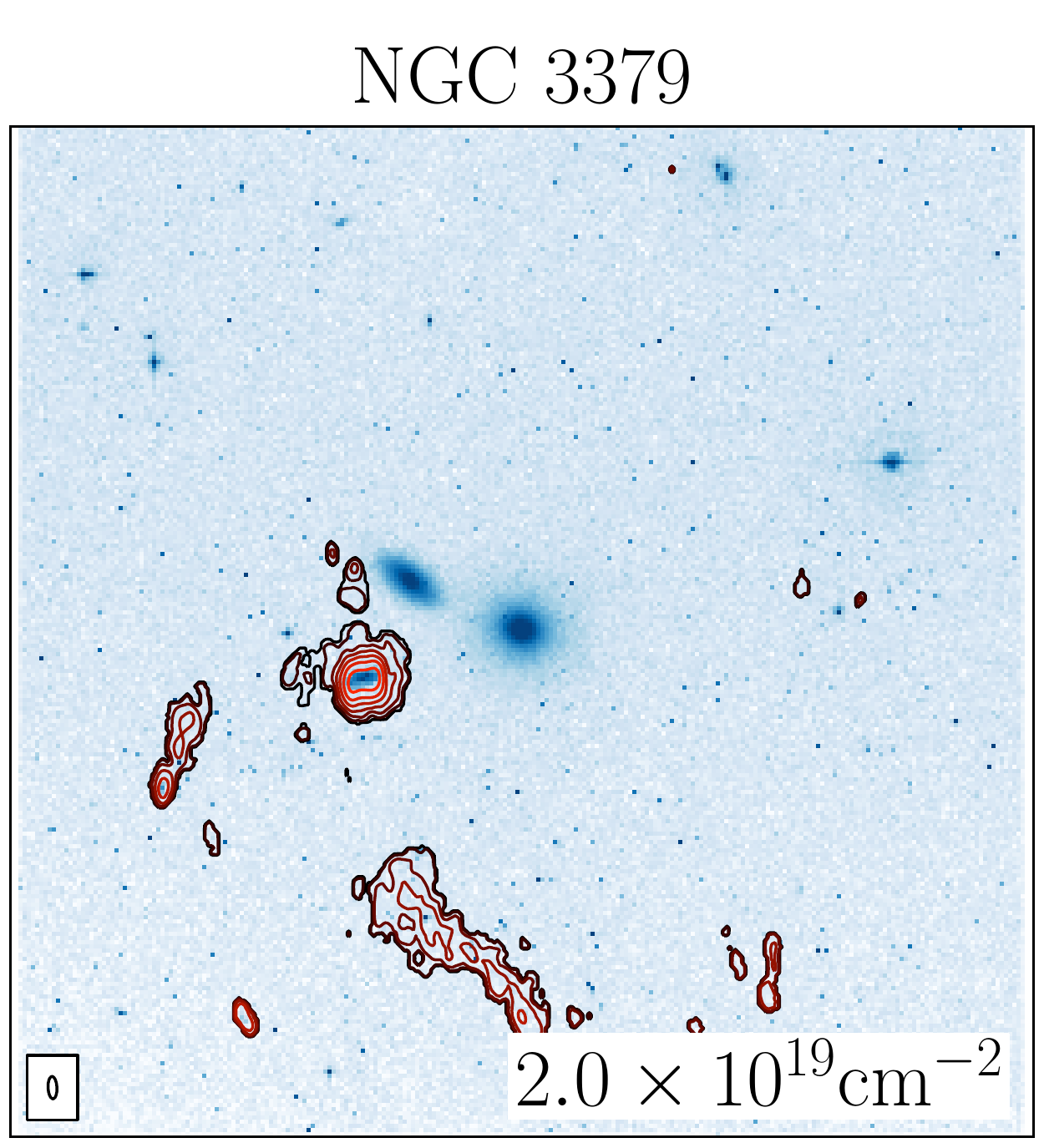}
\includegraphics[width=50mm]{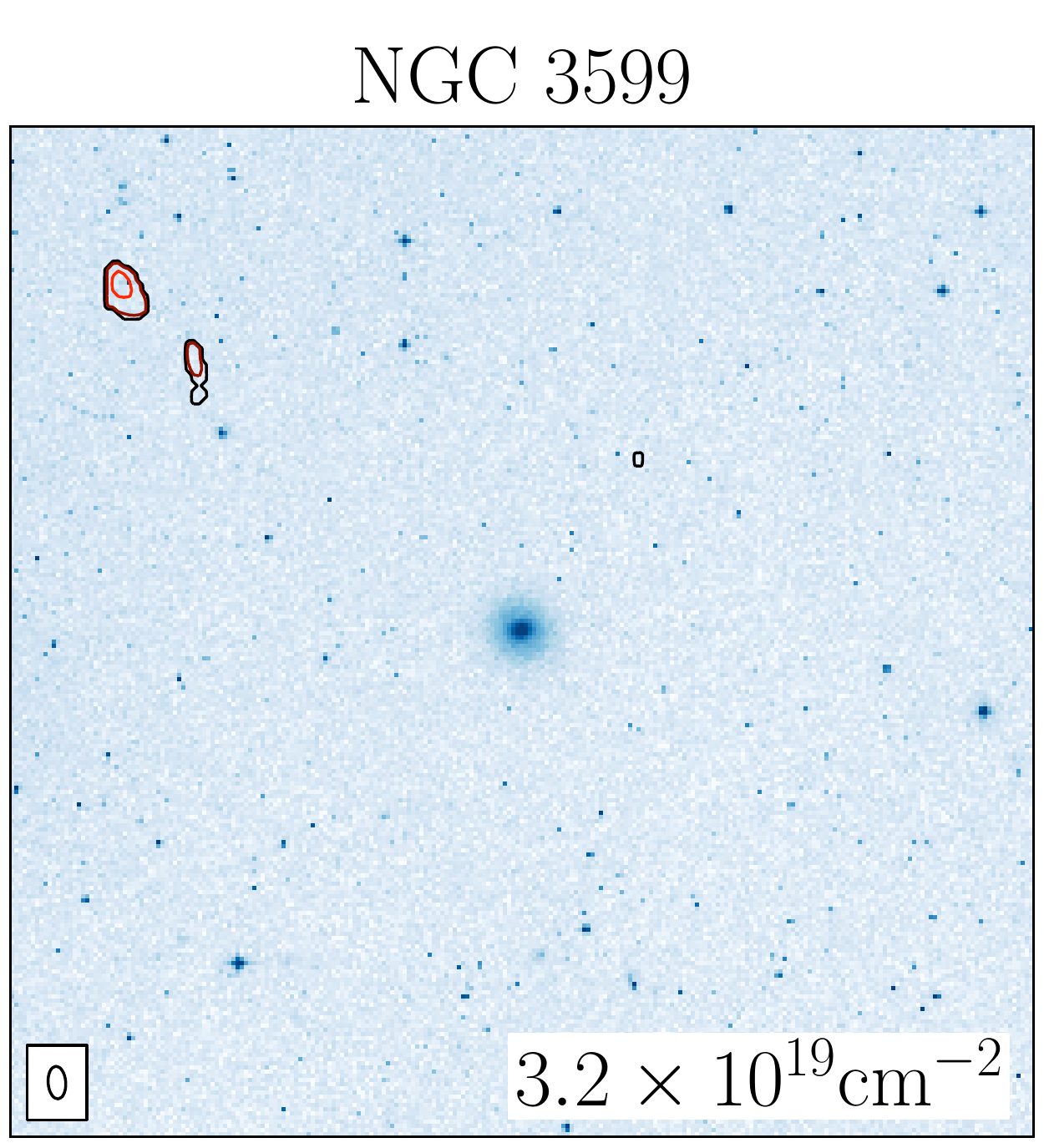}
\includegraphics[width=50mm]{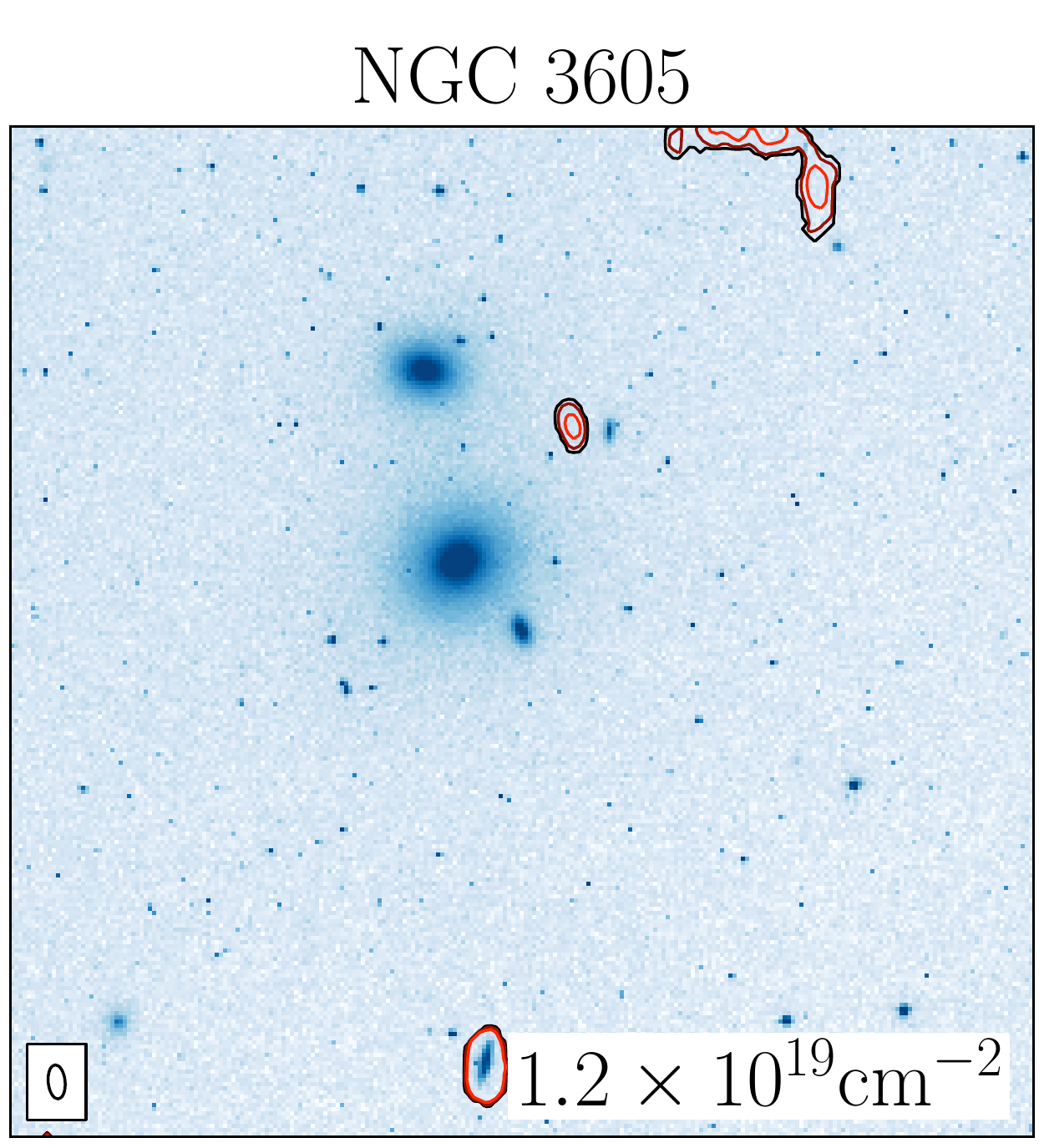}
\includegraphics[width=50mm]{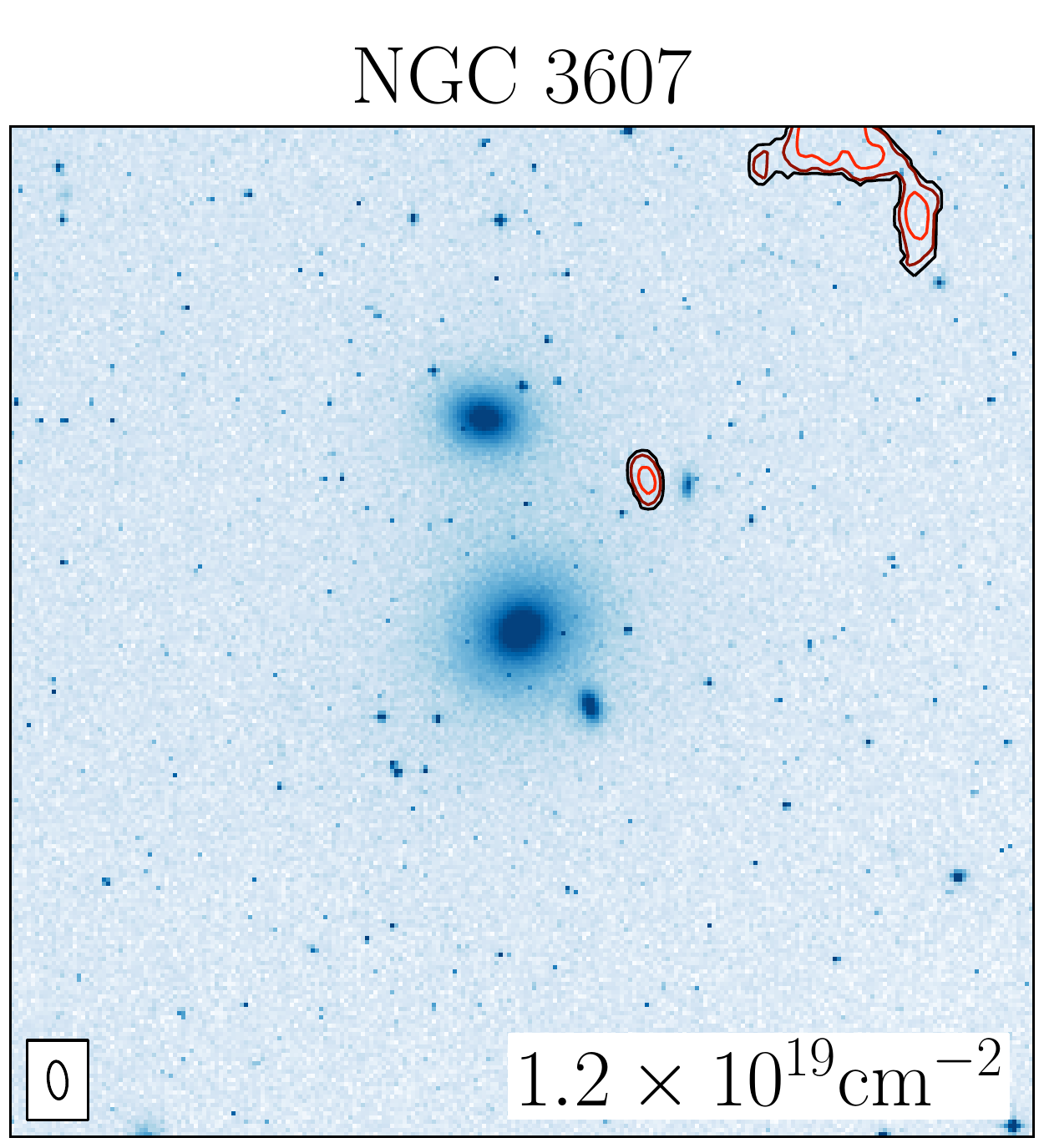}
\includegraphics[width=50mm]{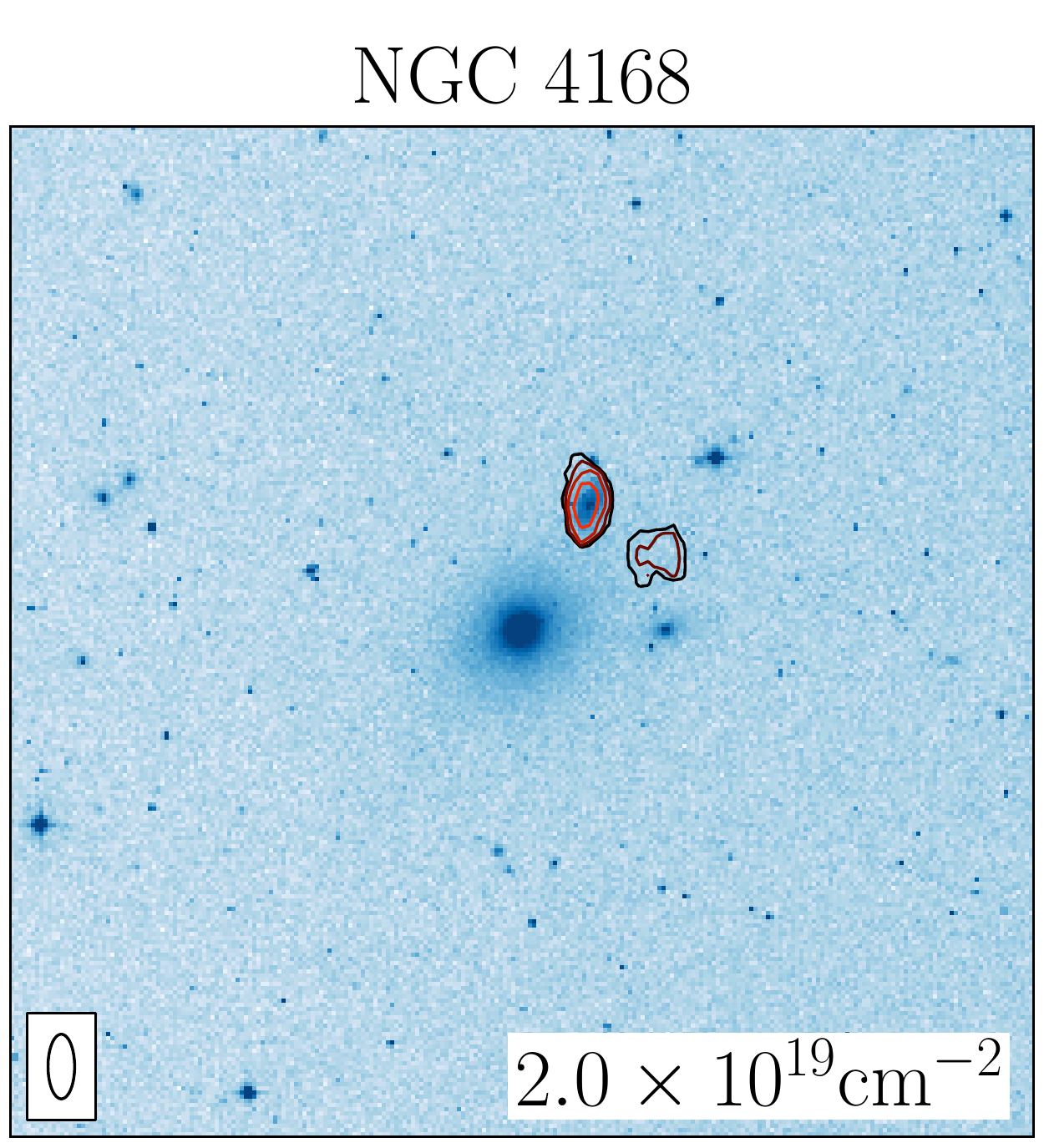}
\includegraphics[width=50mm]{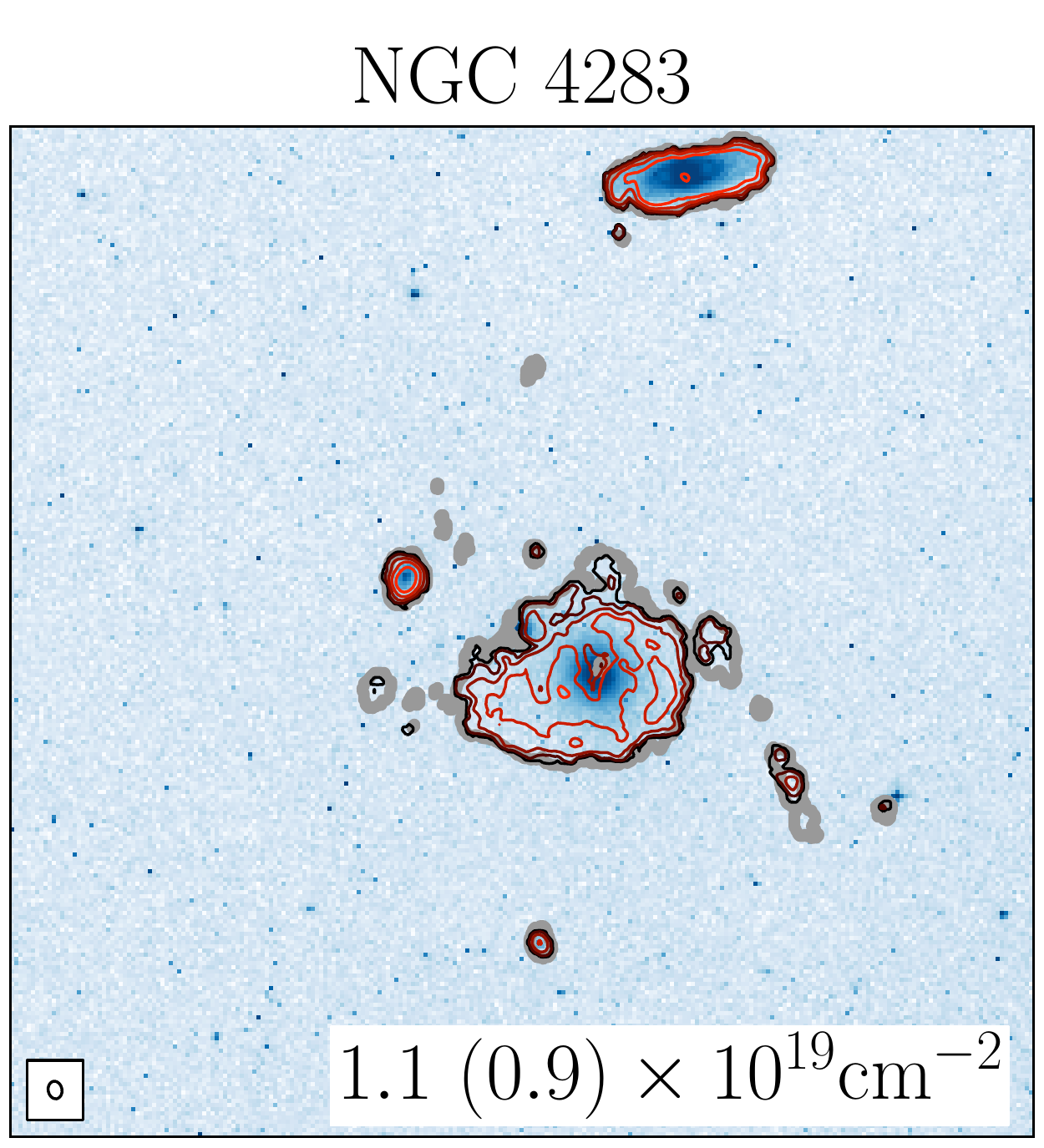}
\includegraphics[width=50mm]{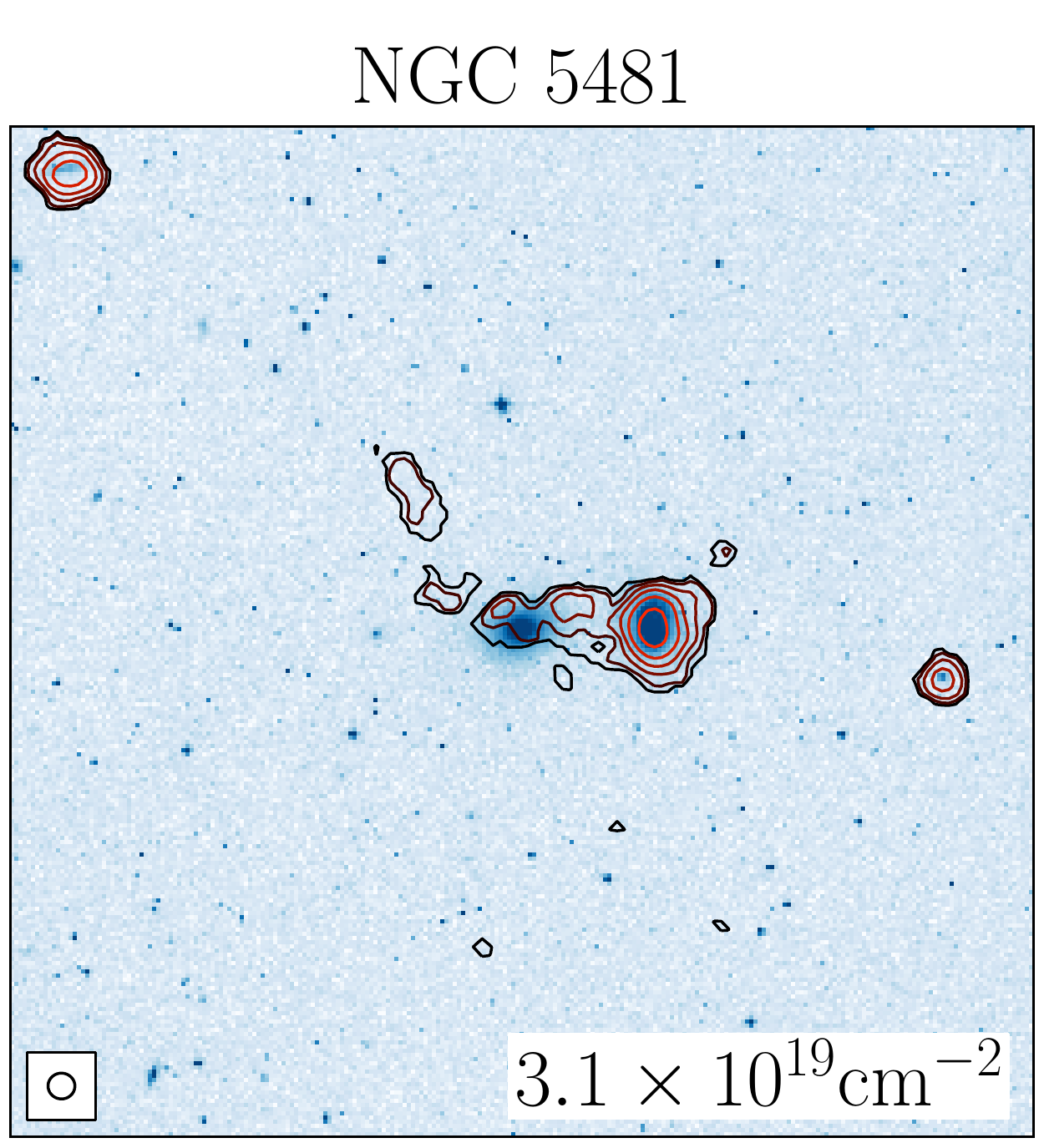}
\includegraphics[width=50mm]{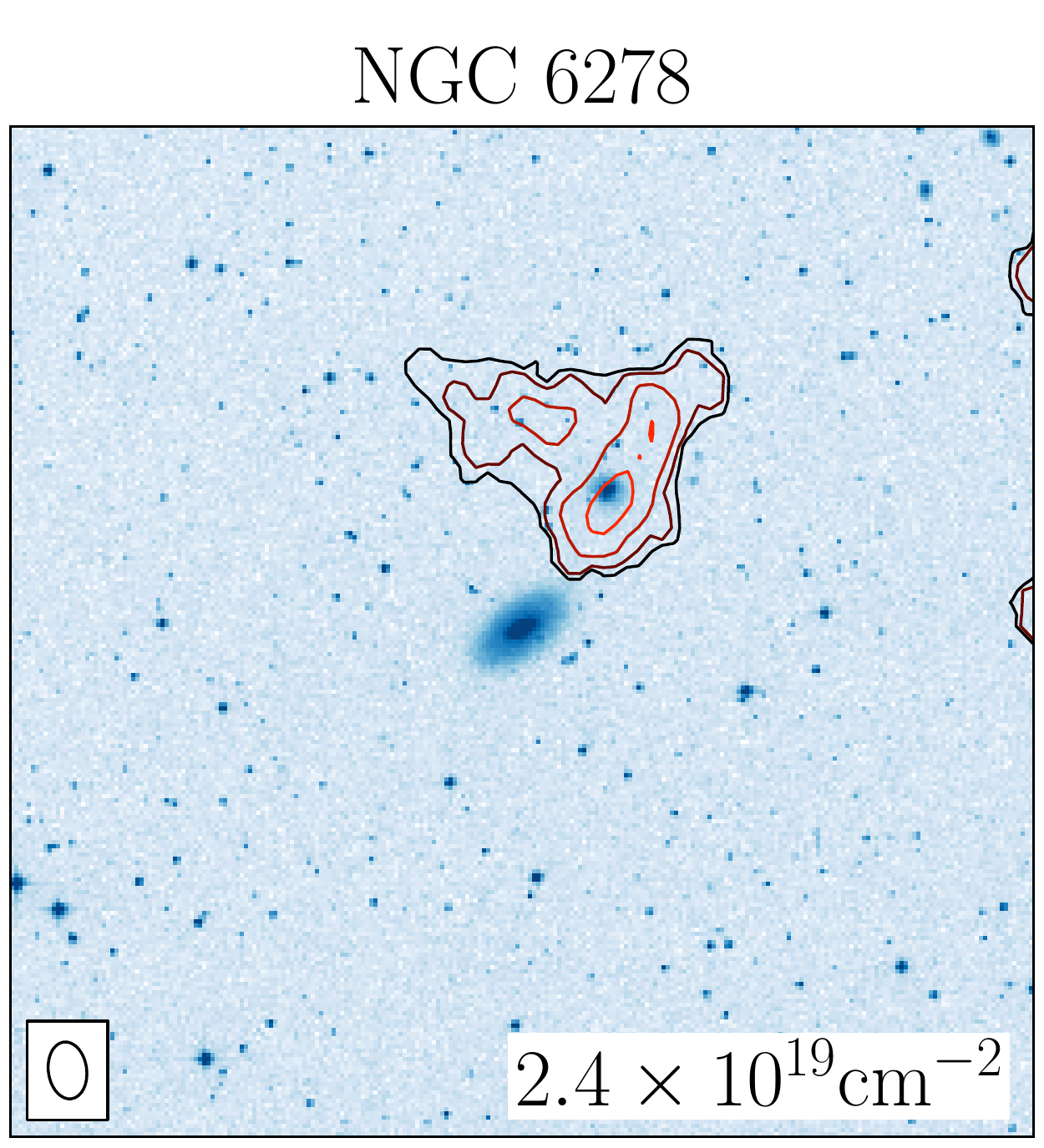}
\caption{Fields where \hi\ is detected in the proximity of an ETG but is either clearly associated to a nearby galaxy or cannot be unambiguously associated to our target galaxy. See the caption of Fig. \ref{fig:gallery} for a description of the images.}
\label{fig:gallery_undet_a}
\end{figure*}

\section{Table of \hi\ properties}

\clearpage
\begin{deluxetable}{r r r r r r r r}
\tablewidth{0pt}
\tabcolsep=5pt

\tablecaption{\hi\ properties of \atlas\ galaxies.}

\tablehead{
\colhead{Name} &
\colhead{noise} &
\colhead{beam} &
\colhead{log$_{10}$ \mhi} &
\colhead{log$_{10}$ \mhil} &
\colhead{\hi\ class} &
\colhead{notes} &
\colhead{source} \\

\colhead{} &
\colhead{(mJy beam$^{-1}$)} &
\colhead{(arcsec$^2$)} &
\colhead{(\msun)} &
\colhead{(\mlsun)} &
\colhead{} &
\colhead{} &
\colhead{} \\

\colhead{(1)} &
\colhead{(2)} &
\colhead{(3)} &
\colhead{(4)} &
\colhead{(5)} &
\colhead{(6)} &
\colhead{(7)} &
\colhead{(8)} \\
}

\startdata
   IC~0598 &   0.46 &  37.7$\times$ 37.3 & $<7.45$ & $<-2.90$ &   - &          - &          - \\ 
   IC~3631 &   0.60 &  77.8$\times$ 31.8 & $<7.71$ & $<-2.40$ &   - &          - &          - \\ 
  NGC~0661 &   0.51 &  48.2$\times$ 32.9 & $<7.37$ & $<-3.22$ &   - &          - &          - \\ 
  NGC~0680 &   0.53 &  51.4$\times$ 33.7 & $ 9.47$ & $ -1.51$ & $u$ &          - &          - \\ 
  NGC~0770 &   0.55 &  66.2$\times$ 33.5 & $<7.56$ & $<-2.78$ &   - &          - &          - \\ 
  NGC~0821 &   0.30 &  91.0$\times$ 31.8 & $<6.91$ & $<-4.00$ &   - &          - &      s (4) \\ 
  NGC~1023 &   0.20 &  39.0$\times$ 34.6 & $ 9.29$ & $ -1.63$ & $u$ &          - &      s (4) \\ 
  NGC~2481 &   0.52 &  49.1$\times$ 34.2 & $<7.42$ & $<-3.25$ &   - &          - &          - \\ 
  NGC~2549 &   0.43 &  39.1$\times$ 36.4 & $<6.51$ & $<-3.78$ &   - &          - &          s \\ 
  NGC~2577 &   0.48 &  51.0$\times$ 34.0 & $<7.35$ & $<-3.33$ &   - &          - &          - \\ 
  NGC~2592 &   0.50 &  46.8$\times$ 34.9 & $<7.18$ & $<-3.28$ &   - &          - &          - \\ 
  NGC~2594 &   0.50 &  46.8$\times$ 34.9 & $ 8.91$ & $ -1.35$ & $D$ &        P,R &          - \\ 
  NGC~2679 &   0.47 &  41.5$\times$ 36.3 & $<7.35$ & $<-3.08$ &   - &          - &          - \\ 
  NGC~2685 &   0.23 &  28.4$\times$ 25.6 & $ 9.33$ & $ -1.10$ & $D$ &          W &      j (4) \\ 
  NGC~2764 &   0.50 &  52.3$\times$ 33.1 & $ 9.28$ & $ -1.31$ & $D$ &        W,L &          - \\ 
  NGC~2768 &   0.41 &  37.5$\times$ 36.8 & $ 7.81$ & $ -3.38$ & $u$ &          $R01$ &          s \\ 
  NGC~2778 &   0.47 &  40.2$\times$ 36.1 & $<7.06$ & $<-3.14$ &   - &          - &          - \\ 
  NGC~2824 &   0.49 &  45.7$\times$ 34.4 & $ 7.59$ & $ -2.89$ & $d$ &          $R01$,A &          - \\ 
  NGC~2852 &   0.47 &  38.4$\times$ 36.3 & $<7.27$ & $<-2.91$ &   - &          - &          - \\ 
  NGC~2859 &   0.48 &  40.5$\times$ 36.2 & $ 8.46$ & $ -2.50$ & $D$ &          R &          - \\ 
  NGC~2880 &   0.48 &  38.4$\times$ 38.3 & $<7.03$ & $<-3.47$ &   - &          - &          - \\ 
  NGC~2950 &   0.47 &  38.5$\times$ 37.1 & $<6.69$ & $<-3.79$ &   - &          - &          - \\ 
  NGC~3032 &   0.24 &  46.3$\times$ 35.0 & $ 8.04$ & $ -2.08$ & $d$ &          C &      s (4) \\ 
  NGC~3073 &   0.47 &  38.6$\times$ 37.0 & $ 8.56$ & $ -1.46$ & $u$ &          - &          - \\ 
  NGC~3098 &   0.50 &  46.7$\times$ 33.5 & $<7.12$ & $<-3.28$ &   - &          - &          - \\ 
  NGC~3182 &   0.47 &  38.3$\times$ 37.4 & $ 6.92$ & $ -3.67$ & $d$ &          $R01$ &          - \\ 
  NGC~3193 &   0.52 &  51.0$\times$ 32.8 & $ 8.19$ & $ -2.98$ & $u$ &          - &          - \\ 
  NGC~3226 &   0.49 &  54.8$\times$ 32.5 & $<7.10$ & $<-3.51$ &   - &          - &          - \\ 
  NGC~3230 &   0.62 &  78.5$\times$ 32.6 & $<7.71$ & $<-3.28$ &   - &          - &          - \\ 
  NGC~3245 &   0.49 &  42.0$\times$ 35.8 & $<7.00$ & $<-3.79$ &   - &          - &          - \\ 
  NGC~3248 &   0.55 &  49.6$\times$ 33.1 & $<7.22$ & $<-3.07$ &   - &          - &          - \\ 
  NGC~3301 &   0.53 &  50.7$\times$ 32.8 & $<7.13$ & $<-3.49$ &   - &          - &          - \\ 
  NGC~3377 &   0.57 &  74.8$\times$ 32.1 & $<6.52$ & $<-3.89$ &   - &          - &          s \\ 
  NGC~3379 &   0.59 &  80.9$\times$ 32.1 & $<6.49$ & $<-4.35$ &   - &          - &          s \\ 
  NGC~3384 &   0.59 &  80.9$\times$ 32.1 & $ 7.25$ & $ -3.47$ & $c$ &          - &          s \\ 
  NGC~3400 &   0.52 &  42.2$\times$ 35.9 & $<7.19$ & $<-2.85$ &   - &          - &          - \\ 
  NGC~3412 &   0.60 &  77.3$\times$ 32.2 & $<6.55$ & $<-3.78$ &   - &          - &          - \\ 
  NGC~3414 &   0.19 &  44.9$\times$ 33.1 & $ 8.28$ & $ -2.63$ & $D$ &      $R01$,P,R,L &      s (4) \\ 
  NGC~3457 &   0.55 &  61.8$\times$ 32.2 & $ 8.07$ & $ -2.00$ & $u$ &          - &          - \\ 
  NGC~3458 &   0.48 &  38.3$\times$ 37.2 & $<7.35$ & $<-3.21$ &   - &          - &          - \\ 
  NGC~3489 &   0.25 &  78.5$\times$ 35.0 & $ 6.87$ & $ -3.63$ & $d$ &          - &      s (4) \\ 
  NGC~3499 &   0.50 &  38.1$\times$ 36.9 & $ 6.81$ & $ -3.25$ & $d$ &          $R01$ &          - \\ 
  NGC~3522 &   0.56 &  58.0$\times$ 33.2 & $ 8.47$ & $ -1.51$ & $D$ &       $R01$,P,R &          - \\ 
  NGC~3530 &   0.49 &  38.4$\times$ 37.3 & $<7.37$ & $<-2.74$ &   - &          - &          - \\ 
  NGC~3595 &   0.45 &  39.5$\times$ 35.5 & $<7.43$ & $<-3.19$ &   - &          - &          - \\ 
  NGC~3599 &   0.55 &  61.0$\times$ 32.8 & $<7.03$ & $<-3.17$ &   - &          - &          - \\ 
  NGC~3605 &   0.34 &  62.9$\times$ 31.0 & $<6.83$ & $<-3.21$ &   - &          - &      s (4) \\ 
  NGC~3607 &   0.34 &  62.9$\times$ 31.0 & $<6.92$ & $<-4.29$ &   - &          - &      s (4) \\ 
  NGC~3608 &   0.34 &  62.9$\times$ 31.0 & $ 7.16$ & $ -3.61$ & $c$ &          - &      s (4) \\ 
  NGC~3610 &   0.49 &  38.7$\times$ 38.0 & $<7.02$ & $<-3.77$ &   - &          - &          - \\ 
  NGC~3613 &   0.49 &  38.5$\times$ 38.2 & $<7.28$ & $<-3.74$ &   - &          - &          - \\ 
  NGC~3619 &   0.49 &  38.5$\times$ 38.2 & $ 9.00$ & $ -1.74$ & $D$ &          M &          - \\ 
  NGC~3626 &   0.55 &  59.4$\times$ 32.4 & $ 8.94$ & $ -1.70$ & $D$ &          C &          - \\ 
  NGC~3648 &   0.48 &  39.2$\times$ 36.8 & $<7.38$ & $<-3.16$ &   - &          - &          - \\ 
  NGC~3658 &   0.50 &  40.5$\times$ 36.5 & $<7.42$ & $<-3.27$ &   - &          - &          - \\ 
  NGC~3665 &   0.50 &  40.5$\times$ 36.5 & $<7.43$ & $<-3.85$ &   - &          - &          - \\ 
  NGC~3674 &   0.47 &  38.3$\times$ 37.1 & $<7.41$ & $<-3.19$ &   - &          - &          - \\ 
  NGC~3694 &   0.51 &  40.4$\times$ 34.3 & $<7.49$ & $<-2.76$ &   - &          - &          - \\ 
  NGC~3757 &   0.51 &  38.4$\times$ 37.1 & $<7.10$ & $<-3.07$ &   - &          - &          - \\ 
  NGC~3796 &   0.49 &  37.8$\times$ 36.3 & $<7.10$ & $<-2.95$ &   - &          - &          - \\ 
  NGC~3838 &   0.51 &  38.7$\times$ 35.5 & $ 8.38$ & $ -1.94$ & $D$ &         R &          - \\ 
  NGC~3941 &   0.51 &  39.1$\times$ 36.0 & $ 8.73$ & $ -1.80$ & $D$ &        C,R &          - \\ 
  NGC~3945 &   0.49 &  37.9$\times$ 37.7 & $ 8.85$ & $ -2.18$ & $D$ &          R &          - \\ 
  NGC~3998 &   0.50 &  37.7$\times$ 37.0 & $ 8.45$ & $ -2.19$ & $D$ &      M,L,A &          - \\ 
  NGC~4026 &   0.50 &  38.4$\times$ 36.7 & $ 8.50$ & $ -2.02$ & $u$ &          R &          - \\ 
  NGC~4036 &   0.52 &  37.8$\times$ 37.6 & $ 8.41$ & $ -2.66$ & $D$ &          R &          - \\ 
  NGC~4078 &   0.62 &  96.9$\times$ 30.7 & $<7.64$ & $<-2.86$ &   - &          - &          - \\ 
  NGC~4111 &   0.50 &  37.2$\times$ 36.6 & $ 8.81$ & $ -1.81$ & $u$ &          - &          - \\ 
  NGC~4119 &   2.82 & 180.0$\times$180.0 & $<7.10$ & $<-3.25$ &   - &          - &   $\alpha$ \\ 
  NGC~4143 &   0.53 &  40.5$\times$ 38.7 & $<6.80$ & $<-3.75$ &   - &          - &          - \\ 
  NGC~4150 &   0.14 &  44.1$\times$ 35.4 & $ 6.26$ & $ -3.72$ & $d$ &          $R01$ &      s (4) \\ 
  NGC~4168 &   0.62 &  77.3$\times$ 31.7 & $<7.46$ & $<-3.46$ &   - &          - &          - \\ 
  NGC~4203 &   0.51 &  40.1$\times$ 35.9 & $ 9.15$ & $ -1.53$ & $D$ &        W,L &          - \\ 
  NGC~4251 &   0.52 &  42.2$\times$ 36.0 & $<6.97$ & $<-3.82$ &   - &          - &          - \\ 
  NGC~4262 &   0.63 &  79.3$\times$ 30.2 & $ 8.69$ & $ -1.66$ & $D$ &          R &          s \\ 
  NGC~4267 &   3.59 & 180.0$\times$180.0 & $<7.17$ & $<-3.42$ &   - &          - &   $\alpha$ \\ 
  NGC~4278 &   0.16 &  43.1$\times$ 33.4 & $ 8.80$ & $ -2.04$ & $D$ &        $R01$,M,L &      s (4) \\ 
  NGC~4283 &   0.16 &  43.1$\times$ 33.4 & $<6.36$ & $<-3.67$ &   - &          - &      s (4) \\ 
  NGC~4340 &   0.65 &  69.6$\times$ 38.0 & $<7.03$ & $<-3.48$ &   - &          - &          - \\ 
  NGC~4346 &   0.48 &  38.0$\times$ 36.5 & $<6.66$ & $<-3.67$ &   - &          - &          - \\ 
  NGC~4350 &   0.65 &  69.6$\times$ 38.0 & $<6.88$ & $<-3.68$ &   - &          - &          - \\ 
  NGC~4371 &   2.65 & 180.0$\times$180.0 & $<7.10$ & $<-3.59$ &   - &          - &   $\alpha$ \\ 
  NGC~4374 &   1.08 &  59.0$\times$ 34.0 & $<7.26$ & $<-4.10$ &   - &          - &          s \\ 
  NGC~4377 &   2.76 & 180.0$\times$180.0 & $<7.16$ & $<-3.13$ &   - &          - &   $\alpha$ \\ 
  NGC~4379 &   2.66 & 180.0$\times$180.0 & $<7.04$ & $<-3.17$ &   - &          - &   $\alpha$ \\ 
  NGC~4382 &   0.60 &  56.7$\times$ 35.3 & $<6.97$ & $<-4.39$ &   - &          - &          s \\ 
  NGC~4387 &   0.68 &  72.7$\times$ 34.0 & $<7.03$ & $<-3.14$ &   - &          - &          s \\ 
  NGC~4406 &   0.35 &  76.0$\times$ 25.0 & $ 8.00$ & $ -3.33$ & $u$ &          - &          o \\ 
  NGC~4425 &   0.35 &  76.0$\times$ 25.0 & $<6.71$ & $<-3.44$ &   - &          - &          o \\ 
  NGC~4429 &   2.93 & 180.0$\times$180.0 & $<7.12$ & $<-3.92$ &   - &          - &   $\alpha$ \\ 
  NGC~4435 &   3.69 & 180.0$\times$180.0 & $<7.23$ & $<-3.62$ &   - &          - &   $\alpha$ \\ 
  NGC~4452 &   4.65 & 180.0$\times$180.0 & $<7.27$ & $<-2.80$ &   - &          - &   $\alpha$ \\ 
  NGC~4458 &   0.62 &  39.4$\times$ 34.0 & $<6.91$ & $<-3.10$ &   - &          - &          s \\ 
  NGC~4459 &   0.64 &  69.7$\times$ 34.6 & $<6.91$ & $<-3.95$ &   - &          - &          s \\ 
  NGC~4461 &   4.76 & 180.0$\times$180.0 & $<7.33$ & $<-3.22$ &   - &          - &   $\alpha$ \\ 
  NGC~4473 &   0.63 &  36.7$\times$ 34.0 & $<6.86$ & $<-3.96$ &   - &          - &          s \\ 
  NGC~4474 &   3.01 & 180.0$\times$180.0 & $<7.08$ & $<-3.14$ &   - &          - &   $\alpha$ \\ 
  NGC~4477 &   0.66 &  71.5$\times$ 34.4 & $<6.95$ & $<-3.86$ &   - &          - &          s \\ 
  NGC~4489 &   0.47 &  64.3$\times$ 32.2 & $<6.74$ & $<-3.21$ &   - &          - &          - \\ 
  NGC~4494 &   0.51 &  45.9$\times$ 33.8 & $<6.84$ & $<-4.12$ &   - &          - &          - \\ 
  NGC~4503 &   3.11 & 180.0$\times$180.0 & $<7.14$ & $<-3.46$ &   - &          - &   $\alpha$ \\ 
  NGC~4521 &   0.48 &  38.5$\times$ 37.4 & $ 7.75$ & $ -3.13$ & $c$ &          - &          - \\ 
  NGC~4528 &   3.74 & 180.0$\times$180.0 & $<7.18$ & $<-2.95$ &   - &          - &   $\alpha$ \\ 
  NGC~4550 &   0.66 &  79.0$\times$ 33.4 & $<6.89$ & $<-3.33$ &   - &          - &          s \\ 
  NGC~4551 &   5.74 & 180.0$\times$180.0 & $<7.39$ & $<-2.80$ &   - &          - &   $\alpha$ \\ 
  NGC~4552 &   0.60 &  81.4$\times$ 32.1 & $<6.87$ & $<-4.16$ &   - &          - &          s \\ 
  NGC~4564 &   0.67 &  82.7$\times$ 34.5 & $<6.91$ & $<-3.63$ &   - &          - &          s \\ 
  NGC~4596 &   3.01 & 180.0$\times$180.0 & $<7.13$ & $<-3.64$ &   - &          - &   $\alpha$ \\ 
  NGC~4608 &   3.69 & 180.0$\times$180.0 & $<7.22$ & $<-3.27$ &   - &          - &   $\alpha$ \\ 
  NGC~4621 &   0.67 &  81.7$\times$ 34.3 & $<6.86$ & $<-4.11$ &   - &          - &          s \\ 
  NGC~4638 &   2.65 & 180.0$\times$180.0 & $<7.12$ & $<-3.39$ &   - &          - &   $\alpha$ \\ 
  NGC~4649 &   3.09 & 180.0$\times$180.0 & $<7.18$ & $<-4.32$ &   - &          - &   $\alpha$ \\ 
  NGC~4660 &   0.69 &  84.4$\times$ 34.3 & $<6.88$ & $<-3.51$ &   - &          - &          s \\ 
  NGC~4694 &   0.37 &  34.7$\times$ 31.2 & $ 8.21$ & $ -1.96$ & $u$ &          - &          c \\ 
  NGC~4710 &   0.57 &  69.5$\times$ 32.1 & $ 6.84$ & $ -3.88$ & $d$ &          $R01$ &          - \\ 
  NGC~4733 &   3.79 & 180.0$\times$180.0 & $<7.12$ & $<-2.92$ &   - &          - &   $\alpha$ \\ 
  NGC~4754 &   3.52 & 180.0$\times$180.0 & $<7.18$ & $<-3.59$ &   - &          - &   $\alpha$ \\ 
  NGC~4762 &   3.01 & 180.0$\times$180.0 & $<7.40$ & $<-3.70$ &   - &          - &   $\alpha$ \\ 
  NGC~5103 &   0.48 &  37.4$\times$ 36.9 & $ 8.57$ & $ -1.69$ & $D$ &        M,L &          - \\ 
  NGC~5173 &   0.46 &  38.5$\times$ 36.5 & $ 9.33$ & $ -1.13$ & $D$ &        M,L &          - \\ 
  NGC~5198 &   0.25 &  37.1$\times$ 35.1 & $ 8.49$ & $ -2.46$ & $u$ &          - &      s (2) \\ 
  NGC~5273 &   0.50 &  40.9$\times$ 34.9 & $<6.81$ & $<-3.45$ &   - &          - &          - \\ 
  NGC~5308 &   0.87 &  34.3$\times$ 34.0 & $<7.63$ & $<-3.34$ &   - &          - &          s \\ 
  NGC~5322 &   0.48 &  38.0$\times$ 37.6 & $<7.34$ & $<-4.08$ &   - &          A &          - \\ 
  NGC~5342 &   0.51 &  38.0$\times$ 37.4 & $<7.50$ & $<-2.85$ &   - &          - &          - \\ 
  NGC~5353 &   0.47 &  38.7$\times$ 36.5 & $<7.45$ & $<-3.90$ &   - &          A &          - \\ 
  NGC~5355 &   0.47 &  38.7$\times$ 36.5 & $<7.50$ & $<-2.77$ &   - &          - &          - \\ 
  NGC~5358 &   0.47 &  38.7$\times$ 36.5 & $<7.52$ & $<-2.60$ &   - &          - &          - \\ 
  NGC~5379 &   0.51 &  39.9$\times$ 39.5 & $<7.36$ & $<-2.79$ &   - &          - &          - \\ 
  NGC~5422 &   0.47 &  38.3$\times$ 36.9 & $ 7.87$ & $ -2.92$ & $d$ &          $R01$ &          - \\ 
  NGC~5473 &   0.46 &  38.4$\times$ 36.9 & $<7.40$ & $<-3.61$ &   - &          - &          - \\ 
  NGC~5475 &   0.47 &  38.4$\times$ 37.0 & $<7.28$ & $<-3.19$ &   - &          - &          - \\ 
  NGC~5481 &   0.49 &  38.4$\times$ 37.0 & $<7.21$ & $<-3.18$ &   - &          - &          - \\ 
  NGC~5485 &   0.47 &  37.8$\times$ 37.5 & $<7.17$ & $<-3.59$ &   - &          - &          - \\ 
  NGC~5500 &   0.46 &  37.9$\times$ 36.9 & $<7.36$ & $<-2.73$ &   - &          - &          - \\ 
  NGC~5557 &   0.47 &  39.2$\times$ 35.9 & $ 8.57$ & $ -2.69$ & $u$ &          - &          - \\ 
  NGC~5582 &   0.48 &  38.5$\times$ 36.1 & $ 9.65$ & $ -0.97$ & $D$ &          R &          - \\ 
  NGC~5611 &   0.48 &  40.0$\times$ 35.6 & $<7.15$ & $<-3.04$ &   - &          - &          - \\ 
  NGC~5631 &   0.48 &  37.2$\times$ 36.4 & $ 8.89$ & $ -1.90$ & $D$ &        M,W,L &          - \\ 
  NGC~5687 &   0.58 &  39.4$\times$ 36.9 & $<7.32$ & $<-3.28$ &   - &          - &          - \\ 
  NGC~5866 &   0.47 &  38.3$\times$ 37.1 & $ 6.96$ & $ -3.95$ & $d$ &          $R01$,A &          - \\ 
  NGC~6149 &   0.53 &  56.0$\times$ 32.4 & $<7.56$ & $<-2.79$ &   - &          - &          - \\ 
  NGC~6278 &   0.52 &  49.2$\times$ 32.9 & $<7.67$ & $<-3.32$ &   - &          - &          - \\ 
  NGC~6547 &   0.52 &  45.9$\times$ 33.3 & $<7.63$ & $<-3.12$ &   - &          - &          - \\ 
  NGC~6548 &   0.53 &  58.6$\times$ 32.3 & $<7.12$ & $<-3.47$ &   - &          - &          - \\ 
  NGC~6703 &   0.46 &  38.3$\times$ 36.4 & $<7.18$ & $<-3.67$ &   - &          - &          - \\ 
  NGC~6798 &   0.60 &  38.1$\times$ 37.0 & $ 9.38$ & $ -1.34$ & $D$ &          C &          - \\ 
  NGC~7280 &   0.60 &  64.5$\times$ 32.7 & $ 7.92$ & $ -2.52$ & $u$ &          $R01$ &          - \\ 
  NGC~7332 &   0.40 &  46.1$\times$ 31.0 & $ 6.62$ & $ -4.19$ & $c$ &          $R01$ &    s (1.5) \\ 
  NGC~7454 &   0.55 &  65.1$\times$ 32.3 & $<7.16$ & $<-3.35$ &   - &          - &          - \\ 
  NGC~7457 &   0.49 &  38.4$\times$ 33.5 & $<6.61$ & $<-3.66$ &   - &          - &          s \\ 
  NGC~7465 &   0.54 &  66.6$\times$ 32.3 & $ 9.98$ & $ -0.46$ & $u$ &          - &          - \\ 
PGC~028887 &   0.57 &  85.4$\times$ 32.1 & $ 7.65$ & $ -2.57$ & $c$ &          - &          - \\ 
PGC~029321 &   0.58 &  78.8$\times$ 31.7 & $<7.68$ & $<-2.30$ &   - &          A &          - \\ 
PGC~035754 &   0.51 &  40.3$\times$ 35.9 & $<7.58$ & $<-2.49$ &   - &          - &          - \\ 
PGC~044433 &   0.58 &  76.4$\times$ 31.9 & $<7.66$ & $<-2.55$ &   - &          - &          - \\ 
PGC~050395 &   0.47 &  37.8$\times$ 37.5 & $<7.51$ & $<-2.57$ &   - &          - &          - \\ 
PGC~051753 &   0.45 &  38.5$\times$ 36.6 & $<7.52$ & $<-2.56$ &   - &          - &          - \\ 
PGC~061468 &   0.54 &  57.1$\times$ 32.4 & $<7.54$ & $<-2.45$ &   - &          - &          - \\ 
PGC~071531 &   0.52 &  54.7$\times$ 32.5 & $<7.37$ & $<-2.64$ &   - &          - &          - \\ 
 UGC~03960 &   0.46 &  50.0$\times$ 33.6 & $ 7.79$ & $ -2.28$ & $D$ &        W,L &          - \\ 
 UGC~04551 &   0.47 &  39.7$\times$ 38.4 & $<7.25$ & $<-3.23$ &   - &          - &          - \\ 
 UGC~05408 &   0.47 &  38.3$\times$ 37.1 & $ 8.52$ & $ -1.60$ & $d$ &          - &          - \\ 
 UGC~06176 &   0.52 &  51.6$\times$ 33.3 & $ 9.02$ & $ -1.36$ & $D$ &          W &          - \\ 
 UGC~08876 &   0.48 &  39.4$\times$ 35.7 & $<7.43$ & $<-2.83$ &   - &          - &          - \\ 
 UGC~09519 &   0.48 &  40.2$\times$ 35.6 & $ 9.27$ & $ -0.83$ & $D$ &          M &          - \\ 
\enddata
\tablecomments{
\\
Column (1): Principal designation from LEDA.\\
Column (2):  Median noise in the \hi\ cube.\\
Column (3): Major and minor axis of the beam.\\
Column (4): Total \hi\ mass calculated assuming galaxy distances given in Paper I.\\
Column (5): Ratio between \mhi\ and the $K$-band luminosity given in Paper I.\\
Column (6): \hi\ class: $D$ = large disc/ring; $d$ =  small disc/ring; $u$ = unsettled; $c$ = cloud.\\
Column (7): Notes on \hi\ data, morphology and kinematics: $R01$ indicates that additional \hi\ is detected in the $R01$ cube, and the \mhi\ value is obtained from it; A = \hi\ detected (also) in absorption; C = \hi\ counter-rotating relative to the stellar kinematics; L = lopsided \hi\ morphology; M = \hi\ misaligned relative to the stellar kinematics; P = polar; R = ring; W = warp.\\
Column (8): Source of \hi\ data if other than the present study: c = \cite{2009AJ....138.1741C}; j = \cite{2009A&A...494..489J}; o = Oosterloo; s = Sauron (M06, O10);  $\alpha$ = ALFALFA. The number in parenthesis indicates the integration time (in units of 12 h) for data obtained with WSRT observations longer than 12 h.\\
}

\label{tab:sample}
\end{deluxetable}

\label{lastpage}

\end{document}